\documentstyle[12pt]{article}
\newcommand{\nc}{\newcommand}
\nc{\renc}{\renewcommand}

\nc{\half}{{\textstyle{1\over2}}}
\nc{\etal}{\mbox{\it et al. }}
\nc{\ie}{{\it i.e.}}
\nc{\eg}{{\it e.g.}}

\renc{\thefootnote}{\arabic{footnote}}
\nc{\capt}[1]{{\bf Figure.} {\small\sl #1}}

\def\ds{\displaystyle}

\nc{\eqs}[2]{\mbox{Eqs.~(\ref{#1},\,\ref{#2})}}
\nc{\eq}[1]{\mbox{Eq.~(\ref{#1})}}
 \nc{\figs}[2]{\mbox{Figs.~(\ref{#1},\,\ref{#2})}}
\nc{\fig}[1]{\mbox{Fig~.(\ref{#1})}}

\nc{\tag}[1]{\label{#1} \marginpar{{\footnotesize #1}}}
\nc{\mtag}[1]{\label{#1} \mbox{\marginpar{{\footnotesize #1}}}}
\renc{\baselinestretch}{1.5}
\newlength{\overeqskip}
\newlength{\undereqskip}
\nc{\be}[1]{\begin{equation} \mbox{$\label{#1}$}}
\nc{\bea}[1]{\begin{eqnarray} \mbox{$\label{#1}$}}
\nc{\Section}[2]{\section{#2}\label{#1}}
\nc{\Bibitem}[1]{\bibitem{#1}}
\nc{\Label}[1]{\label{#1}}
\nc{\eea}{\vspace{\undereqskip}\end{eqnarray}}
\nc{\ee}{\vspace{\undereqskip}\end{equation}}
\nc{\bdm}{\begin{displaymath}}
\nc{\edm}{\end{displaymath}}
\nc{\dpsty}{\displaystyle}
\nc{\bc}{\begin{center}}
\nc{\ec}{\end{center}}
\nc{\ba}{\begin{array}}
\nc{\ea}{\end{array}}
\nc{\bab}{\begin{abstract}}
\nc{\eab}{\end{abstract}}
\nc{\btab}{\begin{tabular}}
\nc{\etab}{\end{tabular}}
\nc{\bit}{\begin{itemize}}
\nc{\eit}{\end{itemize}}
\nc{\ben}{\begin{enumerate}}
\nc{\een}{\end{enumerate}}
\nc{\bfig}{\begin{figure}}
\nc{\efig}{\end{figure}}
\nc{\arreq}{&\!=\!&}
\nc{\arrmi}{&\!-\!&}
\nc{\arrpl}{&\!+\!&}
\nc{\arrap}{&\!\!\!\approx\!\!\!&}
\nc{\non}{\nonumber\\*}
\nc{\align}{\!\!\!\!\!\!\!\!&&}

\def\lsim{\; \raise0.3ex\hbox{$<$\kern-0.75em
       \raise-1.1ex\hbox{$\sim$}}\; }
\def\gsim{\; \raise0.3ex\hbox{$>$\kern-0.75em
       \raise-1.1ex\hbox{$\sim$}}\; }
\nc{\DOT}{\hspace{-0.08in}{\bf .}\hspace{0.1in}}
\nc{\Laada}{\hbox {$\sqcap$ \kern -1em $\sqcup$}}
\nc\loota{{\scriptstyle\sqcap\kern-0.55em\hbox{$\scriptstyle\sqcup$}}}
\nc\Loota{{\sqcap\kern-0.65em\hbox{$\sqcup$}}}
\nc\laada{\Loota}
\nc{\qed}{\hskip 3em \hbox{\BOX} \vskip 2ex}

\nc{\real}{{\rm I \! R}}
\nc{\Z}{{\sf Z \!\!\! Z}}
\nc{\complex}{{\rm C\!\!\! {\sf I}\,\,}}
\def\bigid{\leavevmode\hbox{\small1\kern-3.8pt\normalsize1}}
\def\id{\leavevmode\hbox{\small1\kern-3.3pt\normalsize1}}
\nc{\slask}{\!\!\!/}
\nc{\bis}{{\prime\prime}}
\nc{\pa}{\partial}
\nc{\na}{\nabla}
\nc{\ra}{\rangle}
\nc{\la}{\langle}
\nc{\goto}{\rightarrow}
\nc{\swap}{\leftrightarrow}

\nc{\EE}[1]{ \mbox{$\cdot10^{#1}$} }
\nc{\abs}[1]{\left|#1\right|}
\nc{\at}[2]{\left.#1\right|_{#2}}
\nc{\norm}[1]{\|#1\|}
\nc{\abscut}[2]{\Abs{#1}_{\scriptscriptstyle#2}}
\nc{\vek}[1]{{\rm\bf #1}}
\nc{\integral}[2]{\int\limits_{#1}^{#2}}
\nc{\inv}[1]{\frac{1}{#1}}
\nc{\dd}[2]{{{\partial #1}\over{\partial #2}}}
\nc{\ddd}[2]{{{{\partial}^2 #1}\over{\partial {#2}^2}}}
\nc{\dddd}[3]{{{{\partial}^2 #1}\over
        {\partial #2 \partial #3}}}
\nc{\dder}[2]{{{d #1}\over{d #2}}}
\nc{\ddder}[2]{{{d^2 #1}\over{d {#2}^2}}}
\nc{\dddder}[3]{{d^2 #1}\over
        {d #2 d #3}}
\nc{\dx}[1]{d\,^{#1}x}
\nc{\dy}[1]{d\,^{#1}y}
\nc{\dz}[1]{d\,^{#1}z}
\nc{\dl}[1]{\frac{d\,^{#1}l}{(2\pi)^{#1}}}
\nc{\dk}[1]{\frac{d\,^{#1}k}{(2\pi)^{#1}}}
\nc{\dq}[1]{\frac{d\,^{#1}q}{(2\pi)^{#1}}}

\nc{\cc}{\mbox{$c.c.$ }}
\nc{\hc}{\mbox{$h.c.$ }}
\nc{\cf}{cf.\ }
\nc{\erfc}{{\rm erfc}}
\nc{\Tr}{{\rm Tr\,}}
\nc{\tr}{{\rm tr\,}}
\nc{\pol}{{\rm pol}}
\nc{\sign}{{\rm sign}}
\nc{\bfT}{{\bf T }}

\def\GeV{{\rm\ GeV}}

\def\TeV{{\rm\ TeV}}

\nc{\cA}{{\cal A}}
\nc{\cB}{{\cal B}}
\nc{\cD}{{\cal D}}
\nc{\cE}{{\cal E}}
\nc{\cG}{{\cal G}}
\nc{\cH}{{\cal H}}
\nc{\cL}{{\cal L}}
\nc{\cO}{{\cal O}}
\nc{\cT}{{\cal T}}
\nc{\cN}{{\cal N}}
\nc{\rvac}[1]{|{\cal O}#1\rangle}
\nc{\lvac}[1]{\langle{\cal O}#1|}
\nc{\rvacb}[1]{|{\cal O}_\beta #1\rangle}
\nc{\lvacb}[1]{\langle{\cal O}_\beta #1 |}
\nc{\bb}{\bar{\beta}}
\nc{\bt}{\tilde{\beta}}
\nc{\ctH}{\tilde{\cal H}}
\nc{\chH}{\hat{\cal H}}
\nc{\1}{\aa}
\nc{\2}{\"{a}}
\nc{\3}{\"{o}}
\nc{\4}{\AA}
\nc{\5}{\"{A}}
\nc{\6}{\"{O}}
\nc{\al}{\alpha}
\nc{\g}{\gamma}
\nc{\Del}{\Delta}
\nc{\e}{\epsilon}
\nc{\eps}{\epsilon}
\nc{\lam}{\lambda}
\nc{\om}{\omega}
\nc{\Om}{\Omega}
\nc{\ve}{\varepsilon}
\nc{\mn}{{\mu\nu}}
\nc{\k}{\kappa}
\nc{\vp}{\varphi}

\nc{\advp}[3]{{\it  Adv.\ in\ Phys.\ }{{\bf #1} {(#2)} {#3}}}
\nc{\annp}[3]{{\it  Ann.\ Phys.\ (N.Y.)\ }{{\bf #1} {(#2)} {#3}}}
\nc{\apl}[3]{{\it  Appl. Phys. Lett. }{{\bf #1} {(#2)} {#3}}}
\nc{\apj}[3]{{\it  Ap.\ J.\ }{{\bf #1} {(#2)} {#3}}}
\nc{\apjl}[3]{{\it  Ap.\ J.\ Lett.\ }{{\bf #1} {(#2)} {#3}}}
\nc{\app}[3]{{\it Astropart.\ Phys.\ }{{\bf #1} {(#2)} {#3}}}
\nc{\cmp}[3]{{\it  Comm.\ Math.\ Phys.\ }{{ \bf #1} {(#2)} {#3}}}
\nc{\cqg}[3]{{\it  Class.\ Quant.\ Grav.\ }{{\bf #1} {(#2)} {#3}}}
\nc{\epl}[3]{{\it  Europhys.\ Lett.\ }{{\bf #1} {(#2)} {#3}}}
\nc{\ijmp}[3]{{\it Int.\ J.\ Mod.\ Phys.\ }{{\bf #1} {(#2)} {#3}}}
\nc{\ijtp}[3]{{\it Int.\ J.\ Theor.\ Phys.\ }{{\bf #1} {(#2)} {#3}}}
\nc{\jmp}[3]{{\it  J.\ Math.\ Phys.\ }{{ \bf #1} {(#2)} {#3}}}
\nc{\jpa}[3]{{\it  J.\ Phys.\ A\ }{{\bf #1} {(#2)} {#3}}}
\nc{\jpc}[3]{{\it  J.\ Phys.\ C\ }{{\bf #1} {(#2)} {#3}}}
\nc{\jap}[3]{{\it J.\ Appl.\ Phys.\ }{{\bf #1} {(#2)} {#3}}}
\nc{\jpsj}[3]{{\it J.\ Phys.\ Soc.\ Japan\ }{{\bf #1} {(#2)} {#3}}}
\nc{\lmp}[3]{{\it Lett.\ Math.\ Phys.\ }{{\bf #1} {(#2)} {#3}}}
\nc{\mpl}[3]{{\it  Mod.\ Phys.\ Lett.\ }{{\bf #1} {(#2)} {#3}}}
\nc{\ncim}[3]{{\it  Nuov.\ Cim.\ }{{\bf #1} {(#2)} {#3}}}
\nc{\np}[3]{{\it  Nucl.\ Phys.\ }{{\bf #1} {(#2)} {#3}}}
\nc{\npps}[3]{{\it  Nucl.\ Phys.\ Proc.\ Suppl.\ }{{\bf #1} {(#2)} {#3}}}
\nc{\pr}[3]{{\it Phys.\ Rev.\ }{{\bf #1} {(#2)} {#3}}}
\nc{\pra}[3]{{\it  Phys.\ Rev.\ A\ }{{\bf #1} {(#2)} {#3}}}
\nc{\prb}[3]{{\it  Phys.\ Rev.\ B\ }{{{\bf #1} {(#2)} {#3}}}}
\nc{\prc}[3]{{\it  Phys.\ Rev.\ C\ }{{\bf #1} {(#2)} {#3}}}
\nc{\prd}[3]{{\it  Phys.\ Rev.\ D\ }{{\bf #1} {(#2)} {#3}}}
\nc{\prl}[3]{{\it Phys.\ Rev.\ Lett.\ }{{\bf #1} {(#2)} {#3}}}
\nc{\pl}[3]{{\it  Phys.\ Lett.\ }{{\bf #1} {(#2)} {#3}}}
\nc{\prep}[3]{{\it Phys.\ Rep.\ }{{\bf #1} {(#2)} {#3}}}
\nc{\prsl}[3]{{\it Proc.\ R.\ Soc.\ London\ }{{\bf #1} {(#2)} {#3}}}
\nc{\ptp}[3]{{\it  Prog.\ Theor.\ Phys.\ }{{\bf #1} {(#2)} {#3}}}
\nc{\ptps}[3]{{\it  Prog\ Theor.\ Phys.\ suppl.\ }{{\bf #1} {(#2)} {#3}}}
\nc{\physa}[3]{{\it  Physica\ A\ }{{\bf #1} {(#2)} {#3}}}
\nc{\physb}[3]{{\it  Physica\ B\ }{{\bf #1} {(#2)} {#3}}}
\nc{\phys}[3]{{\it Physica\ }{{\bf #1} {(#2)} {#3}}}
\nc{\rmp}[3]{{\it  Rev.\ Mod.\ Phys.\ }{{\bf #1} {(#2)} {#3}}}
\nc{\rpp}[3]{{\it Rep.\ Prog.\ Phys.\ }{{\bf #1} {(#2)} {#3}}}
\nc{\sjnp}[3]{{\it Sov.\ J.\ Nucl.\ Phys.\ }{{\bf #1} {(#2)} {#3}}}
\nc{\spjetp}[3]{{\it Sov.\ Phys.\ JETP\ }{{\bf #1} {(#2)} {#3}}}
\nc{\yf}[3]{{\it Yad.\ Fiz.\ }{{\bf #1} {(#2)} {#3}}}
\nc{\zetp}[3]{{\it Zh.\ Eksp.\ Teor.\ Fiz.\  }{{\bf #1}  {(#2)} {#3}}}
\nc{\zp}[3]{{\it Z.\ Phys.\ }{{\bf #1} {(#2)} {#3}}}
\nc{\ibid}[3]{{\sl ibid.\ }{{\bf #1} {#2} {#3}}}
\nc{\rf}[1]{(\ref{#1})}
\nc{\nn}{\nonumber \\*}
\nc{\bfB}{\bf{B}}
\nc{\bfv}{\bf{v}}
\nc{\bfx}{\bf{x}}
\nc{\bfy}{\bf{y}}
\nc{\vx}{\vec{x}}
\nc{\vy}{\vec{y}}
\nc{\oB}{\overline{B}}
\nc{\oI}{\overline{I}}
\nc{\oR}{\overline{R}}
\nc{\rar}{\rightarrow}
\nc{\ti}{\times}
\nc{\slsh}{\hskip-5pt/}
\nc{\sm}{Standard~Model~}
\nc{\MP}{M_{\rm Pl}}
\nc{\tp}{t_{\rm Pl}}
\nc{\ave}{\bar{E}}

\nc{\eff}{{\rm eff}}
\nc{\kk}{\vek{k}}
\nc{\pp}{{\rm p}}
\nc{\ga}{g_{a\gamma}}
\nc{\vv}{\\}
\nc{\eee}{{\bf E}}
\nc{\bbb}{{\bf B}}
\nc{\qcd}{T_{\rm QCD}}
\nc{\G}{\rm \ G}
\def\vec#1{{\bf #1}}
\def\lae{\;^{<}_{\sim} \;} \def\gae{\; ^{>}_{\sim} \;}

\begin{document}
{\title{\vskip-2truecm{\hfill {{\small \\
    \hfill HIP-2002-46/TH\\
    }}\vskip 1truecm}
{\bf Cosmological consequences of MSSM flat directions }}
%\vspace{1.2cm}
{\author{
{\sc  Kari Enqvist$^{1}$}\\
{\sl\small Department of Physical Sciences, University of Helsinki,}\\
{\sl\small and Helsinki Institute of Physics}\\
{\sl\small P.O. Box 9, FIN-00014 University of Helsinki, Finland.}\\
{\sc and}\\
{\sc  Anupam Mazumdar$^{2,3}$}\\
{\sl\small The Abdus Salam, International Centre for Theoretical Physics,}\\
{\sl\small Strada Costiera-11, 34100, Trieste, Italy.}
}
\maketitle
\begin{abstract}
We review the cosmological implications of the flat directions
of the Minimally Supersymmetric Standard Model (MSSM). We describe how
field condensates are created along the flat directions because of
inflationary fluctuations. The post-inflationary dynamical evolution
of the field condensate can charge up the condensate with $B$ or $L$ in
a process known as Affleck-Dine baryogenesis. Condensate fluctuations
can give rise to both adiabatic and isocurvature density perturbations
and could be observable in future cosmic microwave experiments. In many cases
the condensate is however not the state of lowest energy but fragments,
with many interesting cosmological consequences. Fragmentation is triggered by
inflation-induced perturbations and the condensate lumps will eventually
form non-topological solitons, known as $Q$-balls. Their properties depend
on how supersymmetry breaking is transmitted to the MSSM; if by gravity,
then the $Q$-balls are semi-stable but long-lived and can be the source
of all the baryons and LSP dark matter; if by gauge interactions, the
$Q$-balls can be absolutely stable and form dark matter that can be
searched for directly. We also discuss some cosmological applications
of generic flat directions and $Q$-balls in the context of self-interacting
dark matter, inflatonic solitons and extra dimensions.
\noindent
\end{abstract}
\footnoterule
{\small $^1$~kari.enqvist@helsinki.fi}\\
{\small $^2$~a.mazumdar@ictp.trieste.it,~anupamm@hep.physics.mcgill.ca}\\
{\small $^3$~Physics Department, McGill University, Canada, from October 2002}

\thispagestyle{empty}
\newpage
\setcounter{page}{1}

\tableofcontents
%%%%%%%%%%%%%%%%%%%%%%%%%%%%%%%%%%%%%%%%%%%%%%%%%%%%%%%%%%%%%%%%%%

\newpage

\section{Introduction}

The interplay between particle physics and cosmology plays an increasing
role in understanding the physics beyond the Standard Model (SM)~\cite{sm}
and the early Universe before the era of Big Bang Nucleosynthesis (BBN)
\cite{sarkar96,olive00333}. On both fronts we currently lack hard data.
Above the electroweak scale $E\sim {\cal O}(100)$~GeV, the particle content
is largely unknown, while beyond the BBN scale $T\sim {\cal O}(1)$~MeV,
there is no direct information about the thermal history of the Universe.
However, there are some observational hints, as well as a number of
theoretical considerations, which seem to be pointing towards a wealth of
new physics both at small distances and in the very early Universe.
Perhaps most importantly, new data is expected soon from accelerator
experiments such as LHC and from cosmological measurements
carried out by satellites such as MAP~\cite{MAP} and Planck \cite{PLANCK}.

In cosmology the recent observations of the cosmic microwave background
(CMB) radiation, which has a temperature $\sim 2.728\pm 0.004$~K
\cite{peebles93}, have given rise to an era of precision cosmology. The
Cosmic Background Explorer (COBE) satellite~\cite{COBE} first detected
in a full-sky map a temperature perturbation of one part in $10^{5}$
at scales larger than $7$~degrees~\cite{smoot91}. The irregularities
are present at a scale larger than the size of the horizon at the time
when the microwave photons were generated and cannot be explained
within the traditional hot Big Bang model~\cite{liddle-lyth00}. The
recent balloon experiments BOOMERANG~\cite{BOOMERANG} and
MAXIMA~\cite{MAXIMA}, together with the ground-based DASI
\cite{pryke01} experiment have established the existence of the first
few acoustic peaks in the positions predicted by cosmic inflation
\cite{guth81,linde82108,linde90,liddle-lyth00}. Inflation, a period of
exponential expansion in the very early Universe, is a direct link to
physics at energy scales that will not be accessible to Earth-bound
experiments for any foreseeable future. Inflation could occur because
a slowly rolling scalar field, the inflaton, dynamically gives rise to
an epoch dominated by a false vacuum. During inflation quantum fluctuation
are imprinted on space-time as energy perturbations which then are
stretched outside the causal horizon. These primordial fluctuations
eventually re-enter our horizon, whence their form can be extracted
from the CMB (for a review, see \cite{mukhanov92,liddle-lyth00}).

Inflation can be considered as a model for the origin of matter
since all matter arises from the vacuum energy stored in the
inflaton field. However the present models do not give clear
predictions as to what sort of matter there is to be found in the
Universe. From observations we know that baryons constitute about
3\%\ of the total mass ~\cite{olive00333}, whereas relic diffuse
cosmic ray background virtually excludes any
domains of anti-baryons in the visible Universe \cite{cohenetal97}.
Almost 30\%\ of the total energy density is in non-luminous, non-baryonic
dark matter~\cite{jungman96}. Its origin and nature is unknown, although
various simulations of large scale structure formation suggest that there
must be at least some {\it cold dark matter} (CDM), comprising of particles
with negligible velocity, although there may also be a component of
{\it hot dark matter} (HDM), comprising of particles with relativistic
velocities~\cite{liddle93}. The rest of the energy density is in the
form of dark energy~\cite{perlmutter97,reiss98}.

The striking asymmetry in the baryonic matter has existed at least
since the time of BBN and plays an important role in providing the
right abundances for the light elements. The present Helium ($^3He$),
Deuterium ($D$) and Lithium abundances suggest a baryon density and
an asymmetry relative to photon density of order $10^{-10}$~\cite{olive00333}.
Such an asymmetry is larger by a factor of $10^{9}$ than what it should
have been by merely assuming a initially baryon symmetric hot Big Bang
\cite{kolbturner90}. Therefore baryon asymmetry must have been created
dynamically in the early Universe.

The origin of baryon asymmetry and dark matter bring cosmology and
particle physics together. Within SM all the three Sakharov conditions for
baryogenesis~\cite{sakharov67} are in principle met; there is baryon number
violation, $C$ and $CP$ violation, and an out-of-equilibrium environment
during a first-order electroweak phase transition. However, it has turned
out that within SM the electroweak phase transition is not strong enough
\cite{cohen93,yaffe95,rubshap96,kajantie96}, and therefore the existence
of baryons requires new physics. Regarding HDM, light neutrinos are a
possible candidate \cite{liddle93,dolgov02}, but there is no
candidate for CDM in the SM. HDM alone cannot lead a successful structure
formation because of HDM free streaming length~\cite{bonometto84,liddle93}.
Therefore one must resort to physics beyond the SM  also to find a
candidate for CDM \cite{jungman96}.

The tangible evidence for small but non-vanishing neutrino masses
as indicated by the neutrino oscillations observed by the
Super-Kamiokande~\cite{superk} and SNO collaborations~\cite{sno} is
definitely another indication for new physics beyond the SM. The main sources
of neutrino mass could be either Dirac or Majorana. A Dirac  neutrino
would require a large fine tuning in the Yukawa sector (one part in
$10^{11}$) while a Majorana mass would appear to require a scale much
above the electroweak scale together with an extension of the SM gauge
group $SU(3)_{C}\times SU(2)_{L}\times U(1)_{Y}$. In the Majorana case
the lightness of the neutrino could be explained via the see-saw mechanism
\cite{see-saw,mohapatra80}.

A theoretical conundrum is that the mass scale of SM is
$\sim{\cal O}(100)$~GeV, much lower that the scale of gravity
$M_{\rm P}=(8\pi G_{N})^{-1/2}=2.436\times 10^{18}$~GeV, and not
protected from quantum corrections. The most popular remedy is of
course supersymmetry~(for a review, see \cite{nilles84,haber85,bailin94}),
despite the fact that so far supersymmetry has evaded all observations
\cite{lepsusy}. The minimal supersymmetric extension of the SM is
called the MSSM. Supersymmetry must be broken at a scale
$\sim {\cal O}(1)$~TeV, presumably in some hidden sector from which
breaking is transmitted to the MSSM, e.g., by gravitational
\cite{nilles84,haber85} or gauge interactions~\cite{giudice98}.

In the MSSM the number of degrees of freedom are increased by virtue
of the supersymmetric counterparts of the SM bosons and fermions. One
of them, known as the lightest supersymmetric particle (LSP), could be
absolutely stable with a mass of the order of supersymmetry breaking
scale. LSP would be a natural candidate for CDM (see e.g. \cite{jungman96}).
In addition, because of the larger parameter space, electroweak baryogenesis
in MSSM in principle has a much better chance to succeed. However, there
are a number of important constraints, and lately Higgs searches at LEP
have narrowed down the parameter space to the point where it
has all but disappeared~\cite{cline00,quiros01,carena02}.

Electroweak baryogenesis within MSSM thus appears to be heading towards
deep trouble. Moreover, although MSSM can provide CDM, there is no
connection between dark matter and electroweak baryogenesis. On the
other hand, by virtue of supersymmetry, MSSM has the intriguing feature
that there are directions in the field space which have virtually no
potential. They are usually known as {\it flat directions}, which are made up
of squarks and sleptons and therefore carry baryon number and/or
lepton number. The MSSM flat directions have been all classified
\cite{gherghetta96}.

Because it does not cost anything in energy, during inflation squarks and
sleptons are free to fluctuate along the flat directions and form
scalar condensates. Because inflation smoothes out all gradients, only
the homogeneous condensate mode survives. However, like any massless
scalar field, the condensate is subject to inflaton-induced zero point
fluctuations which impart a small, and in inflation models a calculable,
spectrum of perturbations on the condensate. After inflation the dynamical
evolution of the condensate can charge the condensate up with a large
baryon or lepton number, which can then released into the Universe
when the condensate decays, as was first discussed by Affleck and Dine
\cite{affleckdine85}.

The potential along the MSSM flat direction is not completely flat because of
supersymmetry breaking. In addition to the usual soft supersymmetry breaking,
the non-zero energy density of the early Universe also breaks supersymmetry,
in particular during inflation when the Hubble expansion dominates over
any low energy supersymmetry breaking scale \cite{dine95,dine96}. Flatness
can also be spoiled by higher-order non-renormalizable terms, and the details
of the condensate dynamics depend on these.

In most cases, the MSSM condensate along a flat direction is however not
the state of lowest energy. The condensate typically has a negative
pressure, which causes the inflation-induced perturbations to grow.
Because of this the condensate fragments, usually when the Hubble scale
equals the supersymmetry breaking scale, into lumps of condensate matter
which eventually settle down to non-topological solitons dubbed as
$Q$-balls by Coleman \cite{coleman85}. $Q$-balls carry a global charge,
which in the case of MSSM is either $B$ or $L$.

The properties of $Q$-balls depend on supersymmetry breaking. If
transmitted to MSSM by gravity, the $Q$-balls turn out to be only
semistable but nevertheless long-lived compared with the time scales
of the very early Universe \cite{enqvist98}. When they decay, they
may provide not only the baryonic matter but also dark matter LSPs
\cite{enqvist99}. If supersymmetry breaking is transmitted from the
hidden sector to MSSM by gauge interactions, the resulting $Q$-balls
would be stable and could exist at present as a form of dark
matter \cite{kusenko97405}. In this case one can make direct searches
for their existence \cite{arafune00}. In both cases there is a prediction
for the relation between the baryon and dark matter densities.
Moreover, the condensate perturbations are inherited by the $Q$-balls,
and can thus be a source of both isocurvature and adiabatic density
perturbations \cite{enqvist9983,enqvist0062,kawasaki01}.

This review is organized as follows. In Section $2$, we recapitulate
some basic cosmology, and in particular baryogenesis. We briefly discuss
various popular schemes of baryogenesis and describe the original
Affleck-Dine baryogenesis. In Section $3$, we present some background
material for inflation, mainly concentrating on supersymmetric  models.
Quantum fluctuations and reheating are also discussed.
In Section $4$, we present the MSSM flat directions and discuss their
properties. Various contributions to the flat direction potential in
the early Universe are listed. Low energy supersymmetry breaking schemes,
such as gravity and gauge mediation, are also discussed. In Section $5$,
we discuss the dynamical properties of flat directions and the running
of the flat direction potential due to gauge and Yukawa interactions.
Leptogenesis along $LH_{u}$ flat direction, and the condensate evaporation
in a thermal bath, is also described. We discuss fragmentation of the
condensates for both gravity and gauge mediated supersymmetry breaking
and present the relevant numerical studies. In Section $6$, $Q$-ball
properties are presented in detail. We describe various types of $Q$-balls,
their interactions and their behavior at finite temperature. We discuss
surface evaporation, diffusion, and dissociation of charge from $Q$-balls
in a thermal bath. In Section $7$, we focus on the cosmological consequences
of $Q$-balls. We consider $Q$-ball baryogenesis and non-thermal dark matter
generation through charge evaporation for different types of $Q$-balls.
We discuss $Q$-balls as self-interacting dark matter and present
experimental and astrophysical constraints on stable $Q$-balls. In
Section $8$, we briefly survey beyond-the-MSSM-condensates by
considering inflatonic $Q$-balls and Affleck-Dine mechanism without
MSSM flat directions. We also describe solitosynthesis, a process of
accumulating large $Q$-balls in a charge asymmetric Universe.

%%%%%%%%%%%%%%%%%%%%%%%%%%%%%%%%%%%%%%%%%%%%%%%%%%%%%%%%%%%%%

\newpage
%%%%%%%%%%%%%%%%%%%%%%%%%%%%%%%%%%%%%%%%%%%%%%%%%%%%%%%%%%%%%
\section{Baryogenesis}

%%%%%%%%%%%%%%%%%%%%%%%%%%%%%%%%%%%%%%%%%%%%%%%%%%%%%%%%%%%%%
\subsection{Baryon asymmetric Universe}

There are only insignificant amounts of anti-particles within the
solar system. Cosmic ray showers contain $\sim 10^{-4}$ anti-protons
for each proton \cite{cosmicrays}, but the anti-protons are by-products
of the interaction of the primary beam with the interstellar
dust medium. This strongly suggest that galaxies and intergalactic
medium is made up of matter rather than anti-matter, and if there were
any anti-matter, the abundance has to be smaller than one part in $10^{4}$.
The absence of annihilation radiation from the Virgo cluster shows that
little anti-matter is to be found within a 20 Mpc sphere, and the relic diffuse
cosmic ray background virtually excludes domains of anti-matter in the
visible Universe \cite{cohenetal97}.

The best present estimation for the baryon density comes from BBN \cite{copi95}
combined with the CMB experiments and it is given by~\cite{olive02}
\begin{equation}
\label{omegab}
0.010 \leq \Omega_{b}h^2\leq 0.022\,,
\end{equation}
where $\Omega_{b}\equiv \rho_{b}/\rho_{c}$ defines the fractional baryon
density $\rho_{b}$ with respect to the critical energy density of the
Universe: $\rho_{c}=1.88~h^2\times 10^{-29}~{\rm g~cm^{-3}}$. The
observational uncertainties in the present value of the Hubble constant;
$H_{0}=100~h~{\rm km \cdot s^{-1}\cdot Mpc^{-1}}\approx (h/3000){\rm Mpc^{-1}}$
are encoded in $h$. Various considerations such as Hubble Space Telescope
observations and type Ia supernova data suggest that $h=0.70$
\cite{riess96}. However, from the age of the globular cluster which comes out
to be $11$~Gyr, $h$ seems to take lower value of about $0.5$~\cite{reid97}.
The present convention is to take $0.5 \leq h \leq 0.8$. In terms of
the baryon and photon number densities we may write
\begin{equation}
\eta\equiv \frac{n_{b}-n_{\bar b}}{n_{\gamma}}=2.68\times 10^{-8}
\Omega_{b}h^2\,,
\end{equation}
where $n_{b}$ is the baryon number density and $n_{\bar b}$ is for
anti-baryons. The photon number density is given by
$n_{\gamma}\equiv (2\zeta(3)/\pi^2)T^3$. Observations of the deuterium
abundance in quasar absorption lines suggest \cite{tytler96}
\begin{equation}
\label{range1}
4(3)\times 10^{-10} \leq \eta \leq 7(10)\times 10^{-10}\,.
\end{equation}
The conservative bounds are in parenthesis.

Often in the literature the baryon asymmetry is given in relation
to the entropy density $s=1.8g_{\ast}n_{\gamma}$, where $g_{\ast}$
measures the effective number of relativistic species which itself
a function of temperature. At the present time $g_{\ast} \approx 3.36$,
while during BBN $g_{\ast} \approx 10.11$, rising up to $106.75$ at
$T\gg 100$~GeV. In the presence of supersymmetry at $T \gg 100$~GeV,
the number of effective relativistic species are doubled to $213.30$.

The baryon asymmetry $B$, defined as the difference of baryon and
anti-baryon number densities relative to the entropy density, is
bounded by
\begin{equation}
\label{range2}
5.7(4.3)\times 10^{-11} \leq B\equiv \frac{n_{b}-n_{\bar b}}{s}\leq
9.9(14)\times 10^{-11}\,,
\end{equation}
where the numbers in parenthesis are conservative bounds \cite{tytler96}.
If at the beginning $\eta =0$, then the origin of this small number can
not be understood in a CPT invariant Universe by a mere thermal decoupling
of nucleons and anti-nucleons at  $T\sim 20$~MeV. The resulting asymmetry
would be too small by at least nine orders of magnitude, see
\cite{kolbturner90}.

%%%%%%%%%%%%%%%%%%%%%%%%%%%%%%%%%%%%%%%%%%%%%%%%%%%%%%%%%%%%%%
\subsection{Thermal history of the Universe}

\subsubsection{Expanding Universe}

The hot Big Bang cosmology assumes that the Universe is spatially homogeneous
and isotropic and can be described by the Friedmann-Robertson-Walker (FRW)
metric
\begin{equation}
ds^2=g_{\mu \nu}dx^{\mu}dx^{\nu}=dt^2-a^2(t)\left[\frac{dr^2}{1-Kr^2}
+r^2(d\theta^2+\sin^2\theta d\phi^2)\right]\,,
\end{equation}
where $a(t)$ is the scale factor that determines the expansion or the
contraction of the Universe; the constant $K$ defines the spatial geometry. If
$K=0$, the Universe is flat and has  Euclidean geometry, otherwise there
is a spatial curvature corresponding either to a closed elliptic ($K=+1$)
or an open hyperbolic ($K=-1$) geometries. The value of $K$ cannot however fix
the global topology; for instance, an Euclidean topology can be flat and
infinite $\rm R^{3}$, or a surface of a 3-torus $\rm T^{3}$. However,
topology has other observable consequences, e.g.,  for the
pattern of CMB temperature fluctuations \cite{levin02}.

There are two characteristic scales corresponding to the homogeneous and
isotropic Universe: the curvature scale $r_{\rm curv}=a(t)|K|^{-1/2}$,
and the Hubble scale
\begin{equation}
H^{-1}=\left[\frac{\dot a(t)}{a(t)}\right]^{-1}\,,
\end{equation}
where dot denotes derivative w.r.t. $t$. The Hubble time is denoted by
\begin{equation}
\label{efolds}
t_{\rm Hub} =\int_{i}^{f}\frac{dt}{H^{-1}}=\ln \left(\frac{a_{f}}{a_{i}}
\right)\,.
\end{equation}

The behavior of the scale factor depends on the energy momentum tensor
of the Universe. For a perfect fluid
\begin{equation}
T_{\mu\nu}=-pg_{\mu\nu}+(\rho+p)u_{\mu}u_{\nu}\,,
\end{equation}
where $\rho$ is the energy density and $p$ is the pressure of a fluid and
the four velocity $u_{\mu}\equiv dx_{\mu}/ds$.  For the FRW metric and for
the perfect fluid the equations of motion gives the Friedmann-Lemaitre equation
\begin{equation}
\label{hubble}
H^2=\frac{\rho}{3M_{\rm P}^2}-\frac{K}{a(t)^2}\,,
\end{equation}
also known as the Hubble equation. The acceleration equation is given by
\begin{equation}
\label{accelerate}
\frac{\ddot a(t)}{a(t)}=-\frac{1}{6M_{\rm P}^2}(\rho+3p)\,,
\end{equation}
and the conservation of the energy momentum tensor $T^{\mu\nu}_{;\nu}=0$
gives
\begin{equation}
\label{em}
\frac{d(\rho a^3)}{da}=-3pa^2\,.
\end{equation}
Note that $\rho a^3$ is constantly decreasing in an expanding Universe for
a positive pressure.

The early Universe is believed to have been radiation dominated  with
$p=\rho/3$ and $a(t) \propto t^{1/2}$, followed by a matter dominated era
with $p=0$ and $a(t)\propto t^{2/3}$. The early Universe might also have
had an era of acceleration, known as the inflationary phase, which could
have happened only if
\begin{equation}
\ddot{a} > 0~\Leftrightarrow ~\rho +3p<0 \,.
\end{equation}
A geometric way of defining inflation is \cite{liddle-lyth00}
\begin{equation}
\label{def}
\frac{d(H^{-1}/a(t))}{dt} <0\,,
\end{equation}
which states that the Hubble length as measured in comoving coordinates
decreases during inflation. We will use this particular definition
of inflation very often while discussing the number of e-foldings and
density perturbations.

The Hubble expansion rate is related to the temperature by
\begin{equation}
H=\sqrt{\frac{\rho}{3M_{\rm P}^2}}=1.66\times g_{\ast}^{1/2}\frac{T^2}
{M_{\rm P}}\,,
\end{equation}
where $g_{\ast}$ is the total number of relativistic
degrees of freedom and it is given by
\begin{equation}
g_{\ast}(T)=\sum_{i=\rm b}g_{i}\left(\frac{T_{i}}{T}\right)^4+\frac{7}{8}
\sum_{i=\rm f}g_{i}\left(\frac{T_{i}}{T}\right)^4\,.
\end{equation}
Here  $T_{i}$ denotes the effective temperature of species $i$, which
decouples at a temperature $T=T_{\rm D}$.

During the radiation era when $H=(1/2t)$, one finds
\begin{equation}
\label{timetemp}
\frac{t}{1~{\rm s}}\approx 2.42 g_{\ast}^{-1/2}\left(\frac{1~{\rm MeV}}{T}
\right)^2\,.
\end{equation}

%%%%%%%%%%%%%%%%%%%%%%%%%%%%%%%%%%%%%%%%%%%%%%%%%%%%%%%%%%%%%%%%%

\subsubsection{Entropy}

An ideal gas of particles respects the Fermi-Dirac or Bose-Einstein
distributions
\begin{equation}
\label{distribution}
f_{i}(\vec{p},\mu,T)=\left[\exp((E_{i}-\mu_{i})/T)\mp 1\right]^{-1}\,,
\end{equation}
where $E_{i}^2=|\vec{p}|^2+m^2$, $\mu_{i}$ represents the chemical potential
of the species $i$, $-/+$ corresponds to Bose/Fermi statistics. The value of
$\mu$ is equal and opposite for particles and anti-particles. Therefore
in the early Universe a finite net chemical potential is proportional to the
particle anti-particle asymmetry.  The bound on charge asymmetry relative
to the photon number density is severe, less than one part in $10^{43}$ at
temperatures close to BBN \cite{masso}, while baryon asymmetry is
comparatively larger, but still small enough for $\mu_{e},~\mu_{b} \approx 0$
to be an excellent approximation. Neutrinos may however carry a net $B-L$
charge which need not be vanishingly small at early times, although a
large enough neutrino chemical potential can affect nucleosynthesis,
for example, see~\cite{dolgov02}.

The number density $n$, energy density $\rho$, and pressure $p$ can be
expressed in terms of temperature, and $g_{i}$ is the number of internal
degrees of freedom \cite{harrison73}
\begin{eqnarray}
\label{eqbm}
n_{i}(T)=\frac{g_{i}}{(2\pi)^3}\int f_{i}(\vec{p},\mu,T) d^{3}{\vec p}
=\frac{g_{i}}{2\pi^2}T^3I_{i}^{11}(\mp)\,, \nonumber \\
\rho_{i}(T)=\frac{g_{i}}{(2\pi)^3}\int E_{i}(\vec{p}) f_{i}
(\vec{p},\mu,T) d^{3}{\vec p}=\frac{g_{i}}{2\pi^2}T^4I_{i}^{21}(\mp)\,,
\nonumber \\
p_{i}(T)=\frac{g_{i}}{(2\pi)^3}\int \frac{|\vec{p}|^2}{3E_{i}(\vec{p})}
f_{i}(\vec{p},\mu,T) d^{3}{\vec p}=\frac{g_{i}}{6\pi^2}T^4I_{i}^{03}(\mp)\,,
\end{eqnarray}
where
\begin{equation}
I_{i}^{ab}(\mp)\equiv \int_{x_{i}}^{\infty}\frac{y^a(y^2-x_{i}^2)^{b/2}}{
(e^{y}\mp 1)}dy\,, \quad \quad x_{i}\equiv \frac{m_{i}}{T}\,.
\end{equation}
For a relativistic case with $x_{i}\ll 1$,
\begin{eqnarray}
I^{11}_{\rm r}(-)&=&2\zeta(3)\,,\quad \quad I_{\rm r}^{21}(-)=I^{03}_{\rm r}(-)
=\frac{\pi^4}{15}\,, ~~~{\rm for~bosons}\,, \nonumber \\
I^{11}_{\rm r}(+)&=&\frac{3\zeta(3)}{2}\,, \quad \quad I^{21}_{\rm r}(+)=
I^{03}_{\rm r}(+)=\frac{7\pi^4}{120}\,, ~~~{\rm for~fermions}\,,
\end{eqnarray}
where $\zeta$ denotes the Riemann Zeta function and $\zeta(3)=1.202$. Thus
the energy density of radiation reads
\begin{equation}
\rho_{\rm r}=\frac{\pi^2}{30}g_{\ast}T^4\,.
\end{equation}

For non-relativistic particles with $x_{i}\gg1$, one obtains for both
bosons and fermions
\begin{equation}
\label{nonrelatdist}
n_{\rm nr}(T)=\frac{\rho_{\rm nr}}{m}=g_{i}\left(
\frac{m_{i}T}{2\pi}\right)^{3/2}e^{-m_{i}/T}\,, \quad \quad \quad
p_{\rm nr}=0\,.
\end{equation}
If the chemical potential is non-zero,  the exponential \eq{nonrelatdist}
also includes a factor $e^{+\mu_{i}/T}$.

The entropy density is defined as
\begin{equation}
s\equiv \frac{S}{T}=\frac{\rho_{i}+p_{i}}{T}\,,
\end{equation}
where $d(sa^3)=0$ is a thermodynamically conserved quantity. The decoupling
temperature can be expressed as \cite{srednicki88}
\begin{equation}
\label{dectemp}
\frac{T_{D}}{T}=\left(\frac{g_{\ast S_{A}}(T_{\rm D})}{g_{\ast S_{A}}(T)}
\frac{g_{\ast S-S_{A}}(T)}{g_{\ast S-S_{A}}(T_{\rm D})}\right)^{1/3}\,,
\end{equation}
where $S$ is the total entropy and $S_A$ the entropy in the degrees
of freedom that have decoupled at $T_D$.

%%%%%%%%%%%%%%%%%%%%%%%%%%%%%%%%%%%%%%%%%%%%%%%%%%%%%%%%%%%%%%%
\subsubsection{Nucleosynthesis}

According to BBN (for reviews see \cite{olive00333,olive02}) the light
elements $^2H$, $^3He$, $^4He$, and $^7Li$ have been synthesized
during the first few hundred seconds. The abundances depend on the
baryon-to-photon ratio
\begin{equation}
\eta\equiv \frac{n_B}{n_\gamma}\,.
\end{equation}
All the relevant physical processes take place essentially
in the range from a few MeV $\sim 0.1$~sec down to $60-70$~KeV
$\sim 10^{3}$~sec. During this period only photons, $e^{\pm}$ pairs,
and the  three neutrino flavors contribute significantly
to the energy density. Any additional energy density may be parameterized
in terms of the effective number of light neutrino species $N_{\nu}$,
so that
\begin{equation}
g_{\ast}=10.75+\frac{7}{4}(N_{\nu}-3)\,.
\end{equation}
Nucleosynthesis starts off with a freezing out of the weak interaction
between neutron and proton at $T_D\approx 0.8$~MeV. Free neutrons keep
decaying until deuterium begins to form through $n+p\rightarrow d+\gamma$.
Deuterium synthesis is over by $T_{D}\approx 0.086~{\rm MeV}$
(assuming $\eta = 5\times 10^{-10}$). At $T_{D}$, neutron abundance
has been depleted to $X_{n}(t_{D})\equiv n/(n+p)\approx 0.122$.
All the surviving neutrons are now captured through
$n+D\rightarrow (^3H,^3He)$, and subsequently by virtue of the process
$(^3H,^3He)+n \rightarrow ^4He$, which has a binding energy of
$28.3$~MeV. The total mass fraction of primordial helium, which
is denoted by $Y_{\rm P}(^4He)$, is given by
\begin{equation}
Y_{\rm P}(^4He)\approx 2X_{n}(T_{D})=0.245\,.
\end{equation}
Adopting the experimentally allowed range of $0.22 < Y_{\rm P}< 0.26$,
one can constraint that number of light neutrino species by \cite{copi95}
\begin{equation}
N_{\nu} \leq 4\,.
\end{equation}
The four LEP experiments combined give the best fit as~\cite{pdg}
\begin{equation}
N_{\nu}=2.994 \pm 0.12\,.
\end{equation}
Nucleosynthesis also constrains many non-conventional ideas, for instance
alternative theories of gravity such as scalar-tensor theories~\cite{will93}.

Besides $^4He$, D and $^3He$ are produced at the level of
$10^{-5}$, and $^7Li$ at the level of $10^{-10}$. The theoretical
prediction has some slight problems in fitting the observed $^4He$ and
$^7Li$ abundances. Both seem to indicate
$1.7\times 10^{-10}< \eta<4.7\times 10^{-10}$, corresponding to
$0.006< \Omega_{b}h^2 < 0.017$ with a central value
$\Omega_{b}h^2=0.009$ \cite{olive02}. The abundance ratio
$D/H$  is comparable with $^4He$ and $^7Li$
abundances at the $2\sigma$ level in the range
$4.7\times10^{-10}<\eta<6.2\times 10^{-10}$, which corresponds to
$0.017<\Omega_{b}h^2<0.023$. The likelihood analysis which includes all
the three elements $({\rm D}, ^4He, {\rm and}~^7Li)$ yields
\cite{olive02}
\begin{equation}
4.7\times 10^{-10} <\eta < 6.2\times 10^{-10}\,, \quad \quad \quad
0.017 < \Omega_{b}h^2<0.023\,.
\end{equation}
Despite the uncertainties there appears to be a general concordance
between theoretical BBN predictions and observations, which is now
being bolstered by the CMB data from several different experiments.
The results from the ground based DASI experiment indicates
$\Omega_{b}h^2=0.022^{+0.004}_{-0.003}$ \cite{pryke01}, while the results
from the BOOMERANG balloon-borne experiment imply
$\Omega_{b}h^2=0.021^{+0.004}_{-0.003}$ \cite{BOOMERANG}.
MAXIMA, another balloon experiment, quotes a somewhat larger
value $\Omega_{b}h^2=0.0325\pm 0.006$~\cite{stompor01}.

%%%%%%%%%%%%%%%%%%%%%%%%%%%%%%%%%%%%%%%%%%%%%%%%%%%%%%%%%%%%
\subsection{Requirements for baryogenesis}

As pointed out by
Sakharov \cite{sakharov67}, baryogenesis requires three ingredients:
$(1)$ baryon number non-conservation, $(2)$ $C$ and $CP$ violation,
and $(3)$ out-of-equilibrium condition. All these three conditions
are believed to be met in the very early Universe\footnote{There have
been attempts \cite{omne69,sato81} for  baryogenesis via a repulsive
interaction between baryons and anti-baryons which would lead to their
spatial separation  before thermal decoupling of nucleons and anti-nucleons.
However at such early times the causal horizon contained only a very small
fraction of the solar mass so that the asymmetry could not be smooth at
distances greater than the galactic size.}.

%%%%%%%%%%%%%%%%%%%%%%%%%%%%%%%%%%%%%%%%%%%%%%%%%%%%%%%%%%%
\subsubsection{Non-conservation of baryonic charge}

In the SM, baryon number $B$ is violated by non-perturbative instanton
processes \cite{thooft76a,thooft76b}. At the quantum level both baryon number
current $J^{\mu}_{B}$ and the lepton number current $J^{\mu}_{L}$ are
not conserved because of chiral anomalies \cite{abj}. However the
anomalous divergences of $J^{\mu}_{B}$ and $J^{\mu}_{L}$ come with an equal
amplitude and an opposite sign. Therefore $B-L$ remains conserved,
while $B+L$ may change via processes which interpolate between the
multiple non-Abelian vacua of $SU(2)$. The probability for the $B+L$
violating transition is however exponentially suppressed
\cite{thooft76a,thooft76b}. As was first pointed out by Manton~\cite{manton83},
at high temperatures the situation is different, so that when
$T\gg M_W$, baryon violating transitions are in fact copious
(see Sect.~$2.4.2$).

In addition to baryogenesis, $B$ violation also leads to proton decay
in GUTs.  For instance, the dimension $6$ operator $(QQQL)/\Lambda$
generates observable proton decay unless $\Lambda \geq 10^{15}$~GeV. In the
MSSM the bound is $\Lambda \geq 10^{26}$~GeV
because the decay can take place via a dimension $5$ operator. In the MSSM
superpotential there are also terms which can lead to $\Delta L =1$ and
$\Delta B=1$. Similarly there are other processes such as neutron-anti-neutron
oscillations in SM and in supersymmetric theories which lead
to $\Delta B =2$ and $\Delta B=1$ transitions~\cite{enqvist86}. These
operators are constrained by the measurements of the proton lifetime,
which yield the bound  $\tau_{p} \geq 10^{33}$~years~\cite{pdg}.

%%%%%%%%%%%%%%%%%%%%%%%%%%%%%%%%%%%%%%%%%%%%%%%%%%%%%%
\subsubsection{$C$ and $CP$ violation}

Weak interactions ensures maximum $C$ violation while neutral Kaon is an
example of $CP$ violation in the quark sector which has a relative
strength $\sim 10^{-3}$ \cite{pdg}. $CP$ violation could also expected
to be found in the neutrino sector. Beyond the SM there are many sources
for $CP$ violation. An example is the axion proposed for solving the
strong $CP$ problem in $QCD$ \cite{pecceiquinn77}. Quantum
fluctuations of light scalars in the early  Universe, in particular
during inflation, can create different domains of various $C$ and $CP$
phases. $C$ and $CP$ can also be spontaneously broken during a phase
transition, so that domains of broken phases form with different $CP$-charges
\cite{vilenkin94}.

%%%%%%%%%%%%%%%%%%%%%%%%%%%%%%%%%%%%%%%%%%%%%%%%%%%%%%%%%%%
\subsubsection{Departure from thermal equilibrium}

Departure from a thermal equilibrium cannot be achieved
by mere particle physics considerations but is coupled to
the dynamical evolution of the Universe. If $B$-violation
processes are in thermal equilibrium, the inverse processes
will wash out the pre-existing asymmetry $(\Delta n_{b})_0$
\cite{weinberg79}. This is a consequence of $S$-matrix unitarity
and $CPT$-theorem \cite{dolgov81}. However there are several
ways of obtaining an out-of-equilibrium process in the early
Universe.

\begin{itemize}

\item{{\it Out-of-equilibrium decay or scattering}:\\
The Universe in a thermal equilibrium can not produce any asymmetry,
rather it tries to equilibrate any pre-existing asymmetry. If
the scattering rate $\Gamma < H$, the process can take
place  out-of-equilibrium. Such a situation is appropriate for e.g.
GUT baryogenesis~\cite{dolgov81,kolb83}.

}

\item{{\it Phase transitions}:\\
Phase transitions are ubiquitous in the early Universe. The
transition could be of {\it first}, or of {\it second} (or of still higher)
order. First order transitions proceed by barrier penetration and
subsequent bubble nucleation resulting in a temporary departure from
equilibrium. Second order phase transitions have no barrier between
the symmetric and the broken phase. They are continuous and equilibrium
is maintained throughout the transition.

Prime examples of first order phase transitions in the early Universe
are the QCD and electroweak phase transitions. The nature and details
of QCD phase transition is still very much an open
debate~\cite{kapusta88,yaffe95,karsch2000}, and although a mechanism
for baryogenesis during QCD phase transition has been
proposed~\cite{zhitnitsky99}, much more effort has been devoted to the
electroweak phase transition~\cite{cohen93,rubshap96} (see Sect.~$2.4.2$).

}

\item{{\it Non-adiabatic motion of a scalar field}:\\
Any complex scalar field carries $C$ and $CP$, but the symmetries
can be broken by terms in the scalar potential. This can
lead to a non-trivial trajectory of a complex scalar field in the
phase space. If a coherent scalar field is trapped in a local minimum
of the potential and if the shape of the potential changes to become a
maximum, then the field may not have enough time to readjust with
the potential and may experience completely non-adiabatic motion. This is
similar to a second order phase transition but it is the non-adiabatic
classical motion which prevails over the quantum fluctuations, and therefore,
departure from equilibrium can be achieved. If the field condensate carries
a global charge such as the baryon number, the motion can charge up the
condensate. This is the basis for the Affleck-Dine
baryogenesis~\cite{affleckdine85} (see Sect.~$2.5$).

}

\end{itemize}

%%%%%%%%%%%%%%%%%%%%%%%%%%%%%%%%%%%%%%%%%%%%%%%%%%%%%%%%%%%%%%%%%%
\subsubsection{Sphalerons}

In the SM $B+L$ is very weakly violated in the vacuum~\cite{thooft76a}.
At finite temperatures violation is large
\cite{manton83,klinkhamer84,kunz92,brihaye93,moreno97}
by virtue of the sphaleron configurations, which mediate transitions between
degenerate gauge vacua with different Chern-Simons numbers related to the
net change of $B+L$. Thermal scattering produces sphalerons which in
effect decay in $B+L$ non-conserving ways below $10^{12}$~GeV
\cite{bocharev87}, and thus can exponentially wash away $B+L$ asymmetry.
Sphalerons and associated electroweak baryogenesis has been reviewed in
\cite{dolgov92,cohen93,rubshap96,trodden98,riotto99,riotto98}. Let us
here just give a brief summary of the main ingredients.

\begin{itemize}
\item{{\it Chiral anomalies}\\
An anomaly means that a classical current conservation no longer holds
at the  quantum level; an example is the chiral anomaly~\cite{abj}.
In the SM there is classical conservation of the baryon and
lepton number currents $J^{\mu}_{B}$ and $J^{\mu}_{L}$, but
because of chiral anomaly the currents are not conserved. Instead
\cite{thooft76a},
\begin{eqnarray}
\partial_{\mu}J^{\mu}_{B}&=&-\frac{\alpha_{2}}{8\pi}N_{g}
W^{\mu\nu}_{i}\tilde W_{i\mu\nu}+
\frac{\alpha_{1}}{8\pi}N_{g}\left(\frac{4}{9}+\frac{1}{9}-\frac{2}{36}
\right)F^{\alpha\beta}\tilde F_{\alpha\beta}\,, \nonumber \\
\partial_{\mu}J^{\mu}_{L}&=&-\frac{\alpha_{2}}{8\pi}N_{g}
W^{\mu\nu}_{i}\tilde W_{i\mu\nu}+
\frac{\alpha_{1}}{8\pi}N_{g}\left(1-\frac{1}{2}\right)F^{\alpha\beta}
\tilde F_{\alpha\beta}\,,
\end{eqnarray}
where $N_{g}$ is the number of generations, $\alpha_2$ and $\alpha_1$
($W_{i\mu\nu}$ and $F_{\mu\nu}$) are respectively the $SU(2)$ and $U(1)$
gauge couplings (field strengths), and the various numbers inside the
brackets correspond to the squares of the hypercharges multiplied by the
number of states. Note that while at the quantum level $B+L$ is violated,
$B-L$ is still conserved.

}

\item{{\it Gauge theory vacua}\\
The vacuum structure of the gauge theories is very rich
\cite{thooft76b,polyakov77}. In case of $SU(2)$, the vacua are
classified by their homotopy class $\{\Omega_{n}({\rm r})\}$,
characterized by the winding number $n$ which labels the so called
$\theta$-vacua \cite{thooft76b,polyakov77}. A gauge invariant quantity
is the difference in the winding number (Chern-Simons number)
\begin{equation}
\label{cs1}
N_{CS}\equiv n_{+}-n_{-}=\frac{\alpha_{2}}{8\pi}\int d^4x W^{\mu\nu}_{a}
\tilde W_{a\mu\nu}\,.
\end{equation}
In the electroweak sector the field density $W\tilde W$ is related
to the divergence of $B+L$ current. Therefore a change in $B+L$ reflects
a change in the vacuum configuration and is determined by the
difference in the winding number
\begin{equation}
\Delta{(B+L)}=\int d^4x\partial_{\mu}J^{\mu}_{B+L}=-\frac{\alpha_{2}}{4\pi}
N_{g}\int d^4x W_{a}^{\mu\nu}\tilde W_{a\mu\nu}=-2N_{g}N_{CS}\,.
\end{equation}
For three generations of SM leptons and quarks the minimal violation is
$\Delta(B+L)=6$. Note that the proton decay $p\rightarrow e^{+}\pi^{0}$
requires $\Delta(B+L)=2$, so that despite $B$-violation, proton decay is
completely forbidden in the SM.

The probability amplitude for tunneling  from an $n$ vacuum at
$t\rightarrow -\infty$ to an $n+N_{CS}$ vacuum at $t\rightarrow +\infty$
can be estimated by the WKB method~\cite{thooft76a}
\begin{equation}
P(N_{CS})_{B+L}\sim \exp\left(\frac{-4\pi N_{CS}}{\alpha_{2}(M_{Z})}
\right)\sim 10^{-162N_{CS}}\,.
\end{equation}
Therefore, as advertised, the baryon number violation rate is totally
negligible in the SM at zero temperature, but as argued in
a seminal paper by Kuzmin, Rubakov and Shaposhnikov \cite{kuzmin85}, at
finite temperatures the situation is completely different.
}

\item{{\it Thermal tunneling}\\
The sphaleron is a field configuration sitting at the top of the
potential barrier between two vacua with different Chern-Simons numbers
and can be reached simply because of thermal fluctuations \cite{kuzmin85}.
Neglecting $U(1)_Y$, the zero temperature sphaleron solution was first
found by Manton and Klinkhamer \cite{manton83,klinkhamer84}.
At finite temperature the energy obeys an approximate scaling law
\cite{brihaye93,moreno97}
$E_{sph}(T)=E_{sph}(0)\langle\Phi(T)\rangle/\langle\Phi(0)\rangle$:
\begin{equation}
\label{sphe00}
E_{sph}(T)=\frac{2m_{W}(T)}{\alpha_2}B\left({\lambda}/{g_{2}}\right)\,,
\end{equation}
where $m_{W}(T)=(1/2)g_{2}\langle\Phi(T)\rangle$ is the mass of the W-boson
and the function $B$ has a weak dependence on $\lambda/g_2$, where $\lambda$
is the quartic self coupling of the Higgs.
Below the critical temperature of the electroweak phase transition,
the sphaleron rate is exponentially suppressed
\cite{carson90}:
\begin{equation}
\label{gammalow}
\Gamma \sim 2.8\times 10^{5}\kappa T^4\left(\frac{\alpha_2}{4\pi}\right)^4
\left(\frac{E_{sph}(T)}{B(\lambda^2/g)}\right)^7e^{-E_{sph}/T}\,.
\end{equation}
where $\kappa$ is the functional determinant which can take the values
$10^{-4}\leq \kappa \leq 10^{-1}$ \cite{dine92}.
Above the critical temperature the rate is however unsuppressed.
Requiring that the Chern-Simons number changes at most by
$\Delta N_{CS} \sim 1$, one can estimate from Eq.~(\ref{cs1}) that
$\Delta N_{CS}\sim g_{2}^{2}l_{sph}^{2}W_{i}^{2}\sim 1\rightarrow W_{i}\sim\frac{1}{g_{2}l_{sph}}$.
Therefore a typical energy of the sphaleron configuration is given by
\begin{equation}
E_{sph}\sim l_{sph}^3(\partial W_{i})^2 \sim \frac{1}{g_{2}^2l_{sph}}\,.
\end{equation}
At temperatures greater than the critical temperature there is no
Boltzmann suppression, so that the thermal energy $\propto T\geq E_{sph}$.
This determines the size of the sphaleron as
\begin{equation}
l_{sph}\geq \frac{1}{g_2^2T}\,.
\end{equation}
This is exactly the infrared cut-off generated by the magnetic mass of
order $\sim g_2^2T$. Therefore, based on this coherence length scale one
can estimate the baryon number violation per volume $\sim l_{sph}^3$,
and per unit time $\sim l_{sph}$. On dimensional grounds the transition
probability would then be given by
\begin{equation}
\label{sphr}
\Gamma_{sph}\sim \frac{1}{l_{sph}^3 t} \sim \kappa(\alpha_{2}T)^4\,.
\end{equation}
where $\kappa$ is a constant which incorporates various uncertainties.
However, the process is inherently non-perturbative, and it has been
argued that damping of the magnetic field in a plasma suppresses the
sphaleron rate by an extra power of $\alpha_2$~\cite{arnold97}, with
the consequence that $\Gamma_{sph}\sim \alpha_{2}^5T^4$. Lattice simulations
with hard thermal loops also give
$\Gamma_{sph}\sim {\cal O}(10)\alpha_{2}^5T^4$~\cite{moore9858}

}

\end{itemize}

%%%%%%%%%%%%%%%%%%%%%%%%%%%%%%%%%%%%%%%%%%%%%%%%%%%%%%%%%
\subsubsection{Washing out $B+L$}

In the early Universe the transitions $\Delta N_{CS}=+1$ and
$\Delta N_{CS}=-1$ are equally probable. The ratio of rates for the two
transitions is given by
\begin{equation}
\label{rsph}
\frac{\Gamma_{sph~+}}{\Gamma_{sph~-}}=e^{-\Delta F/T}\,,
\end{equation}
where $\Delta F$ is the free energy difference between the two vacua.
Because of a finite $B+L$ density, there is a net chemical potential
$\mu_{B+L}$. Therefore
\begin{equation}
\label{old}
\Delta F\sim \mu^2_{B+L}T^2 +
{\cal O}(T^4) \equiv \frac{n^2_{B+L}}{T^2}+{\cal O}(T^4) \,.
\end{equation}
One then obtains \cite{bocharev87}
\begin{equation}
\label{old1}
\frac{dn_{B+L}}{dt}=\Gamma_{sph~+}-\Gamma_{sph~-}\sim N_{g}\frac{
\Gamma_{sph}}{T^3}n_{B+L}\,.
\end{equation}
It then follows that an exponential depletion of $n_{B+L}$ due to sphaleron
transitions remains active as long as
\begin{equation}
\label{active}
\frac{\Gamma_{sph}}{T^3} \geq H ~~\Rightarrow ~~T \leq \alpha_{2}^{4}
\frac{M_{\rm P}}{g_{\ast}^{1/2}}\sim 10^{12}~{\rm GeV}\,.
\end{equation}
This result is important because it suggests that below $T =10^{12}$~GeV,
the sphaleron transitions can wash out any $B+L$ asymmetry being produced
earlier in a time scale $\tau \sim (T^3/N_{g}\Gamma_{sph})$. This seems to
wreck GUT baryogenesis based on $B-L$ conserving groups such as the
minimal $SU(5)$.

%%%%%%%%%%%%%%%%%%%%%%%%%%%%%%%%%%%%%%%%%%%%%%%%%%%%%%%%
%%%%%%%%%%%%%%%%%%%%%%%%%%%%%%%%%%%%%%%%%%%%%%%%%%%%%%%%

\subsection{Alternatives for baryogenesis}

There are several scenarios for baryogenesis (for reviews on
baryogenesis, see \cite{dolgov81,riotto99,riotto98}), the main contenders
being GUT baryogenesis, electroweak baryogenesis, leptogenesis,
and baryogenesis through the decay of a field condensate, or
Affleck-Dine baryogenesis. Here we give a brief description of
these various alternatives.

%%%%%%%%%%%%%%%%%%%%%%%%%%%%%%%%%%%%%%%%%%%%%%%%%%%%%%
\subsubsection{GUT-baryogenesis}

This was the first concrete attempt of model building on baryogenesis
which incorporates out-of-equilibrium decays of heavy GUT gauge bosons
$X,Y\rightarrow qq$, and $X,Y\rightarrow \bar q\bar l$ (see e.g.,
\cite{dolgov81,kolb83,dolgov92,olive90c}).
The decay rate of the gauge boson goes as $\Gamma_{X}\sim \alpha_{X}M_{X}$,
where $M_{X}$ is the mass of the gauge boson and $\alpha^{1/2}_{X}$ is the
GUT gauge coupling. Assuming that the Universe was in thermal equilibrium
at the GUT scale, the decay temperature is given by
\begin{equation}
T_{D}\approx g_{\ast}^{-1/4}\alpha_{X}^{1/2}(M_{X}M_{\rm P})^{1/2}\,,
\end{equation}
which is smaller than the gauge boson mass. Thus, at $T\approx T_{D}$,
one expects $n_{X}\approx n_{\bar X}\approx n_{\gamma}$, and hence the
net baryon density is proportional to the photon number density
$n_{B}=\Delta B n_{\gamma}$. However below $T_{D}$ the gauge boson abundances
decrease and eventually they go out-of-equilibrium. The net
entropy generated due to their decay heats up the Universe with a temperature
which we denote here by $T_{rh}$. Let us naively assume that the energy
density of the Universe at $T_{D}$ is dominated by the $X$ bosons with
$\rho_{X} \approx M_{X}n_{X}$, and their decay products lead to radiation with
an energy density $\rho =(\pi^2/30)g_{\ast}T_{rh}^4$, where
$g_{\ast}\sim {\cal O}(100)$ for $T\geq M_{GUT}$. Equating
the expressions for the two energy densities one obtains
\begin{equation}
n_{X}\approx \frac{\pi^2}{30}g_{\ast}\frac{T_{rh}^4}{M_{X}}\,.
\end{equation}
Therefore the net baryon number comes out to be
\begin{equation}
\label{gutb}
B\equiv \frac{n_{B}}{s}\approx \frac{\Delta B n_{X}}{g_{\ast}n_{\gamma}}
\approx \frac{3}{4}\frac{T_{rh}}{M_{X}}\Delta B\,.
\end{equation}
$T_{rh}$ is determined from the relation
$\Gamma^2_{X}\approx H^2(T_{D})\sim (\pi^2/90)g_{\ast}T^4_{rh}/M_{\rm P}^2$.
Thus,
\begin{equation}
B\approx \left(\frac{g_{\ast}^{-1/2}\Gamma_{X}M_{\rm P}}{M_{X}^2}\right)^{1/2}
\Delta B \equiv \left(\frac{g_{\ast}^{-1/2}\alpha_{X}M_{\rm P}}{M_{X}}
\right)^{1/2}\Delta B\,.
\end{equation}
Uncertainties in $C$ and $CP$ violation are now hidden in $\Delta B$,
but can be tuned to yield total $B\sim 10^{-10}$ in many models.

Above we have tacitly assumed that the Universe is in thermal equilibrium
when $T\geq M_{X}$. This might not be true, since for $2\leftrightarrow 2$
processes the scattering rate is given by $\Gamma\sim \alpha^2 T$, which
becomes smaller than $H$ at sufficiently high temperatures.
Elastic $2\rightarrow 2$ processes maintain thermal contact typically
only up to a maximum temperature $\sim 10^{14}$~GeV, while chemical
equilibrium is lost already at $T\sim 10^{12}$~GeV \cite{kari90,elmfors94}.
It has been argued that QCD gas, which becomes asymptotically free at
high temperatures, never reaches a chemical equilibrium above
$\sim 10^{14}$~GeV \cite{kari93}. In supergravity the maximum temperature
of the thermal bath should not exceed $10^{10}$~GeV~\cite{ellis84b}
(see Sect.~$3.6.3$).

%%%%%%%%%%%%%%%%%%%%%%%%%%%%%%%%%%%%%%%%%%%%%%%%%%%%%%%%

\subsubsection{Electroweak baryogenesis}

A popular baryogenesis candidate is based on the
electroweak phase transition, during which one can in principle  meet all
the Sakharov conditions. There is the sphaleron-induced baryon number
violation above the critical temperature, various sources of $CP$
violation, and an out-of-equilibrium environment if the phase transition
is of the first order. In that case bubbles of broken $SU(2)\times U(1)_Y$
are nucleated into a symmetric background with a Higgs field profile
that changes through the bubble wall~\cite{kuzmin85,kuzmin87,cohen93}.

There are two possible mechanisms which work in a different regime;
local and non-local baryogenesis. In the local case both $CP$ violation
and baryon number violation takes place near the bubble wall. This requires
the velocity of the bubble wall to be greater than the speed of the sound
in the plasma \cite{ambjorn89,turok90}, and the electroweak phase transition
to be strongly first order with thin bubble walls.

The second alternative, where the bubble wall velocity speed is small
compared to the sound speed in the plasma, appears to be more realistic.
In this mechanism the fermions, mainly the top quark and the tau-lepton,
undergo $CP$ violating interactions with the bubble wall, which results
in a difference in the reflection and the transmission probabilities for the
left and right chiral fermions. The net outcome is an overall chiral flux
into the unbroken phase from the broken phase. The flux is then converted
into baryons via sphaleron transitions inside the unbroken phase.
The interactions are taking place in a thermal equilibrium except for
the sphaleron transitions, the rate of which is slower than the rate
at which the bubble sweeps the space.

One great difficulty with the electroweak baryogenesis is the smallness
of $CP$ violation in the SM. It has been pointed out that
an additional Higgs doublet \cite{mclerran89,gunion89,turok90,cohen92}
would provide an extra source for $CP$ violation in the Higgs sector.
However, the situation is much improved in the MSSM where there are
two Higgs doublets $H_{u}$ and $H_{d}$, and two important sources of
$CP$ violation~\cite{ellis82}. The Higgses couple to the charginos
and neutralinos at one loop level leading to a $CP$ violating contribution.
There is also a new source of $CP$ violation in the mass matrix of the
top squarks which can give rise to considerable $CP$ violation \cite{huet96}.

Bubble nucleation depends on the thermal tunneling rate, and the expansion
rate of the Universe. The tunneling rate has to overcome the expansion rate
in order to have a successful phase transition via bubble
nucleation at a given critical temperature $T_{c}>T_{t}>T_0$.
The actual value of the baryon asymmetry produced at the electroweak
baryogenesis is still an open
debate~\cite{carena01,cline00,huber01,cline01,carena02}, but in general
it is hard to generate a large baryon asymmetry. For $T_{c}\sim 100$~GeV,
$N=3$, $\alpha_{2}=0.033$, and $B(\lambda/g_2)\sim 1.87$, one obtains the
condition~\cite{shaposhnikov87,bocharev87,bocharev91,rubshap96}
\begin{equation}
\frac{E_{sph}(T_c)}{T_c}\geq 7\log\left[\frac{E_{sph}(T_c)}{T_c}\right]+
9\log(10)+\log(\kappa)\,.
\end{equation}
which implies~\cite{bocharev91}
\begin{eqnarray}
\label{b1}
\frac{E_{sph}(T_{c})}{T_{c}}&\geq& 45\,~{\rm for}~\kappa =10^{-1}\,,\\
\label{b2}
&\geq& 37\,~{\rm for}~\kappa =10^{-4}\,.
\end{eqnarray}
The standard bound is often taken to be that of Eq.~(\ref{b1}). In terms of
the Higgs field value at $T_c$, one then obtains from Eq.~(\ref{sphe00})
\begin{equation}
\frac{\Phi(T_{c})}{T_{c}}=\frac{g_{2}}{4\pi B(\lambda/g_2)}
\frac{E_{sph}(T_c)}{T_c}\sim \frac{1}{36}\frac{E_{sph}(T_c)}{T_c}\,,
\end{equation}
for the above values of $\alpha_2,B$. Then the bounds in
Eqs.~(\ref{b1},\ref{b2}) translate to
\begin{eqnarray}
\label{b3}
\frac{\Phi(T_{c})}{T_{c}}\geq 1.3~~~(1),
\end{eqnarray}
where the number in parenthesis is for Eq.~(\ref{b2}). Eq.~(\ref{b3})
respectively, implies that the phase transition should be strongly first
order in order that sphalerons do not wash away all the produced baryon
asymmetry. This result is the main constraint on electroweak baryogenesis.

In order to save the SM electroweak baryogenesis some attempts such as   
matter induced effects have also been invoked; exciting the SM gauge 
degrees of freedom in a time varying Higgs background
\cite{Garcia-Bellido:1999sv,Krauss:1999ng}, or via dynamical scalar
field which couples to the SM fields~\cite{Joyce:2000ag}.

Lattice studies suggest that in the SM the phase transition
is strongly first order only below Higgs mass $m_{H}\sim 72$~GeV
\cite{kajantie96,gurtler97,rummukainen98,csikor98}. Above this scale
the transition is just a cross-over. Such a Higgs mass is clearly
excluded by the LEP measurements~\cite{pdg}, thus excluding electroweak
baryogenesis within the SM.

%%%%%%%%%%%%%%%%%%%%%%%%%%%%%%%%%%%%%%%%%%%%%%%%%%%%%

\subsubsection{Electroweak baryogenesis in MSSM}

In the MSSM the ratio $\Phi(T_{c})/T_{c}$ can increase considerably. The
MSSM Higgs sector at finite temperature has been considered in
\cite{giudice92,carena96,carena98,cline98},
for lattice studies see \cite{laine96,cline96,laine98535}.
In the MSSM the right handed stop $\bar t_{R}$
couples to the Higgs with a large Yukawa coupling. This leads to
a strong first order phase transition \cite{carena96,carena98,cline98}.
The LEP precision tests then indicate that the
lightest left handed stop should be heavy heavy with $m_{Q}\geq 500$~GeV.
This implies the for lightest right handed stop mass
\begin{equation}
m^2_{\tilde t}\approx m^2_{U}+0.15M^2_{Z}\cos(2\beta)+m^2_{t}\left(1-
\frac{{\tilde A}^2_{t}}{m^2_{Q}}\right)\,,
\end{equation}
where ${\tilde A}_{t}=A_{t}-\mu/\tan(\beta)$ is the stop mixing parameter,
and $\mu$ is the soft-SUSY breaking mass parameter for the right-handed stop.
The coefficient $\beta $ of the cubic term $\beta TH^3$ in the effective
potential reads
\begin{equation}
\beta_{MSSM} \approx \beta_{SM}+\frac{h_{t}^3\sin^3(\beta)}{4\sqrt{2}\pi}
\left(1-\frac{{\tilde A}^2_{t}}{m^2_{Q}}\right)^{3/2}\,,
\end{equation}
and can be at least one order of magnitude larger than
$\beta_{SM}$. In principle this modification can give rise to a
strong enough first order phase transition.

The sphaleron bound implies Higgs and stop masses in the
range \cite{quiros01,carena02}
\begin{equation}
110~{\rm GeV}\leq m_{H} \leq 115~{\rm GeV}\,, \quad {\rm and}\quad
105~{\rm GeV}\leq m_{\widetilde t_{R}}\leq 165~{\rm GeV}\,.
\end{equation}
The present LEP constraint on the Higgs mass is
$m_{H}\geq 115$~GeV~\cite{lephiggs}. Hence, even an MSSM-based
electroweak baryogenesis may be at the verge of being ruled out.
The  definitive test of the MSSM based electroweak baryogenesis
will obviously come from the Higgs and the stop searches at the LHC
and the Tevatron~\cite{cline00,cline0085,quiros01,carena01,carena02}.

%%%%%%%%%%%%%%%%%%%%%%%%%%%%%%%%%%%%%%%%%%%%%%%%%%%

\subsubsection{Leptogenesis}

Even if $B+L$ is completely erased by the sphaleron transitions, a net
baryon asymmetry in the Universe can still be generated from a non-vanishing
$B-L$ \cite{harvey81}, even if there were no baryon number violating
interactions. Lepton number violation alone can produce baryon asymmetry
$B\sim -L$ \cite{fukugita86}, a process which is known as leptogenesis
(for a recent review \cite{buchmuller00}, and references therein).
For lepton number violation one however has to go beyond the SM.

A popular example is $SO(10)$ GUT model, which can either be broken
into $SU(5)$ and then subsequently to the SM, or into the SM gauge
group directly. The most attractive aspect of $SO(10)$ is that it is
left-right symmetric (for details, see~\cite{langacker81,mohapatra91}),
and has a natural foundation for the see-saw mechanism
\cite{see-saw,mohapatra80} as it incorporates a singlet right-handed
neutrino $N_{R}$ with a mass $M_{R}$. A lepton number violation appears
when the Majorana right handed neutrino decays into the SM lepton doublet
and Higgs doublet, and their $CP$ conjugate state through
\begin{equation}
\label{ndecay}
N_{R}\rightarrow \Phi+ l\,, \quad \quad N_{R}\rightarrow \bar \Phi+\bar l\,,
\end{equation}
There also exist $\Delta L=0$, and $\Delta L=2$ processes mediated by
the right handed neutrino through
\begin{equation}
\label{nscatt}
\frac{(l\Phi)(l\Phi)}{M_{R}}\,,\quad \quad \frac{l l\Phi\Phi}{M_{R}}\,,
\end{equation}
which are dimension $5$ operators \cite{giudice99,zurab01}.
(There are other processes involving t-quarks which may also be
important \cite{luty92,plumacher97}). $CP$ asymmetry is generated
through the interference between tree level and one-loop diagrams.

The total baryon asymmetry and total lepton asymmetry can be found in terms
of the chemical potentials as~\cite{khlebnikov88}
\begin{equation}
\label{mu5}
B=\sum_{i}(2\mu_{qi}+\mu_{u_{R}i}+\mu_{d_{R}i})\,, \quad \quad
L=\sum_{i}(2\mu_{li}+\mu_{e_{R}i})\,,
\end{equation}
where $i$ denotes three leptonic generations. The Yukawa interactions
establish an equilibrium between the different generations
($\mu_{li}=\mu_{l}$ and $\mu_{qi}=\mu_{q}$, etc.), and one obtains
expressions for $B$ and $L$ in terms of the number of colors $N=3$, and the
number of charged Higgs fields $N_H$
\begin{equation}
B=-\frac{4N}{3}\mu_{l}\,,~~L=\frac{14N^2+9NN_{H}}{6N+3N_{H}}\mu_{l}\,,
\end{equation}
together with a relationship between $B$ and $B-L$ \cite{khlebnikov88}
\begin{equation}
B=\left(\frac{8N+4N_{H}}{22N+13N_{H}}\right)(B-L)\,.
\end{equation}
A similar expression has also been found in
\cite{aoki86,dolgov92,buchmuller00}, although there seems to be small
of order one differences. The baryon asymmetry based on the decays of
right handed neutrinos in a thermal bath has been computed in
\cite{luty92,langacker86,buchmuller96,pilaftsis99,buchmuller02}.
In a recent analysis \cite{buchmuller02}, it was pointed out that the
baryogenesis scale is tightly constrained together with with the
heavy right handed neutrino mass $T_{B}\sim M_{1,R}={\cal O}(10^{10})$~GeV,
with an upper bound on the light neutrino masses $\sum_{i}m_{i}<\sqrt{3}$~eV.
The current bound on the right handed neutrino mass is
around $M_{R}\sim {\cal O}(10^{11})$~GeV for light neutrino masses
$m_{1\nu}\approx m_{2\nu}\approx m_{3\nu}\sim {\cal O}(0.1)$~eV.

High scale leptogenesis is  ruled out in a supersymmetric
theory because of the gravitino problem (see Sect. $3.7.1$).
However, if the masses of the right handed neutrinos are such
that the mass splitting is comparable to their decay
widths,  it is possible to obtain an enhancement in the $CP$ phase
of order one~\cite{pilaftsis99}, and possibly a low scale thermal leptogenesis
\cite{hambye02}. Otherwise, one could resort to non-thermal leptogenesis
\cite{kumekawa94,giudice99,zurab01,allahverdim02},
or, to the scattering process discussed in \cite{bentozurab01}, or to
sneutrino driven leptogenesis~\cite{campbell93,zurab01}.

%%%%%%%%%%%%%%%%%%%%%%%%%%%%%%%%%%%%%%%%%%%%%%%%%%%%%%%%%%%
\subsubsection{Baryogenesis through field condensate decay}

Scalar condensates may have formed in the course of the
evolution of the early Universe. In particular, during inflation all
scalar fields are subject to fluctuations driven by the non-zero
inflaton energy density so that fields with very shallow potentials may easily
take non-zero values. An example is the MSSM, where for the squark
and slepton fields there are several directions in the field space
where the potential vanishes completely~\cite{dine96,gherghetta96}. These
directions are called (perturbatively) flat. Field fluctuations
along such flat directions will soon be smoothed out by inflation,
which effectively stretches out any gradients, and only the zero mode,
or the scalar condensate, remains. This mechanism is quite general
and applicable to any order parameter with flat enough potential.
Baryogenesis can then be achieved by the decay of a condensate that
carries baryonic charge, as was first pointed out by Affleck and Dine
\cite{affleckdine85}. As we will discuss, the flat direction condensate
can get dynamically charged with a large $B$ and/or $L$ by virtue
of $CP$-violating self-couplings.

Baryogenesis from MSSM flat directions has the virtue that it
only requires two already quite popular paradigms: supersymmetry
and inflation. In the old version \cite{affleckdine85} baryons were
produced by  a direct decay of the condensate, to be discussed in
Sect.~$2.5.2$. It was however pointed out first by Kusenko and Shaposhnikov
\cite{kusenko98418} in the case of gauge mediated supersymmetry
breaking, and then by Enqvist and McDonald in~\cite{enqvist98} in the
case of gravity mediated supersymmetry breaking, that the MSSM flat direction
condensate most often is not stable but fragments and eventually forms
non-topological solitons called $Q$-balls \cite{coleman85}. These issues
will be dealt in Sects.~$6$ and $7$.

%%%%%%%%%%%%%%%%%%%%%%%%%%%%%%%%%%%%%%%%%%%%%%%%%%%%%%%%%%%%%%%

\subsection{Old Affleck-Dine baryogenesis}

%%%%%%%%%%%%%%%%%%%%%%%%%%%%%%%%%%%%%%%%%%%%%%%%%%%%%%%%%%%%%%%%
\subsubsection{Classical motion of the order parameter}

In the original Affleck-Dine baryogenesis \cite{affleckdine85} it
was assumed that the order parameter along the flat direction is
displaced from the origin because of inflationary fluctuations. Because of
inflation, only the long wave-length model of the order parameter
will survive so that a spatially constant condensate field is formed along
the flat direction. This we shall sometimes call the Affleck-Dine (AD) field.
In an expanding Universe the coherent AD field $\phi$ obeys the
usual equation of motion,
\be{eqmotion}
\ddot\phi+3H\dot\phi +{\partial V\over \partial \phi}=0~,
\ee
where $H$ is the Hubble parameter.

To follow the time evolution of the AD field, let us consider a toy
model with the potential
\be{toy}
V(\phi)=m^2|\phi|^2+\lambda(\phi^4+\phi^{*4})+{|\phi|^6\over M^2}+
\dots\,.
\ee
Although this potential is unrealistic in that it does not take
correctly into account of supersymmetry breaking induced by the non-zero
cosmological constant of the inflationary era, it nevertheless
captures the main features of the initial cosmological evolution
of the AD field.

The theory \eq{toy} has a partially conserved current
$j_\mu=i\phi^*\partial_\mu\phi$, with
\be{toycons}
\partial_\mu j^\mu=\partial_\mu(i\phi^*\partial^\mu\phi-i
\partial^\mu\phi^*\phi)=i\lambda(\phi^{*4}-\phi^4)~.
\ee
The current is conserved for small $\phi$. The role of the higher
order term $|\phi|^6$ is just to stabilize the
potential. In the toy model \eq{toy}, we identify the baryon
number density $n_B$ with $j_0$. The model  also has a $CP$
invariance under which $\phi \leftrightarrow \phi^*$ but which
is violated by the initial conditions, which are taken to be
\be{toyinit}
\phi=i\phi_0,~\dot\phi=0~,
\ee
where $\phi_0$ is real. Writing $\phi=\phi_R+i\phi_I$ one finds the
coupled equations of motion (see e.g., \cite{kinwang88})
\bea{toyeqs}
\ddot\phi_I+3H\dot\phi_I+\left[m^2+12\lambda\phi_R\phi_I+{3|\phi|^4\over M^2}
\right]\phi_I&=&
4\lambda\phi^3_R \nn
\ddot\phi_R+3H\dot\phi_R+\left[m^2+{3|\phi|^4\over M^2}\right]\phi_R&=&
4\lambda(3\phi_I\phi_R^2-\phi^3_I)~.
\eea
In a matter dominated Universe $H=2/(3t)$, so that for large times
$t\gg m^{-1}$ the motion is damped and  \eq{toyeqs} has then
oscillatory solutions of the form
\be{toylarget}
\phi_k={A_k\over mt}\sin(mt+\delta_k)~,~~~ k=I,R~,
\ee
where the amplitudes $A_k$ and the phases $\delta_k$ depend on the
parameters $m$, $\lambda$, $M$, and the initial conditions \eq{toyinit}.
For large times the baryon number is then found to be
\be{toyB}
n_B=2(\phi_I\dot\phi_R-\phi_R\dot\phi_I)={2A_IA_R\over mt^2}
\sin(\delta_i-\delta_R)~.
\ee

If $\phi_0^2\ll mM$, as was tacitly assumed by Affleck and Dine
\cite{affleckdine85}, one may disregard the higher-order terms. In
that case one obtains \cite{dine95} $A_I=\phi_0$ and
$A_R=a_R\lambda\phi^3_0/m^2$, where $a_R=0.85$ is
determined numerically. Likewise, numerically one finds that
$\delta_I-\delta_R=1.54$. Thus,
\be{toyADnB}
n_B={1.7\lambda\phi^4_0\over m^3t^2}\,,
\ee
and the generated baryon number per particle is
\be{toyADB}
R={mn_B\over \rho_\phi}={1.7\lambda\phi_0^2\over m^2}~.
\ee
\eq{toyADB} is true for matter dominated Universe; for radiation dominated
Universe one obtains a similar result, but the numerical prefactor
$1.7$ should be replaced by $-1.3$.

If $\phi_0^2> mM$, one finds $A_I=a_I(mM)^{1/2}$ and
$A_R=a_R(M^3/m)^{1/2}$ with $a_I=0.94,~a_R=-2.86,~\delta_I=011$,
and $\delta_R=-0.41$. It then follows from \eq{toyB}, that
\be{toyng}
n_B=-{2.7\lambda M^2\over mt^2}~.
\ee
Thus, the baryon generation mechanism is remarkably robust. The initial
conditions do not matter, nor the actual expansion rate of the Universe.
The baryon number generated per $\phi$-particle is always large and,
with $\lambda \sim m^2/\langle\phi\rangle^2$, typically $n_{B}\gg 1$.
Although these conclusions were derived in a toy model, similar
results hold true also for the MSSM flat directions.

Thus, to summarize, along a flat direction where squarks and sleptons
have non-zero expectation values, evolution of the AD field condensate,
starting from a $CP$ violating initial value, will dynamically generate large
baryon number density and charge the condensate with $B$ and/or $L$.

%%%%%%%%%%%%%%%%%%%%%%%%%%%%%%%%%%%%%%%%%%%%%%%%%%%%%%%%%%%%%%%%%%%
\subsubsection{Condensate decay}

To provide the Universe with the observed baryon to entropy
ratio, $n_B/s\sim 10^{-10}$, the AD condensate must eventually
transform itself into ordinary quarks. Originally \cite{affleckdine85},
it was thought that this could happen via the decay of the AD field
components (squarks and sleptons) to ordinary quarks and leptons. The AD
condensate can be thought of as a coherent state of $\phi$-particles
where $\phi=\phi_0e^{imt}$ and  $|\phi_0|\gg m$. When
supersymmetry breaking is switched on, the AD field starts
to oscillate about the old vacuum $\langle \phi \rangle  \gg m$.
Writing $\phi=\langle \phi \rangle +\phi'$, one observes
that all the fields to which the excitations $\phi'$ couple
are heavy with masses ${\cal O}(\langle \phi \rangle)$. The
field $\phi'$ itself has a mass of the order of supersymmetry
breaking, ${\cal O}(m)$. Therefore, $\phi'$ can decay to light
fields only through loop diagrams involving heavy fields, with
an effective coupling of the type
$(g^2/\langle\phi\rangle)\phi'\psi\partial\psi^\dagger$,
where $\psi$ is a light fermion and $g$ some coupling constant.
The decay rate is thus,~\cite{affleckdine85}
\be{addecay}
\Gamma\sim g^4 {m^3\over \phi^2~}\,.
\ee

Because of the oscillations of the AD field, the Universe will
eventually become dominated by the energy density in the oscillations,
$\rho_\phi\simeq m^2\phi^2$, so that  $H\sim \rho_\phi^{1/2}/M_P$.
The AD field will decay when $\Gamma\simeq H$, or $\phi\simeq(m^2M_P)^{1/3}$.
This implies a reheating temperature
$T_{rh}\simeq s^{1/3}\simeq \rho_\phi^{1/4}$ while the baryon number
density is $n_B=R\rho_\phi/m$, where $R$ is given in \eq{toyADB}.
Therefore one finally obtains
\be{nBstoy}
{n_B\over s}\simeq{\lambda\phi_0^2\over m^2}\left( \frac Mm\right)^{1/6}\,.
\ee
Depending on $\lambda$, and the size of the initial fluctuation
$\phi_0$ of the AD condensate, $n_B/s$ can be either small or large.
Therefore determining the initial value is of utmost importance
\cite{ellis87184}. This requires us to consider theories of inflation
in more detail, which will be done in the next Section. Following that,
we shall discuss the disappearance of the AD condensate by fragmentation
into (quasi)stable lumps of condensate matter, whose state of lowest
energy is a spherical non-topological soliton, a $Q$-ball \cite{coleman85}.

%%%%%%%%%%%%%%%%%%%%%%%%%%%%%%%%%%%%%%%%%%%%%%%%%%%%%%%%%%

\newpage

\newpage

%%%%%%%%%%%%%%%%%%%%%%%%%%%%%%%%%%%%%%%%%%%%%%%%%%%%%%%%%%%
%%%%%%%%%%%%%%%%%%%%%%%%%%%%%%%%%%%%%%%%%%%%%%%%%%%%%%%%%%%
%%%%%%%%%%%%%%%%%%%%%%%%%%%%%%%%%%%%%%%%%%%%%%%%%%%%%%%%%%%

\section{Field fluctuations during inflation}

Apart from explaining the initial condition for the hot Big Bang model,
the flatness problem, and the horizon problem, cosmological inflation
\cite{starobinsky80,guth81,linde82108} is one of the most favored candidate
for the origin of structure in the Universe (for reviews on inflation,
see~\cite{linde90,lyth98}).
There are many models of inflation, but by far the  simplest is
one in which inflation is generated by the large energy density
of a scalar field. The scalar field driven inflation not only explains
the homogeneity and the flatness problems but also
the observed scale invariance of the density perturbations.

Inflation based on a scalar field theory is described by the
following Lagrangian:
\begin{equation}
{\cal L}=\frac{M_{\rm P}^2}{2}R+\frac{1}{2}\partial_{\mu}\phi
\partial^{\mu}\phi-V(\phi)\,,
\end{equation}
where $R$ is the curvature scalar. The energy-momentum tensor reads
\begin{equation}
T_{\mu\nu}=\partial_{\mu}\phi\partial_{\nu}\phi-\frac{1}{2}g_{\mu\nu}
\partial_{\rho}\phi\partial^{\rho}\phi-g_{\mu\nu}V(\phi)\,
\end{equation}
so that the energy density and the pressure are given by
\begin{eqnarray}
\rho \equiv T_{00} &=&\frac{1}{2}\dot\phi^2+\frac{1}{2a^2(t)}(\nabla\phi)^2
+V(\phi)\,,\\
p \equiv \frac{T_{ii}}{a^2(t)}&=& \frac{1}{2}\dot\phi^2-\frac{1}{6a^2(t)}
(\nabla\phi)^2-V(\phi)\,.
\end{eqnarray}
One of the initial conditions for inflation is that there
must be a homogeneous patch of the Universe which is bigger than the
size of the Hubble horizon \cite{vachaspati98}  (also supported
by numerical studies, see \cite{albrecht85}). However
such a stringent condition can be evaded in a chaotic inflation beginning at
the Planck scale \cite{linde83,linde86,linde90,linde9449}.
More complicated situation can be obtained if there are several
fields that participate in inflation; the classic example is
assisted inflation \cite{liddle98,malik99}.

%%%%%%%%%%%%%%%%%%%%%%%%%%%%%%%%%%%%%%%%%%%%%%%%%%%%%%%%%%%%%%%%%%
\subsection{Fluctuation spectrum in de Sitter space}

The plane wave solution of  a massive scalar field $\phi(\vec x,t)$
in a spatially flat Robertson-Walker metric can be decomposed into
Fourier modes by
\begin{equation}
\phi=\frac{1}{(2\pi)^{3/2}}\int d^3k \left(\phi_{k}(t)e^{ik\cdot\vec x}+
{\rm h.c.}\right)\,.
\end{equation}
Solving the Klein-Gordon equation for the scalar field in a conformal metric:
$ds^2=g_{\mu\nu}dx^{\mu}dx^{\nu}=a^2(\tau,x)(d\tau^2-dx^2)$,
the mode function can be given by
\cite{bunch78,vilenkin82,vilenkinford82,linde82}
\begin{eqnarray}
\phi_{k}(\tau)&=&\left(\frac{\pi}{4}\right)^{1/2}H|\tau|^{3/2}\left(
c_{1}H^{(1)}_{\nu}(k\tau)+c_{2}H^{(2)}_{\nu}(k\tau)\right)\,, \nonumber \\
\tau &=&-H^{-1}e^{-Ht}\,,~~{\rm and}~~\nu^2=\frac{9}{4}-\frac{m^2}{H^2}\,,
\end{eqnarray}
where $m$ is the mass of the scalar field, $H^{(1)}_{\nu}$ and $H^{(2)}_{\nu}$
are the Hankel functions and $c_1,c_2$ are constants. The readers might be
tempted to take the limit $\tau \ll 0$, in order to match the above
solution with the plane wave solution in a Minkowski background. However
this leads to a quasi static de Sitter solution \cite{enqvistng88}.
More technically, it has been shown that  using a point splitting
regularization scheme, it is possible to obtain a Bunch-Davies vacuum
for a de Sitter background which actually corresponds to taking $c_1=0$,
and $c_2=1$. A simple but intuitive way
has been developed in \cite{vilenkin83}, where it has been argued
that during a de Sitter phase, the main contribution to the two point
correlation function comes from the long wavelength modes; $k|\tau| \ll 1$
or $k\ll H\exp(Ht)$. Therefore the two point function is defined by
an infrared cutoff which is determined by the Hubble expansion
\cite{vilenkin83}
\begin{equation}
\label{variance}
\langle \phi^2\rangle\approx \frac{1}{(2\pi)^{3}}\int_{H}^{He^{Ht}}
d^3k|\phi_{k}|^2\,.
\end{equation}
The result of the integration yields
\cite{linde82,vilenkinford82,enqvistng88,vilenkin83} an
indefinite increase in the variance with time
\begin{equation}
\label{variance1}
\langle \phi^2\rangle \approx \frac{H^3}{4\pi^2}t\,.
\end{equation}
This result can also be obtained by considering the Brownian motion
of the scalar field \cite{linde94}.

For a massive field with $m\ll H$, and $\nu\neq 3/2$, one does not obtain an
indefinite growth of the variance of the long wavelength fluctuations, but
\cite{linde82,vilenkinford82,enqvistng88,vilenkin83}
\begin{equation}
\langle \phi^2\rangle=\frac{3H^4}{8\pi^2m^2}\left(1-e^{-(2m^2/3H^2)t}
\right)\,.
\end{equation}
In the limiting case when $m\rightarrow H$, the variance goes as
$\langle \phi^2\rangle \approx H^2$. In the limit $m\gg H$,
the variance goes as $\langle \phi^2\rangle \approx (H^3/12\pi^2m)$
\cite{enqvistng88}. Only in a massless case $\langle \phi^2\rangle$ can
be treated as a homogeneous background field with a long wavelength mode.
This result plays an important role for the rest of this review as it
implies that in a de Sitter phase any scalar field, including the AD
condensate, are subject to quantum fluctuations.

%%%%%%%%%%%%%%%%%%%%%%%%%%%%%%%%%%%%%%%%%%%%%%%%%%%%%%%%%%%
\subsection{Slow roll inflation}

A completely flat potential can render  inflation eternal,
provided the energy density stored in the flat direction dominates.
The inflaton direction is however not completely flat but has
a potential $V(\phi)$ with some slope. An inflationary phase is
obtained while
\begin{eqnarray}
\label{slowr1}
H^2 &\approx & \frac{1}{3M_{\rm P}^2}V(\phi)\,, \\
\label{slowr2}
3H\dot\phi &\approx &-V^{\prime}(\phi)\,,
\end{eqnarray}
where prime denotes derivative with respect to $\phi$. In the above
the approximations are: $\dot\phi^2 < V(\phi)$, and
$\ddot \phi < V^{\prime}(\phi)$, which lead to the slow
roll conditions (see e.g. \cite{liddle-lyth00})
\begin{eqnarray}
\label{ep1}
\epsilon(\phi)&=&\frac{M_{\rm P}^2}{2}\left(\frac{V^{\prime}}{V}\right)^2
\ll 1\,,\\
\label{eta1}
|\eta(\phi)|&=&{M_{\rm P}^2}\frac{V^{\prime\prime}}{V} \ll 1\,.
\end{eqnarray}
Note that $\epsilon$ is positive by definition.

These conditions are necessary but not sufficient for inflation.
They only constrain the shape of the potential but not the velocity of
the field $\dot\phi$. Therefore a tacit assumption behind the success
of the slow roll conditions is that the inflaton field should not have
a large initial velocity.

Inflation comes to an end when the slow roll conditions are violated,
$\epsilon \sim 1$, and $\eta \sim 1$. However, there are certain models
where this need not be true, for instance in hybrid inflation models
\cite{linde9449}, where inflation comes to an end via a phase transition,
or in oscillatory models of inflation where slow roll conditions are
satisfied only on average \cite{damour98}.

One of the salient features of the slow roll inflation is that there exists a
late time attractor behavior. This means that during inflation the
evolution of a scalar field at a given field value has to be independent
of the initial conditions.  Therefore slow roll inflation should provide
an attractor behavior which at late times leads to an identical field
evolution in the phase space irrespective of the initial conditions
\cite{salopek90}. In fact the slow roll solution does not give
an exact attractor solution to the full equation of motion but is nevertheless
a fairly good approximation \cite{salopek90}. A similar statement
has been proven for multi-field exponential potentials without slow roll
conditions (i.e., assisted inflation) \cite{liddle98}. The attractor
behavior of the inflaton leads to powerful predictions which can be
distinguished from other candidates of galaxy formation~\cite{liddle93}.

The standard definition of the number of e-foldings is given by
\begin{equation}
N\equiv \ln\frac{a(t_{end})}{a(t)}=\int_{t}^{t_{end}}Hdt\approx
\frac{1}{M_{\rm P}^2}\int_{\phi_{end}}^{\phi}\frac{V}{V^{\prime}}d\phi\,,
\end{equation}
where $\phi_{end}$ is defined by $\epsilon(\phi_{end})\sim 1$, provided
inflation comes to an end via a violation of the slow roll conditions.
The number of e-foldings can be related to the Hubble crossing
mode $k=a_{k}H_{k}$ by comparing with the present Hubble length $a_{0}H_{0}$.
The final result is \cite{liddle-lyth00}
\begin{equation}
\label{efoldsk}
N(k)=62~-~\ln\frac{k}{a_0H_0}~-~\ln\frac{10^{16}~{\rm GeV}}{V_{k}^{1/4}}~+~
\ln\frac{V_{k}^{1/4}}{V_{end}^{1/4}}~-~\frac{1}{3}\ln\frac{V_{end}^{1/4}}
{\rho_{rh}^{1/4}}\,
\end{equation}
where the subscripts $end$ ($rh$) refer to the end of inflation (onset of
reheating). The details of  the thermal history of the Universe determine the
precise number of e-foldings, but for most practical purposes it is
sufficient to assume that $N(k)\approx 50$, keeping all the uncertainties
such as the scale of inflation and the end of inflation within a margin of
$10$~e-foldings. A significant modification can take place only if there
is an epoch of late inflation such as thermal inflation
\cite{lythstewart}, or in theories with a low quantum gravity scale
\cite{anupam99}.

%%%%%%%%%%%%%%%%%%%%%%%%%%%%%%%%%%%%%%%%%%%%%%%%%%%%%%%%%%%%%
\subsection{Primordial density perturbations}

Initially, the theory of cosmological perturbations has been developed
in the context of FRW cosmology \cite{lifshitz46},
and for models of inflation in
\cite{mukhanov81,guthpi82,hawking82,starobinsky82,bardeen83}.
For a complete review on this topic, see \cite{mukhanov92}.
For a real single scalar field there arise only adiabatic density
perturbations. In case of several fluctuating fields there will in
general also be isocurvature perturbations. We briefly describe the two
perturbations and their observational differences.

%%%%%%%%%%%%%%%%%%%%%%%%%%%%%%%%%%%%%%%%%%%%%%%%%%%%%%%%%%%

\subsubsection{Adiabatic perturbations and the Sachs-Wolfe effect}

Let us consider small inhomogeneities
$\phi(\vec x,t) =\phi(t)+\delta\phi(\vec x,t)$ such that $\delta\phi\ll\phi$.
Perturbations in matter densities automatically induce perturbations in
the background metric, but the separation between the background metric
and a perturbed one
is not unique. One needs to choose a gauge. A simple choice would be to
fix the observer to the unperturbed matter particles, where the observer
will detect a velocity of matter field falling under gravity; this is
known as the Newtonian or the longitudinal gauge because the observer
in the Newtonian gravity limit measures the gravitational potential
well where matter is falling in and clumping. The induced metric
can be written as
\begin{equation}
\label{gauge}
ds^2=a^2(\tau)\left[(1+2\Phi)d\tau^2-(1-2\Psi)\delta_{ik}dx^{i}dx^{k}\right]\,,
\end{equation}
where $\Phi$ has a complete analogue of Newtonian gravitational potential.
In the case when the spatial part of the energy momentum tensor is diagonal,
i.e. $\delta T^{i}_{j}=\delta^{i}_{j}$, it follows that $\Phi=\Psi$
\cite{mukhanov92}. Right at the time of horizon crossing one finds a
solution for $\delta\phi$ as
\begin{equation}
\label{pertphi}
\langle |\delta\phi_{k}|^2\rangle=\frac{H(t_{\ast})^2}{2k^3}\,,
\end{equation}
where $t_\ast$ denotes the instance of horizon crossing. Correspondingly,
we can also define a power spectrum
\begin{equation}
\label{spect}
{\cal P}_{\phi}(k)=\frac{k^3}{2\pi^2}\langle |\delta\phi_{k}|^2\rangle=
\left[\frac{H(t_{\ast})}{2\pi}\right]^2 \equiv
\left. \left[\frac{H}{2\pi}\right]^2\right|_{k=aH}\,.
\end{equation}
Note that the phase of $\delta\phi_{k}$ can be arbitrary, and therefore,
inflation has generated a Gaussian perturbation.

In the limit $k\rightarrow 0$, one can find an exact solution for the long
wavelength inhomogeneities $k\ll aH$ \cite{starobinsky85,mukhanov92}, which
reads
\begin{eqnarray}
\label{pertPhi}
\Phi_{k}&\approx&c_{1}\left(\frac{1}{a}\int_{0}^{t}a~dt^{\prime}\right)^{\cdot}
+ c_{2}\frac{H}{a}\,, \\
\label{pertphi1}
\frac{\delta\phi_{k}}{\dot\phi}&=&\frac{1}{a}\left(c_{1}\int_{0}^{t}
a~dt^{\prime}-c_2\right)\,,
\end{eqnarray}
where  the dot denotes derivative with respect to physical time. The growing
solutions are proportional to $c_1$, the decaying  proportional
to $c_2$. Concentrating upon the growing solution, it is possible to obtain
a leading order term in an expansion with the help of the slow roll conditions:
\begin{eqnarray}
\label{pertPhi1}
\Phi_{k}&\approx &-c_{1}\frac{\dot H}{H^2}\,, \\
\label{pertphi2}
\frac{\delta\phi_{k}}{\dot\phi}&\approx &\frac{c_{1}}{H}\,.
\end{eqnarray}
Note that at the end of inflation, which is indicated by $\ddot a =0$, or
equivalently by $\dot H =-H^2$, one obtains a constant Newtonian
potential $\Phi_{k}\approx c_{1}$. This is perhaps the most
significant result for a single field perturbation.

In a long wavelength limit one obtains a constant of motion $\zeta$
\cite{bardeen83,brandenberger85,mukhanov92} defined as
\begin{equation}
\label{zeta1}
\zeta=\frac{2}{3}\frac{H^{-1}\dot\Phi_{k}+\Phi_{k}}{1+w}+\Phi_{k}\,,~~
w=\frac{p}{\rho}\,.
\end{equation}
If the equation of state for matter remains constant there is
a simple relationship which connects the metric perturbations
at two different times \cite{bardeen83,brandenberger85,mukhanov92}
\begin{equation}
\Phi_{k}(t_f)= \frac{1+\frac{2}{3}\left(1+w(t_f)\right)^{-1}}{1+\frac{2}{3}
\left(1+w(t_i)\right)^{-1}}\Phi_{k}(t_i)\,.
\end{equation}

The comoving curvature perturbation \cite{lukash80} reads in the
longitudinal gauge \cite{mukhanov92} for the slow roll inflation as
\begin{equation}
\label{curvpert}
{\cal R}_{k}=\Phi_{k}-\frac{H^2}{\dot H}\left(H^{-1}\dot\Phi_{k}+\Phi_{k}
\right)\,.
\end{equation}
For CMB and structure formation we need to know the metric perturbation
during the matter dominated era when the metric perturbation is
$\Phi(t_f)\approx (3/5)c_{1}$. Substituting the value of $c_{1}$
from Eq.~(\ref{pertphi2}), we obtain
\begin{equation}
\label{curv1}
\Phi_{k}(t_f) \approx \left.\frac{3}{5}H\frac{\delta \phi_{k}}{\dot\phi}
\right|_{k=aH}\,.
\end{equation}
In a similar way it is also possible to show that the comoving curvature
perturbations is given by
\begin{equation}
{\cal R}_{k}\approx \left.\frac{H}{\dot \phi}\delta \phi \right|_{k=aH}\,,
\end{equation}
where $\delta_{\phi}$ denotes the field perturbation on a spatially flat
hypersurfaces, because on a comoving hypersurface $\delta\phi=0$,
by definition. Therefore, on flat hypersurfaces
\begin{equation}
\delta\phi_{k}=\dot\phi\delta t\,,
\end{equation}
where $\delta t$ is the time displacement going from flat to comoving
hypersurfaces \cite{liddle93,liddle-lyth00}. As a result
\begin{equation}
\label{curvpert1}
{\cal R}_{k} \equiv H \delta t\,.
\end{equation}
Note that during matter dominated era the curvature perturbation and the
metric perturbations are related to each other
\begin{equation}
\label{curmet}
\Phi_{k}=-\frac{3}{5}{\cal R}_{k}\,.
\end{equation}

In the matter dominated era the photon sees this potential well
created by the primordial fluctuation and the redshift in the
emitted photon is given by
\begin{equation}
\frac{\Delta T_{k}}{T}=-\Phi_{k}\,.
\end{equation}
At the same time, the proper time scale inside the fluctuation becomes slower
by an amount $\delta t/t =\Phi_{k}$. Therefore, for the scale factor
$a\propto t^{2/3}$, decoupling  occurs earlier with
\begin{equation}
\frac{\delta a}{a}=\frac{2}{3}\frac{\delta t}{t}=\frac{2}{3}\Phi_{k}\,.
\end{equation}
By virtue of $T\propto a^{-1}$ this results in a temperature which is hotter by
\begin{equation}
\label{sachs}
\frac{\Delta T_{k}}{T} =-\Phi_{k} +\frac{2}{3}\Phi_{k} =-\frac{\Phi_{k}}{3}\,.
\end{equation}
This is the celebrated Sachs-Wolfe effect \cite{sachs67}, which we shall
revisit when discussing isocurvature fluctuations.

%%%%%%%%%%%%%%%%%%%%%%%%%%%%%%%%%%%%%%%%%%%%%%%%%%%%%%%%%%%%%%%%%%%

\subsubsection{Spectrum of adiabatic perturbations}

Now, one can immediately calculate the spectrum of the metric perturbations.
For a critical density Universe
\begin{equation}
\delta_{k}\equiv \left.\frac{\delta\rho}{\rho}\right|_{k}=-\frac{2}{3}
\left(\frac{k}{aH}\right)^2\Phi_{k}\,,
\end{equation}
where $\nabla^2\rightarrow -k^2$, in the Fourier domain. Therefore,
with the help of Eqs.~(\ref{spect},\ref{curv1}), one obtains
\begin{equation}
\label{pspect}
\delta_{k}^2\equiv \frac{4}{9}{\cal P}_{\Phi}(k)\, =\frac{4}{9}\frac{9}{25}
\left(\frac{H}{\dot\phi}\right)^2\left(\frac{H}{2\pi}\right)^2\,,
\end{equation}
where the right hand side can be evaluated at the time of horizon exit
$k=aH$. In fact the above expression can also be expressed in terms of
curvature perturbations \cite{liddle93,liddle-lyth00}
\begin{equation}
\delta_{k}=\frac{2}{5}\left(\frac{k}{aH}\right)^2{\cal R}_{k}\,,
\end{equation}
and following Eq.~(\ref{curvpert}), we obtain
$\delta_{k}^2={4}/{25}{\cal P}_{\cal R}(k)=({4}/{25})(H/\dot\phi)^2(H/2\pi)^2$,
exactly the same expression as in Eq.~(\ref{pspect}).
With the help of the slow roll equation $3H\dot\phi=-V^{\prime}$,
and the critical density formula $3H^2M_{\rm P}=V$, one obtains
\begin{equation}
\label{pspect1}
\delta_{k}^2\approx \frac{1}{75\pi^2 M_{\rm P}^6}\frac{V^3}{V^{\prime 2}}\,
=\frac{1}{150\pi^2 M_{\rm P}^4}\frac{V}{\epsilon}\,,
\end{equation}
where we have used the slow roll parameter
$\epsilon\equiv (M_{\rm P}^2/2)(V^{\prime}/V)^2$.
The COBE satellite measured the CMB anisotropy and fixes the normalization
of $\delta_{\Phi}(k)$ on a very large scale. For a critical density Universe,
if we assume that the primordial spectrum can be approximated by a power law
and ignoring  gravitational waves:
\begin{equation}
\delta_{\Phi}(k)=1.91\times 10^{-5}\left(\frac{k}{k_{pivot}}\right)^{(n-1)/2}
\,,
\end{equation}
where $n$ is the spectral index and $k_{pivot}=7.5a_{0}H_{0}$ is the scale
at which the normalization is independent of the spectral index.

The spectral index $n(k)$ is defined as
\begin{equation}
\label{spectind}
n(k)-1\equiv \frac{d\ln{\cal P}_{\Phi}}{d\ln k}\,.
\end{equation}
This definition is equivalent to the power law behavior if $n(k)$ is fairly
a constant quantity over a range of $k$ of interest. The power spectrum
can  then be written as
\begin{equation}
\label{spectind1}
{\cal P}_{\Phi}(k)\propto k^{n-1}\,.
\end{equation}
If $n=1$, the spectrum is flat and known as Harrison-Zeldovich spectrum
\cite{harrison70}.
For $n\neq 1$, the spectrum is tilted and $n>1$ is known as blue spectrum.
In terms of the slow roll parameters, one can write
\cite{salopek90}
\begin{equation}
\label{spectind3}
\frac{d\epsilon}{d\ln k}=2\epsilon\eta-4\epsilon^2\,,~~\frac{d\eta}{d\ln k}
=-2\epsilon\eta +\xi^2\,~~\frac{d\xi^2}{d\ln k}=-2\epsilon\xi^2+\eta\xi^2
+\sigma^3\,,
\end{equation}
where
\begin{equation}
\xi^2\equiv M_{\rm P}^4\frac{V^{\prime}(d^3V/d\phi^3)}{V^2}\,,
~~~~\sigma^3\equiv M_{\rm P}^6\frac{V^{\prime 2}(d^4V/d\phi^4)}{V^3}\,.
\end{equation}
Thus one finds\cite{liddle92291}
\begin{equation}
\label{spectind4}
n-1=-6\epsilon +2\eta\,.
\end{equation}

Slow roll inflation requires that
$\epsilon \ll 1, |\eta|\ll 1$, and therefore naturally predicts small
variation in the spectral index within $\Delta \ln k\approx 1$. The recent
Boomerang data suggest \cite{BOOMERANG}
\begin{equation}
|n-1| \leq 0.1\,.
\end{equation}
The rate of change in $\eta$ is also very small, and can be estimated in a
similar way \cite{kosowsky95}
\begin{equation}
\frac{dn}{d\ln k}=-16\epsilon\eta+24\epsilon^2+2\xi^2\,.
\end{equation}
It is possible to extend the calculation of metric perturbation beyond the
slow roll approximation basing on a formalism similar to that developed in
\cite{mukhanov85,sasaki86,mukhanov89,lidsey97}.

%%%%%%%%%%%%%%%%%%%%%%%%%%%%%%%%%%%%%%%%%%%%%%%%%%%%%%%%%%%%%
\subsubsection{Gravitational waves}

Gravitational waves are linearized tensor
perturbations of the metric and do not couple to the energy
momentum tensor. Therefore, they do not give rise a gravitational
instability, but carry the underlying geometric structure of the space-time.
The first calculation of the gravitational wave production was made in
\cite{grishchuk74}, and the topic has been considered by many authors
\cite{grishchuk89}. For reviews on gravitational waves, see
\cite{mukhanov92,majjore00}.

The gravitational wave perturbations are described by a line element
$ds^2+\delta ds^2$, where
\begin{equation}
ds^2=a^2(\tau)(d\tau^2-dx^{i}dx_{i})\,, ~~~~\delta ds^2=-a^2(\tau)h_{ij}
dx^{i}dx^{j}\,.
\end{equation}
The gauge invariant and conformally invariant $3$-tensor $h_{ij}$ is symmetric,
traceless $\delta^{ij}h_{ij}=0$, and divergenceless $\nabla_{i}h_{ij}=0$
($\nabla_{i}$ is a covariant derivative). Massless spin
$2$ gravitons have two degrees of freedom and as a result are also
transverse. This means that in a Fourier domain the gravitational wave
has a form
\begin{equation}
h_{ij}=h_{+}e^{+}_{ij}+h_{\times}e^{\times}_{ij}\,.
\end{equation}
For the Einstein gravity, the gravitational wave equation of motion follows
that of a massless Klein Gordon equation \cite{mukhanov92}. Especially,
for a flat Universe
\begin{equation}
\ddot h^{i}_{j}+3H\dot h^{i}_{j}+\left(\frac{k^2}{a^2}\right)h^{i}_{j}=0\,,
\end{equation}
As  any massless field, the gravitational waves also feel the quantum
fluctuations in an expanding background. The spectrum mimics that of
Eq.~(\ref{spect})
\begin{equation}
{\cal P}_{grav}(k)=\left.\frac{2}{M_{\rm P}^2}\left(\frac{H}{2\pi}\right)^2
\right|_{k=aH}\,.
\end{equation}
Note that the spectrum has a Planck mass suppression, which suggests that the
amplitude of the gravitational waves is smaller compared to that of the
adiabatic perturbations. Therefore it is usually assumed that their
contribution to the CMB anisotropy is small. The corresponding
spectral index can be calculated as
\cite{liddle92291}
\begin{equation}
 n_{grav}=\frac{d\ln{\cal P}_{grav}(k)}{d\ln k}~=-2\epsilon\,.
\end{equation}
Note that the spectral index is negative.

%%%%%%%%%%%%%%%%%%%%%%%%%%%%%%%%%%%%%%%%%%%%%%%%%%%%%%
\subsection{Multi-field perturbations}

In multi-field inflation models contributions to the density
perturbations come from all the fields. However unlike in a single scalar
case, in the multi-field case there might not be a unique late time
trajectory corresponding to all the fields. This is true in particular
for those fields that are effectively massless during inflation, such as the
MSSM flat direction fields. Therefore, in these cases scalar perturbations
will depend on the field trajectories and thus on the choice of initial
conditions, with an ensuing loss of predictivity. In a very few cases
it is possible to obtain a late time attractor behavior of all the
fields; an example is assisted inflation
\cite{liddle98}. Let us here nevertheless assume that there
is an underlying unique late time trajectory resulting in a
simple expression for the amplitude of the density
perturbations and the spectral index \cite{sasaki96,lyth98}.

%%%%%%%%%%%%%%%%%%%%%%%%%%%%%%%%%%%%%%%%%%%%%%%%%%%%%%%%%%%%%%%
\subsubsection{Adiabatic and isocurvature conditions}

There are only two kinds of perturbations that can be generated.
The first one is the adiabatic perturbation discussed previously;
it is a perturbation along the late time classical trajectories
of the scalar fields during inflation.
When the primordial perturbations enter our horizon they
perturb the matter density with a generic {\it adiabatic condition}, which is
satisfied when the density contrast of the individual species is related
to the total density contrast $\delta_{k}$
\begin{equation}
\label{adia}
\frac{1}{3}\delta_{k b}=\frac{1}{3}\delta_{k c}=\frac{1}{4}\delta_{k\nu}=
\frac{1}{4}\delta_{k\gamma}=\frac{1}{4}\delta_{k}\,,
\end{equation}
where $b$ stands for baryons, $c$  for cold dark matter, $\gamma$
for photons and $\nu$  for neutrinos.

The other type is the isocurvature perturbation. During
inflation this can be viewed as a perturbation orthogonal to the unique
late time classical trajectory. Therefore, if there were $N$ fluctuating
scalar fields during inflation, there would be $N-1$ degrees of freedom
which would contribute to the isocurvature perturbation.

The {\it isocurvature condition} is known as $\delta\rho=0$:
the sum total of all the energy contrasts must be zero. The
most general density perturbations is then given by a linear
combination of an adiabatic and an isocurvature density perturbations.

%%%%%%%%%%%%%%%%%%%%%%%%%%%%%%%%%%%%%%%%%%%%%%%%%%%%

\subsubsection{Adiabatic perturbations due to multi-field}

In a comoving gauge Eq.~(\ref{curvpert}) with
${\cal R}=-H\delta\phi/\dot\phi$ holds good even for multi-field
inflation models, provided we identify each field component of $\phi$ along
the slow roll direction. There also exists a relationship between
the comoving curvature perturbations and the number of e-foldings $N$
\cite{starobinsky85,salopek95,sasaki96,lyth98}
\begin{equation}
\label{multi1}
{\cal R} =\delta N=\frac{\partial N}{\partial \phi_{a}}\delta\phi_{a}\,,
\end{equation}
where $N$ is measured by a comoving observer while passing from flat
hypersurface (which defines $\delta\phi$) to the comoving hypersurface
(which determines $\cal R$) \cite{sasaki96,kodama84}. The repeated indices
are summed over and the subscript $a$ denotes a component of the inflaton.
A more intuitive derivation has been given in
\cite{lyth98,liddle-lyth00}.

If again one assumes that the perturbations in $\delta\phi_{a}$ have random
phases with an amplitude $(H/2\pi)^2$, one obtains
\begin{equation}
\label{multi5}
\delta_{k}^2=\frac{V}{75\pi^2~M_{\rm P}^2}\frac{\partial N}{\partial\phi_{a}}
\frac{\partial N}{\partial\phi_{a}}\,.
\end{equation}
For a single component
$\partial N/\partial \phi\equiv (M_{\rm P}^{-2}V/ V^{\prime})$, and then
Eq.~(\ref{multi5}) reduces to Eq.~(\ref{pspect1}). By using slow roll
equations we can again define the spectral index
\begin{equation}
\label{multi6}
n-1=-\frac{M_{\rm P}^2V_{,a}V_{,a}}{V^2}-\frac{2}{M_{\rm P}^2 N_{,a}
N_{,a}}+2\frac{M_{\rm P}^2N_{,a}N_{,b}V_{,ab}}{V N_{,c}N_{,c}}\,,
\end{equation}
where $V_{,a}\equiv \partial V/\partial \phi_{a}$, and similarly
$N_{,a}\equiv\partial N/\partial \phi_{a}$. For a 
single component we recover Eq.~(\ref{spectind4}) from Eq.~(\ref{multi6}). 
These results prove useful in constraining the AD potential by cosmological 
density perturbations, as will be discussed in Sect.~$5.3.$

%%%%%%%%%%%%%%%%%%%%%%%%%%%%%%%%%%%%%%%%%%%%%%%%%%%%%%%%%%
\subsubsection{Isocurvature perturbations and CMB}

One may of course simply assume a purely
isocurvature initial condition. For any species the entropy
perturbation is defined by
\begin{equation}
S_{i}=\frac{\delta n_{i}}{n_{i}}-\frac{\delta n_{\gamma}}{n_{\gamma}}\,,
\end{equation}
Thus, if initially there is a radiation bath with
a common radiation density contrast $\delta_{r}$, a baryon-density contrast
$\delta_{b}=3\delta_{r}/4$, and a CDM density contrast $\delta_{c}$, then
\begin{equation}
\label{cdm}
S=\delta_{c}-\frac{3}{4}\delta_{r}=\frac{\rho_{r}\delta\rho_{c}-(3/4)
\rho_{c}\delta\rho_{r}}{\rho_{r}\rho_{c}}=\frac{\rho_{r}+(3/4)\rho_{c}}
{\rho_{r}\rho_{c}}\delta\rho_{c}\approx \delta_{c}\,,
\end{equation}
where we have used the isocurvature condition
$\delta\rho_{r}+\delta\rho_{c}=0$, and the last equality holds in a
radiation dominated Universe.

However a pure isocurvature perturbation
gives five times larger contribution to the Sachs-Wolfe effect compared
to the adiabatic case \cite{starobinsky84,liddle93,liddle-lyth00}.
This result can be derived very easily in a matter dominated era with
an isocurvature condition $\delta\rho_{c}=-\delta\rho_{r}$, which gives
a contribution ${\cal R}_{k}=(1/3)S_{k}$. Therefore from
Eqs.~(\ref{curmet},\ref{sachs}), we obtain $\Delta T_{k}/T =-S_{k}/15$.
There is an additional contribution from radiation because we are in a
matter dominated era, see Eq.~(\ref{cdm}),
$S\approx \delta_{c} \equiv -(3/4)\delta_{r}$. The sum total isocurvature
perturbation $\Delta T_{k}/T = -S/15 -S/3= -6S/15$, where $S$ is measured
on the last scattering surface. The Sachs-Wolfe effect for isocurvature
perturbations fixes the {\em slope} of the perturbations, rather than
the amplitude \cite{liddle00,linde97}. Present CMB data rules out pure
isocurvature perturbation spectrum \cite{stompor96,eksvaliviita},
although a mixture of adiabatic and isocurvature perturbations remains
a possibility~\cite{stompor96,kawasaki9654,enqvist0061,pierpaoli9909420}.
In the latter case it has been argued that the adiabatic and
isocurvature perturbations might naturally turn out to be correlated
\cite{langlois99,bucher00}. The most general power
spectrum is not a single function but a $5\times 5$ matrix, which
contains all possible adiabatic and isocurvature perturbations
together with their cross-correlations. As discussed in \cite{bucher00},
resolving the perturbation spectrum in all its generality would be an
observational challenge that probably would have to wait for the determination
of the polarization spectrum by the Planck Surveyor Mission.

It is sometimes useful to consider the ratio $\alpha$ of the total
power spectra, defined as ${\cal P}_{tot}={\cal P}_{ad}+{\cal P}_{iso}$
\cite{kanazawa99,enqvist0061}, where $\alpha$ is defined as
\cite{enqvist0061}
\begin{equation}
\label{isoalpha}
\alpha =\left.\frac{16}{25}\frac{{\cal P}_{iso}}{{\cal P}_{ad}}
\right |_{k=aH}=\left| \frac{\delta_{\gamma}^{i}}{\delta_{\gamma}^{a}}\right|
\,.
\end{equation}
where $\delta_{\gamma}^{i}$ is the perturbation in the photon energy
density due to isocurvature perturbations and $\delta_{\gamma}^{a}$ is
the perturbation due to adiabatic perturbations.

%%%%%%%%%%%%%%%%%%%%%%%%%%%%%%%%%%%%%%%%%%%%%%%%

%%%%%%%%%%%%%%%%%%%%%%%%%%%%%%%%%%%%%%%%%%%%%%%%%%%%%%%%%%%
\subsection{Inflation models}

A detailed account on inflation model building can be found in many
reviews \cite{linde90,olive90c,lyth98,liddle-lyth00}. Here we briefly
recall some of the popular models with a particular emphasis on
supersymmetric inflation. First we recapitulate some aspects of
non-supersymmetric models.

%%%%%%%%%%%%%%%%%%%%%%%%%%%%%%%%%%%%%%%%%%%%%%
\subsubsection{Non-supersymmetric inflation}

The very first attempt to build an inflation model was made in
\cite{starobinsky80}, where one loop quantum correction to the
energy momentum tensor due to the space-time curvature were taken into
account, resulting in terms of higher order in curvature invariants.
Such corrections to the Einstein equation admit a de Sitter solution
\cite{dowker76}, which was presented in \cite{starobinsky80,vilenkin85}.
Inflation in Einstein gravity with an additional $R^2$ term
was considered in \cite{starobinsky83}
(for a discussion of inflation in pure $R^2$ gravity, see
\cite{buchdahl62}). Such a theory is conformally equivalent to a
theory with a canonical gravity \cite{whitt84}
with a scalar field having a potential term. A similar situation arises
in theories with a variable Planck mass, i.e., in scalar tensor
theories \cite{will93}. Inflation in these models has been studied
extensively~\cite{la89}.

The simplest single field inflation model is arguably chaotic inflation
\cite{linde83,linde91} with a generic potential
\begin{equation}
V=\frac{\lambda}{M_{\rm P}^{\alpha-4}}\phi^{\alpha}\,,
\end{equation}
where $\alpha$ is a positive even integer. In chaotic inflation slow
roll takes place for $\phi \gg M_{\rm P}$, and the two slow roll
parameters are given by \cite{liddle-lyth00}
\begin{equation}
\epsilon\equiv \frac{\alpha^2}{2}\frac{M_{\rm P}^2}{\phi^2}\,, \quad \quad
\eta =\alpha(\alpha-1)\frac{M_{\rm P}^2}{\phi^2}\,.
\end{equation}
Inflation ends when $\epsilon\equiv 1$, or, $\phi\approx\alpha M_{\rm P}$.
The cosmological scales leave the horizon when
$\phi=\sqrt{2N\alpha}M_{\rm P}$, and the spectral indices for scalar
and tensor perturbations turn out to be \cite{liddle-lyth00}
\begin{equation}
n=1-\frac{2+\alpha}{2N}\,, \quad \quad \quad r=\frac{3.1\alpha}{N}\,.
\end{equation}
The amplitude of the density perturbations, if normalized at the COBE
scale, yields the constraint $\lambda \simeq 4\times 10^{-14}$.

An exponential potential, such as might arise in string theories and
theories with extra dimensions,
\begin{equation}
V(\phi)=V_{0}\exp\left(-\sqrt{\frac{2}{p}}\frac{\phi}{M_{\rm P}}\right)\,.
\end{equation}
would give rise to a power law $a(t)\propto t^{p}$ for the scale factor,
so that inflation occurs when $p>1$. Multiple exponentials with differing
slopes give rise to what has been dubbed as assisted inflation
\cite{liddle98}.

%%%%%%%%%%%%%%%%%%%%%%%%%%%%%%%%%%%%%%%%%%%%%%%%%%%%%%%%%%%%
\subsubsection{F-term inflation}

In four dimensions the $N=1$ supersymmetric potential receives two
contributions: one from the F-term, which is related to the
chiral supermultiplets, and the second from the D-term, which
contains the gauge interactions. For a detailed discussion of
supersymmetric inflation we refer to the review by Lyth and Riotto
\cite{lyth98}. Here we give a brief resume of the two types of inflation.

Historically, supersymmetric inflation was first introduced to
cure some of the problems associated with the fine tuning of new
inflation \cite{ellis82a}, but since then utilizing supersymmetry
as a tool for inflation has gained in popularity (we describe
supersymmetry in Sect.~$4.2$, and for supergravity, see Sect.~$4.5.2$.).
The F-term potential can be derived from the superpotential $W$
\begin{equation}
\label{spot1}
V(\phi,\phi^{\ast})=F^{\ast i}F_{i}\,, \quad \quad F_{i}=-\left(
\frac{\partial W}{\partial\phi_{i}}\right)^{\ast}\,,
\end{equation}
where for renormalizable interactions $W$ has a mass dimension three.

Supersymmetry is broken whenever $|F|^2\neq 0$. A simple working
example is to consider the superpotential
\begin{equation}
\label{suppot1}
W=\lambda S(\phi^{2}-\phi_0^{2})\,
\end{equation}
which is invariant under a global $R$ symmetry with the superfields
$S$ and $N$ carrying respectively the $R$ charges $1$ and $0$. The
scalar components of these superfields can be written in the form
\begin{equation}
S=\frac{\sigma}{\sqrt{2}}\,, \quad \quad \quad \quad
\phi=\frac{\phi_1+i\phi_2}{\sqrt{2}}\,,
\end{equation}
where we have used $R$-transformation in order to make $S$ real. The potential
follows from Eq.~(\ref{spot1}):
\begin{equation}
\label{spot2}
V= \lambda^2\phi_0^4-\lambda^2 \phi_0^2(\phi_1^2-\phi_2^2)+\frac{\lambda^2}{4}
(\phi_1^2+\phi_2^2)^2+\lambda^2\sigma^2(\phi_1^2+\phi_2^2)\,.
\end{equation}
The supersymmetric vacuum is located at $\sigma=0$, $\phi_1=\phi_0$,
and $\phi_2=0$.
Note that the potential has a flat direction along $\sigma$-axis when
$\sigma >\sigma_{inst}=\phi_0$. When $\sigma < \sigma_{inst}$, the mass
squared of $\phi_1$  becomes negative and suggests a phase transition
along the $\phi_1$ direction. When this happens $\sigma$, $\phi_1$,
and $\phi_2$  begin to oscillate around their supersymmetry preserving
vacua. If $\phi_1=\phi_2=0$, the height of the potential is given by
$V=\lambda^2 \phi_0^4$, and as a consequence one obtains a period of
inflation. This is the simplest example of a flat direction giving
rise to an inflation potential, and it is known as the hybrid model,
first described in a non-supersymmetric context in~\cite{linde91,linde94}
and in a supersymmetric context in~\cite{copeland94}.

In order to have a graceful exit from inflation one requires a slope
for the flat direction such that $\sigma$ can roll down and approach
$\sigma_{inst}$. The flatness of the potential can be lifted in two
ways: by radiative corrections \cite{dvali94}, or by the low energy
soft supersymmetry breaking effects.

Due to supersymmetry breaking the fermions obtain a mass of the order
$(\partial^2W/\partial \phi^2)=\lambda S$, while the two complex
scalars receive a
mass squared  $\lambda^2S^2\pm \lambda^2\phi_0^2$. The one-loop radiative
correction to the potential is given by~\cite{coleman73}
\begin{equation}
\label{cole}
\delta V=\frac{1}{64\pi^2}\sum_{i}(-)^{f_{i}}M_{i}^4\ln \frac{M_{i}^2}{M^2}\,,
\end{equation}
where $f_{i}$ denotes the number of fermions, $M_{i}^2$ is the fermion
mass squared, and $M$ the cut-off or the renormalization scale. The
summation should be taken over all helicity states $i$. In the present
example the effective potential along the flat direction is given by
\begin{equation}
V=\lambda^2\phi_0^4\left(1+\frac{C\lambda^2}{8\pi^2}\ln\frac{\sigma}{\sqrt{2}M}
\right)\,,
\end{equation}
where $C$ is a constant essentially counting the states running in the loops.
If the loop correction dominates over the tree level potential, there is
a period of inflation which typically ends when
\begin{equation}
\sigma =\lambda\sqrt{\frac{CN}{4\pi^2}}M_{\rm P}\,,
\end{equation}
The COBE normalization sets the scale of inflation to
\begin{equation}
V^{1/4} \sim \left(\frac{50}{N}\right)^{1/4}C^{1/4}\lambda \times 10^{15}
~{\rm GeV}\,
\end{equation}
while the spectral index is given by
\begin{equation}
n=1-\frac{1}{N}\left(1+\frac{3C\lambda^2}{16\pi^2}\right)\,.
\end{equation}
Therefore, depending on the coupling constant $\lambda$ and the number of
e-foldings $N$, it is possible to have a wide range of inflation
energy scales which all provide a spectral index $n \sim 0.96-0.98$.

Soft supersymmetry breaking contributions induce
$m_{\sigma}\sim {\cal O}(\rm TeV)$.
One could also imagine that the mass of $\sigma$  appears dynamically
if $\sigma$ has couplings to bosons and fermions; these may induce a
typical running mass $\propto \sigma^2\ln(\sigma/M)$
\cite{stewart97a,lyth98,covi99}.

%%%%%%%%%%%%%%%%%%%%%%%%%%%%%%%%%%%%%%%%%%%%%%%%%%%%%%%%%%%%%%%
\subsubsection{D-term inflation}

In the above discussion we have neglected the gauge contribution.
The D-term $D^{a}=-g_{a}(\phi^{\ast}T^{a}\phi)$ gives rise to
a scalar potential (see ~\cite{bailin94,nilles84})
\begin{equation}
V(\phi,\phi^{\ast})=\frac{1}{2}\sum D^{a}D_{a}\,
\end{equation}
where  $(T^{a})_{i}~^{j}$ satisfy $[T^{a},T^{b}]=if^{abc}T^{c}$
($f^{abc}$ is the structure constant).

The simplest realization of D-term inflation reproduces the hybrid
potential with three chiral superfields, $S$, $\phi_{+}$, and $\phi_{-}$
with (non-anomalous) $U(1)$ charges $0,+1,-1$ \cite{dvali96}. The
superpotential can be written as
\begin{equation}
\label{dtspot}
W= \lambda S\phi_{+}\phi_{-}\,.
\end{equation}
The scalar potential then reads \cite{dvali96}
\begin{equation}
V=\lambda^2|S|^2\left(|\phi_{+}|^2+|\phi_{-}|^2\right)+\lambda^2|\phi_{+}
\phi_{-}|^2+\frac{g^2}{2}\left(|\phi_{+}|^2-|\phi_{-}|^2+\xi^2 \right)^2\,,
\end{equation}
where $g$ is the gauge coupling and $\xi$ is the Fayet-Iliopoulos D-term.
Note that the potential allows unique supersymmetry preserving vacua with
a broken gauge symmetry $S=\phi_{+}=0$, and $\phi_{-}={\xi}$. By virtue of
the coupling, when $|S|>S_{inst}=g{\xi}/\lambda$, the fields
$\phi_{+}, \phi_{-} \rightarrow 0$, and therefore inflation occurs because
of the Fayet-Iliopoulos D-term $V=g^2\xi^4/2$. The slope along the
inflaton direction $S$ can be generated by the one-loop  contribution
and reads
\begin{equation}
\label{dterm1}
V =\frac{g^2 \xi^4}{2}\left(1+\frac{g^2}{16\pi^2}\ln \frac{\lambda^2|S|^2}
{M_{\rm P}^2}\right)\,.
\end{equation}
Inflation ends when slow roll condition breaks down for
$S\sim (g/2\pi\sqrt{2})M_{\rm P}$, and the predictions for the inflationary
parameters are similar to the previous discussion.
D-term inflation based on an anomalous $U(1)$ symmetry (which could appear
in string theory \cite{dine87}) is no different.

Hybrid inflation is successful but has also problems that are related to
the initial conditions. In \cite{zurab98}, it was pointed out that
hybrid inflation requires an extremely homogeneous field configuration for
the fields orthogonal to the inflaton.  In our example the orthogonal
fields to the inflaton must be set to zero with a high accuracy over a
region much larger than the initial size of the horizon. It is possible to
solve this impasse by having a pre-inflationary matter dominated phase
when the field orthogonal to the inflaton direction  oscillates
and decays into lighter degrees of freedom, gradually
settling down to the bottom of its potential~\cite{zurab98}.

%%%%%%%%%%%%%%%%%%%%%%%%%%%%%%%%%%%%%%%%%%%%%%%%%%%%%%%%%%%%%%%%%
\subsubsection{Supergravity corrections}

When the field values are close to the Planck scale, supergravity (SUGRA)
effects become important and may ruin the flatness of the inflaton potential.
The soft breaking mass of the scalar fields are typically
\cite{dine84,copeland94,stewart95,dine95,dine96}
\begin{equation}
\label{msoftH}
m^2_{soft} \sim \frac{V}{3M_{\rm P}^2} \sim {\cal O}(1)H^2\,.
\end{equation}
Once the inflaton gains a mass $\sim H$, the field simply rolls down to
the minimum of the potential and inflation stops. Indeed, in SUGRA the
slow roll parameter
\begin{equation}
|\eta| \equiv {M_{\rm P}^2}\frac{V^{\prime \prime}}{V} \sim
\frac{m^2_{SUGRA}}{H^2} \sim {\cal O}(1) \,,
\end{equation}
where
$m^2_{SUGRA}\approx m^2_{SUSY}+(V_{SUSY}/3M_{\rm P}^2)\sim m^2_{SUSY}+{\cal O}(1)H^2$.
Note that the latter contribution dominates in an expanding
Universe and violates the slow roll condition. For field values
smaller than Planck scale it is always possible to obtain
$\epsilon \ll 1$, but in supergravity $\eta$ can never be made
less than one for a single chiral field with a minimal kinetic term.
This is known as the $\eta$ problem in SUGRA models of inflation
\cite{copeland94}.

When there are more than one chiral superfields, it might be possible
to cancel the dominant ${\cal O}(1)H$ correction to the inflaton mass
by choosing an appropriate K\"ahler term \cite{stewart95,copeland94}
(see also discussion in \cite{lyth98}). In hybrid inflation
models derived from an F-term the dominant ${\cal O}(1)H$ correction
in the mass term can be canceled if $|N|=0$ exactly, which however seems
to lead to an initial condition problem, as  discussed above.
The fact that the superpotential is linear in $S$ in
Eqs.~(\ref{suppot1},\ref{dtspot}) guarantees the cancellation of the dominant
contribution in the mass term for a minimal K\"ahler term
$\sim |S|^2$. For non-minimal K\"ahler potential such as
$K =|S|^2 +\beta |S|^4/M_{\rm P}^2+...$, one obtains
$(\partial^2 K/\partial S\partial S^{*})^{-1}\sim 1-4\beta |S|^2/M_{\rm P}^2$.
These contributions again lead to a problematic $\beta\times {\cal O}(1)H$
contribution to the inflaton mass unless the value of the unknown constant
$\beta $ is suppressed.

In \cite{dvali96}, it was shown that the $\eta$ problem does not
appear for D-term inflation even for the minimal K\"ahler potential because
the main contribution to the inflation potential does not come from the
vev of the inflaton field alone, but from the Fayet-Iliopoulos term which
belongs to the D-sector of the potential. Based on this fact many D-term
inflation models have been written down \cite{jeannerot97,kolda98}.
Therefore, hybrid inflation, whether realized as an effective potential
coming from F-sector or from D-sector, appears to be among the most
promising models for supersymmetric inflation.

%%%%%%%%%%%%%%%%%%%%%%%%%%%%%%%%%%%%%%%%%%%%%%%%%%%%%%%%%%
\subsection{Reheating of the Universe}

\subsubsection{Perturbative inflaton decay}

Traditionally reheating has been assumed to be a consequence
of the perturbative decay of the inflaton
\cite{albrecht82,starobinsky84,kolbturner90}.
After the end of inflation,  when $H \leq m_{\phi}$, the inflaton field
oscillates about the minimum of the potential. Averaging
over one oscillation results in \cite{turner83}
pressureless equation of state where
$\langle p\rangle =\langle \dot\phi^2/2 -V(\phi)\rangle$
vanishes\footnote{This will be discussed in a more detail in Sect.~$5.8$.},
so that the energy density redshifts as during matter domination with
$\rho_{\phi}=\rho_{i}(a_{i}/a)^{3}$ (subscript $i$ denotes the
quantities right after the end of inflation). If $\Gamma_{\phi}$ represents
the decay width of the inflaton to a pair of fermions, then the inflaton
decays when
$H(a)=\sqrt{(1/3M_{\rm P}^2)\rho_{i}}(a_{i}/a)^{3/2}\approx \Gamma_{\phi}$.
When the inflaton  decays, it releases its energy into the thermal bath
of relativistic particles whose energy density is determined by the reheat
temperature $T_{rh}$, given by
\begin{equation}
T_{rh}=\left(\frac{90}{\pi^2 g_{\ast}}\right)^{1/4}\sqrt{\Gamma_{\phi}
M_{\rm P}}=0.3\left(\frac{200}{g_{\ast}}\right)^{1/4}\sqrt{\Gamma_{\phi}
M_{\rm P}}\,.
\end{equation}

However the inflaton might not decay instantaneously. In such a case
there might already exist a thermal plasma of some relativistic species
at a temperature higher than the reheat temperature already before the
end of reheating~\cite{kolbturner90}. If the inflaton decays with a
rate $\Gamma_\phi$, then the instantaneous plasma temperature is found
to be~\cite{kolbturner90}
\begin{equation}
\label{instT}
T_{inst}\sim \left(g_{\ast}^{-1/2}H\Gamma_{\phi}M_{\rm P}^2\right)^{1/4}\,,
\end{equation}
where $g_{\ast}$ denotes the effective relativistic degrees of freedom in the
plasma. The temperatures reaches its maximum $T_{max}$ soon after the inflaton
field starts oscillating around the minimum. Once the maximum temperature is
reached, then $\rho_{\psi} \sim a^{-3/2}$, and $T\sim a^{-3/8}$ until
reheating and thermalization is completely over
\cite{ellis87,chung99,davidson00,allahverdidrees02,jaikumar02}.

The process of thermalization has two aspects; achieving kinetic equilibrium,
and achieving chemical equilibrium. Kinetic equilibrium can be reached by
$2\rightarrow 2$ scattering and annihilation. For chemical equilibrium
one requires particle number changing interactions such as $2\rightarrow 3$
processes. In \cite{chung99}, soft processes which
allow for small momentum transfer with a larger cross-section have been
advocated for chemical equilibration, while in
\cite{ellis87}, hard processes have been invoked.
Therefore, depending on the interactions, thermalization time scale
could be short, such as in the case of soft scattering processes,
or, it could be long compared to the Hubble time if only hard processes are
operative. Recently it has been argued 
\cite{davidson00,allahverdidrees02,jaikumar02},
that thermalization time scale can be as long as the time it takes for the
inflaton decay products with typical energies ${\cal O}(m_{\phi})$ to
lose the energy $\sim (m_{\phi}-T_{rh})$. The main conclusion is that
inelastic scattering interactions $2\rightarrow 3$ can thermalize the
Universe faster compared to elastic interaction $2\rightarrow 2$.
Inelastic interactions can achieve the kinetic and chemical equilibrium
both, and therefore, $\Gamma_{inel}^{-1}$ could be considered as the true
thermalization time scale. In \cite{jaikumar02}, the authors have 
studied thermalization in QCD based approach with emphasis upon late 
thermalization and hadronization.
%%%%%%%%%%%%%%%%%%%%%%%%%%%%%%%%%%%%%%%%%%%%%%%%%%%%%%%%%%%

\subsubsection{Non-perturbative inflaton decay}

Much effort has lately been devoted to non-perturbative effects which
are essentially non-thermal. These may lead to a rapid transfer of the
inflaton energy to other degrees of freedom by the process known as
preheating.  The requirement is that  the inflaton quanta  couple to other
(essentially massless) fields $\chi$ through e.g. terms like $\phi^2\chi^2$.
The quantum modes of $\chi$ may then be excited during the inflaton
oscillations via a parametric resonance. Preheating has been treated
both analytically
\cite{traschen90,kofman94,cooper94,boyanovsky95,kofman96,allahverdi97,bastero-gil99,heitmann00},
and on lattice \cite{khlebnikov96}.

Like bosons, fermions can also be excited through preheating
\cite{dolgov90,heitmann99,giudice99,maroto99}. In fact it has been
argued that fermionic preheating  is perhaps more effective than
bosonic preheating \cite{heitmann99,giudice99,bastero-gil00}.
However note that supersymmetry is effectively broken during the
inflaton oscillations \cite{maroto99,kallosh99,giudice99a}.
As a consequence, one naturally expects corrections to the inflaton
potential during the oscillations \cite{jeannerot00}. Therefore in most
supersymmetric models of inflation preheating might not turn out
to be very relevant.

%%%%%%%%%%%%%%%%%%%%%%%%%%%%%%%%%%%%%%%%%%%%%%%%%%%%%%%
\subsubsection{Gravitino and inflatino problems}

The reheat temperature should  certainly be above the BBN temperature
$T\geq {\cal O}(1)$~MeV, but there also exists an upper bound
from gravitino overproduction. In supergravity the
superpartner of the graviton is a spin-$3/2$ gravitino, which gets a mass from
the super-Higgs mechanism~\cite{deser77} when supersymmetry is spontaneously
broken. Typically supergravity is broken in a hidden sector by some
non-perturbative dynamics. Supersymmetry breaking  is then
mediated via gravitational (or possibly other) couplings to the observable
sector in such a way that sfermions and gauginos get masses of order
electroweak scale \cite{nath84,nilles84}. In addition, the gravitino also
gets a mass which in the simplest gravity mediated models is of order $1$~TeV
\cite{barbieri82} (see Sect.~$4.4$).

If the gravitino is not the lightest supersymmetric particle (LSP), it
will decay. Gravitino has two helicity states $\pm 3/2$ and $\pm 1/2$.
The latter one is mainly the goldstino mode which is eaten by
the super-Higgs mechanism. The goldstino coupling strength is inversely
proportional to the momentum, so that at low energies the gravitino coupling
is mainly dictated by the goldstino mode~\cite{fayet79}. At temperatures
much above the sparticle masses, it is the massless $\pm 3/2$ mode
that governs the gravitino interactions. The helicity $\pm 3/2$ mode
can decay into gauge bosons and gauginos through a dimension $5$-operator
with a lifetime
\begin{equation}
\tau_{3/2 \rightarrow A_{\mu}\lambda}\approx \frac{4 M_{\rm P}^2}{m_{3/2}^3}\,.
\end{equation}
Typically $\tau \sim 10^{2}-10^{5}$ s for a gravitino mass in the range
$10~{\rm TeV}\leq m_{3/2} \leq 100~{\rm GeV}$.

Although the gravitino interactions with matter are suppressed by the Planck
mass, they can be generated in great abundances very close to the Planck
scale \cite{weinberg82}. Inflation would dilute their number density
\cite{ellis83}, but  during reheating they would be regenerated
though scattering of gauge and gaugino quanta, with adverse consequences
\cite{ellis84b,ellis85b,ellis92,kawasaki95a,fischler94,leigh95,ellis96,boltz01}. The resulting gravitino abundance has been
estimated to be \cite{ellis84b}
\begin{equation}
\label{gravitinoab}
\frac{n_{3/2}}{s} \approx 2.4\times 10^{-13}\left(\frac{T_{rh}}
{10^{9}~{\rm GeV}}\right)\left[1-0.018\ln\left(\frac{T_{rh}}{10^{9}~{\rm GeV}}
\right)\right]\,,
\end{equation}
where $s$ defines the entropy density and $T_{rh}$ denotes the reheating
temperature of the Universe. The abundance \eq{gravitinoab} could well
be increased by an order of magnitude if gravitino interactions with
other chiral multiplets are included \cite{kawasaki95a}. In \cite{fischler94}
it was argued that at  finite temperatures gravitino overproduction could
be enhanced, but the calculation was criticized in \cite{leigh95,ellis96};
for a recent discussion on this topic, see \cite{boltz01}.

Since the gravitino is a late decaying particle, BBN yields a restriction on
the reheat temperature \cite{ellis85b,sarkar96}. For instance,
gravitino decay products can enhance the abundance of $D+^{3}He$ due to
photo fission of $^{4}He$ which implies~\cite{sarkar96}
\begin{eqnarray}
\frac{n_{3/2}}{s} &\leq &(10^{-14}-10^{-11})\Rightarrow \, \nonumber \\
T_{rh} &\leq &(10^7-10^{10})~{\rm GeV}\,, ~~100~{\rm GeV}\leq m_{3/2}
\leq 10~{\rm TeV}\,.
\end{eqnarray}
The constraint on the reheating temperature is~\cite{sarkar96}
\begin{equation}
T_{rh}\leq 2.5\times 10^{8}\left(\frac{m_{3/2}}{100~{\rm GeV}}\right)^{-1}
{\rm GeV}\,,
\end{equation}
for $m_{3/2} \leq 1.6$~TeV.

In gauge mediated supersymmetry breaking scenarios, to be
discussed in Sect. 4.4., the gravitino can have a very light
mass $\sim 10^{-6}$~GeV \cite{dine93} and can be  a hot dark matter candidate
\cite{pierpaoli98}. Small (or large) gravitino masses can also be
obtained in SUGRA models with non-minimal K\"ahler terms, such as
the no-scale model~\cite{ellis84147}. In anomaly mediation
the gravitino mass is large with $m_{3/2}\sim m_{soft}/\alpha \gg m_{soft}$
\cite{randall99}. In general, if the gravitino is not LSP and heavier
than  $10$~TeV, it decays before nucleosynthesis and thus
does not cause any cosmological problems~\cite{khlopov84}.

Gravitinos could also be produced by non-perturbative processes,
as was first described in \cite{maroto99}, where the formalism for
exciting the helicity $\pm 3/2$ component of the gravitino was
developed. Later the production of the helicity $\pm 1/2$ state,
which for a single chiral multiplet is the superpartner of the
inflaton known as inflatino, has been studied  by several authors
\cite{kallosh99,giudice99a,bastero-gil00,kallosh00,nsp01}.
The decay channels of the inflatino have been first discussed in
\cite{rouzbeh00,nilles01}. It has been suggested \cite{rouzbeh00}
and also explicitly shown  \cite{nsp01} that in realistic models with
several chiral multiplets, the helicity $\pm 1/2$ gravitino production
is not a problem for nucleosynthesis as long as the inflationary scale
is sufficiently higher than the scale of supersymmetry breaking in
the hidden sector and the two sectors are gravitationally coupled.
A very late decay of inflatino could however be possible, as argued
in \cite{nilles01,rouzbeh02}. In \cite{rouzbeh02}, it was argued that
if the inflatino and gravitino were not LSP, then late off-shell
inflatino and gravitino mediated decays of heavy relics could be
significant.

%%%%%%%%%%%%%%%%%%%%%%%%%%%%%%%%%%%%%%%%%%%%%%%%%%%
\pagebreak

%%%%%%%%%%%%%%%%%%%%%%%%%%%%%%%%%%%%%%%%%%%%%%%%%%%%%%%%%%%%%%%%
%%%%%%%%%%%%%%%%%%%%%%%%%%%%%%%%%%%%%%%%%%%%%%%%%%%%%%%%%%%%%%%%
%%%%%%%%%%%%%%%%%%%%%%%%%%%%%%%%%%%%%%%%%%%%%%%%%%%%%%%%%%%%%%%%

\section{Flat directions}

%%%%%%%%%%%%%%%%%%%%%%%%%%%%%%%%%%%%%%%%%%%%%%%%%%%%%%%%%%%%%
\subsection{Degenerate vacua}

At the level of renormalizable terms, supersymmetric field theories
generically have infinitely degenerate vacua. This is a consequence of
the supersymmetry and the gauge symmetries (and discrete symmetries such as
$R$-parity) of the Lagrangian, which allow for certain types of
interaction terms only. Therefore, in general there are a number of
directions in the space of scalar fields, collectively called the moduli
space, where the scalar potential is identically zero. In low
energy supersymmetric theories such classical degeneracy is accidental
and is protected from perturbative quantum corrections by a
non-renormalization theorem \cite{gwr79}. In principle the
degeneracies could be lifted by non-perturbative effects. However such
effects are likely to be suppressed exponentially and thus unimportant
because all the couplings of low energy theories are typically weak even
at relatively large vevs. Therefore in the supersymmetric limit when
$M_{\rm p}\rightarrow \infty$, the potential for the flat direction
always vanishes.

In the MSSM the moduli fields are quark, lepton and Higgs chiral fields.
In string theories
there are often additional moduli fields associated with the conformal
field theory degrees of freedom and world sheet discreet $R$-symmetries
\cite{dine88}. The moduli space of string theory can also be lifted by
a soft supersymmetry breaking masses of the order of the gravitino mass
$m_{3/2}$. Since the moduli interactions with others fields are
usually Planck mass suppressed, the string moduli are also a cause for
worry because they may decay after nucleosynthesis. This problem has
been dubbed as the moduli problem \cite{dine84,carlos93}. However,
the MSSM flat directions are  made up of condensates of squarks,
Higgses, and sleptons, and  can evaporate much before nucleosynthesis.

However there is an effective potential for the flat direction
condensate fields which arises as a result of supersymmetry breaking
terms and higher dimensional operators in the superpotential. In this
sense the MSSM flat directions are only approximately flat at vevs
larger than the supersymmetry breaking scale.

%%%%%%%%%%%%%%%%%%%%%%%%%%%%%%%%%%%%%%%%%%%%%%%%%%%%%%%%%%%%%%%%%

\subsection{MSSM and its potential}

Let us remind the reader that the matter fields of MSSM are chiral
superfields $\Phi=\phi+\sqrt{2}\theta\bar\psi+\theta\bar\theta F$, which
describe a scalar $\phi$, a fermion $\psi$ and a scalar auxiliary
field F. In addition to the  usual quark and lepton superfields,
MSSM has two Higgs fields, $H_u$ and $H_d$. Two Higgses are needed
because $H^\dagger$, which in the Standard Model gives masses
to the $u$-quarks,  is forbidden in the superpotential.

The superpotential for the MSSM is given by~\cite{nilles84}
\begin{equation}
\label{mssm}
W_{MSSM}=\lambda_uQH_u\bar u+\lambda_dQH_d\bar d+\lambda_eLH_d\bar e~
+\mu H_uH_d\,,
\end{equation}
where $H_{u}, H_{d}, Q, L, \bar u, \bar d, \bar e$ in
Eq.~(\ref{mssm}) are chiral superfields, and the dimensionless Yukawa couplings
$\lambda_{u}, \lambda_{d}, \lambda_{e}$ are $3\times 3$ matrices in the family
space. We have suppressed the gauge and family indices. Unbarred fields are
$SU(2)$ doublets, barred fields $SU(2)$ singlets. The last term is the $\mu$
term, which is a supersymmetric version of the SM Higgs boson mass.
Terms proportional to $H_{u}^{\ast}H_{u}$ or $H^{\ast}_{d}H_{d}$ are
forbidden in the superpotential, since $W_{MSSM}$ must be analytic in the
chiral fields. $H_{u}$ and $H_{d}$ are required not only because they  give
masses to all the quarks and leptons, but also for the cancellation of
gauge anomalies. The Yukawa matrices determine the masses and CKM mixing
angles of the ordinary quarks and leptons through the neutral components of
$H_{u}=(H^{+}_{u},H^{0}_{u})$ and $H_{d}=(H^{0}_{d}H^{-}_{d})$.
Since the top quark, bottom quark and tau lepton are the heaviest fermions
in the SM, we assume that only the $(3,3)$ element of the matrices
$\lambda_{u}, \lambda_{d}, \lambda_{e}$ are important. In this limit
only the third family and the Higgs fields contribute to the MSSM
superpotential.

The SUSY scalar potential $V$ is the sum of the F- and D-terms and reads
\begin{equation}
\label{fplusd}
V= \sum_i |F_i|^2+\frac 12 \sum_a g_a^2D^aD^a
\end{equation}
where
\begin{equation}
F_i\equiv {\partial W_{MSSM}\over \partial \phi_i},~~D^a=\phi^\dagger T^a
\phi~.
\label{fddefs}
\end{equation}
Here we have assumed that $\phi_i$ transforms under a gauge group
$G$ with the generators of the Lie algebra given by
$T^{a}$.

The $\mu$ term provides masses to the Higgsinos
\begin{equation}
{\cal L} \supset -\mu(\tilde H_{u}^{+}\tilde H_{d}^{-}-\tilde H^{0}_{u}
\tilde H^{0}_{d})+{c.c} \,,
\end{equation}
and contributes to the Higgs $(mass)^2$ terms in the scalar potential  through
\begin{equation}
\label{higgsmass}
-{\cal L} \supset V \supset |\mu|^2(|H^{0}_{u}|^2+|H^{+}_{u}|^2+|H^{0}_{d}|^2+
|H^{-}_{d}|^2)\,.
\end{equation}
Note that Eq.~(\ref{higgsmass}) is positive definite. Therefore, it
cannot lead to electroweak symmetry breaking without including
supersymmetry breaking $(mass)^2$ soft terms for the Higgs fields,
which can be negative. Hence, $|\mu|^2$ should almost cancel the
negative soft $(mass)^2$ term in order to allow for a Higgs vev
of order $\sim 174$~GeV. That the two different sources of masses
should be precisely of same order is a puzzle for which
many solutions has been suggested
\cite{kim84,giudice88,kim92,dvali96478}.

Note also that Eq.~(\ref{mssm}) is the minimal superpotential because
we have not included terms which are gauge invariant and analytic in
the chiral superfields but which violate either baryon number $B$ or lepton
number $L$. The most general gauge invariant and renormalizable superpotential
would not only include Eq.~(\ref{mssm}), but also the terms
\begin{eqnarray}
\label{barv}
W_{\Delta L=1}&=&\frac{1}{2}\lambda^{ijk}L_{i}L_{j}\bar e_{k}+
\lambda^{\prime ijk}L_{i}Q_{j}\bar d_{k}+\mu^{\prime i}L_{i}H_{\mu}\,, \\
\label{lepv}
W_{\Delta B=1}&=&\frac{1}{2}\lambda^{\prime \prime ijk}\bar u_{i}\bar d_{j}
\bar d_{k}\,,
\end{eqnarray}
where $i=1,2,3$ represents the family indices. The chiral supermultiplets
carry baryon number assignments $B=+1/3$ for $Q_{i}$, $B=-1/3$ for
$\bar u_{i}, \bar d_{i}$, and $B=0$ for all others. The total lepton number
assignments are $L=+1$ for $L_{i}$, $L=-1$ for $\bar e_{i}$, and $L=0$ for
all the others. The terms in Eq.~(\ref{barv}) violate lepton number
by one unit, while those in Eq.~(\ref{lepv}) violate baryon number by one unit.

Unless $\lambda^{\prime}$ and $\lambda^{\prime\prime}$ terms are very
much suppressed, one would obtain rapid proton decay which violates
both $B$ and $L$ by one unit. Many other processes also give rise to
violation in baryon and lepton number (for a review, see
\cite{enqvist86}). Therefore, there must be a symmetry forbidding the
terms in Eqs.~(\ref{barv},\ref{lepv}), while allowing for the terms
in Eq.~(\ref{mssm}). The symmetry is known as $R$-parity \cite{fayet79},
which is a discrete parity defined for each particle as
\begin{equation}
P_{R}=(-1)^{3(B-L)+2s}\,
\end{equation}
with $P_{R}=+1$ for the SM particles and the Higgs bosons, while $P_{R}=-1$
for all the sleptons, squarks, gauginos, and Higgsinos.
Here $s$ is spin of the particle. Without the product $(-1)^{2s}$, the
expression is known as matter parity \cite{dimopoulos81}, and denoted by
$P_{M}=(-1)^{3(B-L)}$. The quantity $(-1)^{2s}$ is equal to $1$ whenever
conservation of angular momentum holds at a given vertex. In this case matter
parity and $R$-parity are equivalent. If $R$-parity is conserved then there
will be no mixing between the sparticles and the ones which have $P_{R}=+1$.
This completely forbids potentially dangerous terms in
Eqs.~(\ref{barv},\ref{lepv}).

Matter parity is actually a discrete subgroup of the continuous
$U(1)_{B-L}$ group. Therefore, if a gauged $U(1)_{B-L}$ is broken by scalar
vevs which carry even integer values of $3(B-L)$, then $P_{M}$ survives as
an exactly conserved discrete remnant \cite{mohapatra86}.
Besides forbidding $B$ and $L$ violation from the renormalizable interactions,
$R$-parity has  interesting phenomenological and cosmological consequences.
The lightest sparticle with $P_{R}=-1$, the LSP, must be absolutely stable.
If electrically neutral, the LSP is a natural
candidate for non-baryonic dark matter \cite{goldberg83,ellis84}.
It may be possible to produce LSPs in a next generation collider
experiments.

%%%%%%%%%%%%%%%%%%%%%%%%%%%%%%%%%%%%%%%%%%%%%%%%%%%%%%%%%%%%%
\subsubsection{F-and D-renormalizable flat directions of MSSM}

For a general supersymmetric model with $N$ chiral superfields $X_{i}$,
it is possible to find out the  directions where the potential \eq{fplusd}
vanishes identically by solving simultaneously
\begin{equation}
\label{fflatdflat}
D^{a}\equiv X^{\dagger}T^{a}X=0\,, \quad \quad
F_{X_{i}}\equiv \frac{\partial W}{\partial X_{i}}=0\,.
\end{equation}
Field configurations obeying Eq.~(\ref{fflatdflat}) are called respectively
D-flat and F-flat.

D-flat directions are parameterized by gauge invariant monomials of
the chiral superfields. A powerful tool for finding the flat directions
has been developed in
\cite{buccella82,affleck84,affleck85,dine96,luty96,gherghetta96},
where the correspondence between gauge invariance and flat directions has been
employed. The configuration space of the scalar fields of the
MSSM contains $49$ complex dimensions ($18$ for $Q_{i}$, $9$ each for
$\bar u_{i}$ and $\bar d_{i}$, $6$ for $L_{i}$, $3$ for $\bar e_{i}$, and
$2$ each for $H_{u}$ and $H_{d}$), out of which there are $12$ real
D-term constraints ($8$ for $SU(3)_{C}$, $3$ for $SU(2)_{L}$, and $1$
for $U(1)_{Y}$), which leaves a total of $37$ complex dimensions
\cite{dine96,gherghetta96}. The trick is to construct gauge invariant
monomials forming $SU(3)_{C}$ singlets and then using them as  building blocks
to generate $SU(3)_{C}\times SU(2)_{L}$, and subsequently the whole
$SU(3)_{C}\times SU(2)_{L}\times U(1)_{Y}$ invariant polynomials
\cite{dine96,gherghetta96}. However these invariant monomials give only the
D-flat directions. For F-flat directions, one must solve explicitly
the constraint equations $F_{X_{i}}=0$.

A single flat direction necessarily carries a global $U(1)$ quantum
number, which corresponds to an invariance of the effective Lagrangian for the
order parameter $\phi$ under phase rotation $\phi\to e^{i\theta}\phi$.
In the MSSM the global $U(1)$ symmetry is $B-L$. For example, the
$LH_u$-direction (see below) has $B-L=-1$.

A flat direction can be represented by a composite gauge invariant
operator, $X_m$,  formed from the product of $k$ chiral superfields
$\Phi_i$ making up the flat direction: $X_m=\Phi_1\Phi_2\cdots \Phi_m$.
The scalar component of the superfield $X_m$ is related to the
order parameter $\phi$  through
$X_m=c\phi^m$.

%%%%%%%%%%%%%%%%%%%%%%%%%%%%%%%%%%%%%%%%%%%%%%%%%%%%%%%%%%%

\subsubsection{An example of F-and D-flat direction}

The flat directions in the MSSM are tabulated in Table 1.
An example of a D-and F-flat direction is provided by
\begin{equation}
\label{example}
H_u=\frac1{\sqrt{2}}\left(\begin{array}{l}0\\ \phi\end{array}\right),~
L=\frac1{\sqrt{2}}\left(\begin{array}{l}\phi\\ 0\end{array}\right)~,
\end{equation}
where $\phi$ is a complex field parameterizing the flat direction,
or the order parameter, or the AD field. All the other fields are
set to zero. In terms of the composite gauge invariant operators,
we would write $X_m=LH_{u}~(m=2)$.

From \eq{example} one clearly obtains $
F_{H_u}^*=\lambda_uQ\bar u +\mu H_d=F_{L}^*=\lambda_dH_d\bar e\equiv 0$
for all $\phi$. However there exists a non-zero F-component given
by $F^*_{H_d}=\mu H_u$. Since $\mu$ can not be much larger than the
electroweak scale $M_W\sim {\cal O}(1)$~TeV, this contribution is of
the same order as the soft supersymmetry breaking masses, which are
going to lift the degeneracy. Therefore, following \cite{dine96}, one may
nevertheless consider $LH_u$ to correspond to a F-flat direction.

The relevant D-terms read
\begin{equation}
\label{Dterm0}
D^a_{SU(2)}=H_u^\dagger\tau_3H_u+L^\dagger\tau_3L=\frac12\vert\phi\vert^2
-\frac12\vert\phi\vert^2\equiv 0\,.
\end{equation}
Therefore the $LH_u$ direction is also D-flat.

The only other direction involving the Higgs fields and thus soft
terms of the order of $\mu$ is $H_uH_d$. The rest are purely leptonic,
such as $LL\bar e$, or baryonic, such as $\bar u\bar d\bar d$, or
mixtures of leptons and baryons, such as $QL\bar d$. These combinations
give rise to several independent flat directions that can be obtained
by permuting the flavor indices. For instance, $LL\bar e$ contains
the directions $L_1L_2\bar e_3$, $L_2L_3\bar e_1$, and $L_1L_3\bar e_2$.

%%%%%%%%%%%%%%%%%%%%%%%%%%%%%%%%%%%%%%%%%%%%%%%%%%%%%%%
%\begin{center}
\begin{table}
\vspace{3mm}
\begin{tabular}{|c|c|c|c|}
\hline & $B-L$ & & $B-L$ \\ \hline
$H_{u}H_{d}$ & 0 & $LH_{u}$ &-1\\
$\bar u\bar d\bar d$ & -1 & $QL\bar d$ & -1\\
$LL\bar e$ & -1 & $QQ\bar u\bar d$ & 0\\
$QQQL$ & 0 & $QL\bar u\bar e$ & 0\\
$\bar u\bar u\bar d\bar e $ & 0 & $QQQQ\bar u$ & 1\\
$QQ\bar u\bar u\bar e$ & 1 & $LL\bar d\bar d\bar d$ & -3\\
$\bar u\bar u\bar u\bar e\bar e$ & 1 & $QLQL\bar d\bar d$ & -2\\
$QQLL\bar d\bar d$ & -2 & $\bar u\bar u\bar d\bar d\bar d\bar d$ & -2\\
$QQQQ\bar dLL$ & -1 & $QLQLQL\bar e$ & -1\\
$QL\bar u QQ\bar d\bar d$ & -1 &
$\bar u\bar u\bar u\bar d\bar d\bar d\bar e$ & -1\\
\hline
\end{tabular}
\caption{\label{table1}
{\bf Renormalizable F and D flat directions in the MSSM }}
\end{table}
%\end{center}

%%%%%%%%%%%%%%%%%%%%%%%%%%%%%%%%%%%%%%%%%%%%%%%%%%%%%%

Along a flat direction gauge symmetries get broken, with the gauge
supermultiplets gaining mass by super-Higgs mechanism with
$m_g=g\langle\phi\rangle$. Several chiral supermultiplets typically
become massive by virtue of Yukawa couplings in the superpotential;
for example, in the $LH_u$ direction one finds the mass terms
$W_{\rm mass}=\lambda_u\langle\phi\rangle Q\bar u+\lambda_e\langle\phi\rangle H_d\bar e$.

Of course, there may simultaneously exist several flat directions. For
the purpose of AD mechanism it is the lowest dimensional
operator which determines the baryonic charge of the eventual condensate.
In what follows we will therefore mostly consider a single flat direction.

%%%%%%%%%%%%%%%%%%%%%%%%%%%%%%%%%%%%%%%%%%%%%%%%%%%%%%%%%%%%%%%%
%%%%%%%%%%%%%%%%%%%%%%%%%%%%%%%%%%%%%%%%%%%%%%%%%%%%%%%%%%%%%%%%

\subsection{Lifting the flat direction}

Vacuum degeneracy along a flat direction can be broken in two ways:
by supersymmetry breaking, or by higher order non-renormalizable
operators appearing in the effective low energy theory. Let us first
consider the latter option. Supersymmetry breaking will then be
discussed in more detail in Sects.~$4.4$ and $4.5$.

%%%%%%%%%%%%%%%%%%%%%%%%%%%%%%%%%%%%%%%%%%%%%%%%%%%%%%%%%%%%%%

\subsubsection{Lifting by non-renormalizable operators}

Non-renormalizable superpotential terms in the MSSM can be viewed
as effective terms that arise after one integrates out fields with very
large mass scales appearing in a more fundamental (say, string) theory.
Here we do not concern ourselves with the possible restrictions
on the effective terms due to discrete symmetries present
in the  fundamental theory, but assume that all operators
consistent with symmetries may arise. Thus in terms of the
invariant operators $X_m$, one can have terms of the type \cite{dine95,dine96}
\begin{equation}
\label{Xton}
W=\frac{h}{d M^{d-3}} X^{k}_{m}=\frac{h}{d M^{d-3}}\phi^d\,,
\end{equation}
where the dimensionality of the effective scalar operator $d=mk$,
and $h$ is a coupling constant which could be complex with
$|h|\sim {\cal O}(1)$. Here $M$ is some large mass, typically of
the order of the Planck mass or the string scale (in the heterotic
case $M \sim M_{GUT}$). The lowest value of $k$ is $1$ or $2$, depending
on whether the flat direction is even or odd under $R$-parity.

A second type of term lifting the flat direction would be of the form
\cite{dine95,dine96}
\begin{equation}
\label{2ndtype}
W={h^{\prime}\over M^{d-3}}\psi\phi^{d-1}~\,,
\end{equation}
where $\psi$ is not contained in $X_m$. The superpotential term
\eq{2ndtype} spoils F-flatness through $F_\psi \neq 0$. An example
is provided by the direction $\bar u_1\bar u_2\bar u_3\bar e_1\bar e_2$,
which is lifted by the non-renormalizable term
$W=(h'/M)\bar u_1\bar u_2\bar d_2\bar e_1$. This superpotential term
gives a non-zero contribution
$F_{\bar d_2}^*=(h'/M)\bar u_1\bar u_2\bar e_1\sim (h^{\prime}/M)\phi^3$
along the flat direction.

Assuming minimal kinetic terms, both types discussed above
in Eqs.~(\ref{Xton},\ref{2ndtype}) yield a generic non-renormalizable
potential contribution that can be written as
\be{nrpot}
V(\phi)={\vert\lambda\vert^2\over M^{2d-6}}(\phi^*\phi)^{d-1}~,
\ee
where we have defined the coupling $|\lambda|^2\equiv |h|^2+|h'|^2$.
By virtue of an accidental $R$-symmetry under which $\phi$ has a charge
$R=2/d$, the potential \eq{nrpot} conserves the $U(1)$ symmetry carried
by the flat direction, in spite of the fact that at the superpotential
level it is violated, see Eqs.~(\ref{Xton},\ref{2ndtype}).
The symmetry can be violated if there are multiple flat directions,
or by higher order operator contributions. However it turns out
\cite{dine96} that the $B-L$ violating terms are always subdominant.
This is of importance for baryogenesis considerations, where the necessary
$B-L$ violation should therefore arise from other sources.

The process of finding all the possible non-renormalizable superpotential
contributions lifting a particular flat direction is similar to finding
the D-flat directions discussed in Sect.~$4.2.1$. All the non-renormalizable
operators can be generated from SM gauge monomials with $R$-parity
constraint which allows only even number of odd matter parity fields
($Q,L,\bar u,\bar d,\bar e$) to be present in each superpotential term.
At each dimension $d$, the various $F=0$ constraints are separately imposed in
order to construct the basis for monomials.

As an example, consider flat directions involving the Higgs fields
such as $H_{u}H_{d}$ and $LH_{u}$ directions. Even though they
are already lifted by the $\mu$ term, since $\mu$
is of the order of supersymmetry breaking scale, for cosmological purposes
they can be considered flat, as was discussed in Sect.~$4.2.2$.
At the $d=4$ level the superpotential reads
\begin{equation}
\label{exnon}
W_{4} \supset \frac{\lambda}{M}(H_{u}H_{d})^2+\frac{\lambda_{ij}}{M}
(L_{i}H_{u})(L_{j}H_{u})\,.
\end{equation}
Let us assume $\lambda, \lambda_{ij}\neq 0$.
Note that $F_{H_{d}}=0$ constraint implies
$\lambda H^{\alpha}_{u}(H_{u}H_{d})=0$, which acts as a basis for the
monomials. An additional constraint can be obtained by contracting
$F_{H_{d}}=0$ by $\epsilon_{\alpha \beta}H^{\beta}_{d}$, which forms the
polynomial $H_{u}H_{d}=0$ in the same monomial basis. Similarly the
constraint $F_{H_{u}}=0$, along with the contraction yields
$\lambda^{ij}(L_{i}H_{u})(L_{j}H_{u})=0$. This implies that $L_{i}H_{u}=0$
for all $i$. Therefore the two monomials $LH_{u}$ and $H_{u}H_{d}$ can
be lifted by $d=4$ terms in the superpotential Eq.~(\ref{exnon}).

The other renormalizable flat directions are
$LLE,\bar u\bar u\bar d,Q\bar dL, QQQL,Q\bar uQ\bar d,\bar u\bar u\bar d\bar e$
and
$Q\bar uL\bar e, \bar d\bar d\bar dLL, \bar u\bar u\bar u\bar e\bar e, Q\bar uQ\bar u\bar e, QQQQ\bar u, \bar u\bar u\bar dQdQ\bar d$,
and $(QQQ)_{4}LLL\bar e$. These are lifted primarily by the
superpotential terms which involve either $H_{u}$ or $H_{d}$ if
$d$ is odd, or those which contain neither $H_{u}$ nor $H_d$. The complete
list of superpotential terms which lift the flat directions can be found in
\cite{gherghetta96}. It was shown that all the MSSM flat directions are 
lifted by $d=4,5,6,7,9$ terms in the superpotential. The unique flat 
directions involving $(Q,u,e)$ is lifted by $d=9$, $(L,d)$ by $d=7$, and
$(L,d,e)$ by $d=5$. The flat directions involving $(L,e),(u,d)$ and $(L,d,e)$
are all lifted by $d=6$ terms in the superpotential, while the rest of the
flat directions are lifted already by $d=4$ superpotential terms.

%%%%%%%%%%%%%%%%%%%%%%%%%%%%%%%%%%%%%%%%%%%%%%%%%%%%%%%%%%%
\subsubsection{Lifting by soft supersymmetry breaking}

Vacuum degeneracy will also be lifted by supersymmetry
breaking, as will be discussed in more detail in Sects.~$4.4$ and $4.5$.
It is induced by the soft terms, which in the simplest case read
\begin{equation}
\label{susybreak}
V(\phi)=m_0^2\vert\phi\vert^2+\left[{\lambda A\phi^d\over dM^{d-3}}
+{\rm h.c.}\right]~,
\end{equation}
where the supersymmetry breaking mass $m_0$ and $A$ are typically of the order
of the gravitino mass $m_{3/2}$. An additional soft source for supersymmetry
breaking are the gaugino masses $m_g$. The $A$-term in \eq{susybreak}
violates the $U(1)$ carried by the flat direction and thus provides the
necessary source for $B-L$ violation in AD baryogenesis. In
general, the coupling $\lambda$ is complex and has an associated  phase
$\theta_\lambda$. Writing
$\phi=\vert\phi\vert \exp(i\theta)$, one obtains a potential
proportional to $\cos(\theta_\lambda+n\theta)$ in the angular
direction. This has $n$ discrete minima for the phase of $\phi$,
at each of which $U(1)$ is broken.

%%%%%%%%%%%%%%%%%%%%%%%%%%%%%%%%%%%%%%%%%%%%%%%%%%%%%%%%%
\subsection{Supersymmetry breaking in the MSSM}

In the MSSM there are several proposals for supersymmetry breaking,
which we shall discuss below. However most of the time it is not
important to know the exact mechanism of low energy supersymmetry
breaking. This ignorance of the origin of supersymmetry breaking
can always be hidden by simply writing down explicitly the soft
breaking terms with arbitrary couplings.

%%%%%%%%%%%%%%%%%%%%%%%%%%%%%%%%%%%%%%%%%%%%%%%%%%%%%%%%%%%%
\subsubsection{Soft supersymmetry breaking Lagrangian}

The most general soft supersymmetry breaking terms in the MSSM
Lagrangian can be written as (see e.g. \cite{haber85})
\begin{equation}
{\cal L}_{soft}=-\frac{1}{2}\left(M_{\lambda}\lambda^{a}\lambda^{a}+
{\rm c.c.}\right)-(m^2)^{i}_{j}\phi^{j\ast}\phi_{i}-\left(\frac{1}{2}
b_{ij}\phi_{i}\phi_{j}+\frac{1}{6}a^{ijk}\phi_{i}\phi_{j}\phi_{k}+{\rm c.c.}
\right)\,,
\end{equation}
where $M_{\lambda}$ is the common gaugino mass $(m^2)^{j}_{i}$
are $3\times 3$ matrices determining the masses for
squarks and sleptons, denoted as
$m^2_{Q},m^2_{\bar u},m^2_{\bar d},m^2_{L},m^2_{\bar e}$;
$b_{ij}$ is the mass term for the combination $H_{u}H_{d}$; and finally,
$a^{ijk}$ are complex $3\times 3$ matrices in the family space which yield the
$A$-terms $a_{u},~a_{d},~a_{e}$. There are a total of $105$ new entries in the
MSSM Lagrangian which have no counterpart in the SM. However the arbitrariness
in the parameters can be partly removed by the experimental constraints on
flavor changing neutral currents (FCNC) and $CP$ violation
\cite{dimopoulos95}. In order to avoid FCNC and excessive $CP$
violation, the squark and slepton $(\rm mass)^2$
matrices are often taken to be flavor blind, so that the squark and slepton
mixing angles can  be rotated away. Similarly, one may assume that
the $\phi^3$ couplings are proportional to the Yukawa coupling matrix, so that
$a_{u}=A_{u0}\lambda_{u},~a_{d}=A_{d0}\lambda_{d},$
and $a_{e}=A_{e0}\lambda_{e}$. Large $CP$ violating effects
can be avoided if the soft parameters do not involve new $CP$ phases in 
addition to the SM CKM phases. One can also fix $\mu$ parameter and $b$ 
to be real by an appropriate phase rotation of $H_{u}$ and $H_{d}$.

There are a number of possibilities for the origin of supersymmetry breaking.
Fayet-Iliopoulos mechanism \cite{fayet74} provides supersymmetry breaking by
virtue of a non-zero D-term but requires a $U(1)$ symmetry. However, this
mechanism does not work in the MSSM because some of the squarks and sleptons
will get non-zero vevs which may break color, electromagnetism, and/or
lepton number without breaking supersymmetry.  Therefore the contribution
from the Fayet-Iliopoulos term should be negligible at low scales.

There are models of supersymmetry breaking by F-terms, known as
O'Raifeartaigh models \cite{raifeartaigh75}, where the idea is to pick
a set of chiral supermultiplets $\Phi_{i} \supset (\phi_{i},\psi_{i}F_{i})$
and a superpotential $W$ in such a way that
$F_{i}=-\delta W/\delta\phi^{\ast}_{i}=0$ have no simultaneous solution.
The model requires a linear gauge singlet superfield in the
superpotential. Such singlet chiral supermultiplet is not present in the
MSSM. The scale of supersymmetry breaking has to be set by hand.

The only mechanism of supersymmetry breaking where the breaking
scale is not introduced either at the level of superpotential or in
the gauge sector is through dynamical supersymmetry breaking
\cite{witten82,affleck81}. In these models a small
supersymmetry breaking scale arises by dimensional transmutation.
It is customary to treat the supersymmetry breaking sector as a hidden
sector which has no direct couplings to the visible sector represented
by the chiral supermultiplets of the MSSM. The only allowed interactions
are those which mediate the supersymmetry breaking in the hidden sector to the
visible sector.

The main contenders are  gravity mediated supersymmetry breaking, which is
associated with new physics which includes gravity at the string scale
or at the Planck scale \cite{nath84,nilles84}, and gauge mediated
supersymmetry breaking, which is transmitted to the visible sector
by the ordinary electroweak and QCD gauge interactions
\cite{dine82,dine93,dine95a,dine96a}.  There
are other variants of supersymmetry breaking based upon ideas on
gravity and gauge mediation with some extensions, such as dynamical
supersymmetry breaking (see \cite{shadmi00}, and references therein),
and anomaly mediation (see \cite{randall99,giudice9898}, and
references therein), which we do not consider here.

%%%%%%%%%%%%%%%%%%%%%%%%%%%%%%%%%%%%%%%%%%%%%%%%%%%%%%%%%%%%%%%%%
%%%%%%%%%%%%%%%%%%%%%%%%%%%%%%%%%%%%%%%%%%%%%%%%%%%%%%%%%%%%%%%%%
%%%%%%%%%%%%%%%%%%%%%%%%%%%%%%%%%%%%%%%%%%%%%%%%%%%%%%%%%%%%%%%%%

\subsubsection{Gravity mediated supersymmetry breaking}

Let us assume that supersymmetry is broken by the vev
$\langle F\rangle \neq 0$ and is communicated to the MSSM by
gravity. On dimensional grounds, the soft terms in the visible
sector should then be of the order \cite{nilles84}
\begin{equation}
m_{soft} \sim \frac{\langle F\rangle}{M_{\rm P}}\,.
\end{equation}
Note that $m_{soft} \rightarrow 0$ as
$M_{\rm P} \rightarrow \infty$.
In order to obtain a phenomenologically acceptable soft supersymmetry
mass $m_{soft}\sim {\cal O}(100)$~GeV, one therefore requires
the scale of supersymmetry breaking in the hidden sector to be
$\sqrt{\langle F\rangle} \sim 10^{10}-10^{11}$~GeV.

Another possibility is that the supersymmetry is broken via
gaugino condensate
$\langle 0|\lambda^{a}\lambda^{b}|0\rangle=\delta^{ab}\Lambda^3\neq 0$,
where $\Lambda$ is the condensation scale \cite{nilles82,nilles84}. If
the composite field $\lambda^{a}\lambda^{b}$ belongs to the
$\langle F\rangle \sim \Lambda^3/M_{\rm P}$-term, then again on dimensional
grounds one would expect the soft supersymmetry mass contribution to be
\cite{nilles84}
\begin{equation}
m_{soft} \sim \frac{\Lambda^3}{M_{\rm P}^2}\,.
\end{equation}
In this case the nature of supersymmetry breaking is dynamical and the scale
is given by $\Lambda \sim 10^{13}$~GeV.

The supergravity Lagrangian must contain the non-renormalizable terms
which communicate between the hidden and the observable sectors. For
the cases where the kinetic terms for the chiral and gauge fields are
minimal, one obtains the following soft terms
\cite{nilles84}
\begin{equation}
m_{1/2}\sim \frac{\langle F\rangle}{M_{\rm P}}\,, \quad m_0^2\sim
\frac{|\langle F\rangle|^2}{M_{\rm P}^2}\,, \quad A_{0}\sim \frac{\langle F
\rangle}{M_{\rm P}}\,, \quad B_0 \sim \frac{\langle F\rangle}{M_{\rm P}}\,.
\end{equation}
The gauginos get a common mass $M_{1}=M_{2}=M_3=m_{1/2}$,
the squark and slepton masses are $m^2_{Q}=m^2_{\bar u}=m^2_{\bar d}=m^2_{L}=
m^2_{\bar e} =m^2_{0}$, and for the Higgses $m^2_{H_{u}}=m^2_{H_{d}}=m^2_{0}$.
The $A$-terms are proportional to the Yukawa couplings
while $b=B_{0}\mu$.

Some particular models of gravity mediated supersymmetry breaking
give more detailed estimates of the soft supersymmetry terms.
They include: {\it Dilaton~dominated} models \cite{kaplunovsky93},
which arise in a particular limit of superstring theories, which
have $m_{0}^2=m_{3/2}^2$, and $m_{1/2}=-A_0=\sqrt{3}m_{3/2}$;
{\it Polonyi} models \cite{polonyi77}, where
$m^2_{0}=m^2_{3/2}$, $A_{0}=(3-\sqrt{3})m_{3/2}$, and
$m_{1/2}={\cal O}(m_{3/2})$;
and {\it No-scale} models ~\cite{lahanas87}, which also arise in the
low energy limit of superstrings and in which the gravitino mass is
undetermined at the tree level while the at the string scale
$m_{1/2}\gg m_{0},A_{0},m_{3/2}$.

The predictions for the mass spectrum and other observable can be found
renormalization group (RG) equations; these will be described in
connection with the dynamical evolution of the AD field. Therefore, a
generic flat direction in gravity mediated supersymmetry breaking has
two important components: the soft supersymmetry breaking terms, and
the RG induced logarithmic dependence of the vev.

%%%%%%%%%%%%%%%%%%%%%%%%%%%%%%%%%%%%%%%%%%%%%%%%%%%%%
\subsubsection{Gauge mediated supersymmetry breaking}

In gauge mediated supersymmetry breaking one employs a heavy
messenger sector which couples directly to the supersymmetry
breaking sector but indirectly to the observable sector via
standard model gauge interactions only~\cite{dine82,nappi82}.
As a result the soft terms in the MSSM arise through ordinary
gauge interactions. There will still be gravitational communication,
but it is a weak effect.

The simplest example is a messenger sector with a pair of $SU(2)$ doublet
chiral fields $l,~\bar l$ and a pair of $SU(3)$ triplet fields
$q,~\bar q$, which couple to a singlet field $z$ with Yukawa
couplings $\lambda_2,~\lambda_3$, respectively. The superpotential
is given by
\begin{equation}
W_{mess}=\lambda_{2} zl\bar l+\lambda_{3}zq \bar q\,.
\end{equation}
The singlet acquires a non-zero vev and a non-zero F-term
$\langle F_z\rangle$. This can be accomplished either substituting $z$ into
an O'Raifeartaigh type model \cite{dine82,nappi82}, or by a dynamical
mechanism \cite{dine93,dine95a,dine96a}. One may parameterize
supersymmetry breaking in a superpotential $W_{break}$ by
$\langle \partial W_{break}/\partial z\rangle =-\langle F_{z}^{\ast}\rangle$.
As a consequence, the messenger fermions acquire masses
\begin{equation}
{\cal L}=-\left(\lambda_{2}\langle z\rangle\psi_{l}\psi_{\bar l}+
\lambda_{3}\langle z\rangle \psi_{q}\psi_{\bar q}+{\rm c.c}\right)\,,
\end{equation}
while the scalar messenger partners have a scalar potential given by
\begin{eqnarray}
V&=&|\lambda_{2}\langle z\rangle|^2\left(|l|^2+|\bar l|^2\right)+
|\lambda_{3}\langle z\rangle|^2\left(|q|^2+|\bar q|^2\right)-
\left(\lambda_{2}\langle F_{z}\rangle l\bar l+\lambda_{3}\langle F_{z}\rangle
q\bar q+{\rm c.c.}\right)\, \nonumber \\
&&+{\rm quartic~terms}\,,
\end{eqnarray}
where we have used $\langle \partial W_{mess} /\partial z\rangle =0$, and
we have replaced $z$ and $F_{z}$ by their vevs. It is easy to read off
the eigenvalues of the squared scalar masses and the fermionic and
bosonic spectrum of the messenger sector; for $(l,\bar l)$,
$m^2_{fermions}=|\lambda_2\langle z\rangle|^2$, and
$m^2_{scalars}=|\lambda_2\langle z\rangle|^2\pm |\lambda_2\langle F_z\rangle|$;
for $(q,\bar q)$, $m^2_{fermions}=|\lambda_3\langle z\rangle|^2$, and
$m^2_{scalars}=|\lambda_3\langle z\rangle|^2\pm |\lambda_3\langle F_z\rangle|$.

Supersymmetry breaking is then mediated to the observable fields by
one-loop corrections, which generate masses for the MSSM gauginos
\cite{dine93}. The $q,\bar q$ messenger loop diagrams provide masses
to the gluino and the bino, while $l,\bar l$ messenger loop diagrams
provide masses to the wino and the bino, i.e.,
$M_{a=1,2,3}=(\alpha_{a}/4\pi)\Lambda$, where
$\Lambda =\langle F_{z}\rangle/\langle z\rangle$.

For squarks and sleptons the leading term comes from two-loop diagrams,
e.g. $m^2_{\phi}\propto \alpha^2$. The $A$-terms get negligible
contribution at two-loop order compared to the gaugino masses, they
come with an extra suppression of $\alpha/4\pi$ compared with the gaugino
mass, therefore $a_{u}=a_{d}=a_{e}=0$ is a good approximation. The Yukawa
couplings at the electroweak scale are generated by evolving the
RG equations.

One can estimate~\cite{dine93} the soft supersymmetry breaking masses
to be of order
\begin{equation}
m_{soft}\sim \frac{\alpha_{a}}{4\pi}\frac{\langle F\rangle}{M_{s}}\,.
\end{equation}
If $M_s\sim \langle z\rangle$ and $\sqrt{\langle F\rangle}$ are 
comparable mass scales, then the supersymmetry breaking can take 
place at about $\sqrt{\langle F\rangle} \sim 10^{4}-10^{6}$~GeV.

%%%%%%%%%%%%%%%%%%%%%%%%%%%%%%%%%%%%%%%%%%%%%%%%%%%%%%%%%%%%%%%

\subsection{Supersymmetry breaking in the early Universe}

Non-zero inflationary potential gives rise to
supersymmetry breaking, the scale of which is given by the Hubble
parameter. At early times this breaking is dominant over breaking from
the hidden sector. After the end of inflation, in most models the inflaton
oscillates and its finite energy density still dominates and breaks
supersymmetry in the visible sector. Supersymmetry is broken also by quantum
mechanical effects but these are negligible compared to the classical
supersymmetry breaking from the non-zero energy density of the Universe.

%%%%%%%%%%%%%%%%%%%%%%%%%%%%%%%%%%%%%%%%%%%%%%%%%%%%%%%%%%%%%%

\subsubsection{Inflaton-induced terms}

The early Universe supersymmetry breaking can be transmitted to
the MSSM flat directions either
by renormalizable or non-renormalizable interactions \cite{dine96}.
However at least for a single flat direction, renormalizable
interactions do not lift the MSSM flat directions. In contrast,
the effective potential generated by non-renormalizable interactions
can induce a mass for the flat direction which is independent of the
field values as long as they are below the Planck scale.

At tree level $N=1$ SUGRA potential in four dimensions is
given by the sum of F and D-terms \cite{bailin94}
\begin{equation}
\label{sugrapot}
V=e^{K/M_{\rm P}^2}\left[\left(K^{-1}\right)^{j}_{i}F_{i}F^{j}-3\frac{
|W|^2}{M_{\rm P}^2}\right]+
\frac{g^2}{2}{\rm Re}f^{-1}_{ab}{\hat D}^{a}{\hat D}^{b}\,,
\end{equation}
where
\begin{equation}
F^{i}=W^{i}+K^{i}\frac{W}{M_{\rm P}^2}\,, \quad \quad
{\hat D}^{a}=-K^{i}(T^{a})^{j}_{i}\phi_{j}+\xi^{a}\,.
\end{equation}
where we have added the Fayet-Iliopoulos contribution $\xi^{a}$ to the
D-term. Here $K$ is the K\"ahler potential, which is a function of the
fields $\phi_i$, and $K^i\equiv \partial K/\partial \phi_i$, and 
${\rm Re}f^{-1}_{ab}$ is the inverse of the real part of the gauge 
kinetic function matrix.

A particular class of non-renormalizable interaction terms induced by the
inflaton arise if
the K\"ahler potential has a form \cite{gaillard95,dine96,bagger95}
\begin{equation}
\label{Iphicoupl}
K =\int d^4\theta \frac{1}{M_{\rm P}^2}(I^{\dagger}I)(\phi^{\dagger}\phi)\,,
\end{equation}
where $I$ is the inflaton whose energy density
$\rho \approx \langle \int d^4\theta I^{\dagger}I\rangle$ dominates
during inflation, and $\phi$ is the flat direction. The interaction
Eq. (\ref{Iphicoupl})
will generate an effective mass term in the Lagrangian in the global
supersymmetric limit, given by
\begin{equation}
{\cal L}= \frac{\rho_{I}}{M_{\rm P}^2}\phi^{\dagger}\phi =3H_I^2
\phi^{\dagger}\phi \,,
\end{equation}
where $H_I$ is the Hubble parameter during inflation.

%%%%%%%%%%%%%%%%%%%%%%%%%%%%%%%%%%%%%%%%%%%%%%%%%%%%%%
\subsubsection{Supergravity corrections}

In addition, there are
also inflaton-induced supergravity corrections
to the flat direction. By inspecting the supergravity potential,
one finds the following terms
\begin{eqnarray}
&&\left(e^{K(\phi^{\dagger}\phi)/M_{\rm P}^2}V(I)\right)\,,~~~~
\left(K_{\phi}K^{\phi \bar\phi}K_{\bar\phi}\frac{|W(I)|^2}{M_{\rm P}^4}
\right)\,, \nonumber \\
&&{\rm and} ~~
\left(K_{\phi}K^{\phi \bar I}D_{I}\frac{W^{\ast}(I)W(I)}{M_{\rm P}^2}
+{\rm h.c.}\right)\,.
\end{eqnarray}
Above
$D_{I}\equiv \partial/\partial I+K_{I}W/M_{\rm P}^2$. All these terms provide
a general contribution to the flat direction potential which
is of the form\cite{dine96}
\begin{equation}
\label{mflat}
V(\phi)=H^2M_{\rm P}^2 f\left(\frac{\phi}{M_{\rm P}}\right)\,,
\end{equation}
where $f$ is some function.
Note that this contribution exists also when
the flat direction is lifted by non-renormalizable
superpotential terms.

For a minimal choice of flat direction K\"ahler potential
$K(\phi^{\dagger},\phi)=\phi^{\dagger}\phi$, during inflation
the effective mass for the flat direction is found  to be \cite{dine96}
\begin{equation}
m^2_{\phi}=\left(2+\frac{F^{\ast}_{I}F_{I}}{V(I)}\right)H^2\,.
\end{equation}
Here it has been assumed that the main contribution to the inflaton
potential comes from the F-term. If there were D-term contributions
 $V_{D}(I)$ to the inflationary potential, then a correction of order
$V_{F}(I)/(V_{F}(I)+V_{D}(I))$ must be taken into account.
In purely D-term inflation there is no Hubble
induced mass correction to the flat direction during inflation
because $F_{I}=0$. However, when D-term inflation ends, the energy density
stored in the D-term is converted to an F-term and to kinetic energy of
the inflaton. Thus again a mass term $m_{\phi}^2 =\pm {\cal O}(1)H^2$
appears naturally, however the overall sign is undetermined \cite{kolda98}.

There are additional inflationary contributions to the potential if
the flat direction is lifted by the non-renormalizable operators discussed
earlier in this Section. These new terms come explicitly from the
superpotential part of the flat direction
\begin{eqnarray}
&&\left(W_{\phi}K^{\phi\bar\phi}K_{\bar\phi}\frac{W^{\ast}(I)}{M_{\rm P}^2} +
{\rm h.c.}\right)\,,~~\left(W_{\phi}K^{\phi \bar I}D_{\bar I}W^{\ast}(I)+
{\rm h.c.}\right)\,, \nonumber \\
&&{\rm and}~~\left(\frac{1}{M_{\rm P}^2}K_{I}K^{I\bar I}K_{\bar I}-
3\right)\left(\frac{W(\phi)^{\ast}W(I)}{M_{\rm P}^2}+{\rm h.c.}\right)\,.
\end{eqnarray}
The first one comes from the cross term  between the
derivative of the flat direction superpotential and the inflaton
superpotential, the second is due to the K\"ahler potential coupling
between the flat direction and the inflaton, and the third term is a
cross term between the two superpotentials. All these
terms give a generalized contribution equivalent to an $A$-term of the
MSSM:
\begin{equation}
V(\phi) =H M_{\rm P}^3 f\left(\frac{\phi^{d}}{M_{\rm P}^{d}}\right)\,,
\end{equation}
where $d$ is the dimensionality at which the flat direction is lifted.
The induced $A$-term has an important role to play during the evolution
of the flat direction. A possible $A$-term can also be generated from the
expansion of the K\"ahler potential for field values $I,\phi < M_{\rm P}$,
which is of the form \cite{dine96}
\begin{eqnarray}
\frac{1}{M_{\rm P}}\int d^4\theta I\phi^{\dagger}\phi+{\rm h.c.}\sim
\frac{F_{I}^{\dagger}F_{\phi}\phi}{M_{\rm P}}+{\rm h.c.}\,, \nonumber \\
\frac{1}{M_{\rm P}}\int d^2\theta I W_{i}+{\rm h.c.}\sim
\frac{F_{I}}{M_{\rm P}}W_{i} +{\rm h.c.}\,
\end{eqnarray}
Note that the $A$-terms arise only from terms with a linear coupling of the
inflaton superfield to a gauge invariant-operator $\phi_{i}$.
If $I$ were a composite field rather than a singlet, then such a term
will not arise and an $A$-term will not be generated.
Also, in the case of D-term inflation, the inflaton cannot
induce an $A$-term because $F_{I}=0$. More generally, if there is
a symmetry preventing a linear coupling of the inflaton, then order $H$
$A$-terms can be eliminated also in F-term inflation.
As long as the thermal bath of
the inflaton decay products dominates over the low energy supersymmetry
breaking scale, we should have Hubble induced corrections to
$m^2_{0}, m_{1/2},A$.

If there is a non-minimal dependence of the gauge superfield
kinetic terms on the inflaton field, a Hubble-induced gaugino mass
can also be produced. Generally the gauge superfield
kinetic terms must depend on the field(s) of the hidden
sector in order to obtain gaugino masses of roughly the same order as
(or larger than) scalar masses, as required by phenomenology. Having
$m_{1/2} \sim H$ thus appears to be quite natural unless an
$R$-symmetry forbids terms which are linear in the inflaton superfield
\cite{dine96}.

Since the $\mu$-term does not break supersymmetry, there is a priori
no reason to assume that a $\mu$-term of order $H$ will be created.
(For a discussion, see \cite{giudice88}). In what follows we will
treat $\mu$ as a free parameter.

So far we have not discussed the sign of the Hubble induced mass
correction. In fact with a general K\"ahler term either sign is
possible. Depending on the sign, the dynamical behavior of the AD
field is completely different and therefore the predictions depend crucially
upon the sign. There are however certain cases where the Hubble-induced terms
might not occur at all. An $R$-symmetry \cite{dine95,dine96}
or a special choice of the K\"ahler potential could forbids the AD field
getting the Hubble-induced mass correction \cite{stewart95,gaillard95}.

%%%%%%%%%%%%%%%%%%%%%%%%%%%%%%%%%%%%%%%%%%%%%%%%%%%%%%%%%%%%%%%%
\subsection{The potential for flat direction}

\subsubsection{F-term inflation}

Let us collect all the terms which contribute to the flat direction potential,
which in the case of F-term inflation can be written as \cite{dine95,dine96}
\begin{eqnarray}
\label{adpot0}
V(\phi)&=&-C_{I} H_{I}^2 {|\phi |}^2 + \left(a {\lambda}_d H {{\phi}^d \over d
M^{d-3}} + {\rm h.c.}\right) + m^{2}_{\phi}{|\phi|}^2
+ \left(A_{\phi} {\lambda}_d \frac{{\phi}^d}{dM^{d-3}} + {\rm h.c.}\right)\,,
\nonumber \\
&&+|\lambda|^2\frac{|\phi|^{2d-2}}{M^{2d-6}} \,.
\end{eqnarray}
The first and the third terms are the Hubble-induced and low-energy soft
mass terms, respectively, while the second and the fourth terms are the
Hubble-induced and low-energy $A$ terms. The last term is
the contribution from the non-renormalizable superpotential. The coefficients
$|C_{I}|,~a,~\lambda_{d}\sim {\cal O}(1)$, and the coupling 
$\lambda \approx 1/(d-1)!$. Note that low-energy $A_{\phi}$ term is 
dimensionful. 

Note here the importance of the relative sign of the coefficient
$C_{I}$. At large field values the first term dictates the
dynamics of the AD field. If $C_{I}<0$ , the absolute minimum
of the potential is   $\phi =0$ and during inflation the AD field will
settle down to the bottom of the potential roughly in one Hubble time.
In such case the AD field will not have any interesting classical dynamics.
Its presence would nevertheless be felt because of  quantum
fluctuations. These would be chi-squared in nature since then the classical
energy density of the AD field would be due to its own fluctuations.

If $C_{I} \ll 1$, the AD field takes some time to reach the bottom of
the potential, and if it has a non-zero amplitude after the end of
inflation, its dynamics is  non-trivial.

The most interesting scenario occurs when $C_{I}>0$.  In this
case the absolute value of the AD field settles during inflation to
the minimum given by
\begin{equation}
\label{veveq}
|\phi| \simeq \left({C_{I}\over (d-1) {\lambda}_d} H_{I}M^{d-3}\right)^{1/d-2}
\,.
\end{equation}
Here we have ignored the potential term $\propto a$;
if $C_{I}> 0$, the $a$-term will not change the vev
qualitatively. On the other hand, even for $C_{I}<0$ the potential
Eq.~(\ref{adpot0}) will have a minimum with a non vanishing vev if
$|a|^2 >4(d-1) C_{I}$. However the origin will also be a minimum in this
case. The dynamics then depends on which minimum
the AD field will choose during inflation.

The $a$-term in Eq.~(\ref{adpot0}) violates the global $U(1)$ symmetry
carried by $\phi$. If $|a|$ is ${\cal O}(1)$, the phase $\theta$ of
$\langle \phi \rangle$ is related to the phase of $a$ through $n
\theta + {\theta}_a = \pi$; otherwise $\theta$ will take some random
value, which will generally be of ${\cal O}(1)$. This is the initial
$CP$-violation which is required for baryogenesis/leptogenesis.
In practice, the superpotential term lifting the flat direction
is also the $B$ and $CP$ violating operator responsible for AD baryogenesis,
inducing a baryon asymmetry in the coherently oscillating $\phi$ condensate.

%%%%%%%%%%%%%%%%%%%%%%%%%%%%%%%%%%%%%%%%%%%%%%%%%%%%
\subsubsection{D-term inflation}

In D-term inflation one does not get the Hubble induced mass
correction to the flat direction so that $C_{I}=0$.
Also the Hubble induced $a$-term is absent. However the Hubble induced mass
correction eventually dominates once D-term induced inflation comes to an
end. The potential for a generic flat direction during
D-term inflation is given by
\begin{equation}
\label{adpot1}
V(\phi)=m^{2}_{\phi}{|\phi|}^2
+ \left(A_{\phi} {\lambda}_d \frac{{\phi}^d}{dM^{d-3}} + {\rm h.c.}\right)
+|\lambda|^2\frac{|\phi|^{2d-2}}{M^{2d-6}} \,,
\end{equation}
and after the end of inflation the flat direction potential
is given by \cite{kolda98}
\begin{equation}
\label{adpot2}
V(\phi)=\left(m^{2}_{\phi}-C H^2\right){|\phi|}^2
+ \left(A_{\phi} {\lambda}_d \frac{{\phi}^d}{dM^{d-3}} + {\rm h.c.}\right)
+|\lambda|^2\frac{|\phi|^{2d-2}}{M^{2d-6}} \,,
\end{equation}
where $C\sim {\cal O}(1)$. For $C$ positive, the flat direction settles
down to one of its minima given by Eq.~(\ref{veveq}) provided
$\phi\geq \sqrt{m_{\phi}M/\lambda}$, otherwise
\begin{equation}
\label{veveq1}
|\phi| \simeq \left(\frac{2C}{\lambda_{d}A_{\lambda}(d-1)}H(t)^2M^{d-3}
\right)^{1/d-2}\,,
\end{equation}
Note that in this case that the $A$-term is also responsible for
$B$ and/or $L$, and $CP$ violation. Another generic point to remember
is that in $R$-parity conserving models the $B$ and/or $L$ violating operators
must have even dimensions, so that $d=4$ yields  the minimal operator
for AD baryogenesis.

%%%%%%%%%%%%%%%%%%%%%%%%%%%%%%%%%%%%%%%%%%%%%%%%%%%%%%%%%%%%%%

\pagebreak

%%%%%%%%%%%%%%%%%%%%%%%%%%%%%%%%%%%%%%%%%%%%%%%%%%%%%%%%%%%%%%%
%%%%%%%%%%%%%%%%%%%%%%%%%%%%%%%%%%%%%%%%%%%%%%%%%%%%%%%%%%%%%%%
%%%%%%%%%%%%%%%%%%%%%%%%%%%%%%%%%%%%%%%%%%%%%%%%%%%%%%%%%%%%%%%

\section{Dynamics of flat directions}

After the end of inflation $\langle \phi \rangle$ continues first
to track the instantaneous local minimum of the scalar potential, obtained by
replacing $H_I$ with $H(t)$ in Eq.(\ref{veveq}) or by following
Eq.~(\ref{veveq1}) in the D-term inflation case.
Once $H \simeq m_0\sim m_{3/2}$, the low-energy soft terms take over.
Then $m_{\phi}^2$ becomes positive and $\langle\phi\rangle$ starts to
move in a non-adiabatic way (the phase of $\langle \phi \rangle$
differs from the phase of $A$-term during inflation). As a result
$\langle \phi\rangle$ begins a spiral motion in a complex plane,
which charges up the flat direction condensate, and eventually leads to
generation of a net baryon and/or lepton asymmetry~\cite{dine96}.

For baryogenesis purposes it is essential that the AD condensate obtains a
non-zero vev during the inflationary epoch. In Sect.~$4$, we pointed out
that a non-zero vev of the flat direction condensate is acquired only when the
negative $(\rm mass)^2$ contribution dominates the potential.
The MSSM flat directions which are made up of squarks and sleptons
have Yukawa and gauge interactions. The couplings render the evolution
of a particular flat direction non-trivially, especially when the flat
direction has a time varying mass due to the Hubble expansion
\cite{gaillard95,dine96,campbell99,allahverdi02}.
Moreover, if thermalization is not instantaneous, thermal effects from
reheating can be substantial and might trigger the motion of the flat
direction at an earlier time, there by changing the evolution of the
flat direction condensate in a significant way
\cite{allahverdi00579,anisimov01}.

%%%%%%%%%%%%%%%%%%%%%%%%%%%%%%%%%%%%%%%%%%%%%%%%%%%%%
\subsection{Running of the couplings}

%%%%%%%%%%%%%%%%%%%%%%%%%%%%%%%%%%%%%%

\subsubsection{Running of gravitational coupling}

Any flat direction has two kinds of interactions: renormalizable gauge
or Yukawa interactions, and a non-trivial coupling to the curvature. Both
types of interactions contribute to the logarithmic running of
$({\rm mass})^2$ of the flat direction condensate. The coupling to
the curvature is generic because in principle any scalar field in an expanding
background receives a contribution from the curvature by virtue
of the Lagrangian term $\xi R\phi^2$, where $\xi$ is a coupling
constant. Note that $R\propto +H^2$ in an expanding background.
Any scalar field always gets an additional positive Hubble induced
mass correction, provided $\xi$ is positive. The fundamental theory
might have a conformal invariance, in which case the coupling strength
$\xi=1/6$~\cite{birrell82}, but it is known that conformal invariance is
not protected by any symmetry,  and that quantum corrections always
break conformal invariance. Especially for the flat direction condensate,
spontaneously broken supersymmetry induces soft supersymmetry breaking
terms which break conformal invariance, and the
value of $\xi$ remains undetermined.

It is of course possible to simply set $\xi=0$. If initially
$\xi=0$ at some high scale, renormalization effects
due to scalar field self-interaction will nevertheless generate a non-zero
$\xi$ at lower scales. In an expanding Universe the value of
$\xi$ also changes under the influence of a varying curvature
(see~\cite{buchbinder92}, and references therein). In the
simplest case of a single scalar field with a quartic
self-interaction strength $\lambda$ leads (at one-loop level)
to a logarithmically running $\xi$~\cite{parker84}
\begin{equation}
\label{curvrun}
\xi_{eff}=\xi +\left(\xi-\frac{1}{6}\right)\frac{12}{4\pi^2}\lambda
\ln\left(\left|\frac{m^2+\left(\xi-\frac{1}{6}\right)R}{m^2}\right|\right)\,.
\end{equation}
It is obvious that $\xi=1/6$ is a fixed point of the RG equation.
If the theory has fermions and gauge fields, then obviously
the coefficient in front of the logarithmic term in Eq.~(\ref{curvrun})
will be modified \cite{buchbinder92}.

As we have seen in Sect.~$4.2$, when
supersymmetry is promoted to a local theory, a supergravity correction
is induced to the flat direction which is proportional to the curvature,
and supergravity theories also allow for $\xi R\phi^2$ (e.g.
superconformal supergravity \cite{kallosh00}).

In the context of  MSSM flat directions we have implicitly assumed
$\xi=0$. This is justified from the very definition of
F- and D-flat directions. The only leading order self
coupling term in the flat direction potential is the Hubble
induced A-term in Eq.~(\ref{adpot0}). The overall self coupling constant
is relatively large when the flat direction is lifted at $d=4$, i.e.
the suppression is proportional to ${\cal O}(1)(H/M_{\rm P})$, where we
have replaced $M$ by $M_{\rm P}$ in Eq.~(\ref{adpot0}). In any inflation
model the ratio $H_{I}/M_{\rm P} \ll 1$, which in conjunction with
Eq.~(\ref{curvrun}), suggests that the effect of running on $\xi$
is minimal. For a running $\xi$ the curvature term in Eq.~(\ref{curvrun})
dominates over the mass term. This might not be the case with the flat
direction condensate because the
condensate also receives a field dependent mass while it is evolving.
As long as the vev dependent mass is larger than the curvature induced
mass, the running of any parameter in the theory will be dictated mainly
by the renormalizable quantum effects. We therefore conclude that the
running of $\xi$ can be neglected.

When the field dependent mass of the flat direction field becomes of order
$m^2\sim {\cal O}(H^2)$, it might be prudent to start worrying about the
curvature induced term, especially during inflation. A simple
inspection of Eq.~(\ref{curvrun}) suggests that $\xi$ is always of order
$\ln ({\cal O}(1))$, with virtually no alteration in $\xi_{eff}$. From
now onwards we fix the non-minimal coupling to be $\xi =0$.

%%%%%%%%%%%%%%%%%%%%%%%%%%%%%%%%%%%%%%%%%%%%%%%%%%%%%%%%

\subsubsection{Renormalization group equations in the MSSM}

Let us consider the running of the flat direction $(\rm mass)^2$
below $M_{\rm GUT}$ by assuming that it is the scale where
supersymmetry breaking is transmitted to the visible sector, in
order to avoid uncertainties about physics between $M_{\rm GUT}$ and
$M_{\rm P}$. The running of low-energy soft breaking masses has
been studied in great detail in the context of MSSM phenomenology
\cite{drees95}, in particular in connection with radiative electroweak
symmetry breaking \cite{ewsb}.

Let us recall some of the salient features of the MSSM one-loop
RG equations. The ones relevant to flat directions involve the Higgs
doublet $H_u$ which couples to the top quark, the right-handed stop
$\widetilde{u}_3$, the left-handed doublet of third generation squarks
$\widetilde{Q}_3$ and the $A-$parameter $A_t$ associated with the top
Yukawa interaction. The RG equations read \cite{nilles84}
\begin{eqnarray}
\label{betafct}
{d \over dq} m^2_{H_u} &=& {3h^2_t \over 8 \pi^2}
\left(m^2_{H_u} + m^2_{\widetilde{Q}_3} + m^2_{\widetilde{u}_3} +
|A_t|^2 \right)
- {1 \over 2 \pi^2} \left({1 \over 4} g^2_1 |m_1|^2 + {3 \over 4}
g^2_2 |m_2|^2 \right) \,, \nonumber \\
{d \over dq}m^2_{\widetilde{u}_3} &=& {2h^2_t \over 8 \pi^2}
\left(m^2_{H_u} + m^2_{\widetilde{Q}_3} + m^2_{\widetilde{u}_3} + |A_t|^2
\right)
- {1 \over 2 \pi^2}\left({4 \over 9} g^2_1 |m_1|^2 + {4 \over 3}
g^2_3 |m_3|^2 \right) \,, \nonumber \\
{d \over dq} m^2_{\widetilde{Q}_3} &=& {h^2_t \over 8 \pi^2}
\left(m^2_{H_u} + m^2_{\widetilde{Q}_3} + m^2_{\widetilde{u}_3} + |A_t|^2
\right)\,
- {1 \over 2 \pi^2}\left({1 \over 36} g^2_1 |m_1|^2 + {3 \over 4}
g^2_2 |m_2|^2 + {4 \over 3} g^2_3 |m_3|^2 \right) \, ,
\nonumber \\
{d \over dq} A_t &=& {3 h^2_t \over 8 \pi^2} A_t - {1 \over 2 \pi^2}
\left( {13 \over 36} g^2_1 m_1 + {3 \over 4} g^2_2 m_2 + {4 \over 3}
g^2_3 m_3 \right) \, .
\end{eqnarray}
Here $q$ denotes the logarithmic scale; this could be an external
energy or momentum scale, but in the case at hand the relevant scale
is set by the vev(s) of the fields themselves. $h_t$ is the
top Yukawa coupling, while $g_i$ and $m_i$ are respectively the gauge
couplings and soft breaking gaugino masses of
$U(1)_Y \times SU(2)\times SU(3)$ . If $h_t$ is the only
large Yukawa coupling (i.e. as long as $\tan\beta$ is not very large),
the beta functions for $({\rm mass})^2$ of squarks of the first and
second generations and sleptons only receive significant
contributions from gauge/gaugino loops. A review of these effects can
be found in \cite{drees95}. Here we only mention the main results for
the case of universal boundary conditions, where at $M_{\rm GUT}$
all the scalar masses are $m^2_0$ and the gauginos have a common
soft breaking mass $m_{1/2}$. For a low value of $\tan\beta =1.65$
\footnote{This value corresponds to the case of maximal top
Yukawa coupling, so called fixed point scenario \cite{hill81,carena95}.
Such a low value of $\tan\beta$ is excluded by Higgs searches at LEP
\cite{lephiggs01}, unless one allows stop masses well above $1$~TeV.
We nevertheless include this scenario in our discussion since it
represents an extreme case.},
\begin{equation}
m^2_{H_u} \simeq - \frac{1}{2} m^{2}_{0} - 2 m^{2}_{1/2}
\end{equation}
at the weak scale, while $m^2_{\widetilde u_3}$ and $m^2_{\widetilde Q_3}$
remain positive. The soft breaking $({\rm mass}) ^2$ of the first and
second generations of squarks is $\simeq m^{2}_{0} + (5-7)m^{2}_{1/2}$,
while for the right-handed and left-handed sleptons one gets
$\simeq m^{2}_{0} + 0.1 m^{2}_{1/2}$ and $\simeq m^{2}_{0} + 0.5m^{2}_{1/2}$,
respectively. The important point is that the sum $m^{2}_{H_u} + m^2_L$,
which describes the mass along the $H_u L$ flat direction, is driven to
negative values at the weak scale only for $m_{1/2} \gsim m_0$. This is
intuitively understandable, since Eqs.(\ref{betafct}) have a fixed point
solution \cite{carena95}
$m^{2}_{H_u} + m^{2}_{\widetilde{u}_3} + m^{2}_{\widetilde{Q}_3} = A_t = 0$
when $m_{1/2} =0$.

%%%%%%%%%%%%%%%%%%%%%%%%%%%%%%%%%%%%%%%%%%%%%%%%%%%%%%%%%%

\subsection{Hubble induced radiative corrections}

Here we describe radiative corrections in a cosmological set-up relevant
for the AD mechanism~\cite{allahverdi02}. When the Hubble induced
supersymmetry breaking is dominant, i.e. for $H > \cal{O}(\rm TeV)$,
the evolution of the soft terms is different from the vacuum RG equations
given in \eq{betafct}. For the low-energy supersymmetry breaking case,
constraints from the weak scale (e.g. realization of electroweak
symmetry breaking, and experimental limits on the sparticle masses) give
information about the soft breaking parameters $m^2_0$ and $m_{1/2}$.
Together with fine tuning arguments, these constraints imply that
$m^2_0 > 0$ and that $m_0,~m_{1/2}$ are $\cal{O}(\rm TeV)$. In the
Hubble induced supersymmetry breaking case $m^2_0$ and $m_{1/2}$ are
determined by the scale of inflation (and the form of the K\"ahler
potential). At low scales the Hubble induced terms are completely
negligible because at temperature $T \sim M_W$,  $H \sim {\cal O}(1)$~eV;
at present the Hubble parameter is tiny, $H_0 \sim {\cal O}(10^{-33})$~eV.

There exists an even more fundamental difference between the Hubble induced
and ordinary radiative corrections. In Minkowski space the loop
contributions to beta functions freeze at a scale of the order of the
mass of the particles in the loop. In an expanding Universe the horizon radius
$\propto H^{-1}$ defines an additional natural infrared cut-off for
the theory. The masses of particles coupled to the flat direction  receive
contributions from two sources. There is a supersymmetry preserving part
proportional to the vev $\langle \phi \rangle$, and the Hubble induced
supersymmetry breaking part. The loop contributions to
beta functions should thus be frozen at a scale given by the largest of
$|\langle \phi \rangle|$ and $H$ (recall that $h_t$ and gauge
couplings are close to one). In particular, if the squared mass of the
flat direction condensate is positive at very large scales but turns
negative at some intermediate scale $Q_c$, the origin of the flat direction
potential will cease to be a minimum, provided the Hubble parameter is
less than $Q_c$. On the
other hand, if $m_\phi^2 < 0$ at the GUT scale, its running should
already be terminated at the scale $|\langle \phi \rangle|$ determined
by Eq.(\ref{veveq}).\footnote{Here we note that the Hubble cut-off
usually plays no role in loop corrections to the inflaton
potential. In most inflation models the masses of the fields which
may run in the loop are larger than the Hubble expansion rate due to
the presence of a finite coupling to the inflaton. This will happen
if the (time varying) inflaton vev is large and the couplings are not
very small. In those cases, which are somewhat similar to our case with
$C_I > 0$, one can trust the usual loop calculation evaluated in a flat
space time background \cite{dvali94}.}

In the following two subsections we discuss  separately the cases of positive
and negative GUT-scale $({\rm mass})^2$ for the flat direction condensate.

%%%%%%%%%%%%%%%%%%%%%%%%%%%%%%%%%%%%%%%%%%%%%%%%%%%%%

\subsubsection{The case with $C_I \approx -1$}

\begin{table} \label{table2}
\vspace{3mm}
\begin{tabular}{|c|c||c|c|}\hline
$A_t/H$ & $m_{1/2}/H$ & $Q_c(h_t = 2)$ & $Q_c(h_t = 0.5)$ \\
\hline
$ +1/3$ ($-1/3$)& $1/3$ & $\times $ & $\times $ \\
$+1/3$ ($-1/3$)& $1$ & $10^6-10^7$ & $10^3$ \\
$+1/3$ ($-1/3$)& $3$ & $10^{11}$ & $10^6-10^7$ \\
\hline
$+1$ ($-1$)& $1/3$ & $\times$  & $\times$  \\
$+1$ ($-1$)& $1$ & $10^6-10^7$ & $10^5$ ~ ($\times$) \\
$+1$ ($-1$)& $3$ &  $10^{11}$ &  $10^8$ ~ ($10^6$) \\
\hline
$+3$ ($-3$)& $1/3$ & $\times$  & $10^7$ ~  \\
$+3$ ($-3$)& $1$ & $10^{14}$ ~ ($10^7$) & $10^9$ ~ ($10^3$) \\
$+3$ ($-3$)& $3$ & $10^{15}$ ~ ($10^{11}$) & $10^{10}$ ~ ($10^6$)\\
\hline
\end{tabular}
\caption{The scale $Q_c$ (in GeV) where the squared mass of
the $H_u L$ flat direction changes sign, shown for
$C_I =-1$ and several values for the ratios $A_t/H$ and $m_{1/2}/H$ as well
as the top Yukawa coupling $h_t$, all taken at scale
$M_{\rm GUT} = 2 \cdot 10^{16}$ GeV, from \cite{allahverdi02} }
\end{table}

In this case all scalar fields roll towards the origin very rapidly
and settle there during inflation, provided radiative corrections to their
masses are negligible. A typical flat direction condensate $\phi$ is a linear
combination $\phi = \sum_{i=1}^N a_i \varphi_i$ of the MSSM scalars
$\varphi_i$, implying that
$m^{2}_{\phi} = \sum_{i=1}^N |a_i|^2 m^{2}_{\varphi}$. In \cite{allahverdi02},
it was noticed that with small values for the $\mu$ parameter,
the running of $m^{2}_{\phi}$ crucially depends on $m_{1/2}$.

Let us consider sample cases with gaugino masses $m_{1/2} = (H;~3H;~H/3$), the
A-term\footnote{The RG equations (\ref{betafct}) for $A_t$ show that the
relative sign between $A_t$ and $m_{1/2}$ matters, since it affects the
running of $|A_t|$, and subsequently, scalar soft masses.
Without loss of generality we take the common gaugino mass $m_{1/2}$ to
be positive.}
$A_t(M_{\rm GUT})=(\pm H;~\pm 3H;\pm H/3~)$, top Yukawa
$h_t(M_{\rm GUT}) = (2, \, 0.5)$ and couplings
$g_1(M_{\rm GUT})=g_2(M_{\rm GUT})=g_3(M_{\rm GUT})=0.71$, and
follow the running of scalar soft masses from $M_{\rm GUT}$ down to
$10^3$ GeV, where  low-energy supersymmetry breaking becomes
dominant. The main result is that only the $LH_{u}$ flat direction can
acquire a negative $({\rm mass})^2$ at low scales. In this case
$m^2_{\phi} = {(m^2_{H_u} + m^2_{L} + \mu^2)/ 2}$, where the last term
is from the Hubble induced $\mu$ term. The results are summarized in
Table 2, where it has been assumed that $\mu (M_{\rm GUT})\lsim H/4$
so that the the $\mu$-term contribution to $m^2_\phi$ is
negligible. In general $m^2_\phi$ changes sign at a higher scale for
$h_t(M_{\rm GUT}) = 2$. This is expected since a large Yukawa
coupling naturally maximizes the running of $m^2_{H_u}$. Furthermore,
the difference between $A_t/m_{1/2} < 0$ and $A_t/m_{1/2} > 0$ becomes
more apparent as $|A_t/m_{1/2}|$ increases and $h_t$ decreases. The
quasi fixed-point value of $A_t/m_{1/2}$ is positive
\cite{hill81,carena95}. Positive input values of $A_t$ will thus lead
to positive $A_t$ at all scales, but a negative $A_t(M_{\rm GUT})$ implies
that $A_t \simeq 0$ for some range of scales, which diminishes its effect
in the RG equations, see Eq.~(\ref{betafct}). The sign of $A_t(M_{\rm GUT})$ is
more important for smaller $h_t$, since then $A_t/m_{1/2}$ will evolve
less rapidly.

It was noticed in~\cite{allahverdi02} that the squared mass of
the $H_u L$ flat direction does not change sign when $m_{1/2} = H/3$,
except for\footnote{For this choice of parameters, $A_t$ runs initially
very slowly. It will therefore remain large for some time and helps
$m^{2}_{H_u}$ to decrease quickly towards lower scales.} $A_t = \pm 3H$
and $h_t = 0.5$. This can be explained by the fact that for small $m_{1/2}$
and small or moderate $|A_t|$ we are generally close to the fixed
point solution
\begin{equation}
\label{fixeq}
m^2_{H_u} \simeq -{1 \over 2} H^2; ~m^2_{\widetilde{u}_3} \simeq
0; ~m^2_{\widetilde{Q}_3} \simeq {1 \over 2} H^2.
\end{equation}
Nevertheless, even for $m_{1/2}\ll H$ the squared mass of the
$LH_{u}$ flat direction as well as $m^2_{\widetilde{u}_3}$ are $< 0.2H^2$
above $1$~TeV, because of the fixed point behavior. This implies that
the $LH_{u}$ flat direction can still be viable for baryogenesis, as
pointed out by McDonald \cite{mcdonald99456}. Flat directions built
out of $\widetilde u_3$ will be marginal at best, since the decrease
in $m^2_{\widetilde u_3}$ will be counteracted by other contributions to
$m^2_\phi$; e.g. for the $\bar u_3 \bar d_1 \bar d_2$ flat direction we
find $m^2_\phi > 2 H^2 / 3$ at all scales.

The AD mechanism for baryogenesis should always work if $Q_c > H_I$,
since in that case the global minimum of the potential during inflation
is located at $|\langle \phi \rangle | \not= 0$. Note that in this case the vev
$|\langle \phi \rangle|$ is usually determined by $Q_c$ rather than by
Eq.~(\ref{veveq}).

For scales close to $Q_c$ the mass term in the
scalar potential Eq.~(\ref{adpot0}) can be written as $\beta_\phi
H^2 |\phi|^2 \log({|\phi|}/{Q_c})$, where the coefficient
$\beta_\phi$ can be obtained from the RG equations. If $\beta_\phi > 0$, which
is true for the $H_u L$ flat direction for $C_I < 0$, this term will
reach a minimum at $ \log({|\phi|}/{Q_c}) = -1$. If $Q_c < (H_I
M^{d-3}_{\rm GUT})^{1/d-2}$ the non-renormalizable contributions to
the scalar potential are negligible for $|\phi| \sim Q_c$, so that the
minimum of the quadratic term essentially coincides with the minimum
of the complete potential given by Eq.~(\ref{adpot0}). In models of
high scale inflation (e.g. chaotic inflation models), the Hubble
constant during inflation $H_I$ can be as large as $10^{13}$ GeV.
This implies that $m^2_\phi$ for the $H_u L$ flat direction can
only become negative during inflation if $m^2_{1/2} \gg H^2$, which
includes the ``no-scale'' scenario studied in~\cite{gaillard95}.

The region of the parameter space safely allowing AD leptogenesis is much
larger in models of intermediate and low scale inflation (e.g. some new
inflation models \cite{linde82108}) where $H_I$ is substantially smaller.
In such models one can easily have $H_I < Q_c$ at least for the $H_u L$ flat
direction, unless $m^2_{1/2} \ll H^2$ or $\mu^2 \gsim m^2_{1/2}$
\cite{allahverdi02}.

If $Q_c < H_I$, the condensate $\phi$ settles at the origin during
inflation and its post-inflationary dynamics will depend on the process of
thermalization. If the inflaton decay products thermalize very slowly,
$m^{2}_{\phi}$ is only subject to zero-temperature radiative
corrections and $\langle \phi \rangle$ can move away from the origin
once $H \lsim Q_c$; a necessary condition for this scenario is that
inflatons do not directly decay into fields that are charged under
$SU(3) \times SU(2) \times U(1)_Y$. If $Q_c \gg 1$ TeV, $\phi$ will
readily settle at the new minimum and AD leptogenesis can work.

The situation will be completely different if inflatons
directly decay into some matter fields. In such a case the plasma of
inflaton decay products has a temperature
$T\sim ({\Gamma}_{d} H M^{2}_{\rm Planck})^{1/4}$ \cite{kolbturner90}
(${\Gamma}_{d}$ is the inflaton decay rate). Fields which
contribute to the running of $m^2_\phi$ are in thermal equilibrium
(recall that the flat direction field is stuck at $\phi= 0$) and their back
reaction results in thermal corrections of order $+T^2$ to $m^2_\phi$.
For generic models of inflation $T > H$, implying that thermal
effects exceed radiative corrections. Therefore $\langle \phi \rangle$
remains at the origin at all times and AD leptogenesis will not work.

%%%%%%%%%%%%%%%%%%%%%%%%%%%%%%%%%%%%%%%%%%%%%%%%%%%%%%%%%%%%%%%%

\subsubsection{The case with $C_I \approx +1$}

In this case all flat directions are viable for baryogenesis purposes
provided the running of $m^{2}_{\phi}$ is negligible. Radiative
corrections may change the sign (in this case to positive) at small vev(s),
resulting in the entrapment of $\phi$ at the origin.

A quantitative study of the sample cases discussed above can be
summarized as follows~\cite{allahverdi02}. The squared mass of the
$LH_{u}$ flat direction is always negative at small
scales, unless $\mu^2 \gsim H^2/2$. For $m_{1/2} = 3H$,
$m_\phi^2$ changes sign twice; it is positive for scales $Q$ between
roughly $10^{14}$ and $10^6$ GeV, the precise values depending on
$h_t$ and $A_t$. Slepton masses only receive positive contributions from
electroweak gauge/gaugino loops. As a result, the squared mass of the
$LL\bar e$ flat direction remains negative down to
1 TeV, unless $m_{1/2} > 2 H$; for $m_{1/2} \gsim 3 H$, $Q_c \gsim 10^9$ GeV
even for this flat direction. The squared masses of all
squarks (except $\widetilde{u}_3$) change sign at $Q_c > 1$ TeV unless
$m_{1/2} \lsim H/3$; we find $Q_c \simeq 10^{10} \ (10^{15})$ GeV for
$m_{1/2} / H = 1 \ (3)$. This is due to the large positive
contribution $\propto m_3^2$ to the squared squark masses at scales
below $M_{\rm GUT}$. The corresponding values for the
$\bar u_3 \bar d_i \bar d_j$ and $L Q \bar d$ flat directions are
usually somewhat smaller, due to the
Yukawa terms in the $\beta-$function and the slower running of the slepton
masses, respectively; however, the values of $Q_c$ listed in Table 2
are still a fair approximation.

The positive contribution to the scalar potential from
the non-renormalizable superpotential term now dominates
$-H^2$ (see Eq. (\ref{veveq})). If $Q_c > (H_I M^{d-3}_{\rm GUT})^{1/d-2}$,
$m^2_\phi$ is positive for all vev(s) and hence the flat direction
will settle at the origin during inflation and remain there.
In such a case the flat direction is not suitable for AD
baryogenesis. This can easily happen for flat directions involving
squarks in models with low scale inflation, but is not likely for high
scale inflation models (unless $m_{1/2} \gsim 3H$). For
$H_I < Q_c <(H_I M^{d-3}_{\rm GUT})^{1/d-2}$, feasible for some
flat directions in both intermediate/high scale and low scale models.
During inflation the potential has two minima, at $\langle \phi \rangle = 0$
and at $|\langle \phi \rangle | \sim (H_I M^{d-3}_{\rm GUT})^{1/d-2}$.
Depending on the initial conditions, $\phi$ can roll
down towards either of them and settle there but only the latter one
will be useful for AD baryogenesis.

If $Q_c < H_I$, the flat direction condensate  will settle at the
value determined by
Eq.~(\ref{veveq}) (the only minimum during inflation) and remain there.
The appearance of another minimum at the origin after inflation,
which is possible once $H < Q_c$, does not change the situation
since these minima are separated by a barrier. In this
case radiative corrections will not change the picture qualitatively;
however, they will still modify the quantitative analysis,
since $C_I$ in Eq.(\ref{veveq}) will become scale-dependent.

In brief, the main conclusion is that among the flat
directions $LH_{u}$ is the only robust one in the sense that it
gives rise to AD leptogenesis  independently of the sign of $C_{I}$.

%%%%%%%%%%%%%%%%%%%%%%%%%%%%%%%%%%%%%%%%%%%%%%%%%%%%%%%%%%%%%%%%%%

\subsubsection{Running of the flat direction field in no-scale supergravity}

So far we have dealt with a minimal choice of the K\"ahler potential.
An alternative is non-flat K\"ahler potential; an example of
this is provided by e.g. no-scale models, for which
$K\sim \ln (z+z^*+\phi_i^\dagger\phi)$, where $z$ belongs to supersymmetry
breaking sector, and $\phi_{i}$ belongs to the matter sector, and both are
measured in terms of reduced Planck mass
(for a review, see~\cite{lahanas87}). In no-scale models there
exists an enhanced symmetry known as the Heisenberg symmetry
\cite{binetruy87}, which is defined on the chiral fields
as $\delta z=\epsilon^{\ast}\phi^{i}$, $\delta\phi^{i}=\epsilon^{i}$,
and $\delta y^{i}=0$, where $y^{i}$ are the hidden sector fields, such
that the combinations $\eta =z+z^{\ast}-\phi_{i}^{\ast}\phi^{i}$,
and $y_{i}=0$ are invariant. For a especial choice
\begin{equation}
K=f(\eta)+\ln[W(\phi)/M_{\rm P}^3]^2+g(y)\,,
\end{equation}
The $N=1$ supergravity potential reads~\cite{gaillard95,campbell99}
\begin{equation}
V=e^{f(\eta)+g(y)}\left[\left(\frac{f^{\prime 2}}{f^{\prime \prime}}-3\right)
\frac{|W|^2}{M_{\rm P}^2}-\frac{1}{f^{\prime 2}}\frac{|W_{i}|^2}{M_{\rm P}^2}
+g_{a}(g^{-1})^{a}_{b}g^{b}\frac{|W|^2}{M_{\rm P}^2}\right]\,.
\end{equation}
Note that there is no cross term in the potential such as
$|\phi^{\ast}_{i}W|^2$. As a consequence any tree level flat direction
remains flat even during inflation \cite{gaillard95} (in fact
it is the Heisenberg symmetry which protects the flat directions from obtaining
Hubble induced masses~\cite{binetruy87}). The symmetry
is broken by gauge interactions or by coupling in the renormalizable
part of the K\"ahler potential. Then the mass of the flat direction
condensate arises from the running of the gauge couplings.

For $f(\eta)=-3\ln\eta$, the one-loop corrected supergravity induced
mass term has been calculated in ~\cite{ellis84134,gaillard94,gaillard95},
which gives an effective mass for the flat direction field in the
presence of finite
energy density stored in the inflaton sector. The typical mass of
the flat direction has been computed and comes out to be
$m^2_{\phi}\sim 10^{-2}H^2$ during inflation. The only constraint
is that flat direction must not involve stops~\cite{gaillard95}.

%%%%%%%%%%%%%%%%%%%%%%%%%%%%%%%%%%%%%%%%%%%%%%%%%%%%%%%%%%%%%%%%

\subsection{Post-inflationary running of the flat direction}

Now we focus on the running of the flat direction after inflation.
Here we must take into account the low energy
supersymmetry breaking effects. In particular, the running of the
condensate mass will depend on how supersymmetry is transmitted to the visible
sector.

%%%%%%%%%%%%%%%%%%%%%%%%%%%%%%%%%%%%%%%%%%%%%%%%%%%%%%%%%%%%%%%%%

\subsubsection{Gravity mediated supersymmetry breaking}

For gravity-mediated supersymmetry breaking the scalar potential along a
flat direction has been evaluated as
\cite{enqvist98,enqvist99}
\be{pot}
U(\Phi) \approx m_{\phi}^{2}\left(1 +K \log\left(\frac{|\Phi|^{2}}{M^{2}}
\right) \right) |\Phi|^{2}+ \frac{\lambda^{2}|\Phi|^{2(d-1)}
}{M_{\rm P}^{2(d-3)}} + \left( \frac{A_{\lambda}
\lambda \Phi^{d}}{d M_{\rm P}^{d-3}} + h.c.\right)~,
\ee
where $m_{\phi}$ is the conventional gravity-mediated soft supersymmetry
breaking scalar mass term ($m_{\phi} \approx 100 \GeV$), $K$ is a parameter
which depends on the flat direction, and the logarithmic contribution
parameterizes the running of the flat direction potential with
$M=(M_{\rm P}^{d-3}m_{3/2}/|\lambda|)^{(1/d-2)}$. In the gravity mediated case
$|A_{\lambda}|< dm_{3/2}$, for $d=4,6$.

$K$ can be computed from the RG equations, which to one loop have
the form
\be{rge}
{\partial m_i^2\over \partial t}=\sum_{g}a_{ig} m_g^2+
\sum_a h^2_a(\sum_j b_{ij}m_j^2+A^2)~,
\ee
where $a_{ig}$ and $b_{ij}$ are constants, $m_g$ is the gaugino mass,
$h_a$ the Yukawa coupling, $A$ is the $A$-term, and $t=\ln M_X/q$. The
full RG equations have been listed in \cite{nilles84}. The potential
along the flat direction is then characterized by the amount
of stop mixture (where appropriate), the values of gluino mass and $A$,
and in the special case of the $d=4$~ $H_uL$ -direction, on the $H_u H_d$
-mixing mass parameter $\mu_H$.

The mass of the AD scalar $\phi$ is the sum of the masses of the
squark and slepton fields $\phi_i$ constituting the flat direction,
$m_S^2 = \sum_{a} p_i^2 m_i^2~,$ where $p_i$ is the projection of
$\phi$ along $\phi_i$, and $\sum p_i^2=1$. The parameter $K$ is then
given simply by
\be{K1}
K=\frac{1}{q^2}{\partial m_S^2\over \partial t}\Big\vert_{t={\log} q}~.
\ee
To compute $K$, one has to choose the scale $q$. The appropriate scale
is given by the value of the AD condensate amplitude when it first begins
to oscillate at $H\approx m_{\phi}$ or
\be{phi}
Q=|\phi_0| = \left[\frac{m_{\phi}^2 M^{2(d-3)}}{(d-1)\lambda^2}\right]
^{\frac{1}{2(d-2)}}~,
\ee
The RG running of the flat directions in the case of gravity
mediated supersymmetry breaking was studied
in~\cite{enqvist00483,enqvist0163}, where unification
at $t=0$ was assumed and all the other Yukawa couplings except the top
Yukawa were neglected.

%%%%%%%%%%%%%%%%%%%%%%%%%%%%%%%%%%%%%%%%%%%%%%%%%%%%%%%%%%%%%%%%%%%%%%%%%
\begin{figure}
\leavevmode
\centering
\vspace*{110mm}
\includegraphics{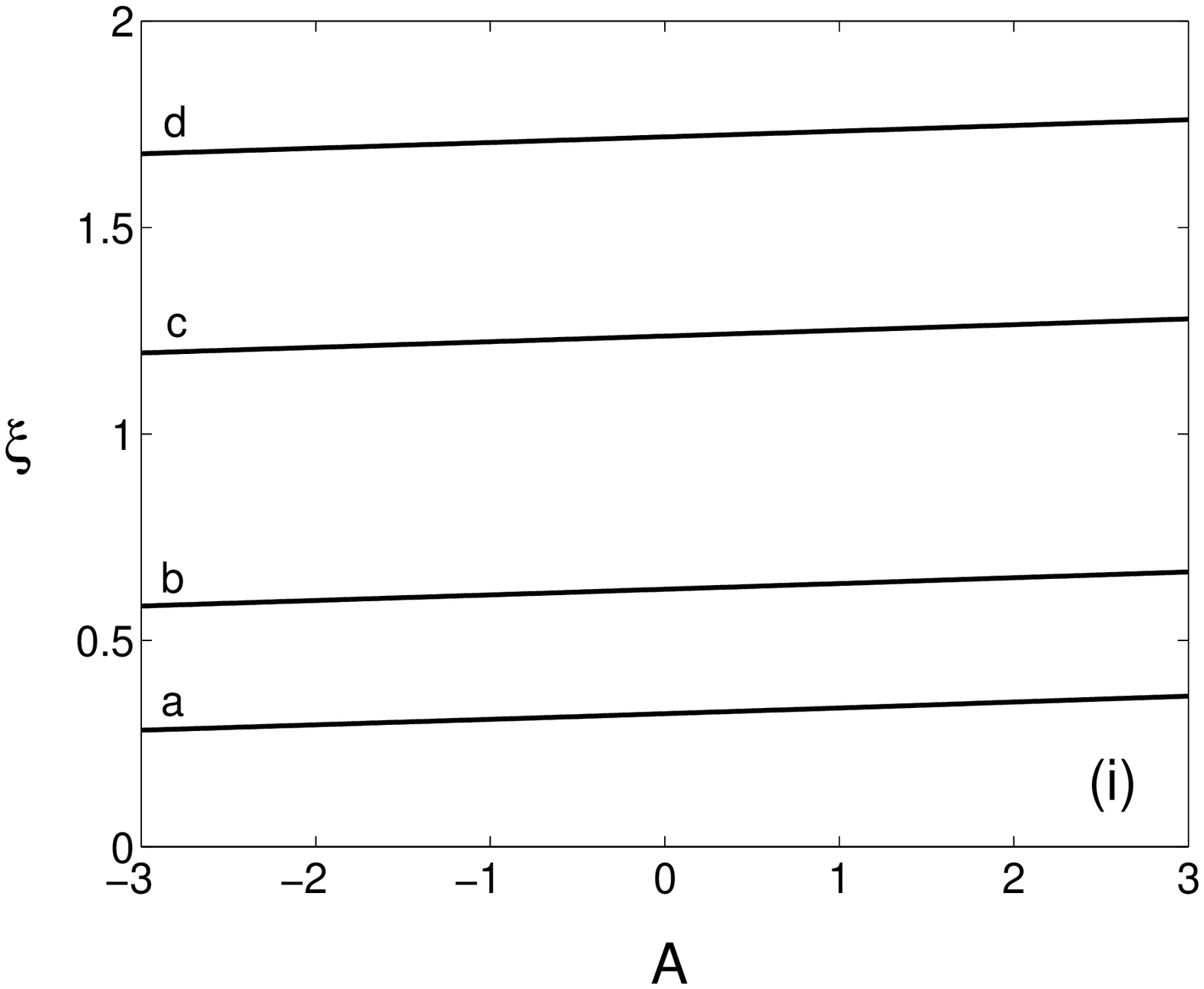}
\includegraphics{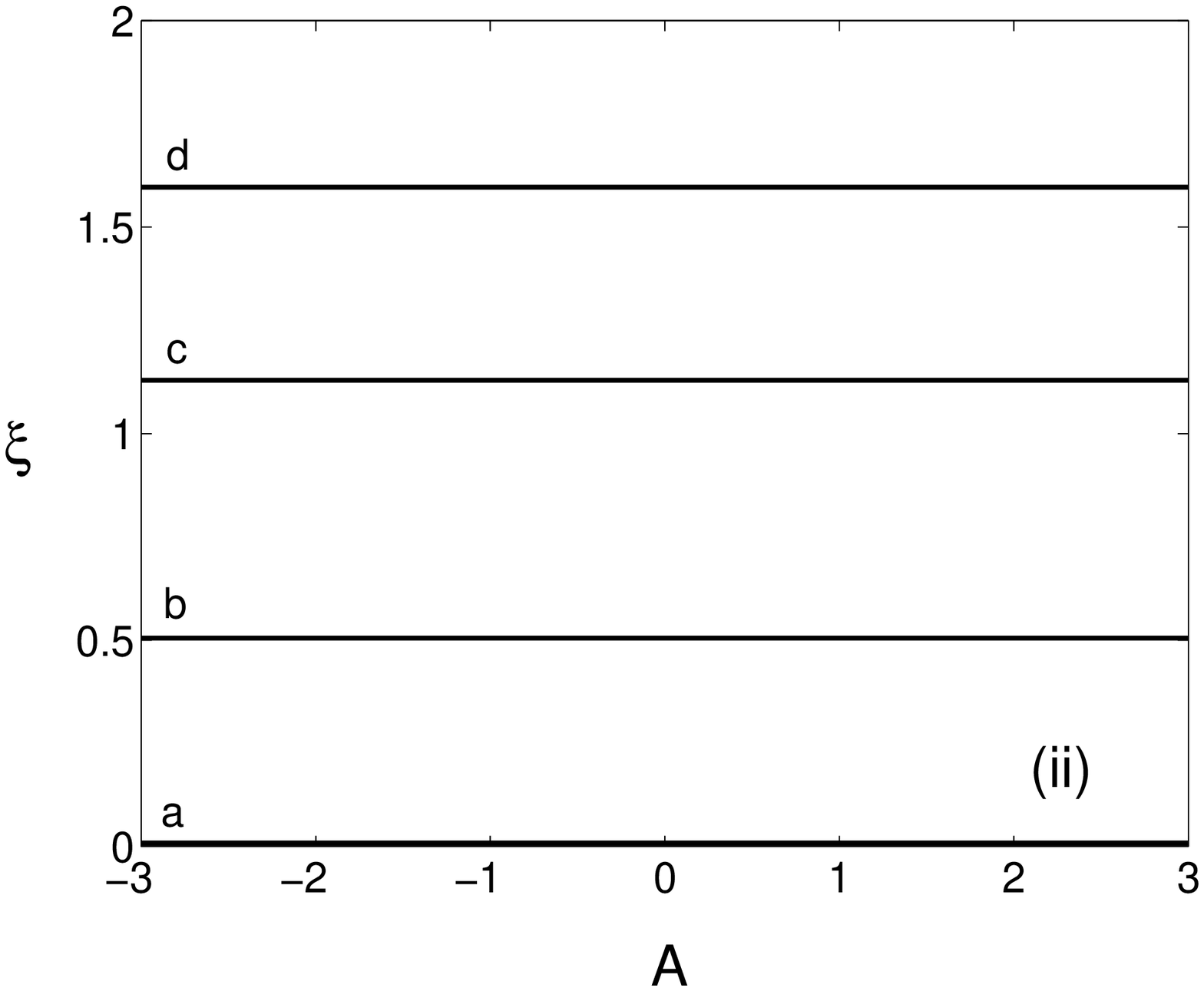}
\includegraphics{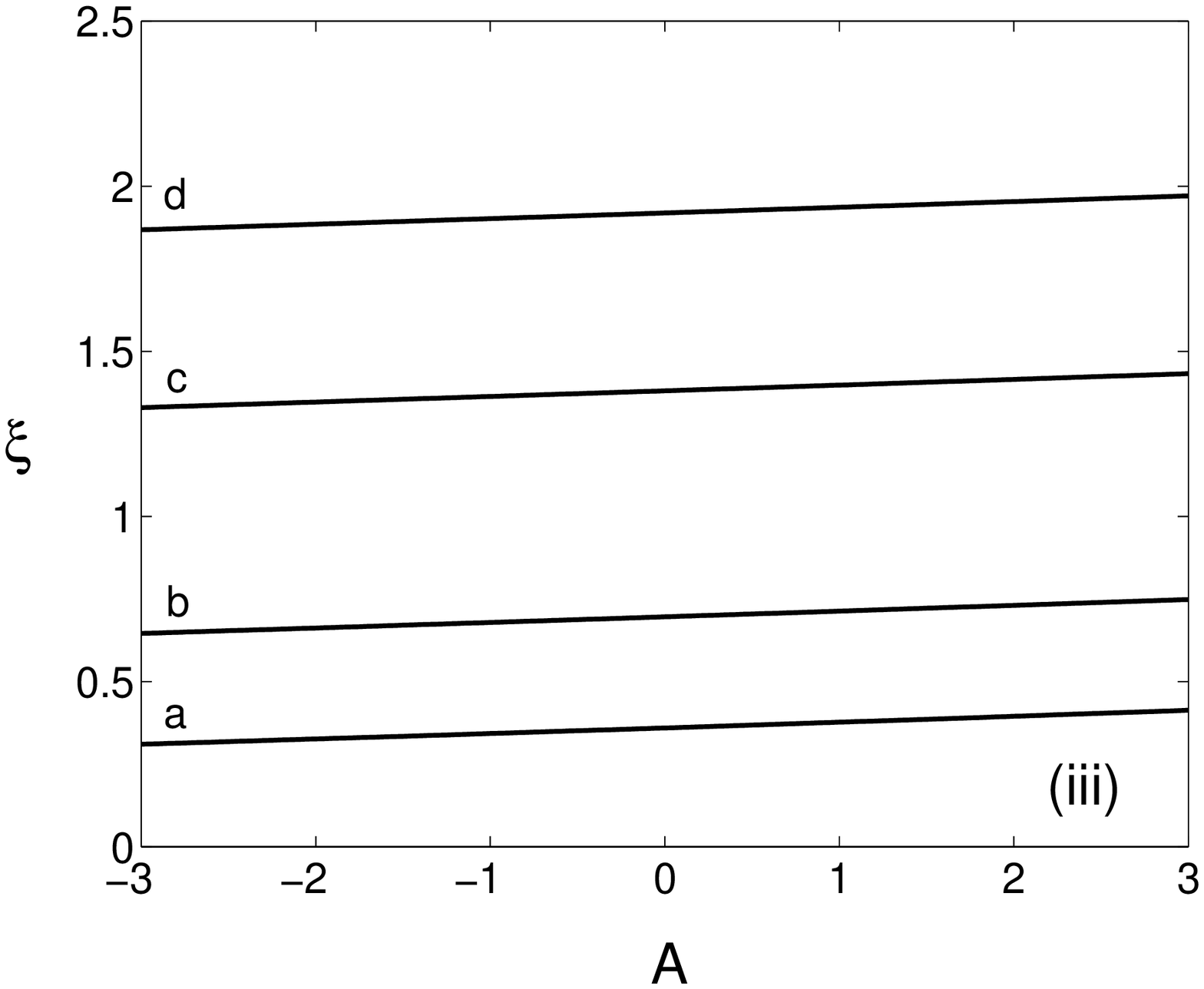}
\includegraphics{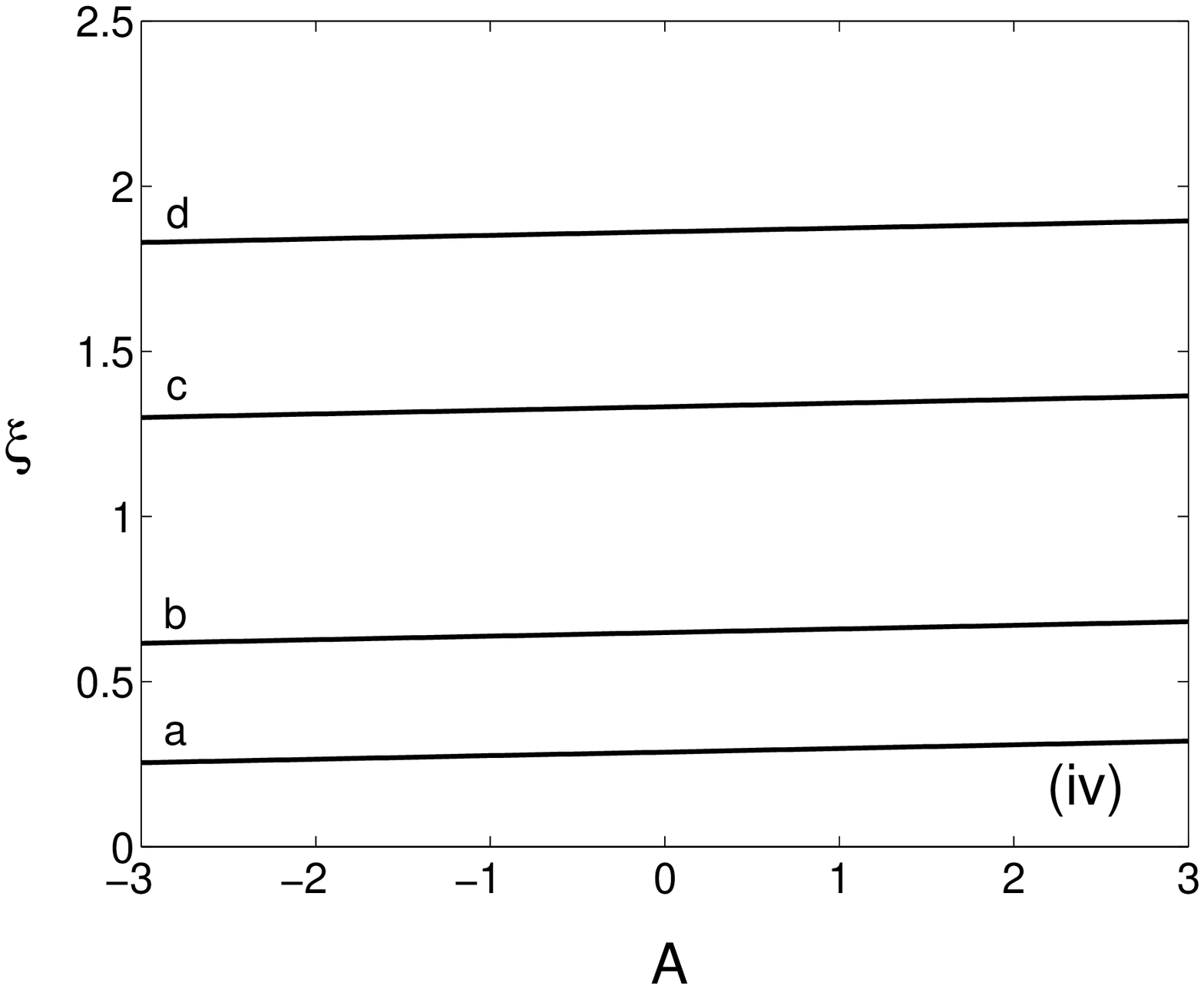}
\caption{Contours of K for two d=4 flat directions in the
$(A, \xi\equiv m_g/m(t=0))$-plane:
(a) $K=0$ (b) $K=-0.01$; (c) $K=-0.05$; (d) $K=-0.1$. The directions are (i)
$Q_3Q_3QL$; (ii) $QQQL$, no stop; (iii) $\bar u_3\bar u\bar d\bar e$;
(iv) $\bar u\bar u\bar d\bar e$ with equal weight for all $\bar u$-squarks,
from~\cite{enqvist00483}.}
\label{kuva1}
\end{figure}
%%%%%%%%%%%%%%%%%%%%%%%%%%%%%%%%%%%%%%%%%%%%%%%%%%%%%%%%%%%%%%%%%%

The contours of $K$ for the $d=4$ $\bar u\bar u\bar d\bar e$ and $QQQL$
directions are shown in Fig.~(\ref{kuva1}) in the $(A, \xi)$-plane,
where $\xi\equiv m_{g}/m(t=0)$
(for $\tan\beta=1$ and $\lambda=1$). These are representative of all the other
directions, too, except for $H_uL$. For $\xi\sim {\cal O}(1)$,
typical value for $K$ is found to be about $-0.05$. Similar
contours can be obtained for the $d=6$ $(\bar u\bar d\bar d)^{2}$ and
$(QL\bar d)^{2}$ directions, see~\cite{enqvist00483}. For all the squark
directions with no stop, as long as $h_b$ and $h_u$ can be neglected,
$K$ is always negative, and the contours of equal $K$ do not depend on $A$.
In the presence of stop mixing $K<0$ is no longer automatic even in
the purely squark directions. The more there is stop, the larger
value of $\xi$ is required for $K<0$. Even for pure stop directions,
positive $K$ is typically obtained only for relatively light gaugino masses
with $\xi\lsim 0.5$.

In contrast to the squark directions, $K$ was found \cite{enqvist00483}
to be always positive in the $H_uL$-direction. This is due to the fact
$H_uL$ does not involve strong interactions which in other directions
are mainly responsible for the decrease of the running scalar masses.
Very roughly, instability is found when
$m_{\widetilde g}\gsim m_{\widetilde t}$,
although the exact condition should be checked case by case. In general,
the sign of $K$ could be deduced from the observation of SUSY parameters
such as $\tan\beta$, the gluino mass and the supersymmetry breaking parameter
$m_{\phi}$~\cite{enqvist00483}.

%%%%%%%%%%%%%%%%%%%%%%%%%%%%%%%%%%%%%%%%%%%%%%%%%%%%%%%%%%%%%%%%%%

\subsubsection{Gauge mediated supersymmetry breaking}

A similar analysis can be made for the gauge mediated case,
where supersymmetry breaking is transmitted to the
observable sector below some relatively low messenger sector scale
$\Lambda_S$, above which the potential is completely flat (see Sect.~$4.4.3$.).
In the gravity mediated scenario the soft masses stay intact, modulo
RG running, up to the Planck scale; in gauge mediation the masses
simply disappear above $\Lambda_S$. For a large condensate
vev, one can integrate out the gauge and chiral fields coupled to
the flat direction in order to obtain an effective low energy theory.
In such a case, as was first pointed out by Kusenko and Shaposhnikov
\cite{kusenko98418}, the potential along the flat direction obtains
a logarithmic correction of the form \cite{kusenko98418,dvali98}
\begin{equation}
\label{potgmsb}
U(\Phi)=m_{\phi}^4\log\left(1+\frac{|\Phi|^2}{m_{\phi}^2}\right)
+\frac{\lambda^2 |\Phi|^{2(d-1)}}{M_{\rm P}^{2(d-3)}}+
\left(\frac{A_{\lambda}\lambda\Phi^{d}}{dM_{\rm P}^{d-3}}+{\rm h.c.}\right)\,.
\end{equation}
where $m_{\phi}\sim 1-100$~TeV. Because of the differences in the
potential, the dynamical evolution of the condensate field will be
markedly different from the gravity mediated case. Because the
messenger sector is not constrained by experiments,
one cannot provide a detailed description of the mass parameters.
Here one should note that in order to
have an AD condensate, the $A$-term is actually constrained. In the
gauge mediated case $|A_{\lambda}| \leq (10^{-4}-10^{-7})m_{\phi}$,
for $d=4,6$.

Eqs.~(\ref{pot}) and (\ref{potgmsb}) are  two book-keeping
equations which are useful for the rest of this review.

%%%%%%%%%%%%%%%%%%%%%%%%%%%%%%%%%%%%%%%%%%%%%%%%%%%%%%%%%%%%%%

\subsection{Density perturbations from the flat direction condensate}

The role of MSSM flat directions is not just limited to generating the
lepton and/or baryon asymmetry in the Universe, but they also play an
interesting role in the dynamics of density perturbations.

%%%%%%%%%%%%%%%%%%%%%%%%%%%%%%%%%%%%%%%%%%%%%%%%%%%%%%%%%%%%%%
\subsubsection{Energetics of flat direction and the inflaton field}

Once the flatness of the flat direction potential is lifted by
non-renormalizable terms, for large field values the condensate
energy density can dominate over the inflaton potential. This could
be disastrous: either inflationary expansion would come to a halt,
or the flat direction condensate fluctuations might ruin the successful
predictions for the angular power spectrum
\cite{enqvist9983,koyama99,enqvist0062,kawasaki01}.

In~\cite{enqvist0062}, the generation of adiabatic density perturbations
was studied for both D-and F-term inflation models. Note that in the
former case there is no Hubble induced mass correction to the flat direction
condensate. The scalar potential for F-and D-flat direction of
dimension $d$ is given by (see Eq.~(\ref{adpot1}) in Sect.~$4.6.2$.)
\begin{equation}
\label{v2}
V(\phi) \approx \frac{\lambda^{2}|\Phi|^{2(d-1)}}{M^{2(d-3)}}\,,
\end{equation}
where only the dominant term from Eq.~(\ref{adpot1})
corresponding to superpotential term of
the form $W = \lambda \Phi^{n}/n M^{d-3}$
has been kept.
Throughout this discussion $R$-parity is conserved and therefore
we deal only with even dimensions $d=4,~6,~8$.

For illustrative purposes let us assume that the D-term inflationary
potential is given by (see Eq.~(\ref{dterm1}) in Sect.~$3.5.3$.)
\begin{equation}
V(S)=\frac{g^2\xi^4}{2}+\frac{g^4\xi^4}{32\pi^2}\ln\left(\frac{S^2}
{Q^2}\right)\,,
\end{equation}
where $S$ is the inflaton component; $Q$ is here the renormalization scale.
For a large initial vev for $\phi, S\sim {\cal O}(M_{\rm P})$,
the dynamics is first dominated by $V(\phi)$. For a sufficiently large
vev of $\phi$ the effective condensate mass squared $V^{''}(\phi)$,
becomes larger than $H^2$. This occurs if $\phi > \phi_{H}$,
where~\cite{enqvist0062}
\begin{equation}
\label{13a}
\phi_{H} = \frac{2^{\frac{(d-1)}{2(d-2)}}}{(6(2d-2)(2d-3))^{\frac{1}{2(d-2)}}}
\left(\frac{g}{\lambda}\right)^{\frac{1}{d-2}} \xi^{\frac{2}{d-2}}
M_{\rm P}^{\frac{d-4}{d-2}}\,.
\end{equation}
If initially $\phi_{i} > \phi_{H}$, then $\phi$ will first oscillate in its
potential with a decreasing amplitude: $\phi(t) \propto a^{-3/d}(t)$
\cite{turner83}. This period ends before the onset of inflaton domination.
The system then enters a regime where both $\phi$ and $S$ are slowly rolling.

The slow rolling dynamics of the scalar fields is given by the solution of
\be{e3}
3 H \dot{\Psi}_{a} = - \frac{\partial V(\Psi_{a})}{\partial \Psi_{a}}
\;\;\; ; \;\;
H = \left(\frac{\sum_{a} V(\Psi_{a})}{3 M_{\rm P}^{2}}\right)^{1/2}~,
\ee
where $\Psi_{a} \equiv S, \; \phi$. By taking the ratio of the
equations for $\phi$ and $S$, one obtains
\be{e4}
\frac{\partial \phi}{\partial S} = \frac{16 \pi^{2} (d-1)
\lambda^{2}\phi^{(2d-3)} S}{2^{d-2}g^{4} \xi^{4} M_{\rm P}^{2(d-3)}}~,
\ee
which has a general solution of the form
\be{e5}
\phi = \phi_{i} \left[ 1 + \alpha_{d} \phi_{i}^{2d-4}
\left( S_{i}^{2} - S^{2}\right) \right]^{-1/(2d-4)} \;\;\; ; \;\;
\alpha_{d}  =  \frac{16 \pi^2(d-2) (d-1) \lambda^{2}}{2^{d-2} M_{\rm P}^
{2(d-3)} g^4 \xi^4}~,
\ee
where $\phi_{i}$ and $S_{i}$ are the initial values at the onset of inflation.
There are two features about this solution. First, since $S_{i}$ is large
compared with the value of $S$ at $N = 50$ e-foldings before inflation, we
see that for sufficiently large $\phi_{i}$ the value of $\phi$ at late times
is {\it fixed} by $S_{i}$,
\be{e6}
\phi \equiv \phi_{*} \approx \left(\frac{1}{\alpha_{d}}\right)^
{\frac{1}{2d-4}} \frac{1}{S_{i}^{1/(d-2)}}~.
\ee
This is true if $\phi_{i} > \phi_{*}$, otherwise, $\phi$ simply remains at
$\phi_{i}$. Second, we can relate $S_{i}$ to the total number of e-foldings
during the $V(S)$ dominated period of inflation. In general, for sufficiently
large $\phi_{i}$, we could have an initial period of $V(\phi)$ dominated
inflation. During this period $S$ does not significantly change
from $S_{i}$. The potential is dominated by $V(\phi)$ once $\phi > \phi_{S}$,
where \cite{enqvist0062}
\be{e7}
\phi_{S} = \frac{\sqrt{2} M_{\rm P}^{\frac{d-3}{d-1}}}
{\lambda^{\frac{1}{d-1}}}
\left( \frac{g^{2} \xi^{4}}{2} \right)^{\frac{1}{2(d-1)}}~.
\ee
$\phi_{S}$ is generally less than $\phi_{H}$ (see Eq.~(\ref{13a})), therefore
$\phi$ will be slow rolling during $V(S)$ domination.

From \eq{e5}, we find that the condition for $S$ to change significantly
from $S_{i}$ at a given value of $\phi$ is given by
\be{e8}
S_{i}  < \frac{1}{\alpha_{d}^{1/2}} \left(\frac{1}{\phi}\right)^{d-2}~.
\ee
The condition for $S$ to change significantly during $V(\phi)$
dominated inflation is given by \eq{e8} with $\phi = \phi_{S}$; one finds
\be{e9}
S_{i} < S_{i\;c} \approx \frac{2^{\frac{d-2}{2(d-1)}}}{4 \pi}
\frac{g^{\frac{d}{d-1}} \xi^{\frac{2}{d-1}} M_{\rm P}^{\frac{d-3}{d-1}}}
{\lambda^{\frac{1}{d-1}}}~.
\ee
Since $S_{i \; c}$ is small compared with $M_{\rm P}$, whereas the value
of $S$ required to generate $50$ e-foldings of inflation,
$S_{50} = g \sqrt{50} M_{\rm P}/(2 \pi)$  is close to $M_{\rm P}$,
it follows that $S_{i}$ ($> S_{50}$) will generally be larger than
$S_{i\;c}$, and so the inflaton will remain at $S_{i}$ until the
Universe becomes inflaton dominated.

In this case the total number of e-foldings during inflaton
domination is given by $N_{S}$ where
$S_{i} = (g/2 \pi) N_{S}^{1/2} M_{\rm P}$. If
$\phi_{i} > \phi_{*}$, then $\phi$ at $N \approx 50$ e-foldings of
inflation will be given by \cite{enqvist0062}
\be{e10}
\phi_{*} \approx \left(
\frac{1}{\alpha_{d}}\right)^{\frac{1}{2d-4}}
\left(\frac{2\pi}{g M_{\rm P} N_{S}^{1/2}}\right)^{\frac{1}{d-2}}~.
\ee
Note that the dependence on $N_{S}$ is quite weak; for the case
of $d=4$ ($d=6$) AD baryogenesis,
$\phi_{*} \propto N_{S}^{-1/4}\; (N_{S}^{-1/8})$. If there is no
large number of inflationary e-foldings one can essentially
fix the value of $\phi_{*}$. In this case one can {\it predict} the
magnitude of the baryonic isocurvature perturbation.

Imposing a chaotic-type initial condition
$V(\phi_{i}) \approx M_{\rm P}^{4}$ yields
\be{e12}
\phi_{i} \approx \frac{\sqrt{2}M_{\rm P}}{\lambda^{\frac{1}{d-1}}}~.
\ee
By directly solving the slow roll equations for $\phi$ and $S$, we
obtain the total number of e-foldings of inflation:
\be{e11}
N_{T} =  N_{\phi} + N_{S} \approx \frac{1}{4(d-1)} \frac{\phi_{i}^2}
{M_{\rm P}^{2}} + \frac{4 \pi^{2} S_{i}^{2}}{g^{2} M_{\rm P}^{2}}~,
\ee
where $N_{\phi}$ is the number of e-foldings during $V(\phi)$ domination,
provided $\phi_{i} > \phi_{S}$. $V(S)$ will dominate the total
number of e-foldings only if
\be{e13}
N_{S} \gae \frac{1}{2 (d-1) \lambda^{2/(d-1)}}~.
\ee
Since $N_{S} > 50$, the above condition will be satisfied
so long as $\lambda$ is not very small (for example, if
$\lambda \approx 1/(d-1)!$). In this case the value of $\phi$
at the time when the CMB perturbations are generated
will be determined mainly by the total number of e-foldings
of inflation, i.e. $N_{T} \approx N_{S}$.

%%%%%%%%%%%%%%%%%%%%%%%%%%%%%%%%%%%%%%%%%%%%%%%%%%%%%%%%%%%%%%%%%

\subsubsection{Adiabatic perturbations during D-term inflation}

The potential for the flat direction condensate is far from flat, and
so if the magnitude of the flat direction condensate is large, it will cause a
large deviation  from scale-invariance to
the adiabatic perturbation.
This will impose an
upper limit on the amplitude of the flat direction condensate at $50$
e-foldings before
the end of inflation. If we assume that the flat direction
follows a late time attractor trajectory together with the inflaton,
then following the analyses in~\cite{enqvist0062},
and Kawasaki and Takahashi~\cite{kawasaki01}, the flat direction induced
adiabatic density perturbation can be estimated from Eq.~(\ref{multi5}).

For a potential of the form $V = V(S) + V(\phi)$, one
obtains (with the help of Eqs.~(\ref{ep1},\ref{eta1},\ref{spectind4},
\ref{multi6})) \cite{enqvist0062}
\be{e20}
\eta = - \frac{M_{\rm P}^{2}}{(V_{S}^{'} + V_{\phi}^{'})V}
\left[ V_{S}^{'}V_{S}^{''} + V_{\phi}^{'} V_{\phi}^{''}
- \frac{2(V_{S}^{'} + V_{\phi}^{'}) (V_{S}^{''}V_{S}^{' \;2}
+ V_{\phi}^{''} V_{\phi}^{'\;2}) }{(V_{S}^{'\;2} + V_{\phi}^{'\;2})}
\right]~\ee
and
\be{e21}  \epsilon = \frac{M_{\rm P}^{2}}{(V_{S}^{'} + V_{\phi}^{'})V}
\left[\frac{(V_{S}^{'} + V_{\phi}^{'}) (V_{S}^{' \;2} +
V_{\phi}^{'\;2}) }{2V}\right]~.
\ee
Particularly, for the case of D-term inflation, if $V_{\phi}^{'}< V_{S}^{'}$
and $V_{\phi}^{''} < V_{S}^{''}$, we obtain the conventional result, see
Eq.~(\ref{spectind4}). Here the isocurvature contribution to
the spectral index has been neglected;  we will discuss it in the
next subsection.

Since the main contribution to the scale-dependence of the perturbations
comes from $\eta$, let us estimate the deviation from scale-invariance
due to the presence of the flat direction condensate. Note that
when $V_{\phi}^{''} > V_{S}^{''}$, with $V_{\phi}^{'} \ll V_{S}^{'}$ and
$V_{\phi} \ll V_{S}$  still satisfied, we can expand
$\eta$ in order to obtain corrections to the conventional D-term
inflation model \cite{enqvist0062}
\be{e24}
\eta \approx M_{\rm P}^{2} \frac{V_{S}^{''}}{V_{S}}  - M_{\rm P}^{2}
\frac{V_{\phi}^{'}V_{\phi}^{''}}{V_{S}V_{S}^{'}}~.
\ee
The condensate scalar induced deviation from the scale invariance
in the spectral index is given by
\be{e25}
\Delta n_{\phi} \approx - \frac{2 V_{\phi}^{''}V_{\phi}^{'}M_{\rm P}^{2}}
{V_{S}V_{S}^{'}}~.
\ee
Requiring that $|\Delta n_{\phi}| < K$, the present CMB observations
imply that $n = 1.2 \pm 0.3$; adopting $K < 0.2$
imposes an upper bound on $\phi$,
\be{e26}
\phi < \phi_{c} = k_{d} \left( \frac{K}{\sqrt{N}}\right)^{\frac{1}{4d-7}}
g^{\frac{5}{4d-7}} \lambda^{\frac{-4}{4d-7}} \xi^{\frac{8}{4d-7}}
M_{\rm P}^{\frac{4d-15}{4d-7}}~,
\ee
where
\be{e27}
k_{d} =  \left(\frac{2^{2(d-1)}}{ 128 \pi (d-1)^{2} (2d-3)}\right)^
{\frac{1}{4d-7}}~.\ee
For the case of minimal $d=4$ AD baryogenesis, one
obtains \cite{enqvist0062}
\be{e28}
\phi_{c} = 0.53 \left( \frac{K}{\sqrt{N}}\right)^{\frac{1}{9}}
\left(g^{{5}} \lambda^{-{4}} \xi^{{8}}M_{\rm P}\right)^{\frac 19}
\sim 10^{16}\GeV~,
\ee
while for $d=6$ AD baryogenesis scenario, one gets
\be{e28a}
\phi_{c} = 0.77 \left( \frac{K}{\sqrt{N}}\right)^{\frac{1}{17}}
\left(g^{{5}} \lambda^{-{4}} \xi^{{8}}
M_{\rm P}^{9}\right)^{\frac{1}{17}}\sim 10^{17}\GeV~.
\ee

%%%%%%%%%%%%%%%%%%%%%%%%%%%%%%%%%%%%%%%%%%%%%%%%%%%%%%%%%

\subsubsection{Adiabatic perturbations during F-term inflation}

During F-term inflation, the dominant part of the flat direction
potential is given by (see Eq.~(\ref{adpot0}), Sect.~$4.6.1$.)
\be{f1}
V_{total}(\phi)\approx \frac{C_{I} H^{2}\phi^{2}}{2} + V(\phi)~,
\ee
where $V(\phi)$ is the usual part from the non-renormalizable
superpotential term. Here we assume that $C_{I}\approx -{\cal O}(1)$.
In such a case the local minimum of the flat direction condensate is given by
Eq.~(\ref{veveq}), denoted here by $\phi_{m}$.

Note that if $\phi$ is close to $\phi_{m}$
($|\delta \phi| \equiv |\phi-\phi_{m}| \lae \phi_{m}$), then
inflation will damp $\delta \phi$ to be close to zero. The equation
of motion for perturbations around this local minimum is given by
\be{f3}
\delta \ddot{\phi} + 3H \delta \dot{\phi} = - k H^{2} \delta \phi \;\; ; \;
k = (2d-4)C_{I} \gae 1  ~,\ee
which has a solution of the form:
\be{f4}
\delta \phi = \delta \phi_{o}e^{\alpha H t} \;\;\; ; \;\,
\alpha = \frac{1}{2} (-3 + \sqrt{9-4k})  ~.
\ee
As long as $Ht \gg1$, i.e. there are a significant
number of e-foldings, the amplitude of the flat direction
condensate will be damped to be exponentially close to the
minimum of its local minimum.

In general, it is likely that the initial value of $\phi$ will not be
close to $\phi_{m}$. It has been shown that the deviation of the
adiabatic perturbation from scale-invariance implies that the value
of the potential at $N \approx 50$ cannot be very much larger
than $\phi_{m}$ \cite{enqvist0062}. The deviation from scale-invariance
due to the flat direction condensate is then
\be{f5}
\Delta n_{\phi} = -\frac{2}{\xi} \frac{d\xi}{dN}
= - \frac{3 V^{'}(\phi)}{V(\phi) + V(S)} \frac{\partial \phi}{\partial N}~.
\ee
For $\phi \gg \phi_{m}$, the $\phi$ field will be rapidly oscillating in
its potential and the change in the amplitude of $\phi$ over an e-folding
due to damping by expansion will be $\partial \phi/\partial N\sim -\phi$.
Requiring that $|\Delta n_{\phi}| < K$ imposes an upper bound
on $\phi$
\be{f6}
\phi \lae \left(\frac{Kd}{6(d-1) \lambda^{2}}\right)^{\frac{1}{2(d-1)}}
\sqrt{2} H^{\frac{1}{d-1}} M^{\frac{d-2}{d-1}}~.
\ee
For $d=4$, one finds \cite{enqvist0062}
\be{f8}
\frac{\phi}{\phi_{m}} \lae \frac{0.8}{C_{I}^{1/4}}
\left(\frac{\lambda M_{\rm P}}{H} \right)^{\frac{1}{6}}  ~,
\ee
while for $d=6$
\be{f9}
\frac{\phi}{\phi_{m}} \lae
\frac{0.9}{C_{I}^{1/8}}\left(\frac{\lambda M_{\rm P}}{H} \right)^
{\frac{1}{20}}~,
\ee
where we have used $K = 0.2$. For typical values of $H_I$,
the scale-invariance of the density perturbations
implies that $\phi$ at $N \approx 50$ e-foldings cannot be much more than an
order of magnitude greater than $\phi_{m}$. Since there is no reason
for $\phi$ to be close to this upper limit when $N \approx 50$, it
is most likely that $\phi$ will be close to $\phi_{m}$ when the primordial
perturbations responsible for large scale structure formation have
left the horizon during inflation.

%%%%%%%%%%%%%%%%%%%%%%%%%%%%%%%%%%%%%%%%%%%%%%%%%%%%%%%%%%%%%%%%%%%

\subsubsection{Isocurvature fluctuations in D-term inflation}

The isocurvature perturbation in the baryon number arises from the AD
scalar if the angular direction is effectively massless, i.e. mass is
small compared with $H$ during and after inflation
\cite{enqvist9983,enqvist0062,kawasaki01}. The resulting
perturbations will be unsuppressed until the baryon number
of the Universe is generated. This in turn requires that
there are no order $H$ corrections to the supersymmetry-breaking $A$-terms.

The baryon number from AD baryogenesis is generated at
$H \approx m_{susy}\sim 100 \GeV$  when the $A$-term can introduce
$B$ and $CP$ violation into the coherently oscillating AD scalar.
If the phase of the AD scalar relative to the real direction (defined
by the $A$-term) is $\theta$, then the baryon number density given by
(see Eq.~(\ref{ap3}), in Sect.~$5.5$.)
\be{b11}
n_{B} \approx m_{susy} \phi_{o}^{2} \sin 2 \theta ~,
\ee
where $\phi_{o}$ is the amplitude of the coherent oscillations at
$H \approx m_{susy}$. One can then obtain fluctuation
in the  baryon number or an isocurvature perturbation as
\be{b22}
\frac{\delta n_{B}}{n_{B}} =  \frac{2 \delta \theta}{\tan(2\theta)}=
\frac{H}{\pi \phi\tan (2\theta)}~\,,
\ee
where $\delta \theta \approx ({H}/{2\pi\phi})$ is generated by quantum
fluctuations of the AD field at the time when the perturbations cross
the horizon. The magnitude of the CMB isocurvature perturbation relative
to the adiabatic perturbation can be written as
\cite{enqvist0062,kawasaki9654,kanazawa99}, (see Eq.~(\ref{isoalpha}),
in Sect.~$3.4.3$.)
\be{e29}
\alpha  = \left| \frac{\delta_{\gamma}^{i}}{\delta_{\gamma}^{a}}\right|
= \frac{\omega}{3} \left( \frac{2 M^{2} V^{'}(S)}{V(S)\tan (2 \theta)
\phi}\right)      ~,
\ee
where $V(S)$ is the inflaton potential (see Sect.~$5.4.1$).
For purely baryonic isocurvature perturbations
$\omega = {\Omega_{B}}/{\Omega_{m}}$, where $\Omega_{B}$ is the
baryon density and $\Omega_{m}$ is the total matter density. For
the case of D-term inflation one obtains
~\cite{enqvist0062}
\be{e31}
\alpha = \frac{1}{6 \pi} \frac{g \omega M}{\phi N^{1/2}\tan(2\theta)} ~,
\ee
where $N \approx 50$.

Requiring that the deviations from the spectral index due to the AD
scalar are acceptably small, for $d=4$, one finds
\be{e32}
\alpha > \alpha_{c}=\frac{3.3 \omega (g\lambda)^{4/9}}{K^{1/9}\tan (2\theta)}~,
\ee
and for $d=6$
\be{e32a}
\alpha>\alpha_{c}=\frac{0.18\omega(g^{3}\lambda)^{4/17}}{K^{1/17}
\tan(2\theta)}  ~.
\ee
The range of $\Omega_{B}$ allowed by nucleosynthesis is
$0.006 \lae \Omega_{B}\lae 0.036$ (for $0.6 \lae h \lae 0.87$)
\cite{olive00333}. For $\Omega_{m} = 0.4$, $K = 0.2$, and
for $d = 4$, we obtain
\be{e33}
\alpha_{c}=(0.06-0.36) \frac{ (g\lambda)^{4/9}}{\tan (2\theta)}~,
\ee
and for $d=6$
\be{e33a}
\alpha_{c}=(3.0 \times 10^{-3} - 0.018) \frac{(g^{3}\lambda)^{4/17}}
{\tan (2\theta)}~.
\ee
(The lower limits should be multiplied by 0.4 for the case
$\Omega_{m} = 1$.) If, for example, $g \sim \lambda \sim 0.1$
and $\tan (2\theta) \lae 1$, one would obtain a lower bound
$\alpha \gae 10^{-2}$ for $d = 4$ and $\alpha \gae 10^{-3}$ for $d=6$.
Such small isocurvature contamination could
be detectable in future CMB experiments.

Present CMB and large-scale structure observations require that
$\alpha \lae 0.1$ \cite{kawasaki9654,kanazawa99}. COBE
normalization combined with the value of $\sigma_{8}$
(the rms of the density field on a scale of $8\ {\rm Mpc}$)
as obtained from X-ray
observations of the local cluster together with the  shape
parameter $\Gamma \approx \Omega_{m}h = 0.25 \pm 0.05$ \cite{peacock94}
from galaxy surveys, which is also consistent with the recent observations of
high-redshift supernovae \cite{perlmutter97,reiss98})
yields the limit $\alpha\lae 0.07$. The limit may however rely too
much on COBE normalization, which is just one experimental result among many.

Future CMB observations by MAP will be able to probe
down to $ \alpha \approx 0.1$, while PLANCK (with CMB polarization
measurements) should be able to see isocurvature perturbations as
small as $0.04$~\cite{enqvist0061} (see also~\cite{pierpaoli9909420}).
For the case of minimal ($d=4$) AD baryogenesis, there is a good
chance that PLANCK will be able to observe isocurvature perturbations
at least if inflation is driven by D-term. For higher dimension AD
baryogenesis ($d \geq 6$) the situation is less certain.

All this assumes that $\phi$ can take any value. This is true if
$\phi_{i} < \phi_{*}$, in which case $\phi$ remains at its initial
value $\phi_{i}$. We have seen that the dynamics of the flat direction
during D-term inflation implies that if $\phi_{i} > \phi_{*}$
then $\phi$ will equal $\phi_{*}$ at $N \approx 50$. In this case we
can fix the magnitude of the isocurvature perturbation.
For $d = 4$, $N\approx 50$ and $\Omega_{m} = 0.4$ it is given by
~\cite{enqvist0062}
\be{e34}
\alpha = \alpha_{*} \approx  (0.17-1.03) \left(\frac{N_{S}}{50}\right)^{1/4}
\frac{(g\lambda)^{1/2}}{\tan(2\theta)}~.
\ee
(For $\Omega_{m} = 1$ this should be multiplied by 0.4.) For $d=6$
and $\Omega_{m} = 0.4$,
\be{e34a}
\alpha = \alpha_{*} \approx  (4.4 \times 10^{-3}-2.6 \times 10^{-2})
\left(\frac{N_{S}}{50}\right)^{1/8}\frac{g^{3/4} \lambda^{1/4}}
{\tan (2\theta)}~.
\ee
If $g, \lambda \gae 0.1$ then for the $d=4$ case one expects
$\alpha_{*} \approx 0.01-0.1$. For the $d=6$ case the isocurvature
perturbation might just be at the observable level.

It is important that one can fix the isocurvature perturbation to be not
much larger than the lower bound coming from adiabatic perturbations.
This is because there is typically a very small range of values of
$\phi$ over which the isocurvature perturbation is less than the present
observational limit, $\alpha \lae 0.1$, but larger than the adiabatic
perturbation lower bound, $\alpha \gae 0.01$ for $d=4$.

%%%%%%%%%%%%%%%%%%%%%%%%%%%%%%%%%%%%%%%%%%%%%%%%%%%%%%%%%%%%%%%%%

\subsubsection{Isocurvature fluctuations in F-term inflation}

If the flat direction condensate is stuck in a local minimum
$\phi \approx \phi_{m}$ given by
Eq.~(\ref{veveq}), the isocurvature perturbation is given by
\cite{enqvist9983,enqvist0062}
\be{f10}
\alpha \approx \frac{2 \omega}{3} \frac{H}{\tan(2 \theta)\delta_{\rho}
\phi_{m}} ~,
\ee
where $\delta_{\rho}= 3\delta T/T \approx 3 \times 10^{-5}$ is the
density perturbation. Given $H$, $d$, and the value of
$\phi_{m}$, the magnitude of the isocurvature perturbation is
essentially fixed. For $d=4$ and $\Omega_{m} = 0.4$, the isocurvature
perturbation has been found to be
\be{f11}
\alpha = (3.1-18.6) \times 10^{2} \; \frac{\lambda^{1/2}}{C_{I}^{1/4}
\tan(2\theta) }\left(\frac{H_{I}}{M_{\rm P}}\right)^{1/2}~
\ee
while for $d=6$
 \be{f12}
\alpha = (2.9-17.4) \times 10^{2} \; \frac{\lambda^{1/4}}{C_{I}^{1/8}
\tan(2\theta) }\left(\frac{H_{I}}{M_{\rm P}}\right)^{3/4}~.
\ee
If we require that $\alpha \lae 0.1$ we find the upper bounds
$H_{I}/M_{\rm P} \lae 10^{-7}/\lambda$ (for $d=4$) and
$H_{I}/M_{\rm P} \lae 10^{-5}/\lambda^{1/3}$ (for $d=6$). For typical
values of $H$ the isocurvature perturbation in the F-term inflation
can be close to the present observational limits. In \cite{kawasaki01},
however, it was pointed out that there would be negligible isocurvature
perturbations produced from the MSSM flat directions, and the present
observations would not be able place any independent constraint upon 
the initial amplitude of the flat directions.

%%%%%%%%%%%%%%%%%%%%%%%%%%%%%%%%%%%%%%%%%%%%%%%%%%%%%%%%%%%%%%%%%

\subsubsection{MSSM flat directions as a source for curvature perturbations}

In a very recent development, a new paradigm has been laid, where 
adiabatic density perturbations were generated from the decay of
the pure isocurvature perturbations. Though it was first suggested in 
\cite{Mollerach:ue}, but it was implemented recently in a pre-Big-Bang
scenario~\cite{Enqvist:2001zp}, where the axion field which generates 
isocurvature perturbations decays late in the Universe.

In order to create pure adiabatic density perturbations it is important 
that the field, known as curvaton \cite{Lyth:2001nq}, is subdominant
during inflation, but becomes dominant during the late phase of the Universe
especially when it is decaying. The curvaton field $\sigma$ generates 
isocurvature perturbations during inflation, assuming that the perturbations
generated from the inflaton field can be negligible, and the curvaton mass
$m^2_{\sigma}\approx V_{\sigma\sigma} \ll H^2_{inf}$. In this limiting
case the power spectrum for the curvaton will be equivalent to a massless
scalar field during inflation; ${\cal P}_{\sigma}=H_{\ast}^2/4\pi^2$, where
$\ast$ denotes the epoch when perturbations are crossing the horizon 
$k=a_{\ast}H_{\ast}$. The curvaton field follows it trajectory after the 
end of inflation, and when $H \sim m_{\sigma}$, the curvaton oscillates 
and eventually decays through its coupling to the SM relativistic degrees 
of freedom. While oscillating it produces density contrast
$\delta_{\sigma}=2\delta\sigma/\sigma$, assuming that $H_{\ast}<\sigma_{\ast}$
\cite{Lyth:2001nq}, and the perturbation spectrum is given by 
\begin{equation}
{\cal P}^{1/2}_{\delta_{\sigma}}=2\frac{{\cal P}_{\sigma}^{1/2}}{\sigma}=
\frac{H_{\ast}}{\pi\sigma_{\ast}}\,.
\end{equation}
When the curvaton decays it converts all its isocurvature perturbations
to the adiabatic ones by following that, before decaying the relativistic 
degrees of freedom due to the inflaton decay products gives rise to density
perturbations in the radiation as $\zeta_{r}=(1/4)\delta_{r}$, and 
$\zeta_{\sigma}=(1/3)\delta_{\sigma}$. With these results the curvature 
perturbations is given by \cite{Lyth:2001nq}
\begin{eqnarray}
\zeta=\frac{4\rho_{r}\zeta_{r}+3\rho_{\sigma}\zeta_{\sigma}}{4\rho_{r}+
4\rho_{\sigma}}\approx\frac{1}{3}\delta_{\sigma}\,,
\end{eqnarray}
supposing that before decaying $\zeta_{r}$ is negligible. If the curvaton 
decay products do not dominate the Universe, then there will adiabatic 
and isocurvature perturbations both \cite{Lyth:2001nq,manycurvaton}.

In \cite{Enqvist:2002rf}, the authors have pointed out that the MSSM 
flat directions which are lifted by the non-renormalizable operators 
such as $d=7,9$ are the best candidate for the curvaton. The flat directions
$d=7,~LL\bar d\bar d\bar d$ (lifted by $H_{u}LLLddd$), and 
$d=9,~Q\bar u Q\bar u Q\bar u\bar e$ 
(lifted by $Q\bar u Q\bar u Q\bar u H_{d}\bar e \bar e$), which are lifted 
by superpotential: $W\sim \sigma^{d-1}\psi/M^{d-3}$. Note that $\psi$ is
the superfield other than the curvaton, does not produce any $A$-term 
in the potential, since $\langle\psi\rangle=0$, and therefore does not give 
rise to any $U(1)$ violating terms in the flat direction potential. It was
shown in \cite{Enqvist:2002rf}, that these flat directions can dominate
the energy density of the Universe and while decaying the squarks and sleptons
can directly decay in the MSSM relativistic degrees of freedom. Therefore
the virtue of this scenario is that the MSSM flat direction is solely 
responsible for reheating the Universe, barring any need for speculation 
from the inflaton coupling to the SM fields. Inflation was supposed to 
happen in the hidden sector of the theory, which does not 
necessarily couple to the SM sector.

%%%%%%%%%%%%%%%%%%%%%%%%%%%%%%%%%%%%%%%%%%%%%%%%%%%%%%%%%%%%%%%%%

\subsection{Baryon number asymmetry}

In both D-and F-term inflation the inflaton and other
scalar fields begin to oscillate coherently about the minimum of
their respective potential after the end of inflation, and the
post-inflationary evolution of the flat direction condensate is no exception.
If $C_{I}$ and $C$ are positive in Eqs.~(\ref{adpot0}) and (\ref{adpot2}),
the corresponding vevs are either given by Eqs.~(\ref{veveq}) or
(\ref{veveq1}). In fact, in D-term inflation models for $|C|$ less
than about $0.5$, it is possible to have a positive $H^2$ correction
and still generate the observed baryon asymmetry as shown by McDonald
\cite{mcdonald99456}. In addition, there has been attempts for AD baryogenesis
in F-term inflation, basing on the low energy effective action of the
heterotic string theory, where inflation is driven by $T$-moduli
\cite{casas97410}. Flat directions beyond MSSM involving a triplet
Higgs has also been considered \cite{senami02}. Here we will mainly
concentrate on the negative $H^2$ correction only.

After inflation, $\langle \phi \rangle$ initially continues to track
the instantaneous local minimum of the scalar potential, which can be
derived by replacing $H_I$ with $H(t)$ in Eq.(\ref{veveq}). Once
$H \simeq m_0$, the low-energy soft terms take over. The condensate
mass squared turns positive, and since the phase of $\langle \phi \rangle$
differs from the phase of $A$, $\langle \phi \rangle$ starts to change
non-adiabatically.

In an absence of $H$ corrections to the $A$-terms, the initial
phase $\theta$ of the AD field relative to the real direction is random
and so typically $\approx 1$. As a result $\langle \phi\rangle$
starts a spiral motion in the complex plane (see Figs.~(\ref{plotreim}),
and forthcoming discussion on the condensate trajectory),
which leads to a generation of a net baryon and/or lepton asymmetry
\cite{dine95,dine96}\footnote{There have been attempts for
AD baryogenesis in local domains, see~\cite{dolgov89}}.

The baryon number density is related to the AD field as
\begin{equation}
n_{B,L}=\beta i(\dot{\phi}^{\dagger}\phi - \phi^{\dagger}\dot{\phi})\,,
\end{equation}
where $\beta$ is corresponding baryon and/or lepton charge of the
AD field. The equations of motion for the AD field are given by
\begin{equation}
\ddot\phi + 3H\dot\phi +\frac{\partial V(\phi)}{\partial \phi^{\ast}}=0\,.
\end{equation}
The above two equations lead to
\begin{eqnarray}
\label{bleq}
\dot n_{B,L}+3Hn_{B,L}&=&2\beta {\rm {\cal I}m}\left[\frac{\partial V(\phi)}
{\partial \phi^{\ast}}\phi\right]\,, \nonumber \\
&=&2\beta\frac{m_{\phi}}{d M^{d-3}}{\rm {\cal I}m}(a\phi^{d})\,.
\end{eqnarray}
By integrating Eq.~(\ref{bleq}), we obtain the baryon and/or lepton number
as
\begin{equation}
a^3(t)n_{B,L}(t)=2\beta |a|\frac{m_{\phi}}{M^{d-3}}
\int^{t}a^3(t^{\prime})|\phi(t^{\prime})|^{d}\sin(\theta)~dt^{\prime}\,,
\end{equation}
Note that $a$ introduces an extra $CP$ phase which we may
parameterize as $\sin(\delta)$. After a few expansion times, the amplitude
of the oscillations will become damped by the expansion of the Universe and
the $A$-term, which is proportional to a large power of $\phi$,
will become gradually negligible. The net baryon and/or lepton
asymmetry is then given by~\cite{dine96}
\begin{eqnarray}
\label{ap3}
n_{B,L}(t_{osc}) &=&  \beta \frac{2(d-2)}{3(d-3)}m_{\phi} \phi_{0}^{2}
\sin2 \theta \; \sin \delta\,, \nonumber \\
&\approx &  \beta \frac{2(d-2)}{3(d-3)}m_{\phi}\left(m_{\phi} M^{d-3}\right)
^{2/(d-2)} \sin2 \theta \; \sin \delta\,,
\end{eqnarray}
where $\sin\delta \sim \sin 2\theta \approx {\cal O}(1)$.

When the inflaton decay products have completely thermalized with a
reheat temperature $T_{rh}$, the baryon and/or lepton asymmetry
is given by~\cite{asaka00}
\begin{eqnarray}
\label{bleq1}
\frac{n_{B,L}}{s} &=&\frac{1}{4}\frac{T_{rh}}{M_{\rm P}^2 H(t_{osc})^2}
n_{B,L}(t_{osc})\,,\nonumber \\
&=&\frac{d-2}{6(d-3)}\beta \frac{T_{rh}}{M_{\rm P}^2 m_{\phi}}
\left(m_{\phi} M^{d-3}\right)^{2/(d-2)} \sin2 \theta \; \sin \delta\,,
\end{eqnarray}
where we have used $H(t_{osc})\approx m_{\phi}$, and $s$ is the entropy
density of the Universe at the time of reheating. For $d=4$, the
baryon-to-entropy ratio turns out to be~\cite{asaka00}
\begin{equation}
\label{assym1}
\frac{n_{B,L}}{s}\approx 1\times 10^{-10}\times \beta \left(\frac{m_{3/2}}
{m_{\phi}}\right)\left(\frac{M}{M_{\rm P}}\right)\left(\frac{T_{rh}}
{10^{9}~{\rm GeV}}\right)\,,
\end{equation}
and for $d=6$
\begin{equation}
\label{assym2}
\frac{n_{B,L}}{s}\approx 5\times 10^{-10}\times\beta\left(\frac{m_{3/2}}
{m_{\phi}}\right)\left(\frac{1~{\rm TeV}}{m_{\phi}}\right)^{1/2}
\left(\frac{M}{M_{\rm P}}\right)^{3/2}\left(\frac{T_{rh}}{100~{\rm GeV}}
\right)\,,
\end{equation}
where we have taken the net $CP$ phase to be $\sim {\cal O}(1)$.
The asymmetry remains frozen unless there is additional entropy
production afterwards. Note that for $d=4$, the required reheat temperature
of the Universe is below the gravitino overproduction bound (see
Sect.~$3.6.2$). For higher dimensional non-renormalizable operators, a
low reheat temperature is favorable, which is indeed a good news.

In this regard low scale inflation, which guarantees a low reheat temperature,
has been given some consideration~\cite{asaka0151} (see also
\cite{stewart9654} where AD baryogenesis after a brief period of thermal
inflation, required to solving the cosmological moduli problem, has been
discussed). Although, in \cite{kasuya0265}, it was pointed out that in
gauge mediated supersymmetry breaking it is hard to reconcile AD
baryogenesis with a moduli problem.

Among the host of MSSM flat directions which are lifted by non-renormalizable
operator and listed in Table 1, the $LH_{u}$ flat direction carrying the
lepton number is the candidate for producing lepton asymmetry in the
Universe (there has been some earlier attempts of direct baryogenesis
via $\bar u\bar d\bar d$ directions, see e.g.
\cite{mcdonald9755}). The lepton asymmetry calculated above in
Eqs.~(\ref{assym1},\ref{assym2}) can be transformed into baryon number
asymmetry via sphalerons $n_{B}/s =(8/23)n_{L}/s$. AD leptogenesis
has important implications in neutrino physics also, because in the
MSSM, the $LH_{u}$ direction is lifted by the $d=4$ non-renormalizable operator
which also gives rise to neutrino masses~\cite{dine96}:
\begin{equation}
W=\frac{1}{2M_{i}}\left(L_{i}H_{u}\right)^2=\frac{m_{\nu~i}}{2\langle H_{u}
\rangle^2}\left(L_{i}H_{u}\right)^2\,,
\end{equation}
where we have assumed the see-saw relation
$m_{\nu~i}=\langle H_{u}\rangle^2/M_{i}$ with diagonal entries for
the neutrinos $\nu_{i}$, $i=1,2,3$. The final $n_{B}/s$ can be
related to the lightest neutrino mass since the flat direction moves
furthest along the eigenvector of $L_{i}L_{j}$ which corresponds to the
smallest eigenvalue of the neutrino mass matrix
~\cite{dine96}.
\begin{equation}
\label{lnb}
\frac{n_{L}}{s}\approx 1\times 10^{-10}\times \beta\left(\frac{m_{3/2}}
{m_{\phi}}\right)\left(\frac{T_{rh}}{10^{8}~{\rm GeV}}\right)\left(
\frac{10^{-6}~{\rm eV}}{m_{\nu l}}\right)\,,
\end{equation}
where $m_{\nu~l}$ denotes the lightest neutrino.

%%%%%%%%%%%%%%%%%%%%%%%%%%%%%%%%%%%%%%%%%%%%%%%%%%%%%%%%%%%%%%%%%%%
%%%%%%%%%%%%%%%%%%%%%%%%%%%%%%%%%%%%%%%%%%%%%%%%%%%%%%%%%%%%%%%%%%%

\subsection{Thermal effects}

In our discussion on the baryon/lepton asymmetry
we have tacitly assumed that the asymmetry has been generated before
the Universe has thermalized and reheated. This might not be the case
if there were light degrees of freedom which have thermalized
with an instantaneous plasma temperature
$T_{inst}\leq(H\Gamma_{d}M_{\rm P})^{1/4}$
before the inflaton has decayed. We remind that
the bulk of energy density is still in the form of inflaton oscillations,
and only a fraction of the energy density has gone into these light
Standard Model degrees of freedom. If the MSSM flat direction couples
with this thermal bath, there arises a modification in the flat
direction potential \cite{allahverdi00579,anisimov01,anisimov02}.

%%%%%%%%%%%%%%%%%%%%%%%%%%%%%%%%%%%%%%%%%%%%%%%%%%%%%%%%
\subsubsection{Thermal corrections to the flat direction potential}

Besides the D-term couplings of the form $g^2\phi^2\alpha^2$,
where $\alpha$ is some field with gauge interactions, there are also
F-term Yukawa couplings to fields $\chi$ which result in a term
$h^2|\phi|^2|\chi|^2$ in the flat direction potential.
$\chi$ and $\alpha$ obtain large masses due to the flat direction vev,
and therefore do  not feel the effect of temperature if the condensate
amplitude $\phi$ is large. As pointed out by Allahverdi, Campbell and
Ellis \cite{allahverdi00579}, the back-reaction effect induces a
mass-squared term $h^2T^2$ for the flat direction. If this exceeds
the negative Hubble induced mass squared term, the flat direction
oscillations starts earlier than otherwise expected. In order for
thermal correction to play
a significant role the couplings $h,g$ must have intermediate strength.
Otherwise, a large coupling would induce a large vev dependent mass
for $\alpha$, which would prevent its thermal excitations, and a very
small coupling would not have significant thermal backreaction at all.

For the inflationary scale $H_I \sim 10^{13}$~GeV and $M=M_{GUT}$,
it has been found
 \cite{allahverdi00579} that a generic MSSM flat direction with
a Yukawa coupling $h\sim 10^{-2}$ starts oscillating at $H\gg 10^{2}$~GeV
for $4\leq d \leq 8$. Since thermal effects induce early oscillations,
baryon asymmetry is also produced much earlier, which could have interesting
consequences.

Another thermal effect has been discussed by Anisimov and Dine
\cite{anisimov01,anisimov02}. All the flat directions which are
lifted at large $d$ give rise to a large mass for $\alpha$, and
consequently one should account for their effect by integrating
out the heavy modes. This would result in terms like
\begin{equation}
\frac{A}{16\pi^2}F^2_{\mu\nu}\ln\frac{|\phi|^2}{M^2}\,.
\end{equation}
In particular  for the flat direction $LH_{u}$ the effective potential thus
obtained has the form \cite{anisimov01}
\begin{equation}
\label{thermlog}
V_{eff}=\alpha_{s}^2(T) a_{g}T^4\ln\frac{|\phi|^2}{M^2}\,,
\end{equation}
where $\alpha_{s}\equiv g_{s}^2/4\pi$, and
$a_{g}=\frac{3N_g}{288}\left(5\frac{N_{f}}{4}+7\frac{N}{2}\right)$
includes leading order contribution of the gluons, gluinos, and quarks
to the free energy for a non-abelian group $SU(N)$. The oscillations 
in the flat direction are induced when
$H_{osc}^2=\partial^2V_{eff}/\partial |\phi|^2=\alpha_{s}^2a_{g}T^4/|\phi|^2$
and one can check that \cite{allahverdi00579} for $d=4$, the thermal mass
correction $\sim h^2T^2$ wins over the logarithmic counterpart
Eq.~(\ref{thermlog}), but for $d=5$ and/or $6$, the logarithmic correction
dominates and the oscillations start earlier than otherwise one would have
expected.

There could also be a thermal enhancement of the $A$-term
\cite{allahverdi00579}, which can arise from the cross terms
\begin{equation}
\label{thermsup1}
W\supset h\phi\alpha \alpha +\lambda_{d}\frac{\phi^{d}}{dM^{d-3}}\, .
\end{equation}
However a symmetry forbids such enhancement \cite{anisimov02}, although
the situation might change if one adds more terms in the superpotential
such as \cite{anisimov02}.
\begin{equation}
\label{thermsup2}
W=\frac{1}{M}\left(aI+b\frac{I^2}{M}\right)\frac{\phi^{d}}{M^{d-3}}\,.
\end{equation}

%%%%%%%%%%%%%%%%%%%%%%%%%%%%%%%%%%%%%%%%%%%%%%%%%%%%%%%%%%%%%%%%%%%%%
\subsubsection{Thermal evaporation of the flat direction}

It has been argued that in general the flat direction condensate
decays as a result
of scattering with the thermalized decay products of the inflaton
\cite{dine96}. Usually the scattering interactions preserve $B$ and $L$,
and therefore the previously produced baryon and/or lepton asymmetry
remains unchanged. In ~\cite{anisimov01} it was assumed that
thermalized fermions scatter with the condensate with a rate
\begin{equation}
\Gamma_{scatt}\simeq yg^2 T\,.
\end{equation}
where $yT$ corresponds to the mass of the condensate. A complete evaporation
was found to be avoided only after reheating if
$T_{rh}\leq (yg^2)^{2/3} M_{\rm P}^{5/6}H^{1/6}$, which is usually
satisfied for a reasonable range of reheat temperatures and Yukawa couplings.

%%%%%%%%%%%%%%%%%%%%%%%%%%%%%%%%%%%%%%%%%%%%%%%%%%%%%%%%%%%%%%%%%%%%%

\subsection{Baryosynthesis and neutrino mass}

As discussed in Sect.~$5.4$., the lepton asymmetry via $LH_{u}$ direction
leads to a prediction on the lightest neutrino mass. It is however
pertinent to include also the finite temperature effects \cite{fujii0163}.
At finite T, the flat direction potential potential for $LH_{u}$
direction can be written as
\cite{fujii0163}
\begin{eqnarray}
\label{thermlept}
V_{total}&=&\left(m_{\phi}^2-C_{I}H^2+\sum_{f_{k}|\phi|<T}c_{k}f^2_{k}
T^2\right)|\phi|^2 +\frac{m_{3/2}}{8M}\left(a_{m}\phi^4+{\rm h.c.}\right)+
\frac{H}{8M}\left(a_{H}\phi^4+{\rm h.c}\right) \nonumber \\
&&+a_{g}\alpha_{s}^2T^{4}\ln\left(\frac{|\phi|^2}{T^2}\right)
+\frac{|\phi|^{6}}{4M^2}\,,
\end{eqnarray}
where $c_{k}=$ are real positive constants and couplings $f_{k}=1-10^{-5}$
in MSSM \cite{asaka00}. The mismatch in phases between $a_{m}$ and $a_{H}$
leads to the helical motion of the flat direction.
Once the inflaton decay products generate a
thermal plasma with a temperature $T=(T_{rh}M_{\rm P}H_{I})^{1/4}$,
thermal corrections take over the Hubble induced term
\begin{equation}
H^2\leq m^2_{\phi}+\sum_{f_{k}|\phi|<T}c_{k}f^2_{k}T^2+a_{g}\alpha^2_{s}(T)
\frac{T^4}{|\phi|^2}\,.
\end{equation}
The flat direction starts to oscillate when \cite{fujii0163}
\begin{equation}
H_{osc}\approx {\rm max}\left[m_{\phi},H_{i},\alpha_{s}T_{rh}
\left(\frac{a_{g}M_{\rm P}}{M}\right)^{1/2}\right]\,,
\end{equation}
where $H_{i}$ is given by \cite{asaka00,fujii0163}
\begin{equation}
H_{i}\approx {\rm min}\left\{\frac{1}{f_{i}^4}\frac{M_{\rm P}T_{rh}^2}{M^2},
(c_{i}^2f_{i}^4M_{\rm P}T_{rh}^2)^{1/3}\right \}\,.
\end{equation}

The lepton asymmetry is then given by \cite{fujii0163}
\begin{equation}
\label{leptonasyfujii}
a^3(t)n_{L}(t)\approx \frac{m_{3/2}}{2M}\int^{t}dt^{\prime}a^3(t^{\prime})
{\cal I}{\rm m}\left(a_{m}\phi^4\right)\,,
\end{equation}
The right hand side of Eq.~(\ref{leptonasyfujii})
initially increases until $H\approx H_{osc}$, after which the
integrand is rapidly damped because $a^3\phi^4\sim t^{-n}$ for $n>1$.
The final lepton asymmetry is determined approximately by
the configuration at the time when the oscillations commence
\cite{fujii0163}
\begin{equation}
n_{L}=\left.\frac{m_{3/2}}{2M}{\cal I}{\rm m}\left(a_{m}\phi^4\right)t
\right|_{H=H_{osc}}=\frac{1}{3}m_{3/2}MH_{osc}\delta_{eff}\,,
\end{equation}
where $\delta_{eff}=\sin(4{\rm arg}(\phi)+{\rm arg}(a_{m}))$ is the net
$CP$ phase. The final baryon to entropy ratio turns out to be \cite{fujii0163}
\begin{equation}
\frac{n_{B}}{s}=10^{-11}\delta_{eff}\times \left(\frac{m_{\nu~l}}{10^{-8}
{\rm eV}}\right)^{-3/2}\left(\frac{m_{3/2}}{1~{\rm TeV}}\right)\,.
\end{equation}
Note that the final expression obtained does not depend upon the
reheat temperature $T_{rh}$, mainly due to the fact that the $H_{osc}$
is determined by thermal correction $\sim T^4\ln(|\phi|^2)$. This is
however true only for
$10^{8}~{\rm GeV}\leq T_{rh}\leq 10^{12}~{\rm GeV}$ as pointed out by
Fujii, Hamaguchi, and Yanagida in~\cite{fujii0163}. For
$10^{5}~{\rm GeV}\leq T_{rh}\leq 10^{8}~{\rm GeV}$,
the dependence on reheat temperature appears as
$n_{B}/s \propto T_{rh}^{1/3}$,
because then the thermal mass term $\sim T^2|\phi|^2$ dominates.

It is possible \cite{fujii0163} to obtain the right amount of baryon asymmetry
with the lightest neutrino mass $m_{\nu l}\simeq (0.1-3)\times 10^{-9}$~eV
and with a $CP$ phase $\delta_{eff}\simeq (0.1-1)$ for a fairly wide
range of reheat temperature
$10^{5}~{\rm GeV}\leq T_{rh}\leq 10^{12}~{\rm GeV}$.

%%%%%%%%%%%%%%%%%%%%%%%%%%%%%%%%%%%%%%%%%%%%%%%%%%%%%%%%%%%%%%%%%%%

\subsection{Trajectory of a flat direction}

Let us now turn our attention to the dynamical evolution of the flat
direction after the end of inflation. Here we assume that the flat
direction is tracking its minimum which is determined by Eq.~(\ref{veveq}).
The trajectory of the flat direction depends upon the potentials
Eqs.~(\ref{pot}) and (\ref{potgmsb}). Here we sketch the main differences
between the gravity and gauge mediated cases.

%%%%%%%%%%%%%%%%%%%%%%%%%%%%%%%%%%%%%%%%%%%%%%%%%%
\begin{figure}
\leavevmode
\centering
\vspace*{4cm}
\includegraphics{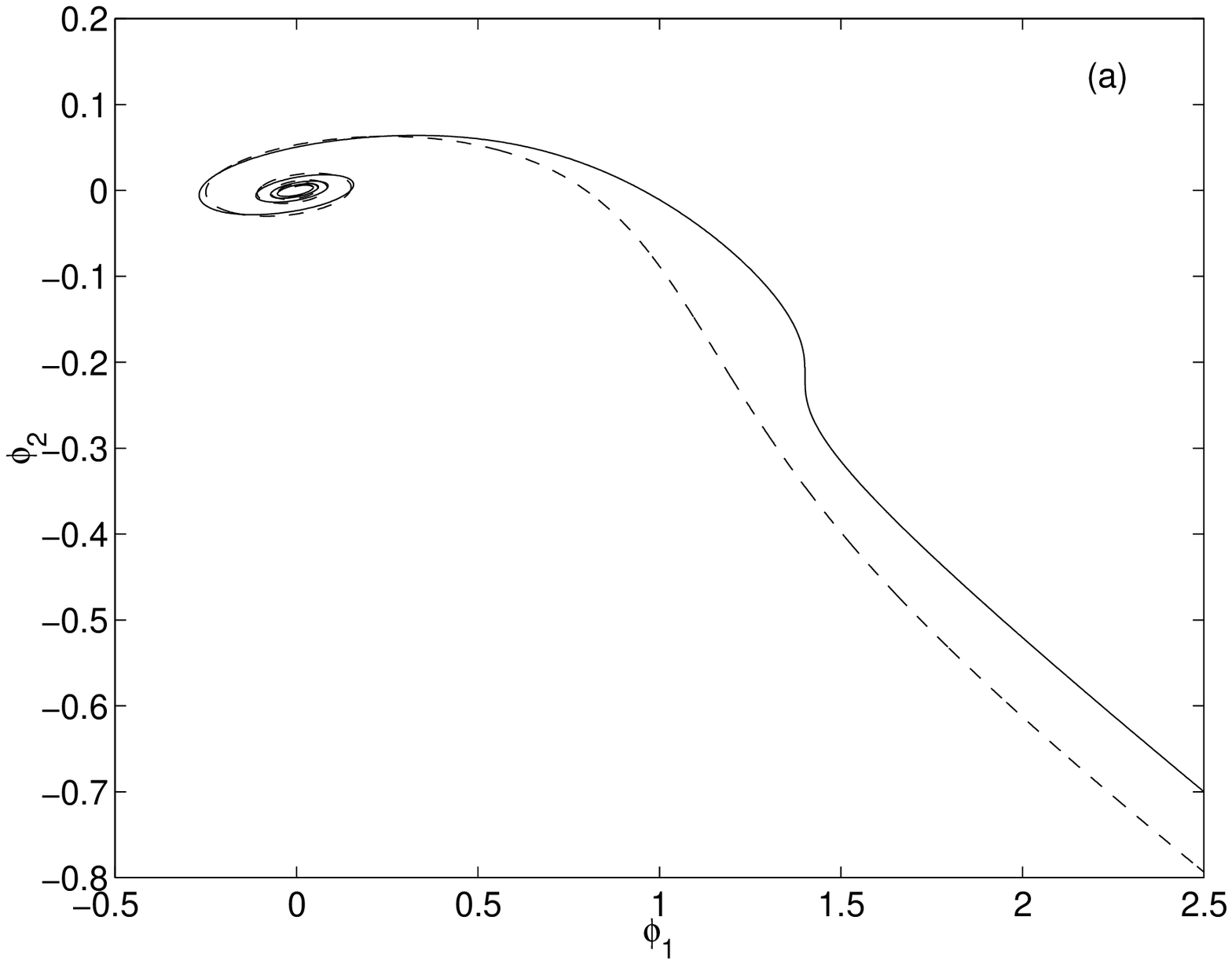}
\includegraphics{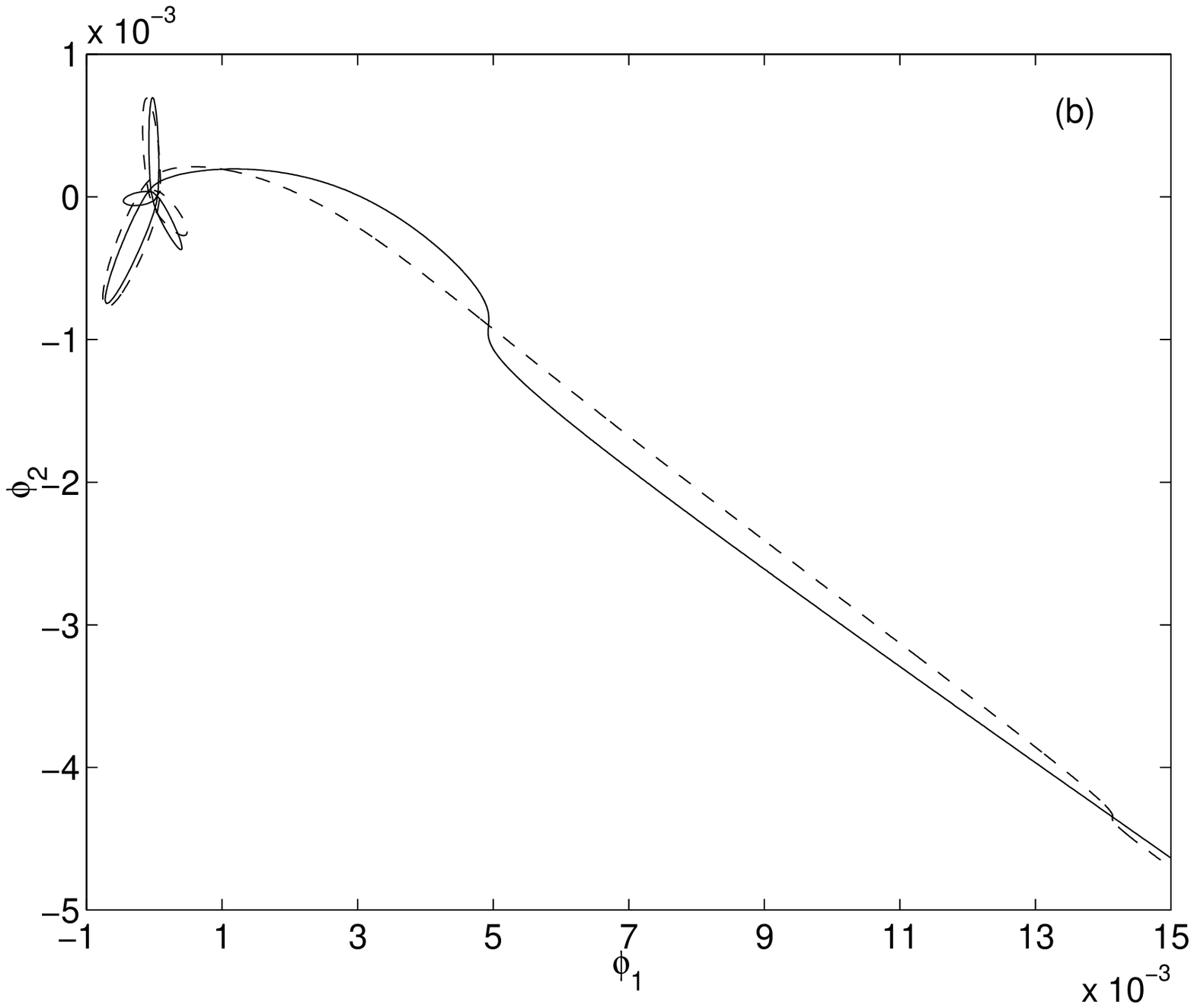}
\caption{\small Affleck-Dine condensate formation with $x=\phi_{1}$
and $y=\phi_{2}$, for (a) gravity mediated
case with $d=4$ (solid) and $d=6$ (dashed), and (b) gauge mediated case
with $d=4$, $m_{\phi}=1\TeV$ (solid) and $m_{\phi}=10\TeV$ (dashed) with
the initial condition $\theta_i=-\pi/10$, from \cite{jokinen02}.}
\label{plotreim}
\end{figure}

%%%%%%%%%%%%%%%%%%%%%%%%%%%%%%%%%%%%%%%%%%%%%%%%%%%%%

Jokinen~\cite{jokinen02} has studied numerically the
trajectories of the flat direction condensate in gravity and gauge mediated
cases, following Eqs.~(\ref{pot}) and (\ref{potgmsb}). The rotation of
the condensate depends on the low energy supersymmetry breaking
mass terms. The classical motion for the condensate
$\phi=(\phi_{1}+i\phi_{2})/\sqrt{2}$ is illustrated in
Fig.~(\ref{plotreim}). In the gravity mediated case,
Fig.~(\ref{plotreim}a), we see that the orbit is a spiraling ellipse and
in the gauge mediated case, Fig.~(\ref{plotreim}b), a precessing
trefoil. From Fig.~(\ref{plotreim}), one can see that there is a
twist on the orbit much before the rotation starts properly. This
is the time of the phase transition, when the condensate $\phi$ starts
to rotate in the pit of the symmetry breaking minimum. The rotation
begins when the symmetry breaking minimum is the vacuum, and ends
when it has become a false vacuum, and twists when the
false vacuum has completely vanished forming a kink on the orbit.
It is possible to produce a condensate through a second order phase
transition, but the charge in that case will be small. It should also
be noted that in the gravity mediated case condensate formation starts when
$C_{I}H^2\sim m_{3/2}^2$ for all values of $d$, $A$ and $a$. In the gauge
mediated case the condensate formation starts at
$C_{I}H^2\sim m_{\phi}^4/|\phi|^2$, so that the formation happens earlier
if the condensate mass $m_{\phi}$ is increased, as can be seen from the
different positions of the kink in Fig.~(\ref{plotreim}b).

%%%%%%%%%%%%%%%%%%%%%%%%%%%%%%%%%%%%%%%%%%%%%%%%%%%%%%%%%%%%
\subsection{Instability of the coherent condensate}

%%%%%%%%%%%%%%%%%%%%%%%%%%%%%%%%%%%%%%%%%%%%%%%%%%%%%%%%%%%%
\subsubsection{Negative pressure}

The effective equation of state of a coherent scalar field
oscillating in a potential $U(\phi)$ with a frequency which
is large compared with $H$ is obtained by averaging
$p/\rho=(\vert\dot\phi\vert^2/\rho)-1$ over one oscillation
cycle $T$. The result is \cite{turner83}
\begin{equation}
p=(\g -1)\rho~,
\end{equation}
where
\be{effgamma}
\g = \frac 2T\int_0^T \left( 1-\frac{U(\phi)}{\rho} \right)dt~.
\end{equation}
For the case  $U\sim m^2\phi^2$, one finds $\gamma =1$, so that one
effectively obtains the usual case of pressureless, non-relativistic
cold matter.

When the motion of the condensate field is not simply oscillatory,
such as in the case for the condensate trajectory, one can generalize
\eq{effgamma} by integrating over the orbit $c$ of the AD field.
In that case
\be{effgamma2}
\g={2\int_cd\vert\phi\vert\left( 1-U(\phi)/\rho\right)^{\frac 12}\over
\int_cd\vert\phi\vert\left( 1-U(\phi)/\rho\right)^{-\frac 12}}~.
\end{equation}
In practice the orbits are nearly elliptical. Then the arc length is
given by
\begin{equation}
\label{major-minor}
d\vert\phi\vert = {d\phi_1\over\sqrt{2}}\sqrt{1+{B^2\phi_1^2\over A^4
(1-\phi_1^2/A^2)}}~,
\end{equation}
where $A$ and $B\le A$ are respectively the semi-major and the semi-minor
axis of the ellipse, and $\phi_1=Re\phi/\sqrt{2}$. For a circular orbit
$B=A$, whereas for pure oscillation (no charge in the condensate) $B=0$.

It is therefore obvious that small corrections to a harmonic
potential of a coherent condensate can easily generate a pressure.
As we have seen, in the gravity mediated case quantum corrections
typically modify the flat direction mass terms by
\begin{equation}
U(\phi)=\frac 12 m_{\phi}^2\phi^2+ Km_{\phi}^2\phi^2\log\left(
{\phi^2\over\mu^2}\right)+\dots\,,
\end{equation}
where $K$ is some constant. If one writes
\begin{equation}
U(\phi)=\frac 12 m_{\phi}^2\phi^2\left({\phi^2\over\mu^2}\right)^x
\end{equation}
one finds that
\begin{equation}
\g = {1+x\over 1+\frac x2}~, ~p={x\over 2+x}~.
\end{equation}
In the case of the logarithmic potential $x\simeq 2K$. There arises
a negative pressure $p=K \rho$ whenever $K<0$ or whenever $x$ is small
and negative. This is a sign of an instability of the condensate under
arbitrarily small perturbations.

This is exactly the situation one finds in the MSSM with flat directions.
The effective mass $m^2_{\rm eff}(\phi)\equiv {dU/ d\phi^2}$
decreases for a range in $\phi$, albeit for different reasons, both for
gravity mediated and gauge mediated supersymmetry breaking. According to
\eq{effgamma2} this results in a negative pressure, which has been computed
numerically by Jokinen \cite{jokinen02}. The results are shown in
Figs.~(\ref{plotpreengr}) and (\ref{plotpreenga}).

%%%%%%%%%%%%%%%%%%%%%%%%%%%%%%%%%%%%%%%%%%%%%%%%%%%%%%%%%%%%%%
\bfig
\leavevmode
\centering
\vspace*{4cm}
\includegraphics{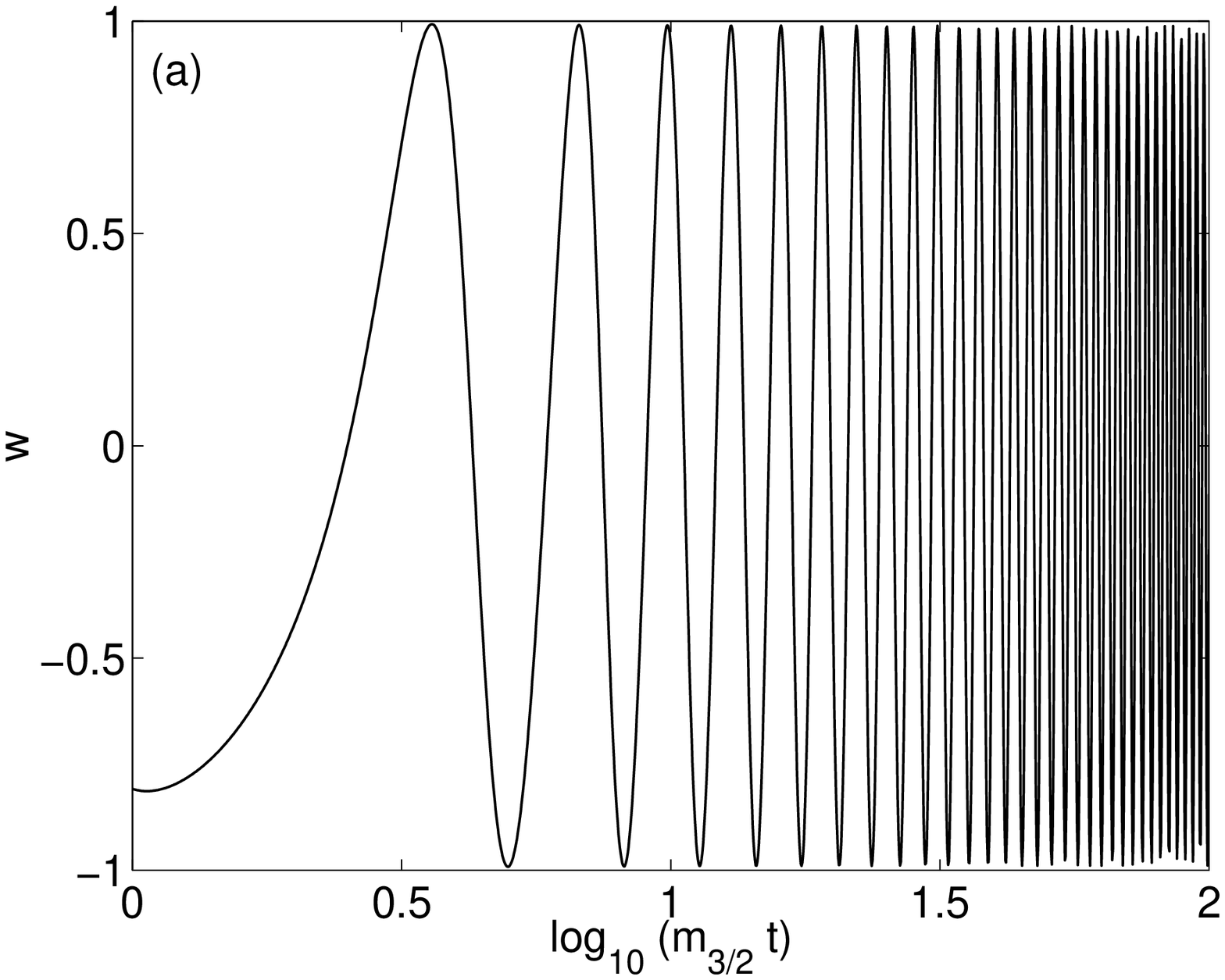}
\includegraphics{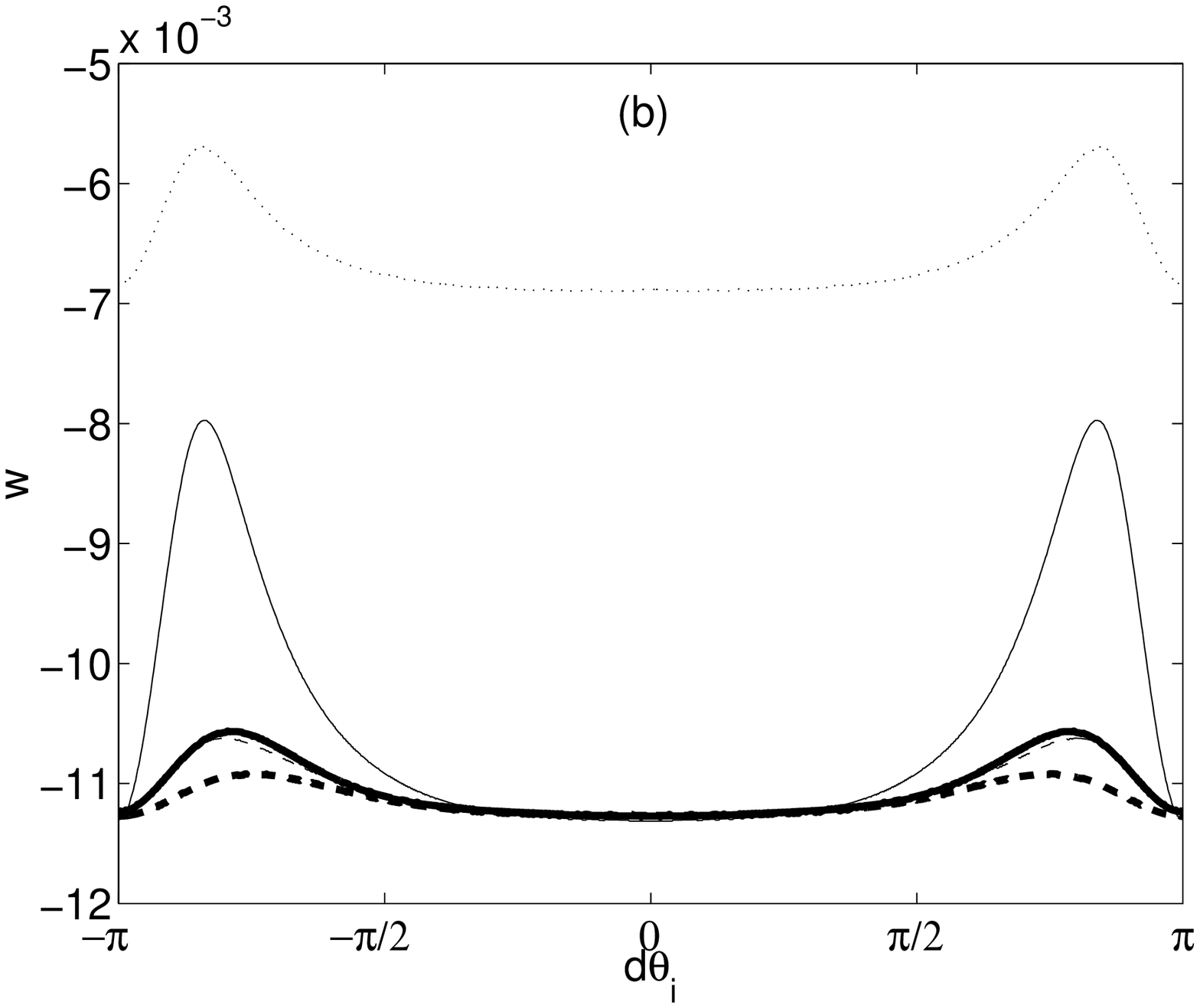}
\includegraphics{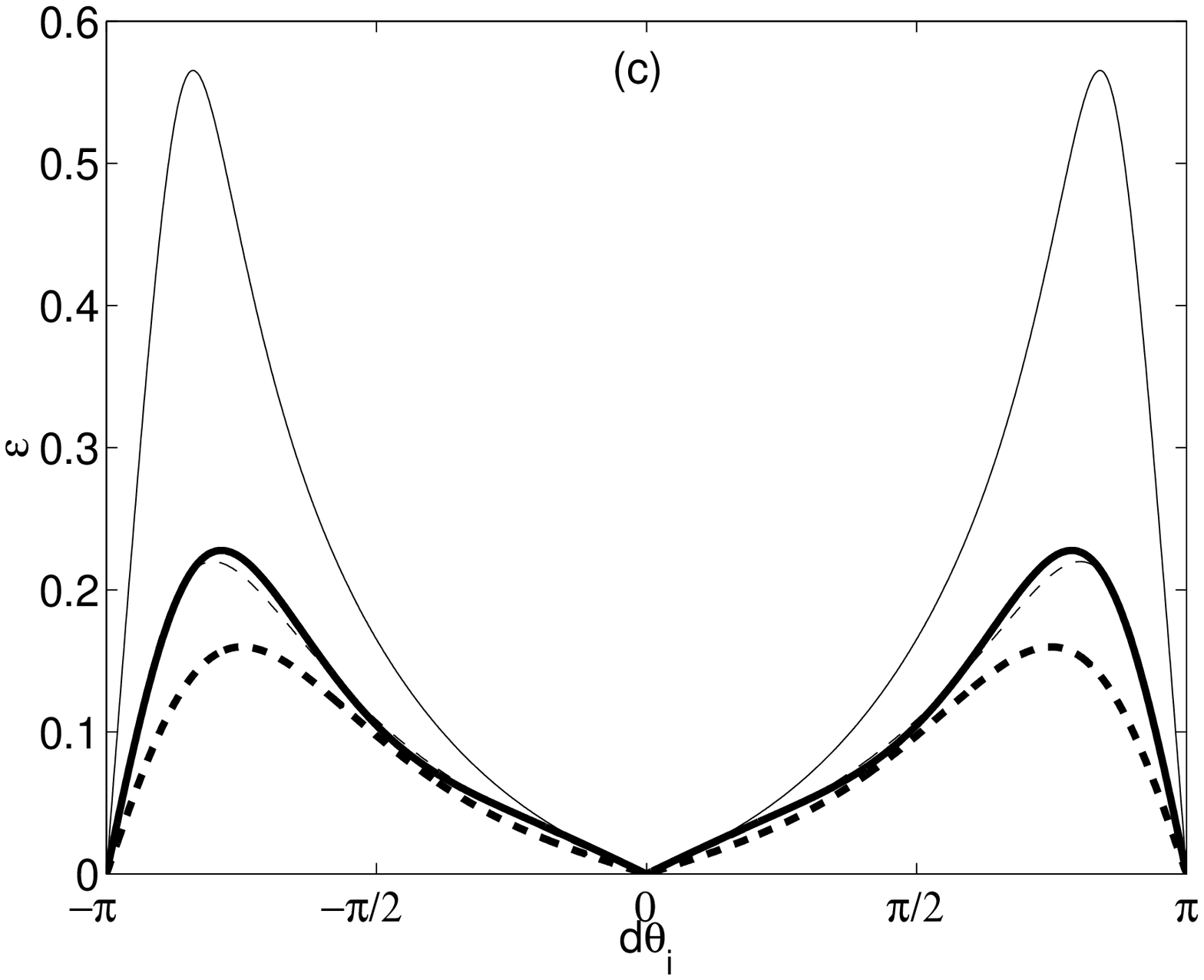}
\caption{\small Pressure-to-energy density ratio, $p/\rho\equiv w=\gamma-1$,
in the gravity mediated case vs. (a) time in logarithmic units for $d=4$,
(b) different initial conditions for $d=4,\,6$; (c) ellipticity
$\varepsilon=B/A$ vs. initial conditions for $d=4$ (thin lines), $d=6$
(thick lines), D-term (solid), F-term (dashed) with  $K=-0.01$ and
$t=100m_{3/2}^{-1}$. In (b) $w$ is shown at $t=300m_{3/2}^{-1}$
with dotted lines for the $d=4$ D-term case, from \cite{jokinen02}.}
\label{plotpreengr}
\efig

%%%%%%%%%%%%%%%%%%%%%%%%%%%%%%%%%%%%%%%%%%%%%%%%%%%%%%%%%%%%%%%

The pressure-to-energy density ratio; $w=\gamma-1$ for the gravity mediated
case is plotted in Fig.~(\ref{plotpreengr}a), which shows that there are
time-dependent oscillations in pressure \cite{jokinen02}. The average
pressure is slightly on the negative side. The average value of $w$ is
shown in Fig.~(\ref{plotpreengr}b) at $t\sim 100m_{3/2}^{-1}$ for a few
different initial conditions. In Fig.~(\ref{plotpreengr}c), the ellipticity
of the orbit, $\varepsilon=B/A$, is plotted to show that $w$ is more negative
if $\varepsilon$ is small. It should be noted that $w$ achieves
values which are more negative than the absolute lower bound coming
from pure oscillation.

A similar analysis has been made for the gauge mediated supersymmetry
breaking case. The results have striking similarities \cite{jokinen02}.
In Figs.~(\ref{plotpreenga}a) and (\ref{plotpreenga}b), the time
development of the pressure-to-energy density ratio, $w$, for $d=4$
and $d=6$ has been depicted. One can see that the pressure is always
negative. The calculation of average pressure is even more involved
than in the gravity mediated case, since the oscillation frequency
becomes very large. In Fig.~(\ref{plotpreenga}c), the ellipticity of the
orbit is shown as a function of different initial conditions.
Jokinen \cite{jokinen02} has pointed it out that quite generically
$\varepsilon \lsim 0.1$.

%%%%%%%%%%%%%%%%%%%%%%%%%%%%%%%%%%%%%%%%%%%%%%%%%%%%
\bfig
\leavevmode
\centering
\vspace*{4cm}
\includegraphics{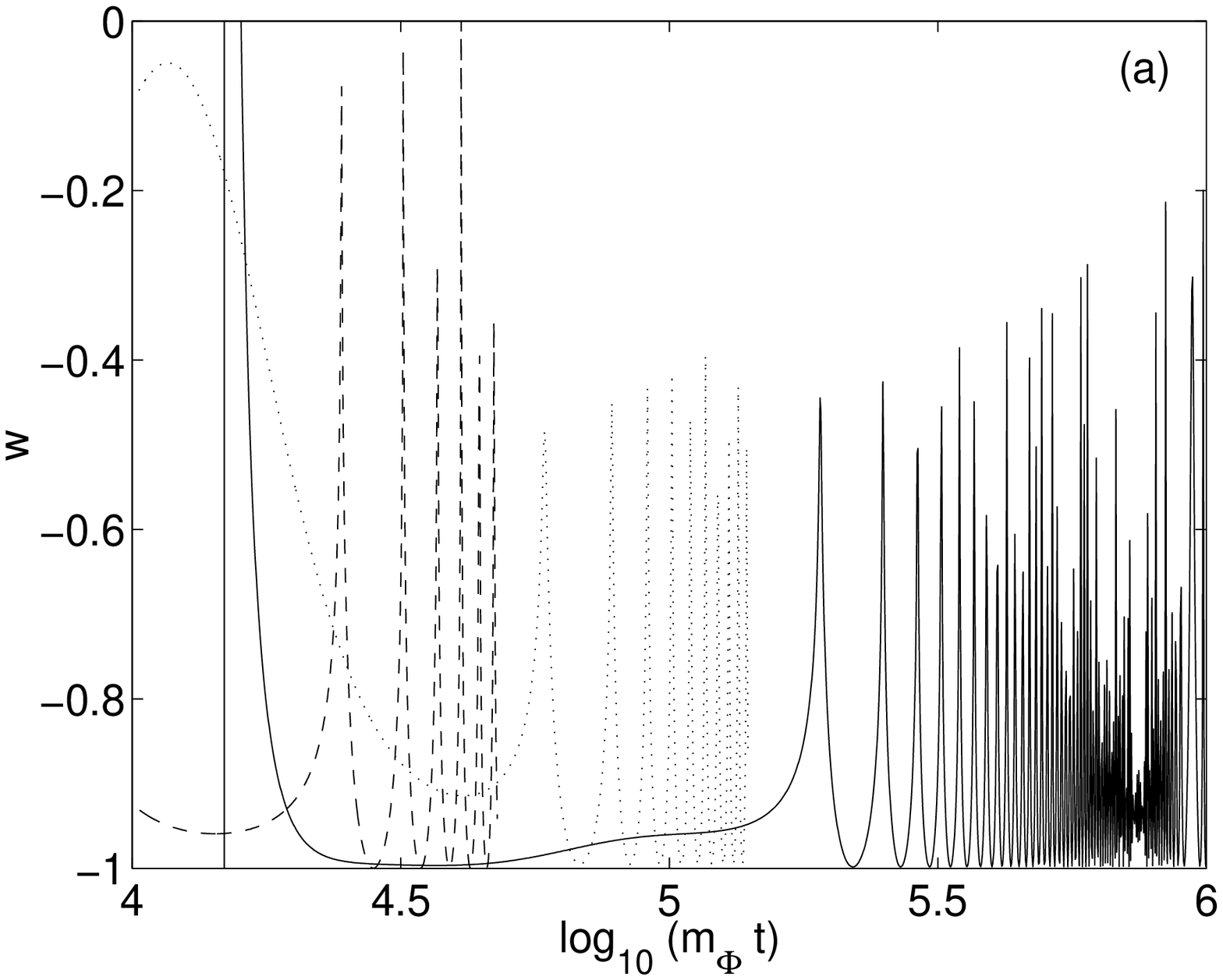}
\includegraphics{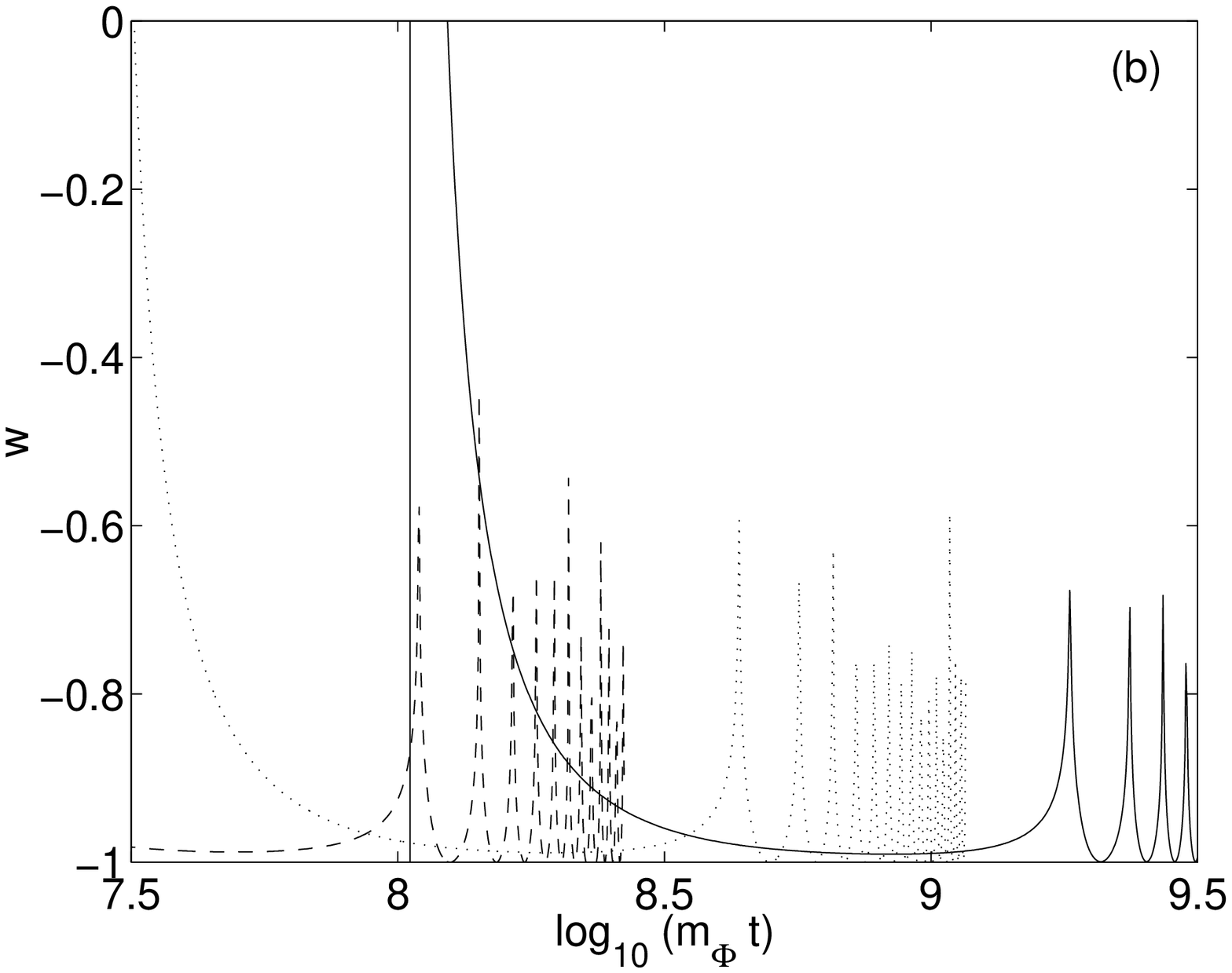}
\includegraphics{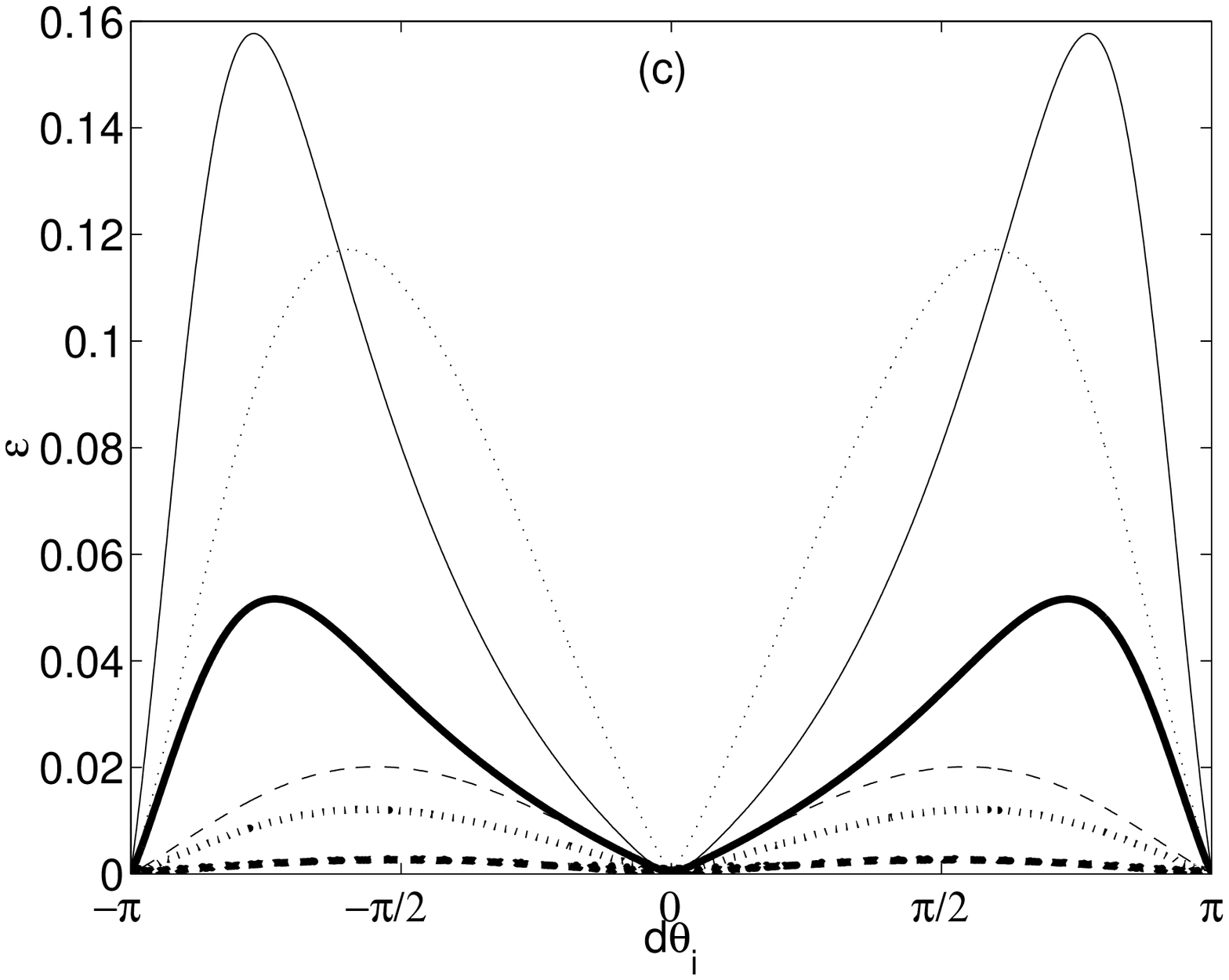}
\caption{\small Pressure-to-energy density ratio, $w$, in the gauge
mediated D-term case (without Hubble induced $A$-term) vs. time
in logarithmic units for (a) $d=4$ and (b) $d=6$; (c) ellipticity
of the orbit where $d=4$ (thin lines) and $d=6$ (thick lines).
The scalar masses
$m_{\phi}=1,\,10,\,100\TeV$ are denoted respectively with solid, dotted
and dashed lines, from \cite{jokinen02}.}\label{plotpreenga}
\efig
%%%%%%%%%%%%%%%%%%%%%%%%%%%%%%%%%%%%%%%%%%%%%%%%%%%%%%%%%%%%%%%%%

%%%%%%%%%%%%%%%%%%%%%%%%%%%%%%%%%%%%%%%%%%%%%%%%%%%%%%%%%%%%%%%%%

\subsubsection{Growth of perturbations in the AD condensate}

%%%%%%%%%%%%%%%%%%%%%%%%%%%%%%%%%%%%%%%%%%%%%%%%%%%%%%%
\bfig
\leavevmode
\centering
\vspace*{3cm}
\includegraphics{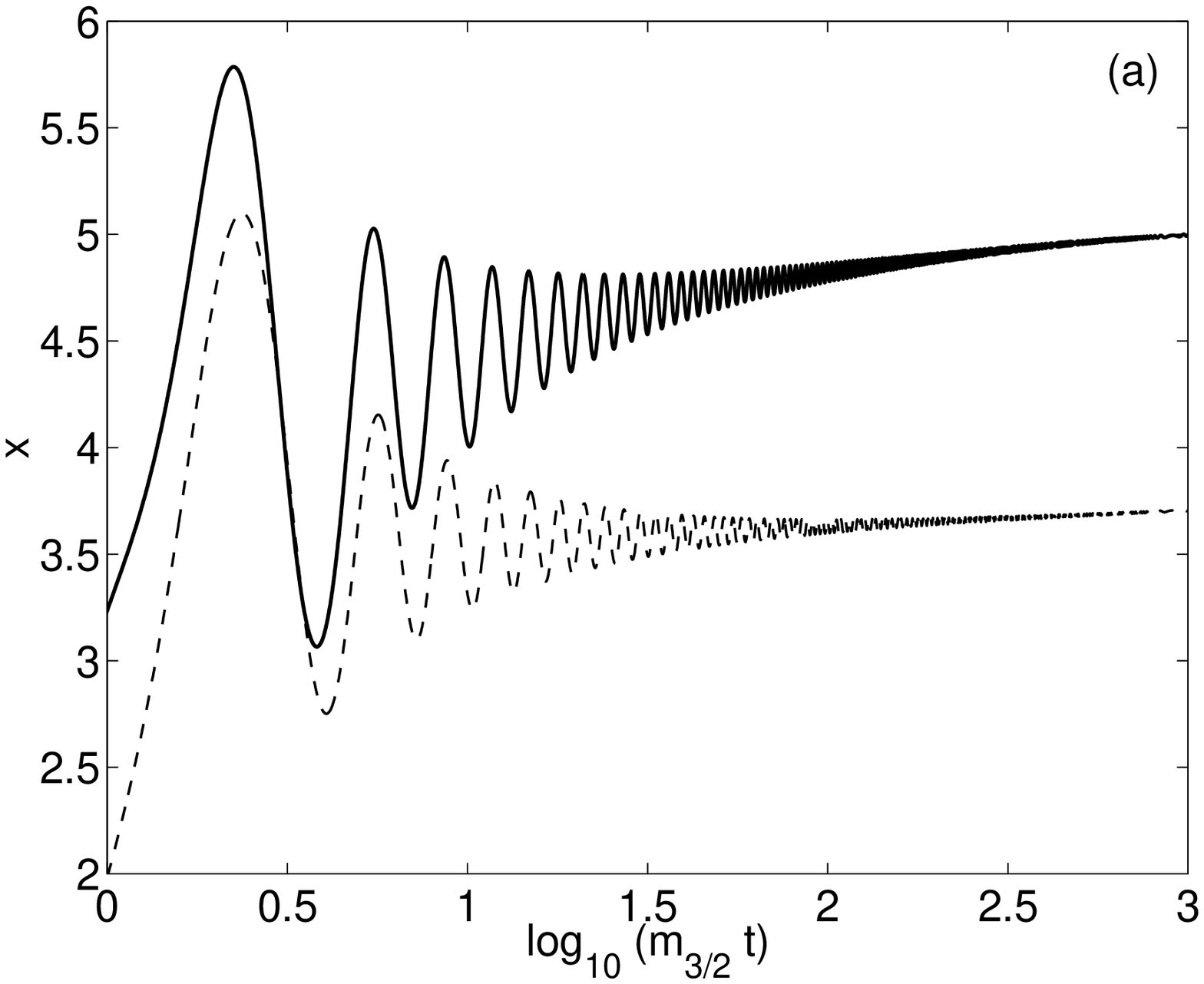}
\includegraphics{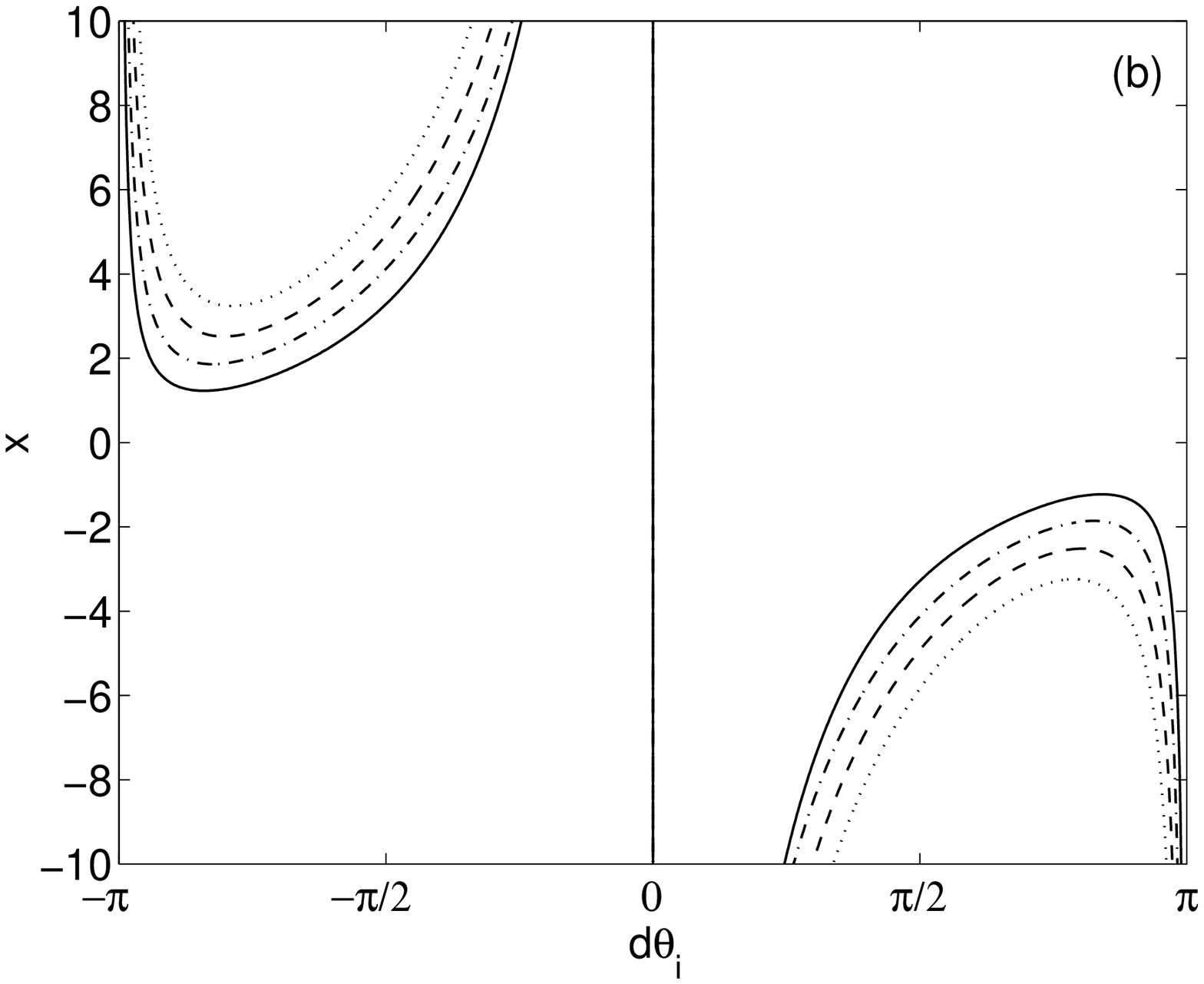}
\includegraphics{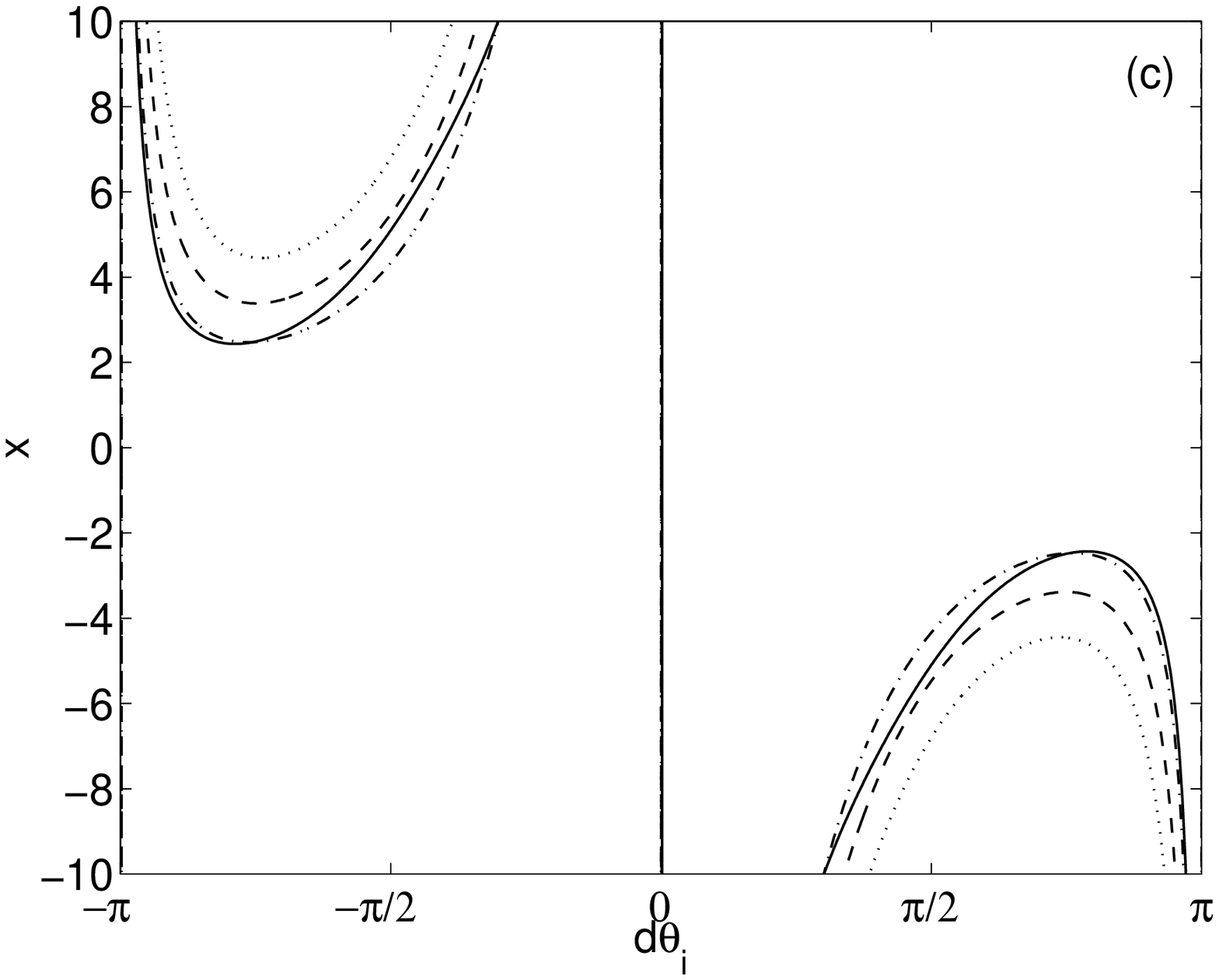}
\caption{\small Energy-to-charge ratio, $x$, in the gravity mediated
case vs. (a) time in logarithmic units for $d=4,\,6$; (b) the D-term case;
(c) the F-term (with Hubble induced $A$-term) case with $d=4,5,6,7$
(solid, dash-dot, dashed and dotted lines), $K=-0.01$ and
$t=100m_{3/2}^{-1}$, from \cite{jokinen02}.}\label{plotenchagr}
\efig

%%%%%%%%%%%%%%%%%%%%%%%%%%%%%%%%%%%%%%%%%%%%%%%%%%%%%%%%%

As a result of internal negative condensate pressure the
quantum fluctuations in the scalar condensate grow according
to~\cite{liddle-lyth00}
\begin{equation}
\ddot{\delta}_{\bf k} = -K{\bf k}^2\delta_{\bf k}~.
\end{equation}
If $K<0$, quantum fluctuations of the condensate field at the
scale $\lambda= 2\pi/\vert{\bf k}\vert$ will grow exponentially in time as
\begin{equation}
\delta\phi_{\bf k}(t)=\delta\phi(0){\rm exp}\left(-K{\bf k}^2 t\right)~.
\end{equation}
In reality the onset of non-linearity sets the scale at which the spatial
coherence of the condensate can no longer be maintained and the condensate
fragments. For the AD condensate the initial perturbation originates
from inflation. Note that since the AD condensate carries a global
charge, due to charge conservation the energy-to-charge ratio
changes as the the condensate fragments.

The energy-to-charge ratio has been estimated numerically for both gravity and
gauge mediated cases by Jokinen \cite{jokinen02}. The time evolution of the
energy-to-charge ratio $x$ is shown in Fig.~\ref{plotenchagr}, where $x$
is also plotted for various initial phases in F-and D-term inflation
models. For the gauge mediation case the plots are quite different from
the gravity mediated case, see Fig.~(\ref{plotenchaga}).

%%%%%%%%%%%%%%%%%%%%%%%%%%%%%%%%%%%%%%%%%%%%%%%%%%%%%%%%%%%%%%%%%%
\bfig
\leavevmode
\centering
\vspace*{4cm}
\includegraphics{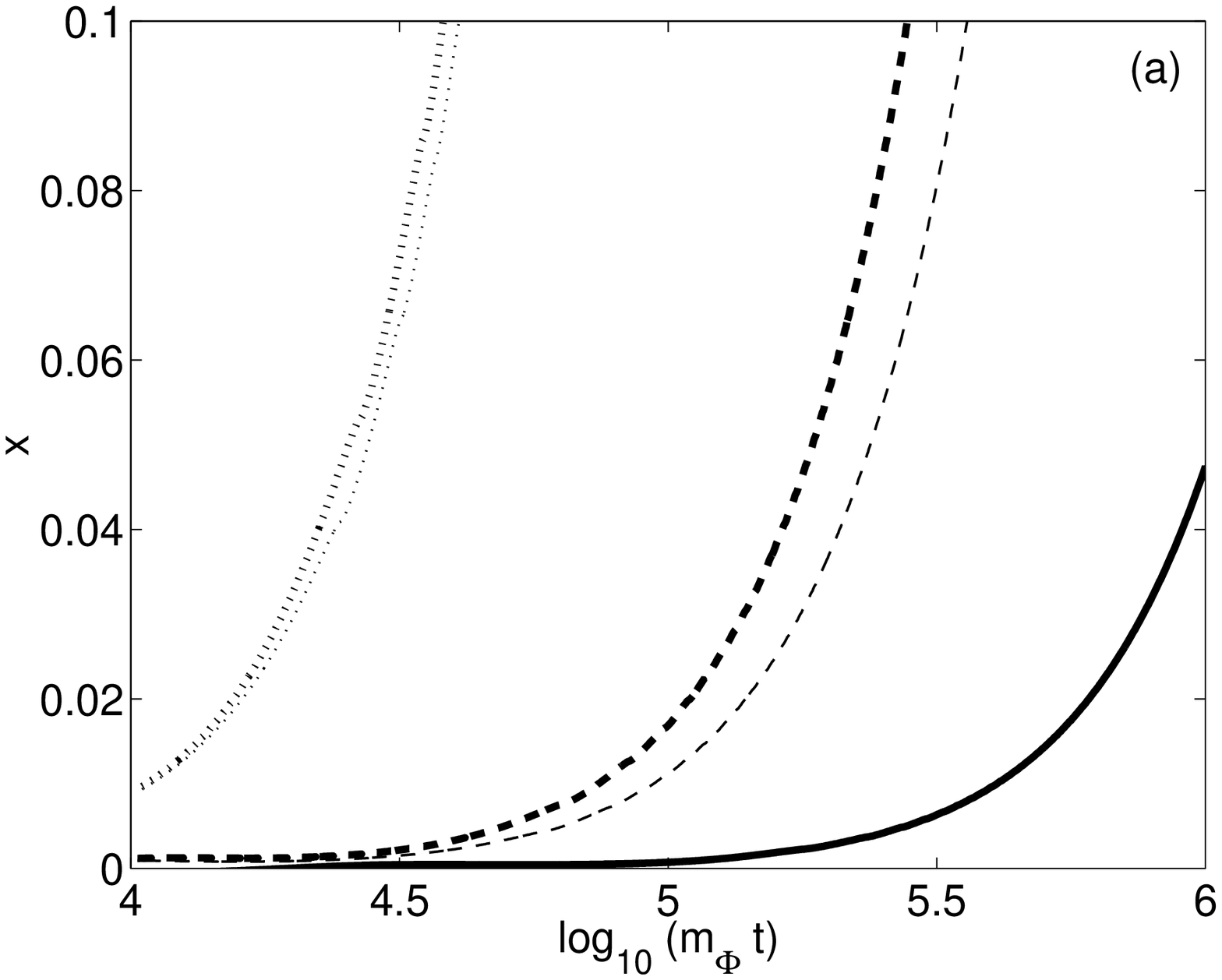}
\includegraphics{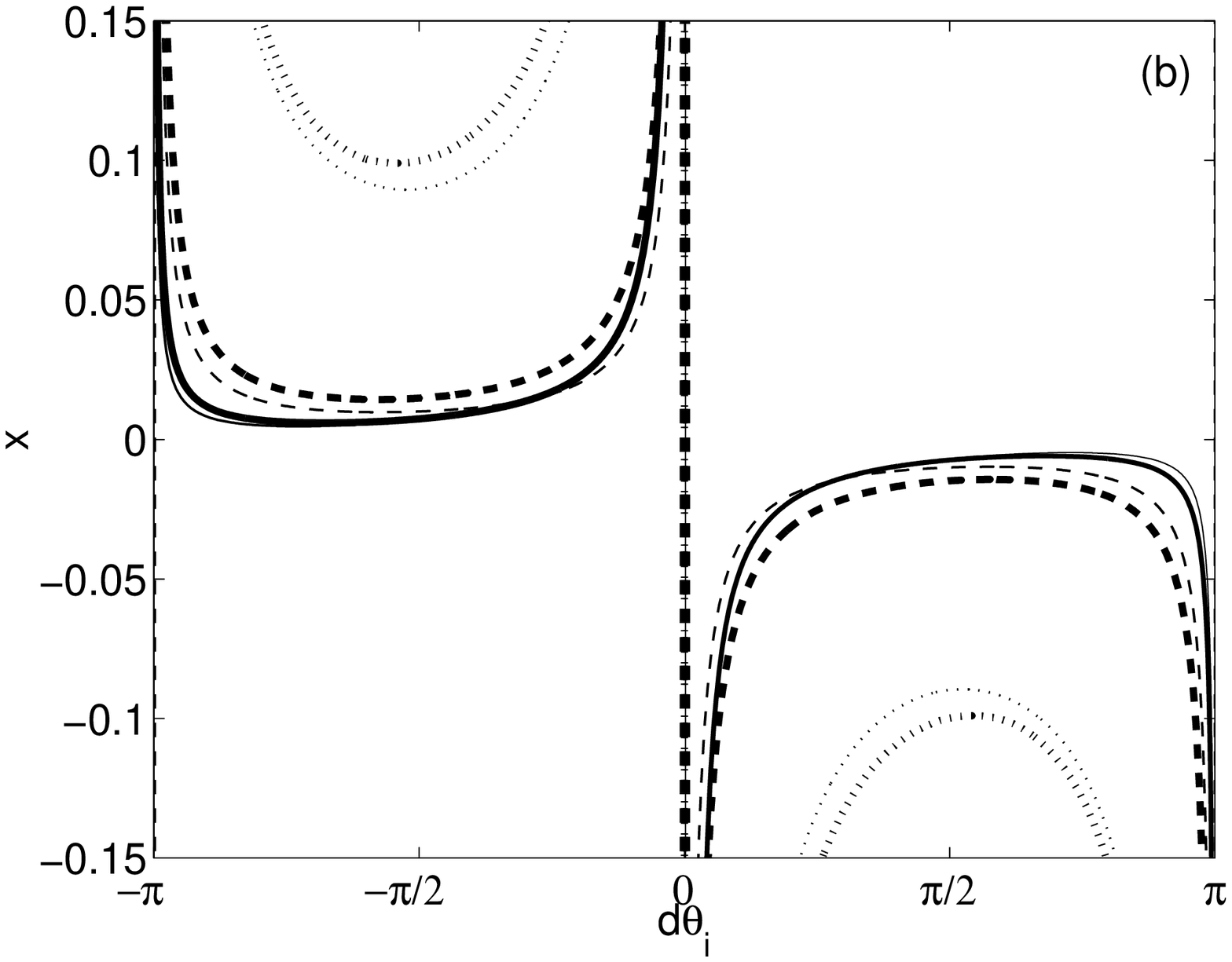}
\includegraphics{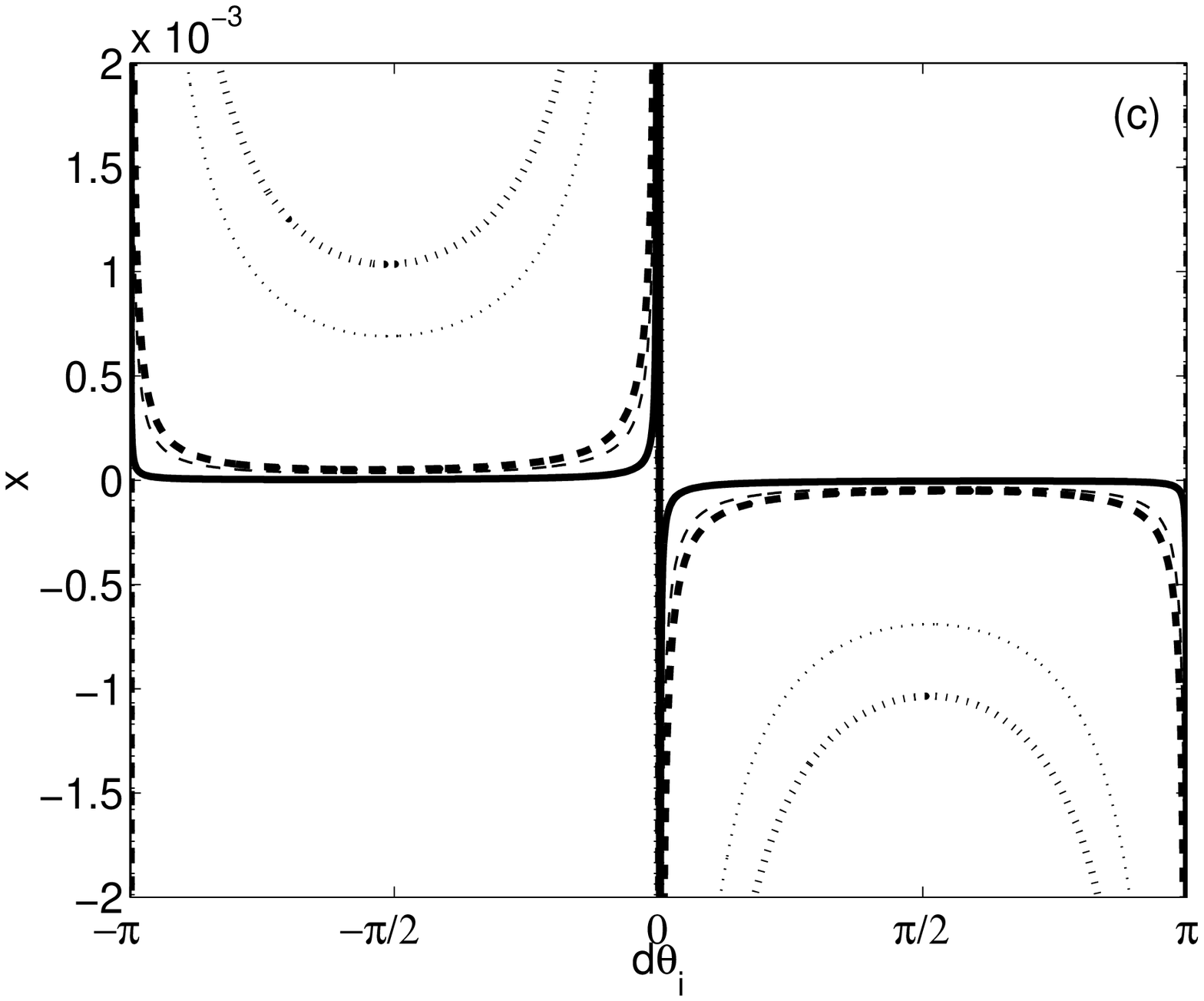}
\caption{\small Energy-to-charge ratio, $x$, in the gauge mediated case
vs. (a) time in logarithmic units of time $d=4$, (b) $d=4$ and (c)
$d=6$ with D-term (thin lines, without Hubble induced A-term)and
F-term (thick lines, with Hubble induced A-term)
and $m_{\phi}=1,\,10,\,100\TeV$ (solid, dashed, dotted lines) at
$t=4\cdot 10^5m_{\phi}^{-1},\,10^5m_{\phi}^{-1},\,4\cdot 10^4m_{\phi}^{-1}$
($d=4$) and $t=4\cdot 10^9,\,10^9,\,4\cdot 10^8m_{\phi}^{-1}$ ($d=6$),
from \cite{jokinen02}.}
\label{plotenchaga}
\efig

%%%%%%%%%%%%%%%%%%%%%%%%%%%%%%%%%%%%%%%%%%%%%%%%%%%%%%%%%%%%%%%

%%%%%%%%%%%%%%%%%%%%%%%%%%%%%%%%%%%%%%%%%%%%%%%%%%%%%%%%%%%%%%%%
\subsubsection{The true ground state}

Under the negative pressure the homogeneous AD condensate fragments and
forms lumps. The question then is, what is the true ground state? The
answer is, a non-topological soliton with a fixed charge, called the
$Q$-ball \cite{rosen68,lee74,coleman85}, which in general is made up of a
complex scalar field with a global U(1) symmetry, for which the
Lagrangian is
\begin{equation}
{\cal L}=\partial_\mu\phi\partial^\mu\phi^* - U(\phi\phi^*)~.
\end{equation}
When supplemented by the CP violating terms, this is the Lagrangian for
the MSSM flat directions.

The conserved current is
$j^\mu = i(\phi\partial^\mu\phi^*-\phi^*\partial^\mu\phi)$, and the conserved
charge, and energy are given by
\begin{eqnarray}
Q &=& \int d^3xj^0~,\\
E &=& \int d^3x[\vert\dot\phi\vert^2+\vert\nabla\phi\vert^2+U(\phi\phi^*)]
\end{eqnarray}
If the charge is kept fixed, the state of lowest energy is found by
minimizing \cite{coleman85}
\begin{equation}
E_\omega=E-\omega(Q-i\int d^3xj^0)~,
\end{equation}
with respect to variations in $\phi$ and the Lagrange multiplier
$\omega$. From $\delta_\phi E_\omega = 0$, it follows that
$\dot\phi-i\omega\phi=0$, so that we may write
\begin{equation}
\label{om0}
\phi(t,{\bf x})=e^{i\omega t}\varphi({\bf x})
\end{equation}
where $\varphi$ may be chosen real by virtue of $U(1)$ invariance.
The charge and energy of such a configuration read
\begin{eqnarray}
\label{chargeq}
Q &=& 2\omega \int d^3x~\varphi^2~,\\
\label{energyq}
E &=&\int d^3x~[(\nabla\varphi)^2+U(\varphi^2)+\omega^2\varphi^2]~,
\end{eqnarray}
and one has to minimize
$E_\omega=\int d^3x  [(\nabla\varphi)^2+\hat U_\omega(\varphi^2)]+\omega Q$
where
\begin{equation}
\label{uomega}
\hat U_\omega(\varphi^2)=U(\varphi^2)-\omega^2\varphi^2~.
\end{equation}
To find a localized configuration that vanishes at spatial infinity one may
make use of the spherical rearrangement theorem, which implies that
$E_\omega$ is minimized by $\varphi({\bf x})$ which is spherically
symmetric and monotonically decreasing. This is equivalent to solving
the equation of motion
\begin{equation}
\label{qbeqm}
{d^2\varphi\over dr^2}+\frac 2r{d\varphi\over dr}-\varphi
{dU_\omega\over d\varphi^2}=0~.
\end{equation}
If by convention we set the globally symmetric minimum to $\varphi=0$ with
$U(0)=0$, one can then show that a non-trivial solution to Eq.~\rf{qbeqm}
is obtained whenever $U(\varphi^2)/\varphi^2$ has a minimum at
$\varphi\ne 0$, i.e. $U(\varphi^2)$ grows more slowly than
$m_{\phi}^2\varphi^2$ over some range. We will discuss
$Q$-balls in detail in the next Sects.~$6$ and $7$.

%%%%%%%%%%%%%%%%%%%%%%%%%%%%%%%%%%%%%%%%%%%%%%%%%%%%%%%%%%%
\subsection{Numerical studies of fragmentation}

Although the homogeneous AD condensate is not the ground state,
it is not obvious that the ground state should always be reached
within cosmic time scales. It is then essential to study the
dynamical evolution of the AD condensate. Since the $Q$-ball formation
is inherently a non-linear phenomenon, analyzing small perturbations
is not sufficient to determine the full dynamical evolution of the AD
condensate. One can nevertheless gather some information about the
gross features of the condensate fragmentation by perturbative
considerations alone.

\nc{\w}{\omega}
%%%%%%%%%%%%%%%%%%%%%%%%%%%%%%%%%%%%%%%%%%%%%%%%%%%%%%%%%
\subsubsection{Perturbation theory}

Negative pressure is equivalent to an attractive force between
the condensate quanta which induces a growing mode in spatial
perturbations. A linearized description of the evolution of
perturbations has been given by Kusenko and Shaposhnikov
in~\cite{kusenko98418}, and by Enqvist and McDonald in
\cite{enqvist98,enqvist9881,enqvist99,enqvist00570}.
For a maximally charged AD condensate ($B=0$ in Eq.~(\ref{major-minor})),
the linearized perturbation
takes the form $\phi = \phi(t) + \delta \phi(x,t)$, and
$\theta = \theta(t)+\delta\theta(x,t)$, where the homogeneous
condensate is described by
\be{le1}
\Phi = \frac{\phi(t)}{\sqrt{2}}e^{i \theta(t)}~,
\ee
with $\phi(t) = (a_{o}/a)^{3/2}\phi_{o}$  and
$\dot{\theta}(t)^{2} \approx m_{\phi}^{2}$.
Initially $\delta \phi(x,t)$ and $\delta \theta (x,t)$ should
satisfy the relationship~\cite{kusenko98418,enqvist98,enqvist99}
\be{ks1}
\delta \theta_{i} \approx \left(\frac{\delta \phi}{\phi}\right)_{i}~.
\ee
The solution of the linear perturbation equations then has the
form \cite{enqvist98}
\be{le2}
\delta \phi \approx \left(\frac{a_{o}}{a}\right)^{3/2}
\delta \phi_{o}\;  \exp \left( \int dt \left(\frac{1}{2}
\frac{\vec{k}^{2}}{a^{2}}\frac{|K| m_{\phi}^{2}}{\dot{\theta}(t)^{2}}
\right)^{1/2} \right)e^{i\vec{k}.\vec{x}}~
\ee
and
\be{le3}
\delta \theta \approx \delta \theta_{i}\; \exp\left( \int dt \left(\frac{1}{2}
\frac{\vec{k}^{2}}{a^{2}}\frac{|K| m_{\phi}^{2}}{\dot{\theta}(t)^{2}}
\right)^{1/2} \right)e^{i\vec{k}\cdot\vec{x}}       ~.
\ee
For the gravity mediation case, the above condition applies if
$\left| \vec{k}^{2}/a^{2} \right|\lae | 2K m_{\phi}^{2}|$, and
$H^{2}$ is small compared with $m_{\phi}^{2}$ and $|K| \ll 1$.
If the first condition is not satisfied, then the gradient energy of the
perturbations produces a positive pressure larger than the negative
pressure due to the attractive force from the logarithmic term,
preventing the growth of the perturbations.

For the case of a matter dominated Universe, the exponential growth
factor is then \cite{enqvist98,enqvist99}
\be{le4}
\int dt \left(\frac{1}{2} \frac{\vec{k}^{2}}{a^{2}}\frac{|K| m_{\phi}^{2}}
{\dot{\theta}(t)^{2}} \right)^{1/2} = \frac{2}{H} \left(\frac{|K|}{2}
\frac{\vec{k}^{2}}{a^{2}}\right)^{1/2}      ~.
\ee
The largest growth factor will correspond to the largest value
of $\vec{k}^{2}$ for which growth can occur,
\begin{equation}
\label{instband}
\left.\frac{\vec{k}^{2}}{a^{2}}\right|_{max} \approx 2 |K| m_{\phi}^{2}\,.
\end{equation}
The value of $H$ at which the first perturbation goes
non-linear is \cite{enqvist98,enqvist99}
\be{le5}
H_{i}\approx \frac{2 |K| m_{\phi}}{\alpha(\lambda)}~,
\ee
with
\be{le6}
\alpha(\lambda) = -\log \left( \frac{\delta \phi_{o}(\lambda)}{\phi_{o}}
\right)~,
\ee
where $\phi_{o}$ is the value of $\phi$ when the condensate oscillations
begin at $H \approx m_{\phi}$. A typical value of $\alpha(\lambda)$
(e.g. with $d=6$) is $\alpha(\lambda) \approx 30$. The initial non-linear
region has a radius $\lambda_{i}$ at $H_{i}$, which is given by
\cite{enqvist98,enqvist99}
\be{le7}
\lambda_{i} \approx \frac{\pi}{|2 K|^{1/2} m_{\phi}}    ~.
\ee

For the case of a non-maximally charged condensate the situation is
slightly different. It is likely that the initial radius and
the time at which the spatial perturbations initially go non-linear will
roughly be the same \cite{enqvist99} as for the maximally charged condensate.
In general, the charge density of the initial non-linear lumps will
essentially be the same as that of the original homogeneous condensate.

The perturbative evolution of a single condensate lump was considered
in \cite{enqvist99}.
In terms of $\phi = (\phi_{1}+i\phi_{2})/\sqrt{2}$, the initial
lumps are described by
\be{g8}
\phi_{1}(r,t) = A \cos(m_{\phi}t) (1 + \cos(\pi r/r_{0}) )       ~
\ee
\be{g9}
\phi_{2}(r,t) = B \sin(m_{\phi}t) (1 + \cos(\pi r/r_{0}) )       ~,
\ee
for $r \leq r_{0}$ and by $\phi_{1,2} = 0$ otherwise. The initial radius
of the lump is $2 r_{0}$, where $r_{0} =  \pi/(\sqrt{2} |K|^{1/2} m_{\phi})$.
The maximally charged condensate lump corresponds to $A = B$, while
the non-maximal lump has $A > B$. The total energy and charge in a
fixed volume are given by~\cite{enqvist00570,enqvist00483}
\be{QE}
E=4\pi\int_V drr^2\rho~~,~Q=4\pi\int_V drr^2 q~\sim AB,
\ee
with $Q_{\rm max}=A^2$.

In \cite{enqvist99}, the behavior of the solutions was found
in a perturbative analysis to depend on $K$, and to a greater
extent on $Q/Q_{\rm max}$. The condensate lump was found to
pulsate while charge is flowing out until the lump reaches a
(quasi-)equilibrium pseudo-breather configuration, also called
$Q$-axiton, with the lump pulsating with only a small difference
between the maximum and minimum field amplitudes. For the $Q$-axiton,
in which the attractive force between the scalars is balanced by the
gradient pressure of the scalar field, the energy per unit charge is much
larger than $m_{\phi}$; indeed, the $Q$-axiton exists even if $Q=0$.
Only for a maximally charged $Q$-axiton are the properties similar to
that of the corresponding $Q$-ball. It is however unclear whether
$Q$-axitons are just an artifact of perturbation expansion.

%%%%%%%%%%%%%%%%%%%%%%%%%%%%%%%%%%%%%%%%%%%%%%%%%%%%%%
\subsubsection{Lattice simulations}

The features of the fragmentation of the AD condensate cannot be fully
captured by studying various mean field theory approaches, such as in
large N-approximation and Hartree-approximation
\cite{boyanovsky95,cooper94,heitmann00}. The formation of
a $Q$-ball is a non-linear process for which various mode-mode interactions
become important. This can be seen by expanding the perturbed
$\phi$ and $\theta$ as shown by Kasuya and Kawasaki in
\cite{kasuya0061,kasuya00,kasuya0062}
\begin{eqnarray}
\label{pertQ}
\delta \ddot \phi +3H\delta \dot\phi-2\dot\theta \phi\delta \phi-
\frac{\nabla^2}{a^2}\delta\phi+U^{\prime\prime}(\phi)\delta\phi&=&0\,,
\nonumber \\
\phi\ddot\theta +3H\phi\delta \dot\theta+2(\dot\phi\delta\dot\theta)-
2\frac{\dot\phi}{\phi}\dot\theta\delta\phi-\phi\frac{\nabla^2}{a^2}\delta
\theta &=&0\,.
\end{eqnarray}
Although the potentials differ in the gauge and gravity mediated cases,
it is nevertheless always possible to identify  the
fastest growing amplified mode. In the gravity mediated
case we have already obtained that by inspecting Eq.~(\ref{le4}).
A similar analysis can be performed for the gauge mediated case by noting that
$U^{\prime\prime}(\phi)\approx -2m_{\phi}^4/\phi^2$. Taking into account
the conservation of charge $\dot\theta\phi^2a^3={\rm const.}$, along with
the approximation of a circular orbit, one may simplify Eq.~(\ref{pertQ})
by seeking a solution of the form
$\delta \phi=\delta\phi_{0}\exp(\alpha t+ikx)$ and
$\delta\theta =\delta\theta_{0}\exp(\alpha t+ikx)$. In order to further
simplify the analysis, one can also assume $a={\rm const.}$ and
$\phi=\phi_{0}={\rm const.}$, so that the phase velocity
$\dot\theta=(U^{\prime}/\phi)^{1/2}\approx \sqrt{2}m_{\phi}^2/\phi_{0}$.
If $\alpha$ is real and positive, the fluctuations grow exponentially
and become non-linear. Solving for $\delta\phi_{0},\delta\theta_{0}$,
Kasuya and Kawasaki finds for gauge mediated case~\cite{kasuya01}
\begin{equation}
\alpha^4+2\left(\frac{\vec k^2}{a^2}+\frac{2m_{\phi}^4}{\phi_{0}^2}\right)
\alpha +\left(\frac{\vec k^2}{a^2}-\frac{4m_{\phi}^4}{\phi_{0}^2}\right)
\vec k^2=0\,.
\end{equation}
Note that in order for $\alpha$ to be positive, one must require the
expression in the second parenthesis to be negative. This means that the
instability band for the fluctuations is given by
\begin{equation}
0<\frac{\vec k}{a}<\frac{2m_{\phi}^2}{\phi_{0}}\,.
\end{equation}
The most amplified mode appears at
$(\vec k/a)_{max}\approx (3/2)^{1/2}m_{\phi}^2/\phi_{0}$
in the gauge mediated case.

Various groups have studied the fragmentation of the AD condensate and
the formation of $Q$-balls numerically. In \cite{enqvist0163}, condensate
fragmentation was simulated numerically on a $2+1$ dimensional
$100\times 100$ lattice, starting with a uniform AD-condensate with
$\phi_0=10^9$~GeV and with an arbitrary phase $\w$,
$\phi=\phi_0 \textrm{e}^{i\w t}+\delta\phi$
with uniformly distributed random noise $\delta\phi\sim \cO(10^{-13})|\phi_0|$
added to the amplitude and phase. The parameter values chosen for the
simulations were $m_{\phi}=10^2\GeV,\ K=-0.1$, and $\lambda=1/2$.
The results indicate that first the charge density
of the condensate decreases uniformly due to the expansion of the Universe. As
time progresses a growing mode can be seen to develop. White noise is
still present but the growing mode soon starts to dominate. This process
continues until lumps of positive charge develop. These are
dynamically arranged in string-like features but the filament texture
is a transient feature which disappears soon, see forthcoming
Figs.~(\ref{figmultamaki1},\ref{figmultamaki2}).

The further evolution of the flat direction depends on the initial
energy-to-charge ratio of the condensate, defined as
\begin{equation}
\label{xdefinition}
x\equiv {E\over m Q}~,
\end{equation}
and hence on the
value of $\w$. If $x=1$, i.e. the
energy-to-charge ratio of the condensate is equal to that of a $Q$-ball,
no negatively charged $Q$-balls, anti-$Q$-balls, are formed. After the
modes grow non-linear, the lumps just evolve into $Q$-balls and finally freeze
due to the expansion of the Universe.

If $\w<1$ so that $x\gg 1$, the fragmentation process has a much more
complicated history \cite{enqvist0163}. After the positively charged
lumps have formed, expanded linearly and then developed non-linearly,
the extra energy stored in them causes the lumps to fragment as they
evolve into $Q$-balls.
In this process a large number of negatively charged $Q$-balls
forms. The total charge in the negative and positive $Q$-balls is
approximately equal so that the initial charge in the condensate is in
fact negligible compared to the amount of charge and anti-charge created.

%%%%%%%%%%%%%%%%%%%%%%%%%%%%%%%%%%%%%%%%%%%%%%%%%%%%%%%%%%%%%%%
\input epsf
\begin{figure}[t!]
\centering
\hspace*{-7mm}
\leavevmode\epsfysize=4cm \epsfbox{relic3D.eps}
\leavevmode\epsfysize=4cm \epsfbox{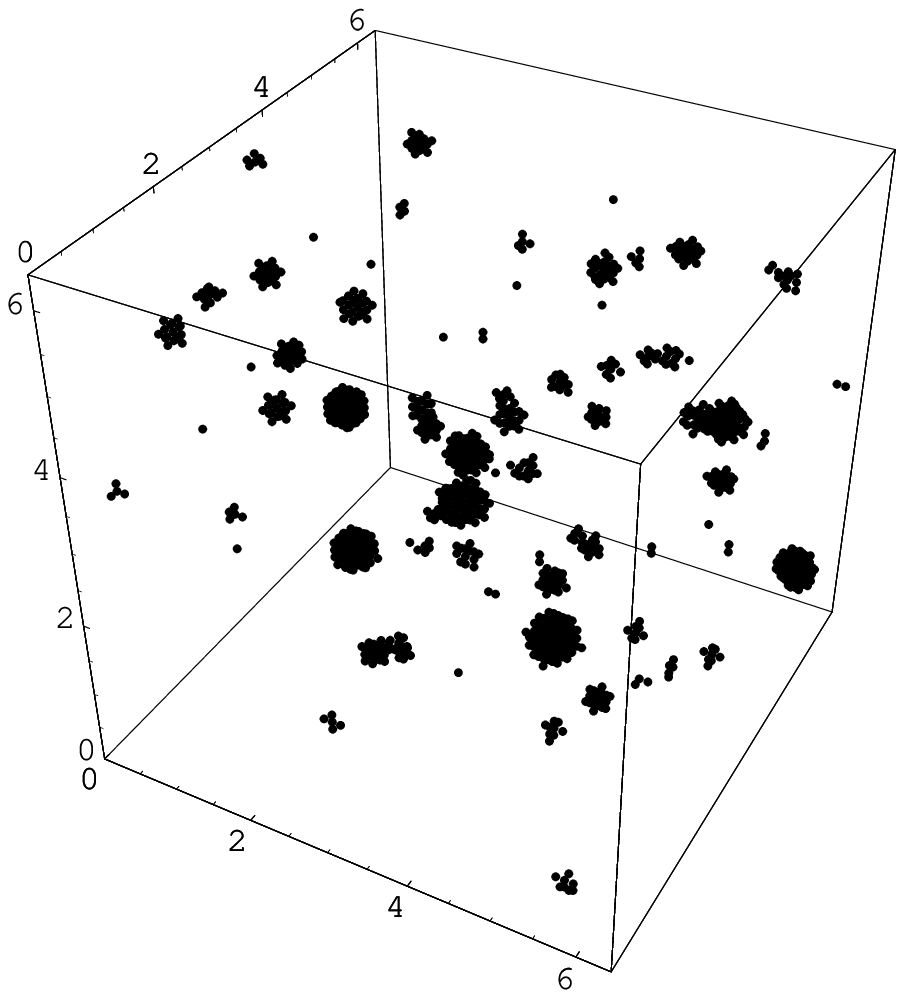}
\leavevmode\epsfysize=6cm \epsfbox{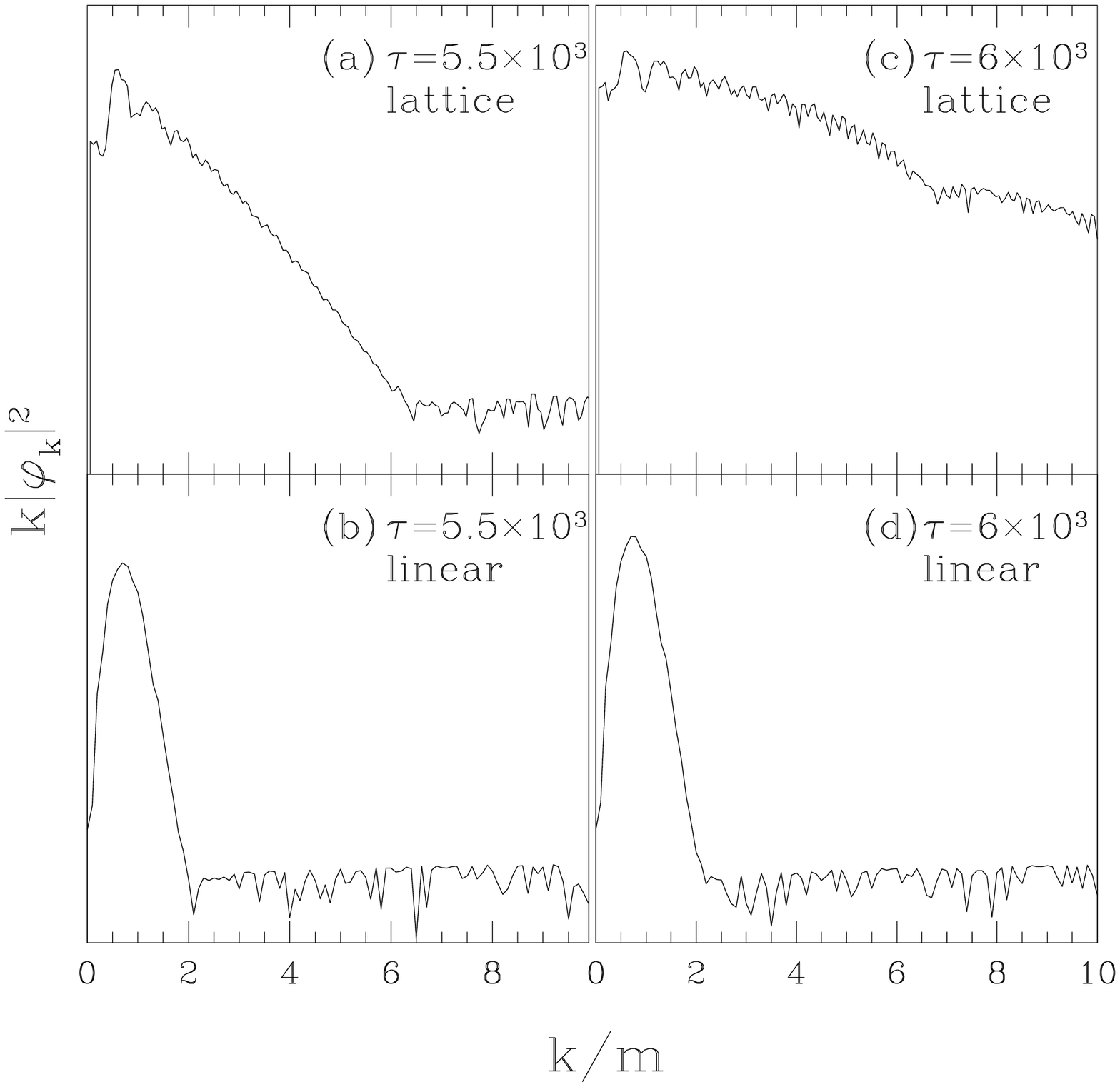}\\[2mm]
\caption{\label{relic3D}
\small $Q$-ball formation in gravity mediated supersymmetry
breaking. From the left, the first
plot shows the amplitude of the condensate after $Q$-balls have formed
at $z=6.3$. The second plot shows around $40$~$Q$-balls
with a largest charge $Q\simeq 5.16\times 10^{16}$, and the third plot
shows the power spectra of the condensate fluctuations
($k|\delta\varphi_k|^2$) when the amplitude of the fluctuations
has become as large as the homogeneous mode: $\langle \delta
\varphi^2 \rangle \sim \varphi^2$. The top panels (a) and (c) show the
full fluctuations calculated on one dimensional lattices, while the
bottom panels (b) and (d) show the linearized fluctuations without
mode mixing, from \cite{kasuya0062}}
\end{figure}
%%%%%%%%%%%%%%%%%%%%%%%%%%%%%%%%%%%%%%%%%%%%%%%%%%%%%%%%%%%%%%%
\begin{figure}[t!]
\centering
\hspace*{-7mm}
\leavevmode\epsfysize=5cm \epsfbox{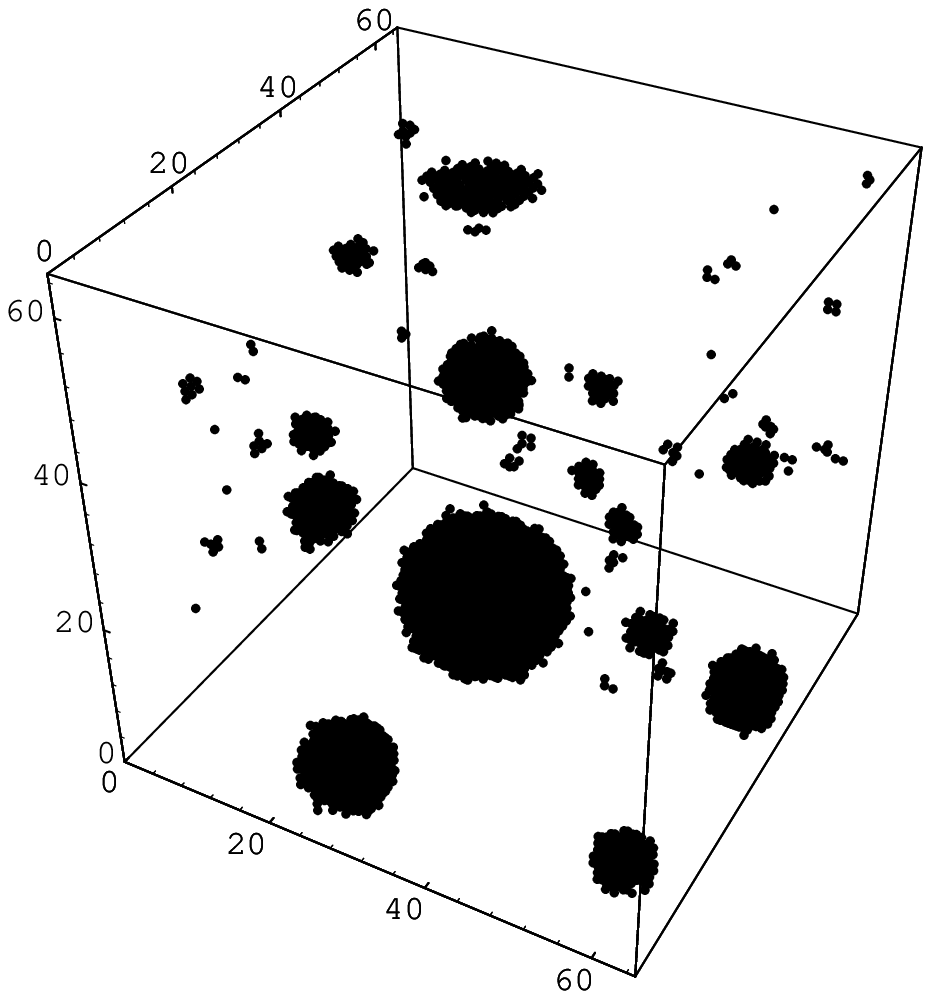}
\leavevmode\epsfysize=7cm \epsfbox{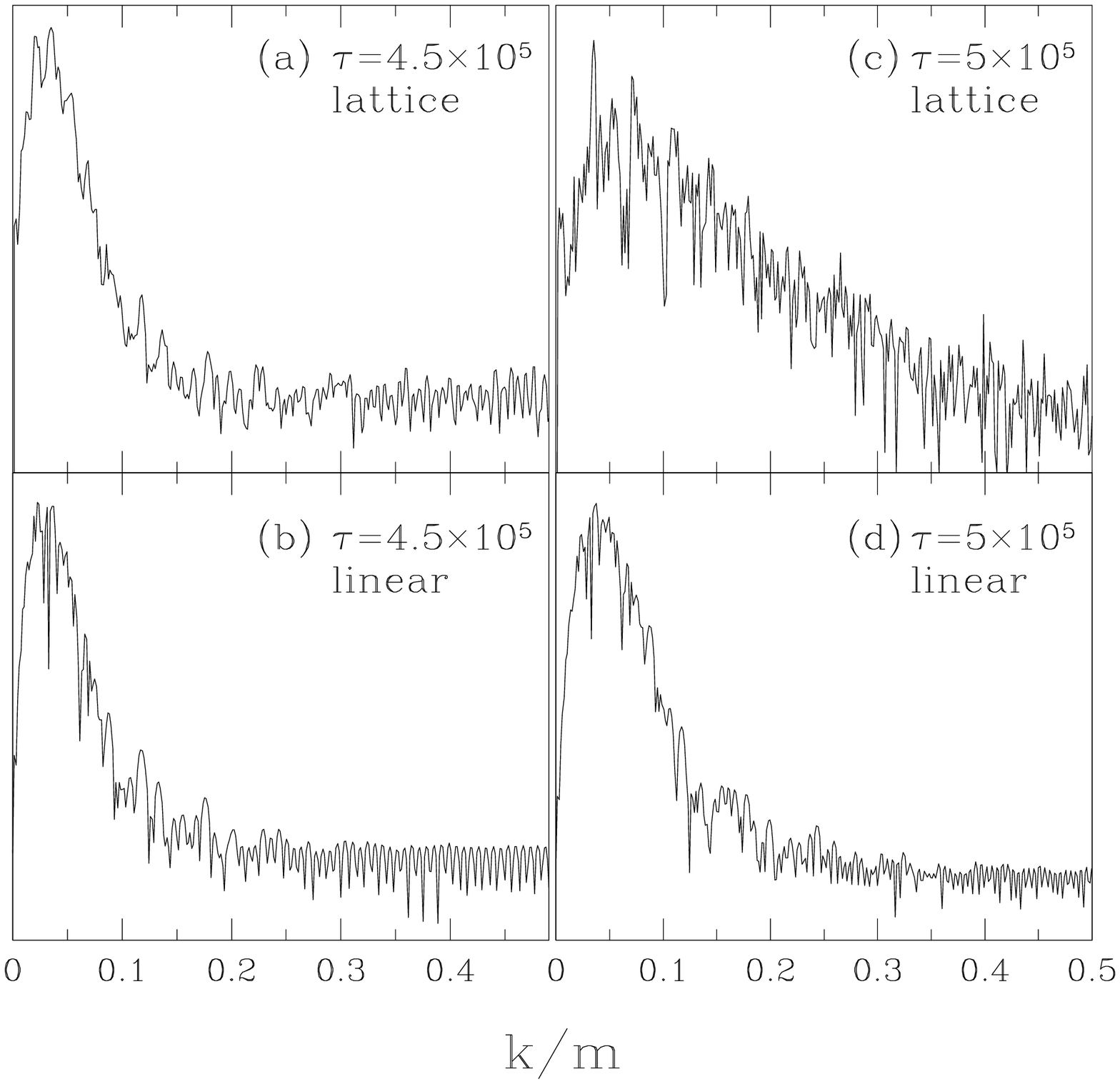}\\[2mm]
\caption{\label{fig-1}
\small Configuration of $Q$-balls on a three dimensional lattice in the gauge
mediated case. More than 30
$Q$-balls have formed, and the largest one has the charge with
$Q\simeq 1.96\times 10^{16}$. The second plot shows the power spectrum
of the condensate fluctuations  when the amplitude
of the fluctuations has become as large as the homogeneous mode. The top
panels (a) and (c) show the full fluctuations calculated on one
dimensional lattices, while the bottom panels (b) and (d) show the
linearized fluctuations without mode mixing, from \cite{kasuya00}.}
\end{figure}

%%%%%%%%%%%%%%%%%%%%%%%%%%%%%%%%%%%%%%%%%%%%%%%%%%%%%%%%%%%%%%%

Full $3+1$ dimensional simulations have been presented by
Kasuya and Kawasaki \cite{kasuya0061,kasuya00,kasuya0062} (for
both gravity mediated and gauge mediated cases), see
Figs.~(\ref{relic3D},\ref{fig-1}), and by Multam\"aki and Vilja
\cite{multamaki02535} (for  a wide range of the energy-to-charge ratio
in the gravity mediated case with), see
Figs.~(\ref{figmultamaki1},\ref{figmultamaki2}).

In \cite{kasuya0061}, the authors simulated $Q$-ball formation
in the gravity mediated scenario on a $(64)^3$ lattice with a
lattice spacing $\Delta \zeta =0.1$. The  results are shown in
Fig.~(\ref{relic3D}). The initial fluctuations in the real and
the imaginary direction were taken to be
$\delta\phi_{1} =\delta \phi_{2} \sim {\cal O}(10^{-7})$, together with
initial field values $\phi_1(0)=\phi_{2}(0)\sim {\cal O}(10^7)$~GeV.
In order to obtain the spectrum the authors relied on a 1d
lattice with $N=1024$ and $\Delta \zeta=0.1$. The result is
depicted in the third plot of Fig.~(\ref{relic3D}) for two different
comoving time scales. One can see the marked difference between the
linearized perturbations and the lattice simulated ones. In the latter
case the spectrum does not fall sharply, which can be attributed
to  mode-mode interactions or  rescattering effects which
kick the lower momentum modes  higher, leading to a broadening and smoothening
of the spectrum. Note that in linearized fluctuations the instability band
is almost the same as Eq.~(\ref{instband}). For example
$\vec k/m_{\phi}=\sqrt{2}a(\tau)|K|^{1/2}\approx 2$ for $|K|=0.01$ and
$\tau=5.5\times 10^{3}$.

In \cite{kasuya00}, Kasuya and Kawasaki repeated their simulation for
the gauge mediated case, which is shown in Fig.~(\ref{fig-1}). Note
that the size of the $Q$-ball is bigger than in the gravity
mediated case.

Multam\"aki and Vilja \cite{multamaki02535} had typical lattice sizes
of $120^3$. They verified that the $2$ dimensional simulations
\cite{enqvist0163} capture all of the essential features of
the AD condensate fragmentation: transient filament structures resulting
in a large number of $Q$-balls and anti-$Q$-balls, which can also be seen in
Fig.~(\ref{figmultamaki2}). Note that when the condensate has the exact
energy-to-charge ratio of a $Q$-ball so that $x=1$, no anti-$Q$-balls
form in Fig.~(\ref{figmultamaki1}), whereas for $x\gg 1$, the number
of $Q$-balls and anti-$Q$-balls are practically equal.

%%%%%%%%%%%%%%%%%%%%%%%%%%%%%%%%%%%%%%%%%%%%%%%%%%%%%%%%%%%%%
\begin{figure}[t!]
\centering
\hspace*{-7mm}
\leavevmode\epsfysize=4cm \epsfbox{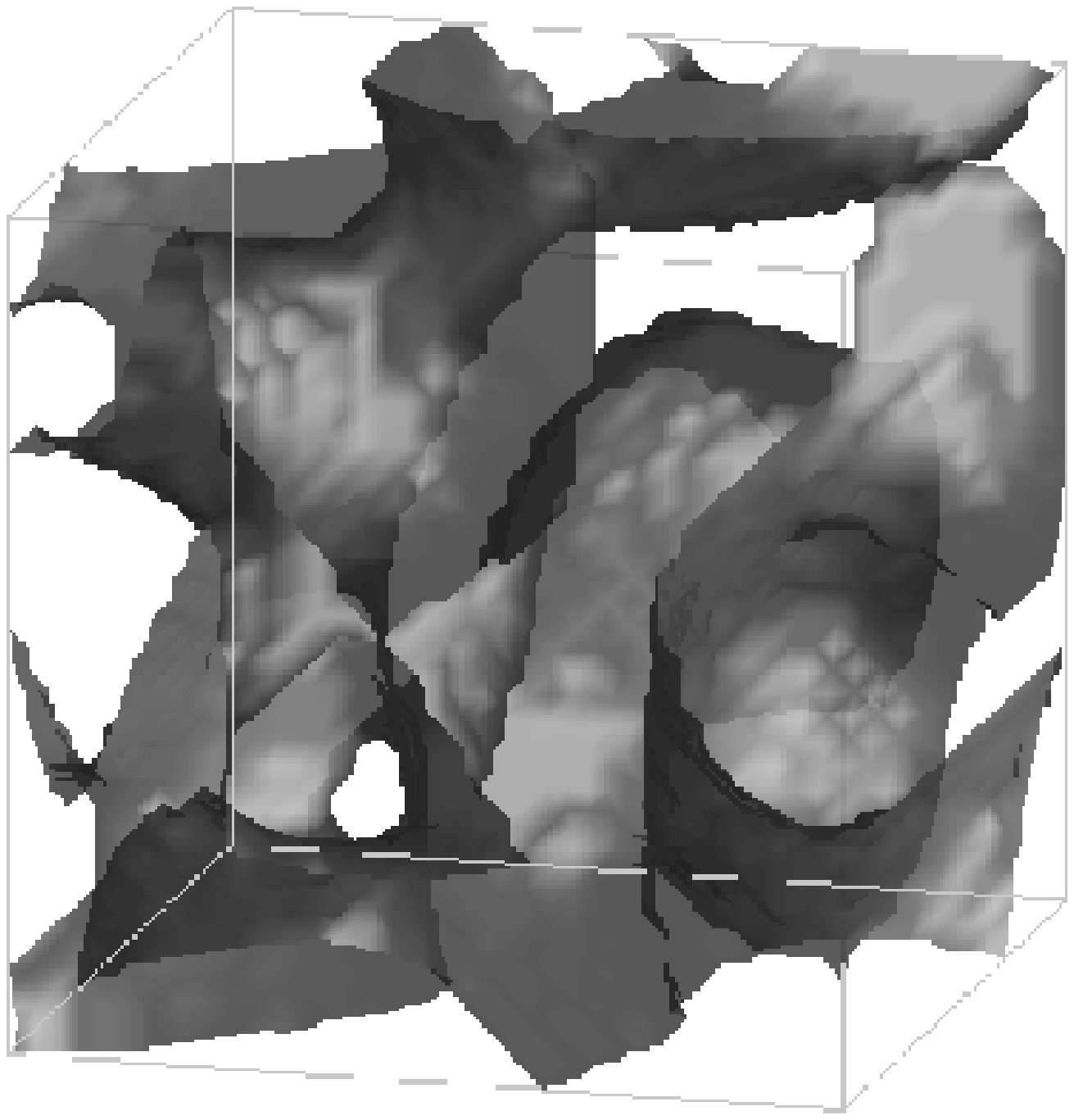}
\leavevmode\epsfysize=4cm \epsfbox{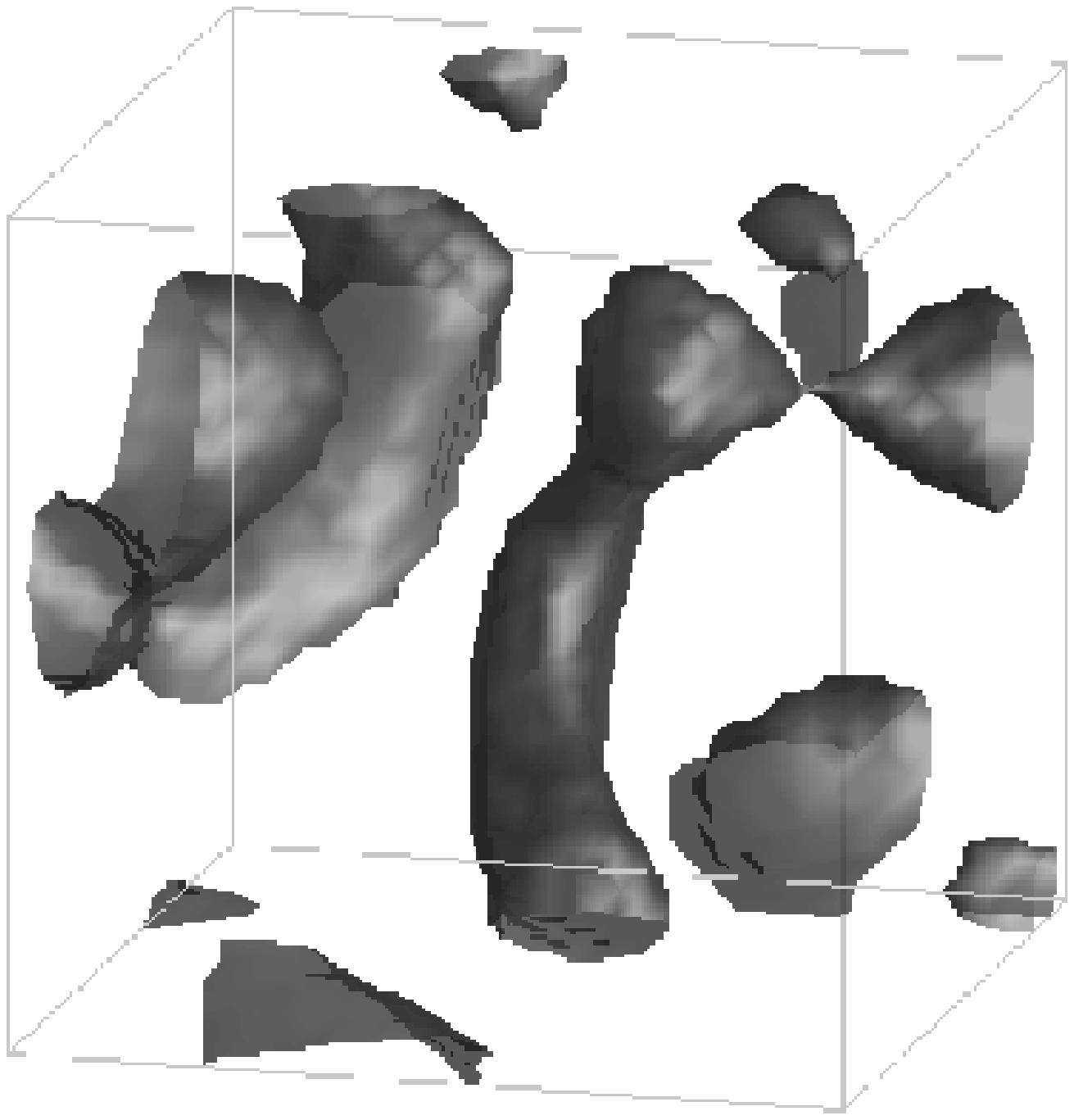}
\leavevmode\epsfysize=4cm \epsfbox{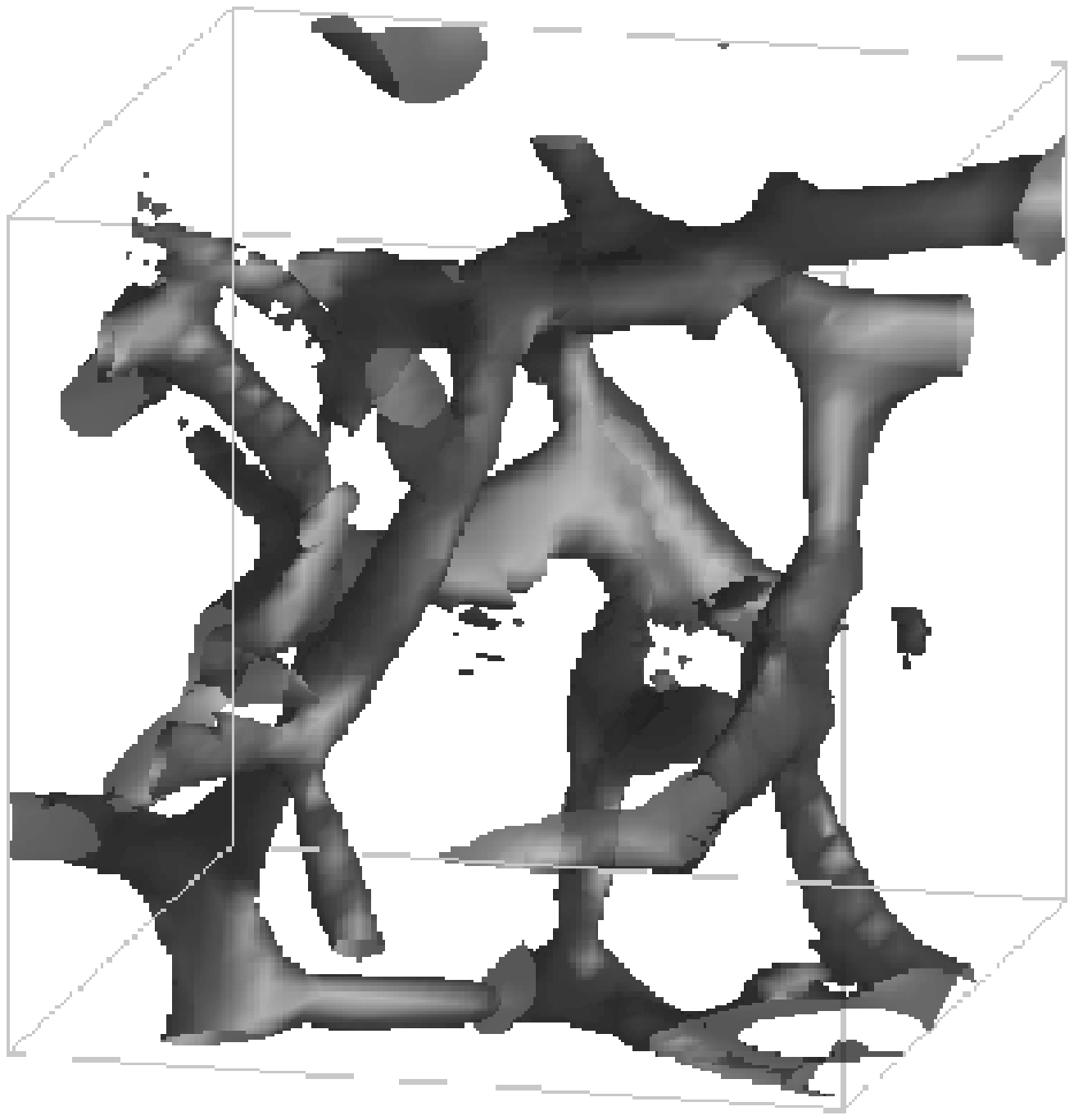}\\[4mm]
\leavevmode\epsfysize=4cm \epsfbox{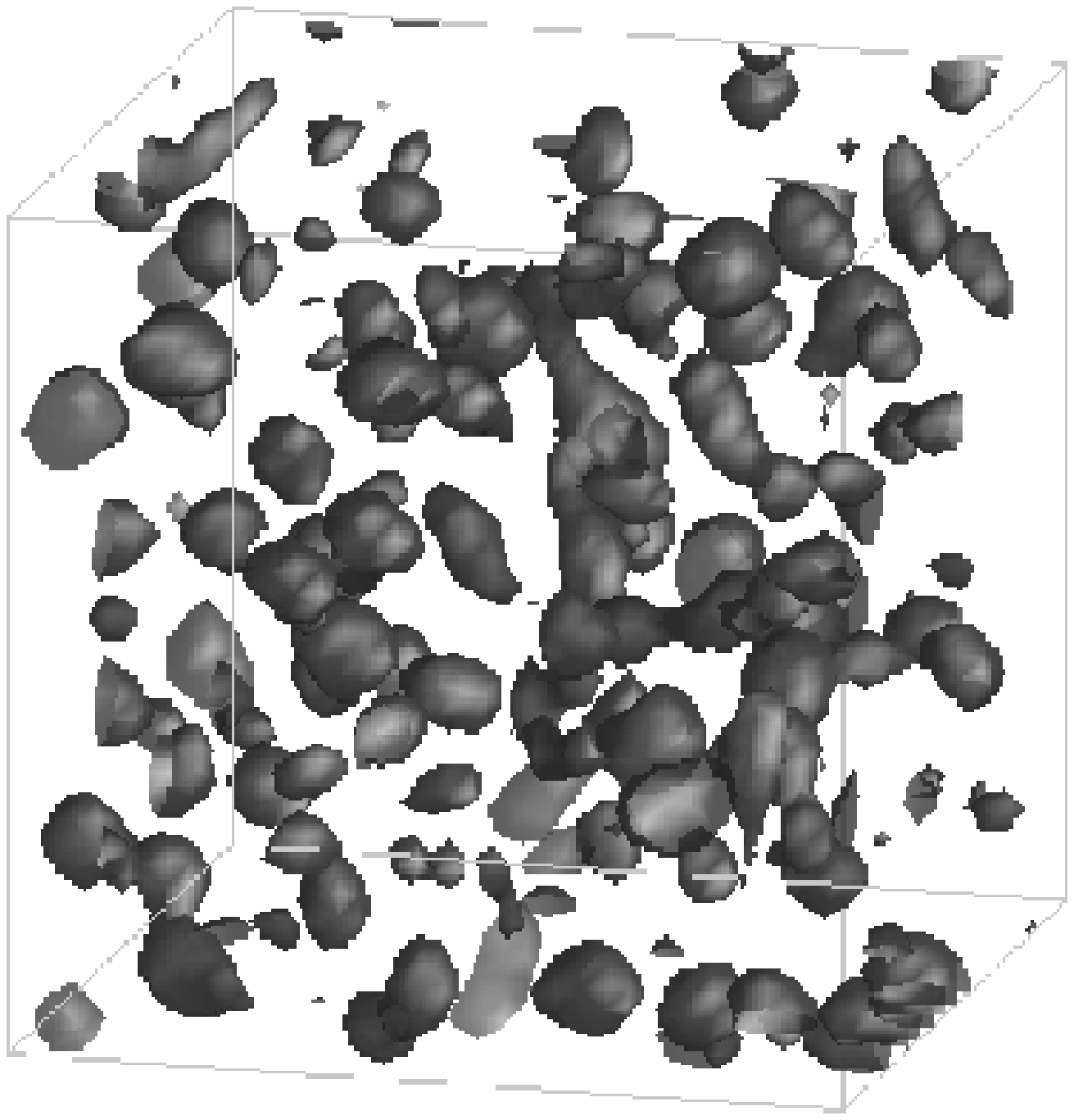}
\leavevmode\epsfysize=4cm \epsfbox{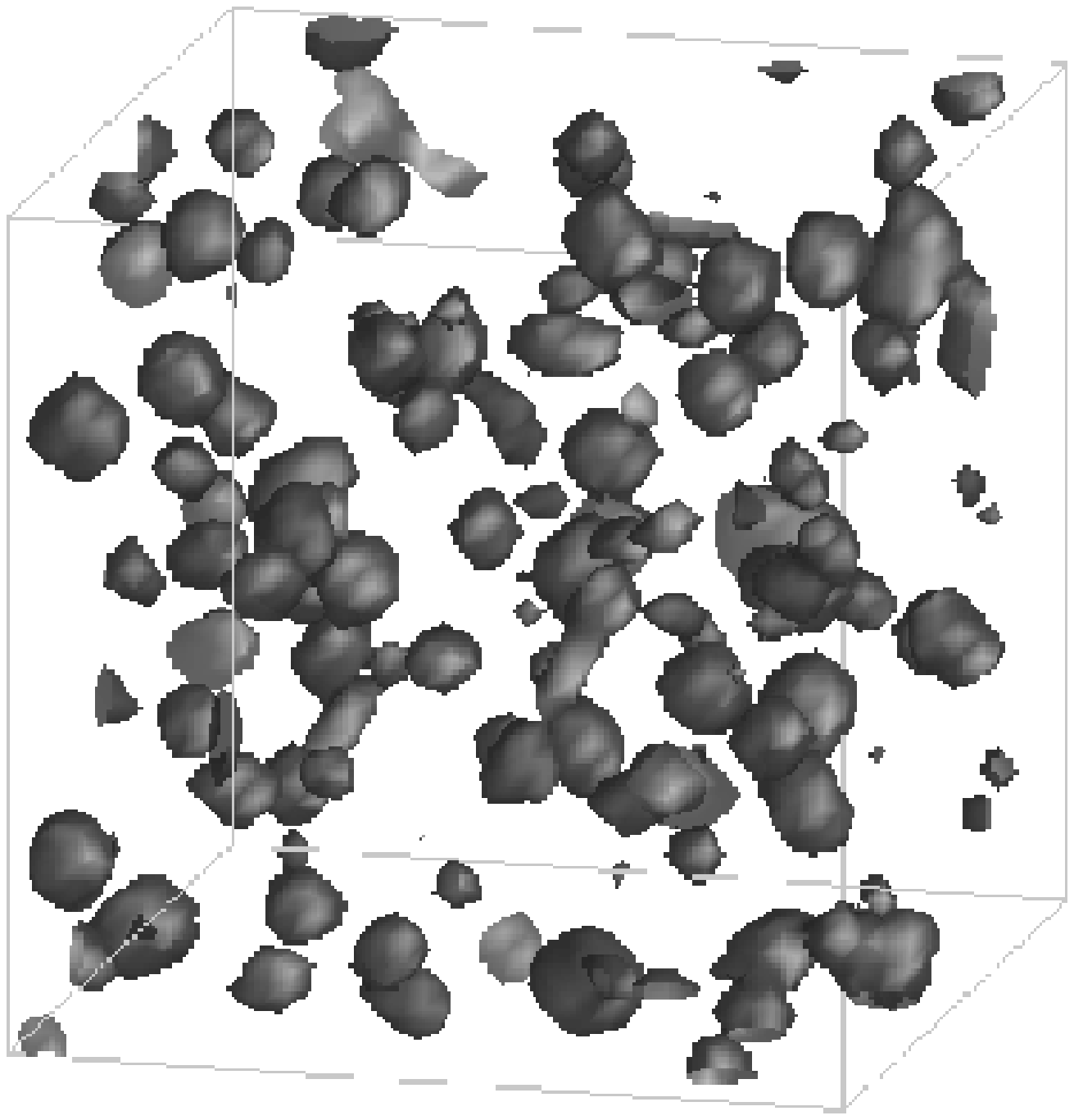}
\leavevmode\epsfysize=4cm \epsfbox{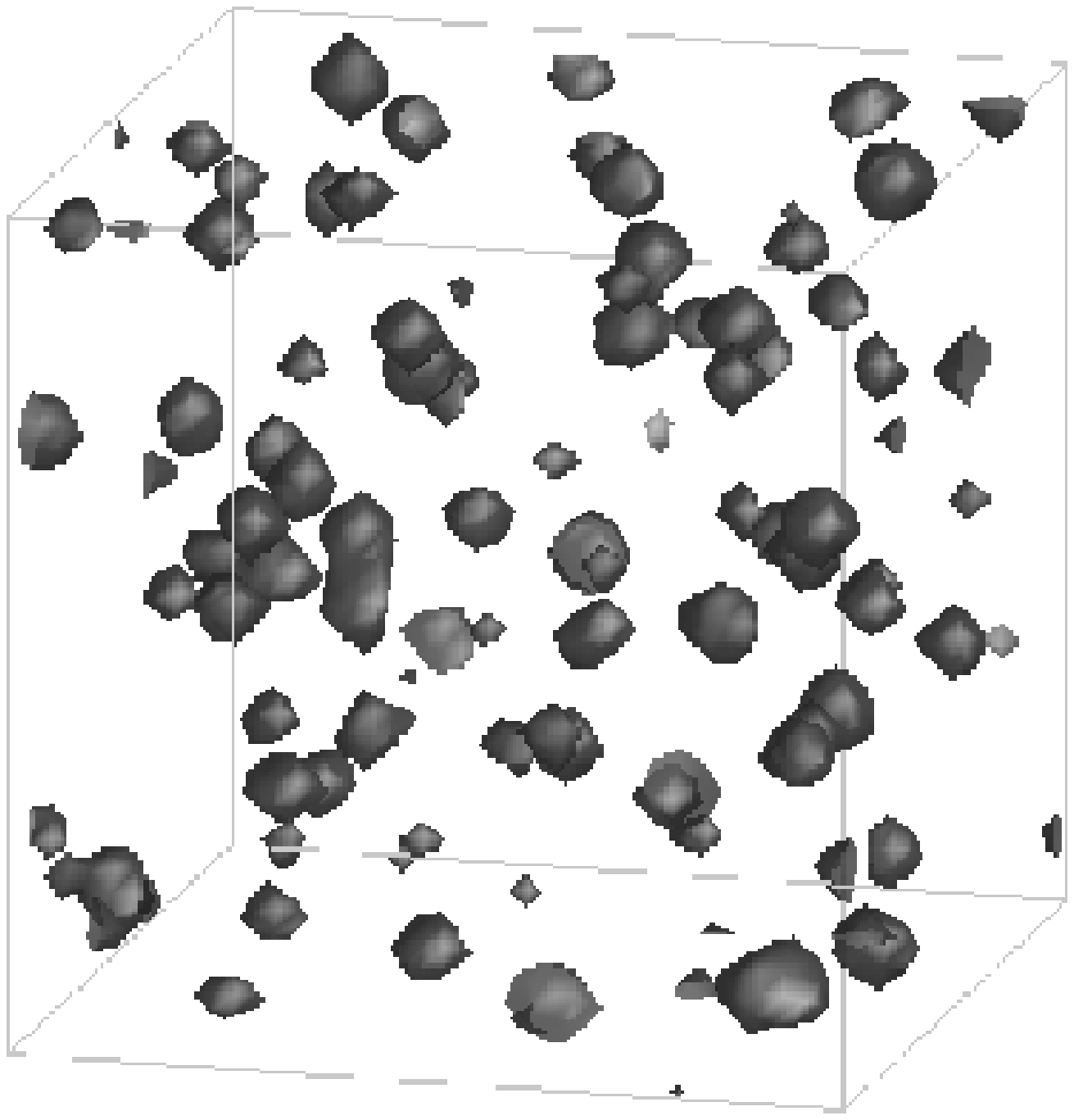}\\[2mm]
\caption{\label{figmultamaki1}
\small Formation and evolution of charged lumps on 3D lattice from
comoving time $\tau=875-3000$, when $x=1$. Only positively charged
$Q$-balls have formed, from \cite{multamaki02535}.}
\end{figure}

\begin{figure}[t!]
\centering
\hspace*{-7mm}
\leavevmode\epsfysize=4cm \epsfbox{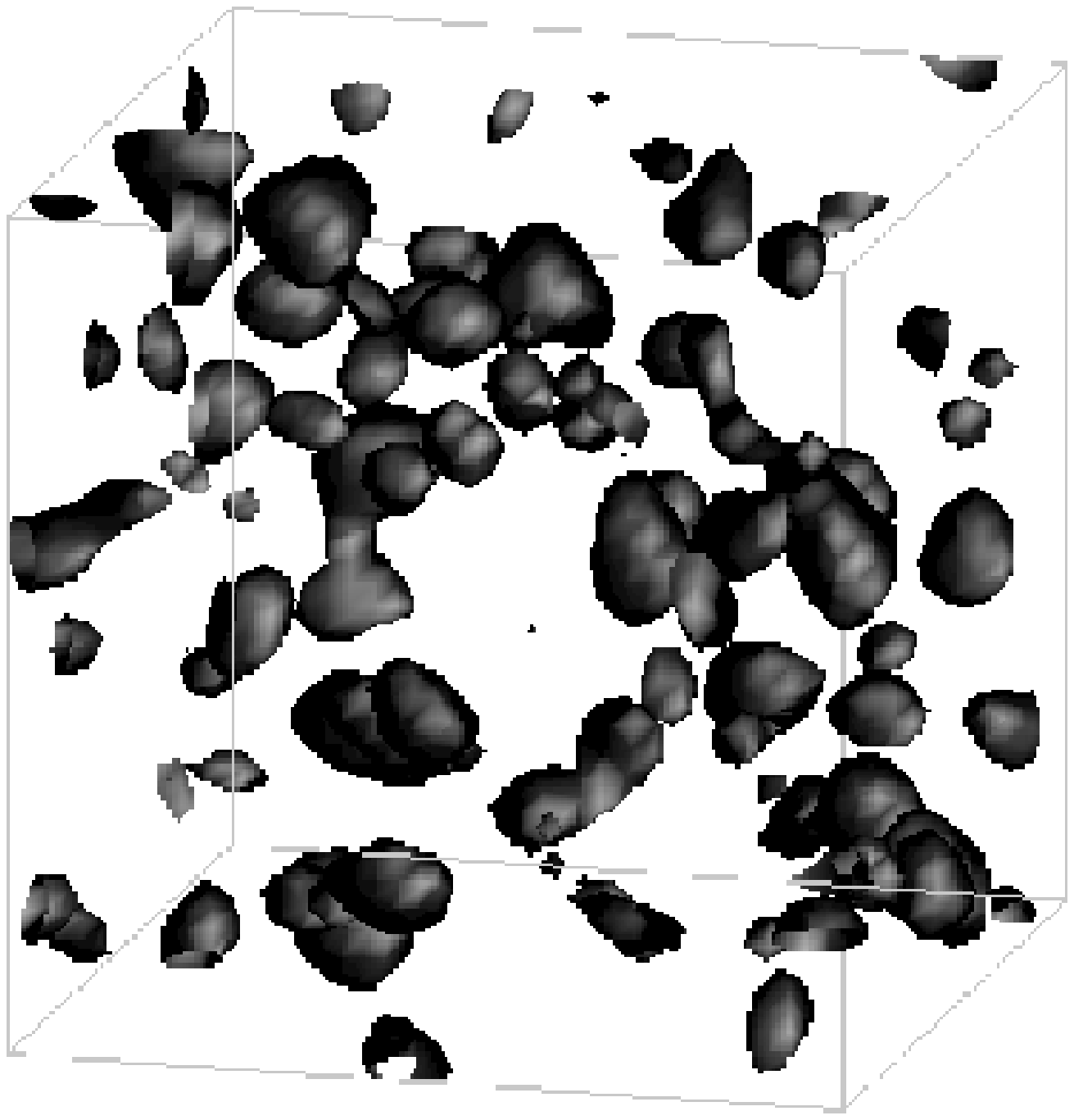}
\leavevmode\epsfysize=4cm \epsfbox{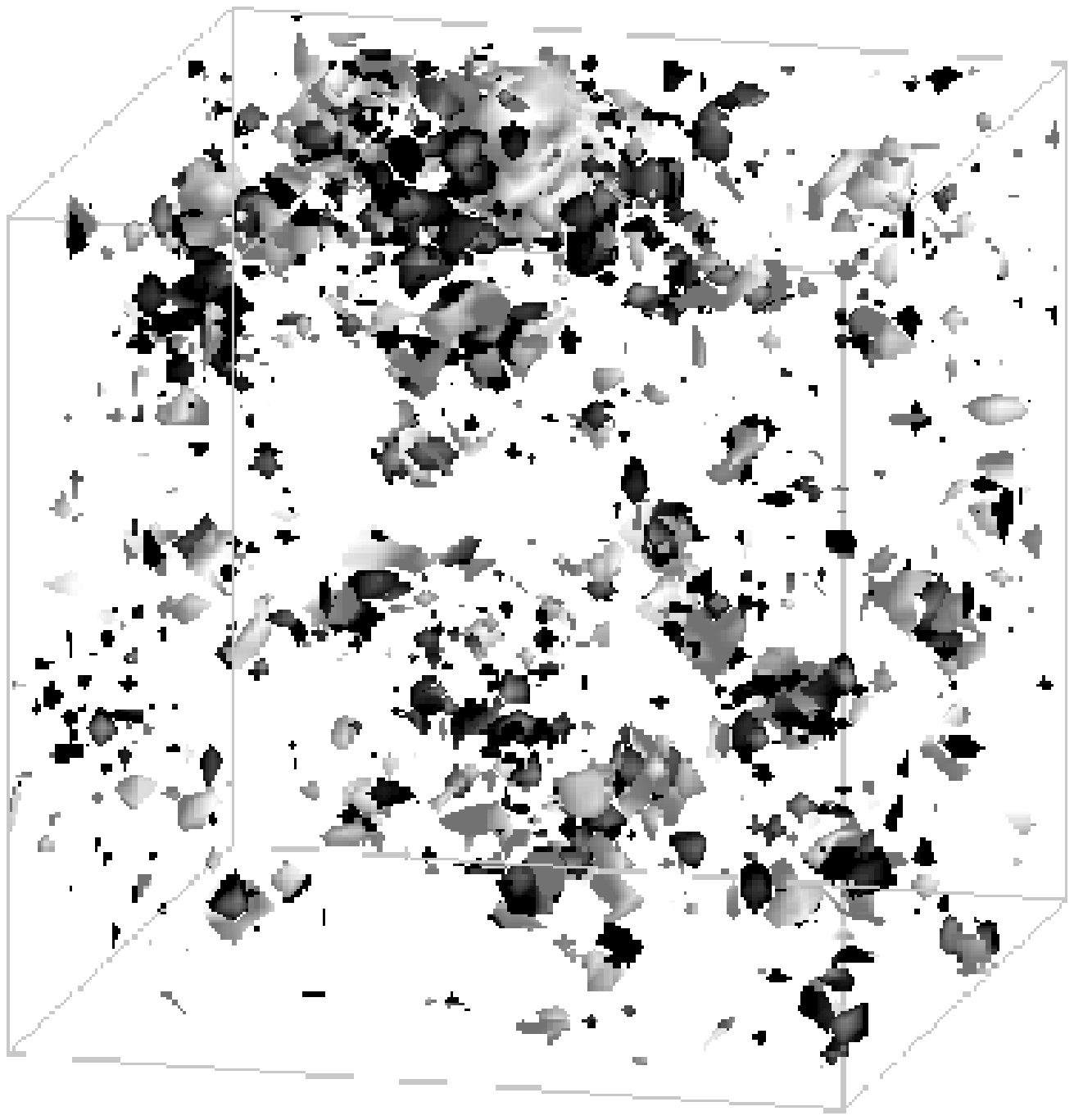}
\leavevmode\epsfysize=4cm \epsfbox{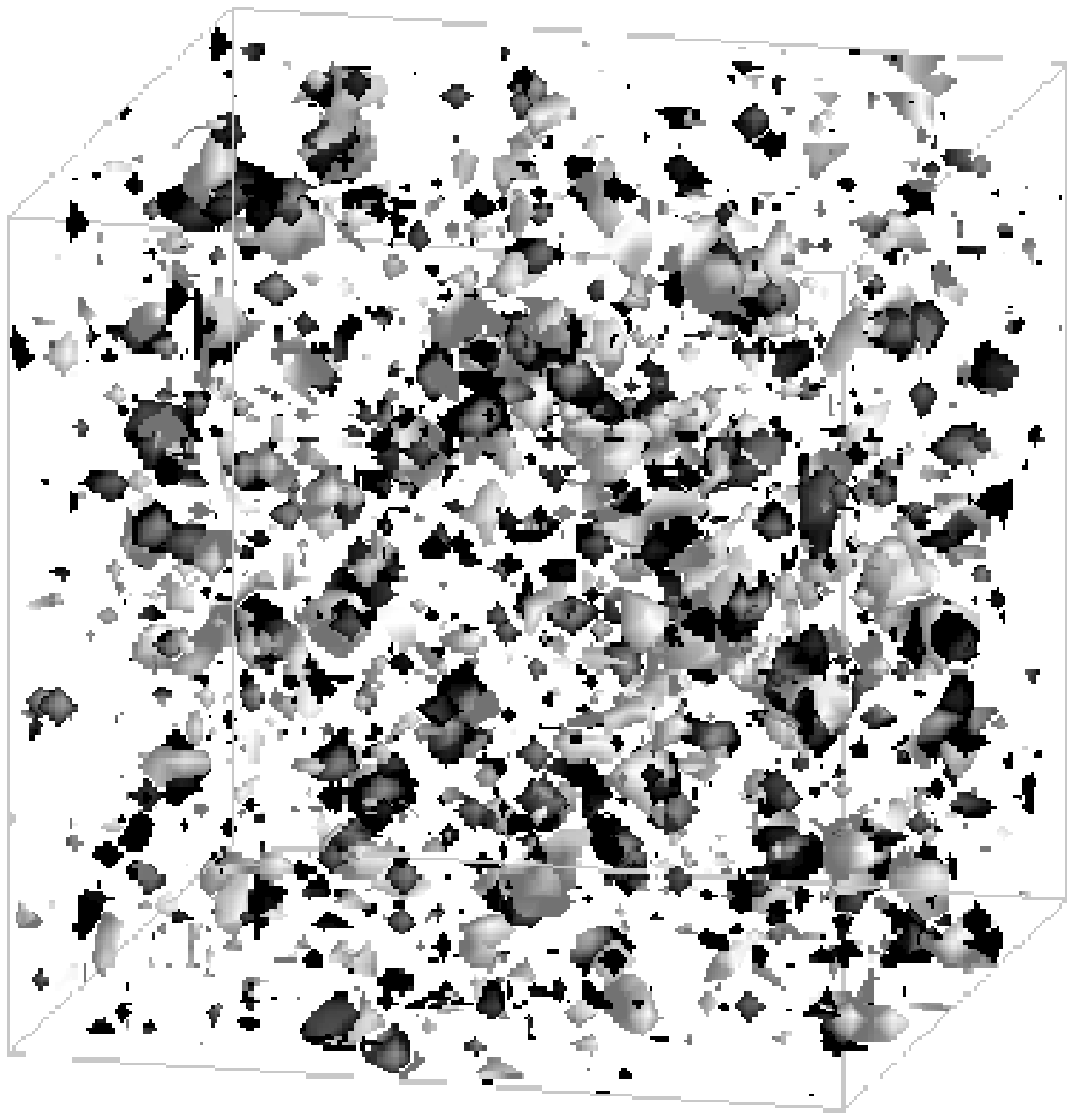}\\[4mm]
\leavevmode\epsfysize=4cm \epsfbox{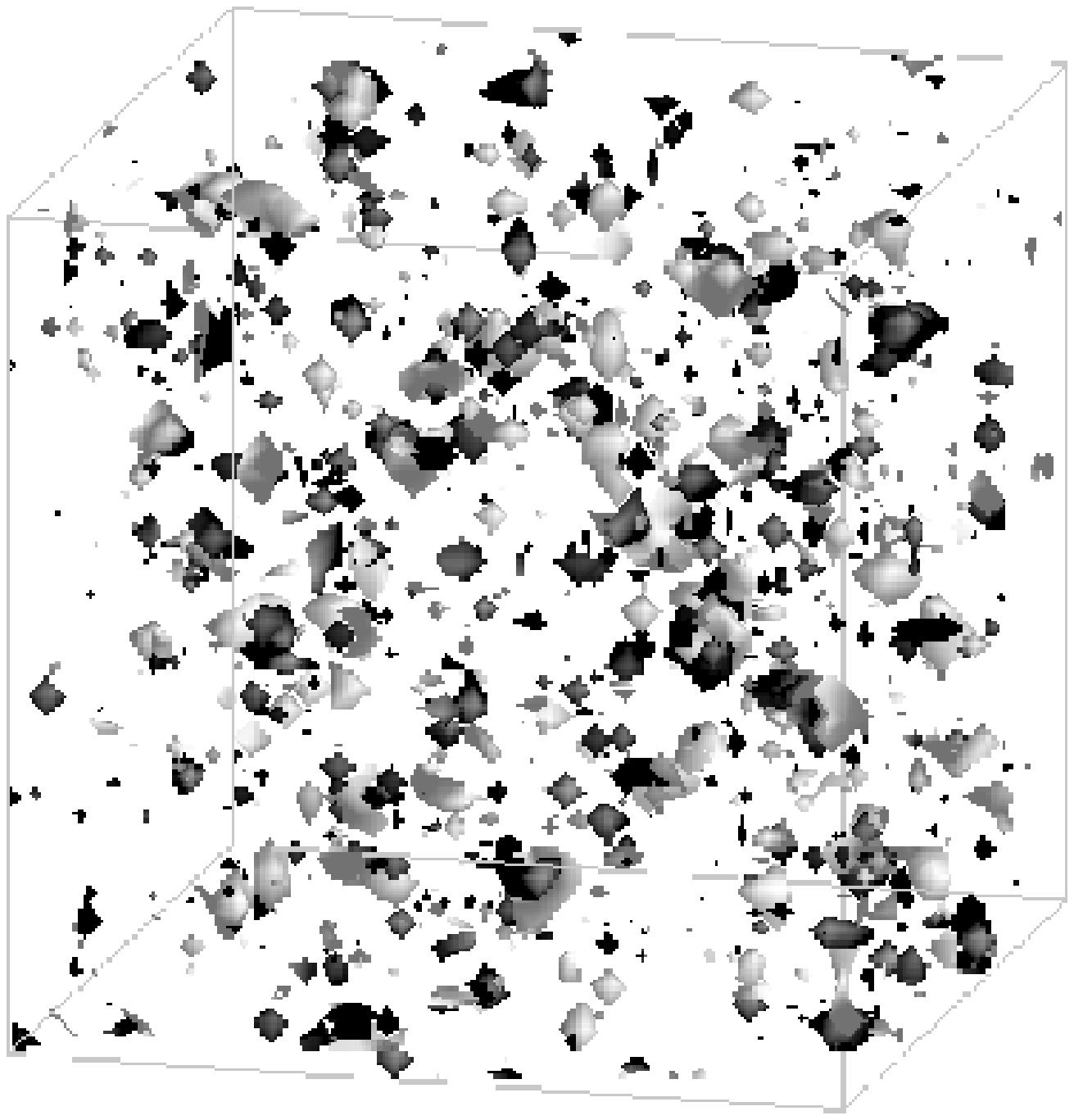}
\leavevmode\epsfysize=4cm \epsfbox{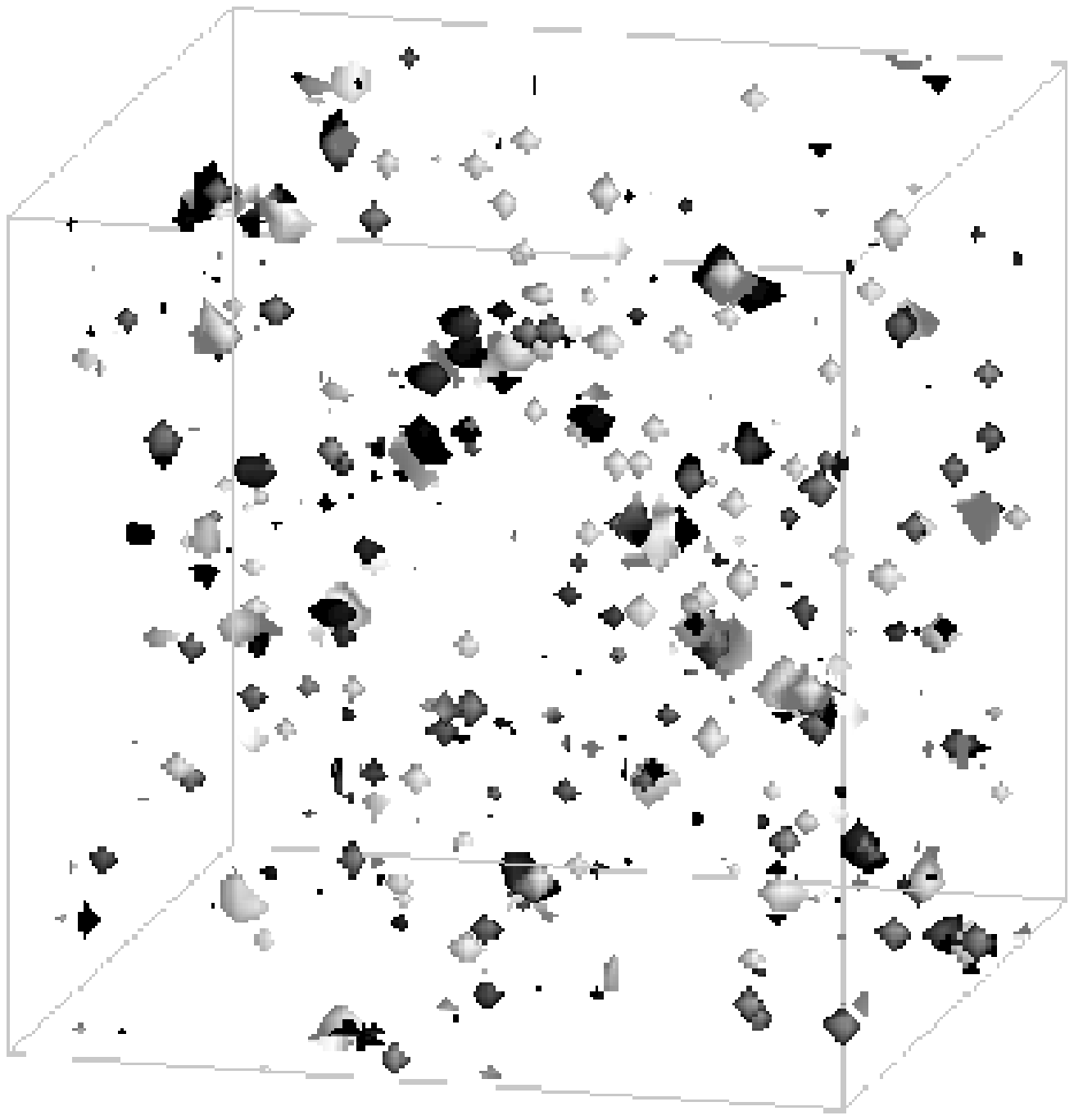}
\leavevmode\epsfysize=4cm \epsfbox{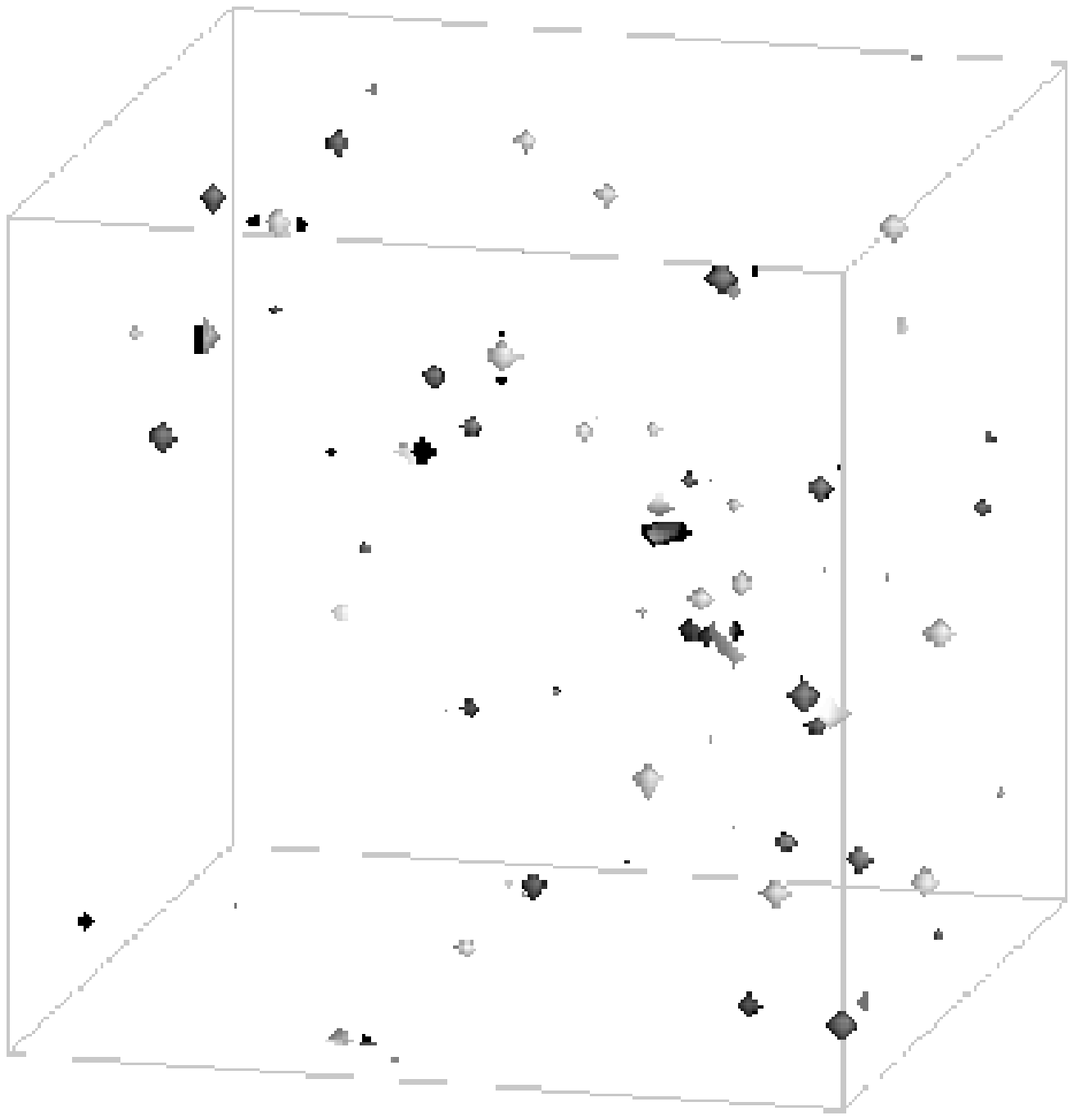}\\[2mm]
\caption{\label{figmultamaki2}
\small Formation and evolution of charged lumps from comoving time
$\tau=1500-7500$, when $x=10^{5}$. Here positive and negatively charged
$Q$-balls have formed, depicted in different shades, from
\cite{multamaki02535}.}
\end{figure}

%%%%%%%%%%%%%%%%%%%%%%%%%%%%%%%%%%%%%%%%%%%%%%%%%%%%%%%%%%%%%

\subsection{Equilibrium ensembles}

%%%%%%%%%%%%%%%%%%%%%%%%%%%%%%%%%%%%%%%%%%%%%%%%%%%%%%%%%%%%%

Some of the lattice results can be understood by analytical
arguments. In particular, the distributions and number densities
of $Q$-balls and anti-$Q$-balls may be obtained simply by maximizing
the entropy. Such approach appears justified, since after fragmentation,
the AD lumps are expected to interact vigorously and the field fragments
will settle to the state of lowest energy by emitting and exchanging
smaller fragments. If the interaction is fast enough compared with
the expansion rate of the Universe, i.e. $\Gamma=n_{tot}\sigma v>H=(2/3t)$,
where $n_{tot}$ is the total number of $Q$-balls and anti-$Q$-balls,
$\sigma\approx \pi R_{Q}^2$ is the geometric cross-section of a
$Q$-ball collision, and $v$ is the average velocity of a $Q$-ball,
then it is naturally expected that the final state should consist
of an equilibrium distribution of $Q$-balls and anti-$Q$-balls in a
state of maximum entropy. Considerations supporting this has been
given in~\cite{enqvist0163}
for the case of gravity mediated supersymmetry breaking.

The $Q$-ball (anti-$Q$-ball) distributions $N_+(Q,p)$ ($N_-(Q,p)$) are
subject to the following constrains:
\bea{eqehto}
E_{tot} = E_++E_- ~, & E_{\pm} = \int dQ~dp~E(Q,p)~N_{\pm}(Q,p)~\nn
Q_{tot} = Q_+-Q_- ~, & Q_{\pm} = \int dQ~dp~Q~N_{\pm}(Q,p)~,
\eea
where $E(Q,p)\approx\sqrt{p^2+m^2Q^2}$ is the energy of a single
$Q$-ball, $E_+$ $(E_-)$ and $Q_+$ $(Q_-)$ are the energy and charge of
$Q$-balls (anti-$Q$-balls), and $E_{tot}$ and $Q_{tot}$ are respectively
the total energy and charge of $Q$-balls and anti-$Q$-balls, which are equal
to the energy and charge of the initial AD-condensate (unless
significant amounts of energy and/or charge are transformed into radiation).
It then follows from \eq{eqehto} that
\be{epayht}
x\equiv E_{tot} \geq {m~(Q_++Q_-)\over m Q_{tot}}\geq 1.
\ee
This condition is independent of the precise form of the $Q$-ball
distributions.

If all the baryon asymmetry resides in $Q$- and anti-$Q$-balls, then
at times earlier than $10^{-6}s$, $Q_{tot}/Q_+\sim \Delta B\sim 10^{-8}$.
(Since $B-L$ is conserved in the MSSM, this holds also for the purely
leptonic flat directions.) From \eq{epayht} it follows
\be{xey}
x \simeq 10^8~.
\ee
Even if all of the baryon asymmetry were not carried by $Q$-balls, a natural
expectation is $x\gg 1$ so that the number of $Q$-balls, $N_+$, and the number
of anti-$Q$-balls, $N_-$, are approximately equal and the total number of
$Q$-balls is $N_{tot}\approx 2N_+$. The main bulk of the $Q$-balls may be
expected to be relativistic \cite{enqvist0163}, and the collision rate much
larger than the Hubble rate.

Indeed, one can verify that the equilibrium assumption is self-consistent,
and it is also supported by numerical studies in $2$ dimensions
\cite{enqvist0163} as well as by the $3$ dimensional simulations
of Multam\"aki and Vilja \cite{multamaki02535}, who actually observe the
$Q$-ball distribution to relax into equilibrium as shown in
Fig.~(\ref{figmultamaki2}).

%%%%%%%%%%%%%%%%%%%%%%%%%%%%%%%%%%%%%%%%%%%%%%%%%%%%%%%%%%%
\pagebreak
%%%%%%%%%%%%%%%%%%%%%%%%%%%%%%%%%%%%%%%%%%%%%%%%%%%%%%%%%%%
%%%%%%%%%%%%%%%%%%%%%%%%%%%%%%%%%%%%%%%%%%%%%%%%%%%%%%%%%%%
%%%%%%%%%%%%%%%%%%%%%%%%%%%%%%%%%%%%%%%%%%%%%%%%%%%%%%%%%%%

\section{$Q$-balls}

$Q$-balls with a global charge have many interesting astrophysical
and cosmological consequences. They may reheat the Universe
\cite{enqvist02a,enqvist02b} and serve as candidates for dark matter
candidate~\cite{kusenko98418,kusenko9880,demir99,kasuya01,kusenko01,enqvist02}.
They may provide baryogenesis and leptogenesis
\cite{enqvist98,enqvist99,enqvist9881} and while decaying, produce
LSP dark matter~\cite{enqvist98440,mcdonald01,kasuya01,fujii02,fujii02b}.
They could be responsible for the generation of cosmic magnetic fields
\cite{shiromizu98}. $Q$-balls could stabilize neutron stars
\cite{kusenko98423} or even form solitonic q-stars
\cite{lee92,lynn89,prikas02}.
They could act as a laboratory for physics beyond the electroweak scale
\cite{dvali98}.

%%%%%%%%%%%%%%%%%%%%%%%%%%%%%%%%%%%%%%%%%%%%%%%%%%%%%%%%%%%%%%
\subsection{$Q$-ball as a non-topological soliton}

%%%%%%%%%%%%%%%%%%%%%%%%%%%%%%%%%%%%%%%%%%%%%%%%%%%%%%%%%%%%%%
\subsubsection{Proofs of existence}

The $Q$-ball is a generic ground state in a broad class of theories
with interacting scalar fields carrying some conserved global charge
\cite{rosen68,lee74,friedberg76,morris78,coleman85,lee92}. The $Q$-ball
is an example of a non-topological soliton whose boundary condition at
infinity is the same as that for the vacuum state, unlike in the case
of topological solitons such as magnetic monopoles \cite{t'hooft74,polyakov74}
(a detailed review on non-topological solitons can be found in, e.g.
\cite{wilets89,lee92}).

$Q$-balls are quite generic solitons in $3+1$ dimensions,
which can be associated with many scalar fields with various
$U(1)$ charges \cite{kusenko97405},  with a non-Abelian symmetries
\cite{safian88a,axenides92a,axenides98}, and also with local gauge symmetries
\cite{klee89,khare89,shiromizu99,demir99,axenides01,fujii0164,levi01}. The main
difference which distinguishes global $Q$-balls from a local $Q$-balls  is that
in the latter case the charge of the stable $Q$-ball is bounded from above.

Recall that in order to find a $Q$-ball solution one must minimize
the energy $E_\omega$ (see Eqs.~(\ref{chargeq},\ref{energyq})) for
a fixed charge with respect to the variations of
$\omega$ and $\varphi(x)$ independently.
Obviously one could try finding directly a $Q$-ball solution by solving
Eq.~(\ref{qbeqm}). A more effective way is to seek a bounce
solution $\bar \varphi(x)$ for tunneling in three Euclidean dimensions
in the potential given by Eq.~(\ref{uomega})
\cite{kusenko97404,kobzarev74,linde83}. Note that the first term in
Eq.~(\ref{energyq}) is then the three dimensional Euclidean action
$S_{3}[\bar\varphi_{\omega}(x)]$ of this bounce solution which
satisfies Eq.~(\ref{qbeqm}) in radial coordinates with the boundary
conditions $\varphi(0)=\varphi^{\prime}(\infty)=0$.

A theorem \cite{coleman85,lee92} (see also \cite{kusenko97404})
 states that if there exists a range of $\varphi$ however small,
and the potential $U(\varphi^2)$ contains an attractive interaction
however weak, then a non-topological soliton solution exists for
\begin{equation}
\label{theorem}
\nu^2 \leq \omega^2 < m_{\phi}^2\,,
\end{equation}
where the mass parameter is defined as
$U(\varphi^2) \rightarrow m_{\phi}^2\varphi^2$ when $\varphi \rightarrow 0$.
The value of $\omega =\sqrt{k^2+m_{\phi}^2}$ determines the frequency of
the $\varphi$ quanta in the field space. For a plane wave solution it
is always true that $\omega^2 > m_{\phi}^2$ while for a solitonic solution
$\omega^2<m_{\phi}^2$. This suggests that there exists a parabola
$\nu^2\varphi^2$ tangent to $U(\varphi^2)$ at $\varphi =\pm \varphi_{0}$, with
$\nu^2 <m^2_{\phi}$. Another useful way of expressing this is through
\begin{equation}
\label{bounceq}
\frac{U(\varphi)}{\varphi^2}={\rm min}\,, ~~~{\rm for}~~
\varphi=\varphi_{0}>0\,.
\end{equation}

For a sufficiently large $Q$ given by Eq.~(\ref{chargeq}), the energy
of a soliton is then given by
\begin{equation}
\label{minen1}
E=|\nu Q|< m_{\phi}|Q|\,,
\end{equation}
which ensures its stability against decay into plane wave solutions
with $\varphi \simeq \varphi_{0}$ inside and $\varphi \simeq 0$ outside
the soliton. Note that the global $U(1)$ symmetry is thus broken inside
the soliton by the vev, however, remains unbroken outside the soliton.
The  most crucial piece is the presence of a global $U(1)$
charge of the $Q$-ball which actually prevents it from decaying and
makes the soliton stable. (For analytical results on $Q$-ball properties,
see \cite{correia}). The above discussion can be repeated in
the presence of several charges $q_{i}$, and an analogue of
Eq.~(\ref{bounceq}) has been established by Kusenko
in ~\cite{kusenko97405}. The main difference between a single
and multi-charged $Q$-ball is that there exists different vevs
corresponding to different charges, which
modifies the appropriate bounce solution Eq.~(\ref{bounceq}).

%%%%%%%%%%%%%%%%%%%%%%%%%%%%%%%%%%%%%%%%%%%%%%%%%%%%%%%%%%%%%%

\subsubsection{Beyond thin wall solution}

The above discussion tacitly assumes a thin wall approximation where
the edge of a soliton is sharply defined, which means that the gradient
energy is small compared to the volume energy. As $\varphi_{0}$
increases, the thin wall limit breaks down for a fixed charge, and one
must seek other methods in order to guarantee the existence of a
$Q$-ball solution as pointed out by Kusenko~\cite{kusenko97404}.
For a flat potential, which mimics the large field value situation,
the equations of motion Eq.~(\ref{qbeqm}) can be solved near the
origin $r=0$, and for large $r$. The $Q$-ball profile can be found to be
\begin{eqnarray}
\label{thick0}
\varphi(r)&=&\varphi_{0}\frac{\sin(\omega r)}{\omega r}\,,~~r< R \,,
\nonumber \\
&=&\varphi_{1}e^{-m_{\phi}r}\,,~~r\geq R\,,
\end{eqnarray}
where the values of $\varphi_{0},\varphi_{1},\omega$ and $R$ are such that
they minimize $E_{\omega}$, while $\varphi(r)$ interpolates $r=R$,
the size of the $Q$-ball, continuously. In a thick wall limit one can
also write down $E_{\omega}$ in terms of dimensionless variable
$\xi =\omega x$ and $\psi=\varphi/\omega$, while neglecting all the
terms in $\hat U_{\omega}$ except the constant term and the
$\omega^2\varphi^2$ term \cite{kusenko97404,linde83}, one obtains
$E_{\omega}\approx a\omega+b/\omega^3+\omega Q$, where $a,b$ are constants
independent of $\omega$.  The size of the $Q$-ball turns out to be
$R\propto 1/\omega$, and the vev of the field in the $Q$-ball interior
is $\varphi_{0}\propto Q^{1/4}$ from Eq.~(\ref{thick0}). Note that there
is no classical limit on the charge of a $Q$-ball. In fact no matter how
small $Q$ is there always exists a value $\omega$ close to $m_{\phi}$, for
which $E_{\omega}$ is minimized. Quantum stability requires that
$Q\geq 1$. When $Q\rightarrow 1$, the quantum corrections will indeed become
important. For $Q\geq 7$ the $Q$-balls are quantum mechanically stable
configurations \cite{graham01}.

The $Q$-ball has been shown to be classically stable by Coleman
\cite{coleman85} (see also \cite{lee92,kusenko97404}). The semiclassical
approach obviously breaks down when quantum fluctuations around the $Q$-ball
are comparable to the energy of the system itself. This happens when
$Q\approx 1$. Even though the charge becomes small, the $Q$-ball
size remains large in comparison to the Compton length $m_{\phi}^{-1}$
\cite{kusenko97404,kusenko9867,graham01}. One can also establish a Virial
theorem, which holds for any $Q$ and does not require any approximation
\cite{kusenko97405}.

%%%%%%%%%%%%%%%%%%%%%%%%%%%%%%%%%%%%%%%%%%%%%%%%%%%%%%%%%
\subsection{Varieties of $Q$-balls}

%%%%%%%%%%%%%%%%%%%%%%%%%%%%%%%%%%%%%%%%%%%%%%%%%%%%%%%%%%
\subsubsection{Thin wall $Q$-balls}

There exist thin wall $Q$-ball solutions, where the boundary is a well
defined edge, as well as thick wall $Q$-balls, where the boundary is not
localized in a narrow region and the soliton is typically described by
a Gaussian profile. Both $Q$-ball types may appear within the same theory.

The thin wall $Q$-ball is the simplest and arises
naturally in any suitable scalar potential that allows for the existence of
a $Q$-ball. As mentioned already, thin wall solution has the profile
$\varphi(r) \approx \varphi_{0}\delta(r-R)$ in the radial direction,
where $R$ is the size of the radius of a spherically symmetric $Q$-ball.
This obviously neglects the surface energy contribution and
yields \cite{coleman85}
\begin{equation}
\label{thin1}
\frac{E}{Q} ={\rm min}~\sqrt{\frac{U(\varphi^2)}{\varphi^2}}
\approx \omega_{c}\,.
\end{equation}
Energy is thus growing linearly with charge. Note that the radius of
such a $Q$-ball can be very large,
\begin{equation}
\label{thin2}
Q=2\omega_{c}\varphi_{0}^2 V=\frac{8\pi}{3}\pi R^3\omega_{c}\varphi_{0}^2\,.
\end{equation}
These are useful relationships for the purposes of this Section.

%%%%%%%%%%%%%%%%%%%%%%%%%%%%%%%%%%%%%%%%%%%%%%%%%%%%%%%%%%%%%

\subsubsection{Thick wall $Q$-balls in the gauge mediated case}

Thick wall $Q$-balls have been widely considered in the literature within
gauge and gravity mediated supersymmetry breaking scenarios. In
both cases the mass of the $Q$-ball grows more slowly compared to a
thin wall case, i.e. the scalar potential grows
slower than $\varphi^2$. In this case $Q$-ball never reaches a thin wall
regime, even if $Q$ is large. The value of $\varphi$ inside a
$Q$-ball extends as far as the gradient terms allow, and the mass of a
$Q$-ball is proportional to $Q^{p}$, where $p<1$.

In the context of gauge mediated supersymmetry breaking the AD potential
takes the form (without the non-renormalizable contributions)
\cite{kusenko98418,dvali98} (see Eq.~(\ref{potgmsb}), in Sect.~$5.3.2$.)
\begin{equation}
\label{gmsbp0}
U(\varphi)\approx m^4_{\phi}\log\left(1+\frac{|\varphi|^2}
{m_{\phi}^2}\right)\,,
\end{equation}
where $m_{\phi} \sim {\cal O}({\rm TeV})$ represents the supersymmetry
breaking scale. In \cite{gouvea97}, the authors have considered the
effective potential of the form
\begin{equation}
\label{exotic}
U \sim F^2\left[\log\left(\frac{|\varphi|^2}{m^2_{\phi}}\right)\right]\,,
\end{equation}
where $F^{1/2}\gg m_{\phi}$. Despite the difference between the
forms of Eqs.~(\ref{gmsbp0}) and (\ref{exotic}), the dynamics of the
flat direction is similar to the one given in Eq.~(\ref{gmsbp0}),
which yields the equation of motion
\begin{equation}
\varphi^{''} + \frac{2}{r} \varphi^{'} = -  \omega^{2} \varphi\,.
\end{equation}
For large $r$, $\varphi(r)\sim \exp(-m_{\phi} r)$, where $m_{\phi}$ is
the mass of $\varphi$ near the origin. The interpolating solution
was already presented in Eq.~(\ref{thick0}). The energy of such a
$Q$-ball grows as \cite{kusenko98418}
\begin{equation}
\label{thick1}
E\approx \frac{4\sqrt{2}}{3}\pi m_{\phi}Q^{3/4}\,.
\end{equation}
The profile of the $Q$-ball is given by $\varphi(r)\sim \exp(-m_{\phi}r)$.
The radius and the value of the vev inside the $Q$-ball is roughly given by
\cite{kusenko98418}
\begin{eqnarray}
\label{thick2}
R &\approx& \frac{1}{\sqrt{2}m_{\phi}}Q^{1/4}\,, \\
\label{thick3}
\varphi_{0} &\approx&\frac{m_{\phi}}{\sqrt{2\pi}}Q^{1/4}\,.
\end{eqnarray}

%%%%%%%%%%%%%%%%%%%%%%%%%%%%%%%%%%%%%%%%%%%%%%%%%%%%%%%%%%%%%%%%%%

\subsubsection{Thick wall $Q$-balls in the gravity mediated case}

If the potential grows only slightly slower than  $\varphi^2$ as in
the case of gravity mediated supersymmetry breaking scenarios, the
potential may be approximately written as
\cite{enqvist98,enqvist9881,enqvist99} (see Eq.~(\ref{pot}), in Sect.~$5.3.1$.)
\begin{equation}
\label{thicklog0}
U(\varphi)\approx m^2_{\phi}\left(1+K\log\left[\frac{\varphi^2}{M^2}\right]
\right)\varphi^2\,,
\end{equation}
where $K<0$, and $M$ is the largest mass scale. Note that at small vevs
we have again neglected the non-renormalizable contributions in the above
potential. The mass scale is given by
$m_{\phi}\sim m_{3/2}\sim {\cal O}({\rm TeV})$.

The $Q$-ball equation of motion is written as
\bea{a1}
\varphi^{''} + \frac{2}{r} \varphi^{'} = -  \omega_{0}^{2} \varphi
+ m_{\phi}^{2} \varphi K \log \left( \frac{\varphi^{2}}{M^{2}}\right)~,
\eea
where $\omega_{0}$ is defined by
\bea{a2}
\omega_{0}^{2} = \omega^{2} -  m_{\phi}^{2} \left( 1 + K \right)\,.
\eea

For thin wall $Q$-balls, the initial value of $\varphi$ is
very close to $\varphi_{c1}$, the value of $\varphi$ for which the
right-hand side of \eq{a1} vanishes. In this case $\varphi$ will remain
close to $\varphi_{c1}$ up to a radius of order
$\omega_{0}^{-1}\log\left(\varphi_{c1}/\delta\varphi(0)\right)$, where
$\delta \varphi(0) = (\varphi_{c1}-\varphi(0))$. It will then decrease
to zero over a distance $\delta r \approx \omega_{0}^{-1}$, corresponding
to the width of the wall of a $Q$-ball. The radius of a thin wall
$Q$-ball can be made arbitrarily large by choosing $\delta \phi(0)$ small
enough.

For a thick wall $Q$-ball, the initial value of $\varphi$
can be much smaller than $\varphi_{c1}$. In this case the non-renormalizable
terms may be neglected. In general, the right-hand side of the
$Q$-ball equation of motion vanishes for three values of $\varphi$,
which correspond to $\varphi_{c1}$, $\varphi_{c2}$ and zero.
$\varphi_{c2}$ corresponds to
the point at which, assuming that the non-renormalizable terms can be
neglected, the first two terms on the  right-hand side of \eq{a1}
cancel, and we obtain~\cite{enqvist99}
\begin{eqnarray}
\label{a3}
\varphi_{c2}(r)&=& Me^{\left( \frac{\omega_{0}^{2}}{2 K m_{\phi}^{2}}
\right)}\,,
\nonumber \\
&=& M e^{-\left(1-\omega^2/m_{\phi}^2-2K\right)}e^{Km^2_{\phi} r^2/2}\,.
\end{eqnarray}
In deriving the last correspondence we have used Eq.~(\ref{a2}). Note that
$\varphi_{c2}$ is an attractor, in a sense that if $\varphi(0)$ is close
to $\varphi_{c2}$ it will tend towards $\varphi_{c2}$ as $r$ increases.

The radius of a thick wall $Q$-ball in the gravity mediated case
is given by \cite{enqvist98,enqvist9881,enqvist99}
\bea{a5}
R \approx \frac{1}{|K|^{1/2}m_{\phi}}\,,
\eea
where $R$ is defined as the radius within which $90\%$ of the $Q$-ball
energy is found, and \cite{enqvist98,enqvist9881,enqvist99}
\bea{a6}
\omega_{0} \approx |K|^{1/2} m_{\phi} ~.
\eea
Since typically $|K|$ is small compared with 1, we find
$\omega \approx m_{\phi}$. In the gravity mediated supersymmetry
case the size of a $Q$-ball does not depend on charge, unlike in the
gauge mediated case, see
Figs.~(\ref{relic3D},\ref{figmultamaki1},\ref{figmultamaki2}), where
the sizes of the $Q$-balls are all equal.

One may take a Gaussian ansatz for the profile of a thick wall $Q$-ball
\cite{enqvist99}
\bea{a7}
\varphi(r) = \varphi(0) e^{-\frac{r^{2}}{R^{2}}}~,
\eea
provided one identifies
$\omega_{0}^{2} \approx 3 |K|m_{\phi}^{2}$
and $R^{2} \approx 2 (|K|m_{\phi}^{2})^{-1}$.

The total charge of the Gaussian thick wall $Q$-ball is given by
\cite{enqvist99}
\bea{a10}
Q = \int dr \; 4\pi r^{2} \omega\varphi_{0}^{2} e^{- \frac{2 r^{2}}{R^{2}} }
 = \left(\frac{\pi}{2}\right)^{3/2} \omega \varphi_{0}^{2} R^{3}~
\eea
while the total energy is given by \cite{enqvist99}
\bea{a11}
E \approx \frac{3}{2} \left(\frac{\pi}{2}\right)^{3/2}\varphi_{0}^{2} R +
\left(\frac{\pi}{2}\right)^{3/2} m_{\phi}^{2}\varphi_{0}^{2} R^{3}~,
\eea
where the second term in the above equation is the combined contribution
from the potential energy and the charge term, where we have used
$\omega^{2}\approx m_{\phi}^{2}$.
Since $R$ is large compared with $m_{\phi}^{-1}$ for small $|K|$,
the potential plus charge term dominates the energy. The radius within
which $90\%$ of the energy lies is then given by
$R_{c} = 1.25 R$. The energy per unit charge is given by \cite{enqvist99}
\bea{a11a}
\frac{E}{Q}=\frac{m^{2}}{\omega}\approx\left(1+\frac{3|K|}{2}\right)
m_{\phi}~,
\eea
where we have used the Gaussian result $\omega_{0}^{2}= 3|K|m_{\phi}^{2}$.
For all practical purposes we can take $E\approx m_{\phi}Q$.
Although the energy per unit charge is larger than $m_{\phi}$, the mass
of the scalar at small values of $\varphi$ will have the form
$m_{\phi}(1 + \alpha |K|)$ (with $\alpha \gae 1$)
once the logarithmic correction to the potential is included,
so that the binding energy per unit charge will be positive and of order
$|K|m_{\phi}$.

The last two examples Eqs.~(\ref{gmsbp0},\ref{thicklog0}) exhibit
two extremes of any thick wall type $Q$-ball. Any
thick wall $Q$-ball should belong somewhere in between, such as
in hybrid case.

%%%%%%%%%%%%%%%%%%%%%%%%%%%%%%%%%%%%%%%%%%%%%%%%%%%%%%%%%%%%%%

\subsubsection{Hybrid case: gauge and gravity mediated $Q$-ball}

In Sect.~$6.2.2$., we discussed the $Q$-ball potential in the gauge mediated
case, but it is  true that any generic flat direction is also lifted by
gravity mediation as well. In the gauge mediated case the full
flat direction potential relevant for  $Q$-ball formation should read as
\cite{kasuya00,kasuya01}
\begin{equation}
\label{gmsbp1}
U(\varphi)=m_{\phi}^4\log\left(1+\frac{|\varphi|^2}{m^2_{\phi}}\right)
+m_{3/2}^2|\varphi|^2\left[1+|K|\log\left(\frac{|\varphi|^2}{M_{\rm P}^2}
\right)\right]\,,
\end{equation}
where $m_{3/2}$ takes values between $100$~KeV and $1$~GeV. The gaugino
loops lead to $K<0$ as  discussed before (see Sect.~$5.3.1$.), but Yukawa
couplings give rise to $K>0$. On the other hand if the Yukawas dominate,
the AD condensate can be stabilized and $Q$-balls can only form for
$\varphi \leq \varphi_{eq}$. The second term in the above
potential dominates when
\begin{equation}
\varphi \geq \varphi_{eq}\equiv \sqrt{2}\frac{m_{\phi}^2}{m_{3/2}}\,.
\end{equation}
There are two distinct regimes. When $\varphi \geq \varphi_{eq}$,
the $Q$-ball properties resemble the gravity mediated thick wall case.
Otherwise, when $\varphi <\varphi_{eq}$, the $Q$-ball properties
are similar to the gauge mediated thick wall case. The energy per
unit charge can be written as \cite{kasuya00,kasuya01}
\begin{equation}
  \frac{E}{Q}\sim \left\{
      \begin{array}{ll}
          \ds{ m_{\phi}Q^{-1/4}}
                       & \ds{\varphi\leq \varphi_{eq}} \\[2mm]
          \ds{ m_{3/2}}
                       & \ds{\varphi \geq \varphi_{eq}.}
      \end{array}
      \right.
\end{equation}
Obviously in between there should be a hybrid regime which
interpolates smoothly between the gauge and gravity mediated cases.

%%%%%%%%%%%%%%%%%%%%%%%%%%%%%%%%%%%%%%%%%%%%%%%%%%%%%%%%%%%%%%%

\subsubsection{Effect of gravity on $Q$-balls}

In principle gravity can give a significant contribution to the $Q$-ball
energy, as shown in a study by Multam\"aki and Vilja \cite{multamaki02}. In a
thin wall case, the interesting result is that gravity limits the
maximal size of a $Q$-ball. The reason is that besides the rotational
motion in the complex field space which generates outward pressure,
there exists a gravitational attraction. Since the gravitational
contribution to the $Q$-ball energy is negative, it is possible that
gravity can render an otherwise unstable $Q$-ball stable.
The effect of gravity on $Q$-balls remains small provided the soliton
is much larger than the Schwarzschild radius and the charge smaller
than the gravitational charge $Q_{g} \sim (M_{\rm P}/m_{\phi})^4$,
which is quite large $Q_{g}\sim 10^{64}$ for $m_{\phi}\sim 100$~GeV.

%%%%%%%%%%%%%%%%%%%%%%%%%%%%%%%%%%%%%%%%%%%%%%%%%%%%%%%%%%%%%%%

\subsubsection{$Q$-balls and local gauge invariance}

So far we have considered $Q$-ball solutions in theories with
a global $U(1)$ symmetry. The symmetry group can however be
extended to include global non-Abelian symmetries
\cite{safian88a,axenides92a,axenides98,fujii0164}.
The existence of $Q$-balls in a supersymmetric Wess-Zumino
model has been demonstrated by Axenides, Floratos and Kehagias
\cite{axenides98}, who showed that $Q$-balls form domains of
manifestly broken supersymmetry.

A gauged $Q$-ball has some additional interesting properties
\cite{klee89,shiromizu99,demir99,axenides01,levi01}.
Taking a complex scalar field
$\phi({\bf r},t)=f({\bf r},t)\exp(-i\theta({\bf r},t))/\sqrt{2}$ coupled
to an Abelian gauge field $A_\mu$, the charge of a given field
configuration is given by \cite{klee89}
\begin{equation}
Q_\phi=\int d^3rf^2(\dot \theta -eA_0)~,
\end{equation}
where $e$ is the gauge charge.
Since the gauge field inside a $Q$-ball is broken by the non-zero vev of
$\varphi$ which couples to the gauge field canonically, the gauge field
is massive and acts as a $U(1)$ superconductor, provided the Compton
wavelength of the gauge field is smaller than the size of a $Q$-ball.
As a result of broken gauge symmetry, there is an extra source of
electrostatic self-energy contribution which comes from the electrostatic
repulsive force due to the presence of a gauge charge. The gauged
charges are also repelled to reside on the boundary of a $Q$-ball.
In a gauged $Q$-ball, for a fixed
charge $Q$, both radius and energy are relatively large. The remarkable feature
is that there is a maximum allowed charge and correspondingly a maximum
radius. This is due to the repulsive electrostatic potential $A_0$ which
tends to destabilize the gauged $Q$-ball. The maximum charge-to-radius
ratio is governed by the gauge charge:
$Q_{max}/R_{max}=4\pi(m_{\phi}-\omega_{c})/e^2$.

In \cite{axenides01}, it has been argued that the presence of fermions
could stabilize the gauged $Q$-ball. During  $Q$-ball formation,
fermions could be trapped inside the $Q$-ball. Those with charges equal
to the charge of the scalar quanta would be repelled from the inside,
whereas fermions with opposite charges would remain and render the
$Q$-ball neutral. Large electric fields inside the $Q$-ball
could also lead to a pair production and to a subsequent screening
of the charge.

An interesting application of a gauged $Q$-ball could be
the hadronic structure of QCD, as in the Friedberg-Lee
model \cite{friedberg77a}, where hadrons are modeled by phenomenological
non-topological solitons. Gauged Q-balls in theories with a
Chern-Simons terms has also been considered in \cite{khare89}.

%%%%%%%%%%%%%%%%%%%%%%%%%%%%%%%%%%%%%%%%%%%%%%%%%%%%%%%%%%%

\subsection{$Q$-ball decay}

%%%%%%%%%%%%%%%%%%%%%%%%%%%%%%%%%%%%%%%%%%%%%%%%%%%%%%%%%%
\subsubsection{Surface evaporation to fermions}

In the MSSM, the scalar field forming a $Q$-ball can interact
with fermions.  Then a $Q$-ball can decay into a pair of fermionic quanta.
$Q$-ball decay has been considered in \cite{cohen86,multamaki00},
where fermion production was studied in a classical background of
a $Q$-ball. As it was first pointed out in \cite{cohen86},
for a large $Q$ the $Q$-ball evaporates through its physical
surface and  there exists an upper bound on the evaporation rate per unit
area. This behavior has been verified by numerical studies
\cite{multamaki00}.

In order to understand this, let us imagine that a region inside a
$Q$-matter with a vanishing $\phi$, which might appear due to
fermion pair production, forms a cavity. Suppose $L$ is the
linear size of the cavity. The energy of $Q$-matter
formerly inside the cavity is given by \cite{cohen86}
\begin{equation}
\label{hand0}
E_{Q}\sim \omega_{0}^2\phi_{0}^2L^3\,,
\end{equation}
while the charge $Q$ within that region is given by \cite{cohen86}
\begin{equation}
\label{hand1}
\Delta Q\sim \omega_{0}\phi_{0}^2L^3\,.
\end{equation}
$N$ massless fermions inside the cavity will have a free Fermi gas
distribution, and therefore, the energy of the fermions is given by
\cite{cohen86}
\begin{equation}
\label{hand3}
E_{\psi}\sim \hbar \frac{N^{4/3}}{L}\sim \left(\frac{\hbar}{L}\right)
\left(\frac{\Delta Q}{\hbar}\right)^{4/3}\,.
\end{equation}
The ratio of energies is given by \cite{cohen86}
\begin{equation}
\label{hand4}
\frac{E_{Q}}{E_{\psi}}\sim \left(\frac{\hbar \omega_{0}^2}{\phi_{0}^2}\right)
^{1/3}\,.
\end{equation}
In the semiclassical limit $\hbar \rightarrow 0$, cavitation is energetically
forbidden irrespective of the size of a cavity. In a sense, it is the
Pauli exclusion principle which keeps the $Q$-ball stable. Fermions are
produced but there exists a Fermi pressure which prevents further production.
Inside $Q$-matter fermions gain mass of order $g\phi_{0}$ and saturate
the Fermi energy so that fermions can be produced only from the surface.

Following~\cite{cohen86}, let us assume that each  fermion
carries the energy $\sim \hbar \omega_{0}/2$. A  simple bound on
the average fermionic pair production from the surface can then be obtained
by noting that maximum current density can be reached only
when the outward moving fermions from the $Q$-wall are occupied while
every level in the inward moving fermions is empty. Then, by assuming
massless Weyl fermions with a single helicity state for each $\vec k$,
one obtains the limit on the outward moving
current~\cite{cohen86}
\begin{equation}
\langle \vec n\cdot \vec j\rangle _{\omega \leq \omega_{0}/2}\leq
\frac{1}{(2\pi)^3}\int_{0}^{\omega_{0}/2}k^2 dk\int_{0}^{1} \cos(\theta)
d(\cos(\theta))\int_{0}^{2\pi}d\phi =\frac{\omega_{0}^3}{192\pi^2}\,.
\end{equation}
Integrating this expression over the enclosing surface yields the maximum
evaporation rate per unit area in the limit
$R\rightarrow\infty$ \cite{cohen86}
\begin{equation}
\label{qevap0}
\frac{dQ}{dt dA}\leq \frac{\omega_{0}^3}{192 \pi^2}\,.
\end{equation}
A detailed calculation relies on estimating the reflection
and transmission coefficients of the ingoing and outgoing waves outside the
$Q$-ball and matching these with the solutions obtained
inside the $Q$-ball. In order to calculate the transmission
coefficient, one has to sum over infinite sequence of scatterings.
In \cite{cohen86}, it was shown  that the maximum transmission
coefficient is 1 when $\omega_{0}\leq g\phi_{0}$. The weak coupling limit
$\omega_{0}>g\phi_{0}$ leads to a different transmission coefficient and
the rate of evaporation is given by \cite{cohen86}
\begin{equation}
\label{qevap1}
\frac{dQ}{dtdA}\simeq 3\pi\left(\frac{g\phi_{0}}{\omega_{0}}\right)
\left(\frac{\omega_{0}^3}{192 \pi^2}\right)=
\frac{g^2\omega_{0}^2\phi_{0}}{32\pi^2}\,.
\end{equation}
The factor $g\phi_{0}$ determines the penetration width of the
fermions. In other words, in the weak coupling limit fermions
can penetrate deep inside a $Q$-ball without completely filling
the Fermi sphere. Within Pauli exclusion principle,
$Q$-matter could decay into very weakly coupled fermions
within the whole $Q$-ball volume. This will be discussed in the
context of $L$-balls in Sect.~$7.1$, which can
decay into massless neutrinos throughout the interior.

In \cite{cohen86,multamaki00}, it was shown numerically
that the evaporation rate is strongly dependent on $R$ but
approaches the limiting profile given in \cite{cohen86}.
In a realistic case a step-function is not always a good
approximation, in particular for a thick wall
$Q$-ball. In the thick wall case the problem has been investigated
numerically  by Multam\"aki and Vilja \cite{multamaki00}, who found
that for a sufficiently large $Q$ the evaporation rate decreases
with increasing $Q$. As a result the $Q$-ball evaporates faster when its
size decreases.

%%%%%%%%%%%%%%%%%%%%%%%%%%%%%%%%%%%%%%%%%%%%%%%%%%%%%%%%%%%%

\subsubsection{The decay temperature}

For a thin wall $Q$-ball the surface area is related to the charge via
$A = (36 \pi)^{1/3}Q^{2/3}/(\kappa^{2/3})$, where
$\kappa = (2 \varphi_{0}^{2} U(\varphi_{0}))^{1/2}$.
The lifetime of a $Q$-ball in this case is given by
\cite{enqvist99}
\be{lttw}
\tau = 144 \pi \left(\frac{4 \pi}{3}\right)^{2/3}\frac{\kappa^{2/3}Q^{1/3}}
{\omega^{3}} ~.
\ee
For a thick wall case, the area of a $Q$-ball is independent of
its charge, being fixed by its radius $R \approx (|K|^{1/2}m_{\phi})^{-1}$.
The $Q$-ball lifetime in this case is then given by \cite{enqvist99}
\be{ltfw}
\tau = \frac{48 \pi Q}{R^{2} \omega^{3}} ~.
\ee
From the above expression, one can estimate the temperature at which
the $Q$-balls decay. By assuming radiation domination, the decay
temperature is defined as
\cite{enqvist99}
\be{tempd}
T_{d} = \left(\frac{1}{k_{T}}\right)^{1/2}
\left(\frac{M_{\rm P}}{2 \tau}\right)^{1/2}~,
\ee
where $k_{T} = \left(\frac{4 \pi^{3} g(T)}{45}\right)^{1/2}$.
By taking $\omega \approx m_{\phi} \approx 100\GeV$,
thick wall $Q$-balls will decay at a temperature \cite{enqvist99}
\be{ttw}
T_{d} \approx \frac{15}{|K|^{1/2}}\left(\frac{\omega}{100 \GeV}\right)^{1/2}
\left(\frac{10^{15}}{Q}\right)^{1/2} \GeV ~,
\ee
where we have set $k_{T} \approx 17$. The $Q$-balls will
decay at a temperature less than $100$~GeV if $Q\gae 2\times 10^{13}|K|^{-1}$.
For thin wall $Q$-balls, the right-hand side of  \eq{ttw} has
an additional factor $\left({Q}/{Q_{c}}\right)^{1/3}$, where $Q_{c}$ is
the value at which the thin wall limit becomes valid.

%%%%%%%%%%%%%%%%%%%%%%%%%%%%%%%%%%%%%%%%%%%%%%%%%%%%%%%%%%%%

\subsubsection{$Q$-ball decay into a pair of bosons}

If the $Q$-ball forming scalar is not the lightest, as might be the
case in the MSSM, it is also possible that the $Q$-ball decays into lighter
bosons. The decay into scalar fields within the $Q$-ball volume is
not blocked by the Pauli exclusion principle. In the case of bosons
one should replace $N^{4/3}$ by $N$ in Eq.~(\ref{hand3}), because the bosons
can condense in the lowest mode of the cavity. The ratio
Eq.~(\ref{hand4}) becomes
\begin{equation}
\label{hand5}
\frac{E_{Q}}{E_{b}}\sim \omega_{0}L\,.
\end{equation}
Cavitation in the case of bosons is always energetically favorable
for sufficiently large $L$. $Q$-ball decaying into light
scalars could be significantly enhanced relative to the decay into
fermions.

The decay into light scalars will only be possible near
the edge of a thick wall $Q$-ball. This is because particles coupling
directly to the condensate scalars will gain a large effective mass
from $\langle \phi \rangle$ inside the $Q$-ball. As a result, decay into
light scalars will only occur via loop diagrams with rates suppressed
by the large effective mass. This tends to make MSSM  $B$-balls long-lived,
as will be discussed later.

Not all the flat directions would have a scalar decay mode. An example is
$\bar u\bar d\bar d$ direction, which is lifted at $d=6$
and contains the right-handed squarks. For the universal boundary condition
for squarks at large scales, RG flow analysis suggests that the left-handed
squark masses are typically heavier than  the right-handed ones. The Higgs
scalar mass could also be heavy compared to the right-handed
squark mass and the slepton masses usually come out to be lighter than
right-handed squarks. Hence the decay of such baryonic direction would be
kinematically forbidden. Even if the right-handed squark decays to
Higgses and to sleptons, it would certainly involve either a pair of
(light) quarks, gauginos, Higgsinos or leptons in the final state.
As a result the core of a $\bar u\bar d\bar d$ $Q$-ball
would be Fermi suppressed. Any final state involving a pair
of fermions can arise only from the surface.

Generically $Q$-balls made up of left-handed squarks and sleptons, such as
the $d=6$ MSSM flat directions $\bar dQL$ or $\bar eLL$ are expected to
decay into a pair of light bosons. Note that these particular directions are
good candidates for carrying $B$ charge. One may parameterize by $f_{s}$
the possible enhancement factor of the scalar decay rate over the fermion
decay rate, so that \cite{enqvist99}
\be{d2}
\left(\frac{dQ}{dt}\right)_{boson}
 =f_{s} \left(\frac{dQ}{dt}\right)_{fermion}
~.\ee
This gives
for the decay temperature \cite{enqvist99}
\be{d3}
T_{d} \approx
\left( \frac{f_{s}\omega^{3}R^{2}M_{\rm P}}{48 \pi k_{T} Q}\right)^{1/2}
 \approx 0.06  \left(\frac{f_{s}}{|K|}\right)^{1/2}
\left( \frac{m_{\phi}}{100\GeV} \right)^{1/2}
\left( \frac{10^{20}}{Q} \right)^{1/2} \GeV\,.
\ee

For $\bar dQL$ or $\bar eLL$ one may estimate the largest possible
enhancement factor using the Gaussian thick wall ansatz as shown
in \cite{enqvist99}. Within the
$Q$-ball, for  $\phi \gg m_{\phi}$, the lowest possible dimension
operator which could allow the condensate scalars to decay at one loop level
to light particles is the one lifted by the $d=5$ operator
\be{d5}
\frac{1}{M} \int d^{4}\theta \varphi \chi^{\dagger} \eta ~,
\ee
where $\chi$ and $\eta$ represent the light particles and
$M \approx g\varphi$, where $g$ is the coupling of the heavy particles
to $\varphi$. The decay rate of the condensate scalars to light scalars
will then be given by~\cite{enqvist99}
\be{d6}
\frac{dQ}{dt} = -\int \omega \varphi^{2}(r) \Gamma(r) 4 \pi r^{2} dr ~,
\ee
where $\omega\varphi^{2}(r)$ is the charge density within the $Q$-ball,
and \cite{enqvist99}
\bea{d7}
\Gamma(r) &\approx& \frac{\alpha^{2} m_{\phi}^{3}}{\phi^{2}}
\; ; \;\;\; g\phi > m_{\phi} ~\nn
&\approx& \alpha m_{\phi} \; ; \;\;\; g\phi < m_{\phi}~,
\eea
with $\alpha = g^{2}/(4 \pi)$ (for simplicity we consider a single
coupling constant $g$). Let $r_{*}$ be the radius at which
$\varphi(r) = m/g$. Then the largest contribution to the decay rate
will come from a region of width $\delta r \approx R^{2}/(4 r_{*})$
around $r_{*}$, over which $\varphi$ has a roughly constant
value $\varphi \approx m/g$, where
\be{d9}
r^{*} = \gamma R  \;\; ; \;\;\;\;\; \gamma = \ln^{1/2}\left(
\frac{g\varphi(0)}{m_{\phi}}\right)     ~.
\ee
From \eq{d6}, the rate can be deduced as \cite{enqvist99}
\be{d10}
\frac{dQ}{dt} \approx
- 4 \pi \alpha \omega m_{\phi} \left(m_{\phi}^{2}\alpha
\int_{0}^{r_{*}} dr r^{2}
+ \phi^{2}(0) \int_{r_{*}}^{\infty} dr r^{2} e^{-\frac{2 r^{2}}{R^{2}}}
\right) ~
\ee
\be{d10a} \approx -\frac{12 \pi \gamma}{|K|^{1/2}}
\left(1 + \frac{\gamma^{2} g^{4}}{3 \pi}\right)
\left(\frac{dQ}{dt}\right)_{fermion}
~.\ee
where we have used $\omega \approx m_{\phi}$. For a thick wall
$Q$-ball and for typical values of the parameters,
$g\varphi(0)/m \approx (0.1-0.01)Q^{1/2}$, so that
$\gamma \approx 4.5$ in Eq.~(\ref{d10a}). One finds that
the enhancement factor is typically \cite{enqvist99}
\be{d11}
f_{s} \approx \frac{170}{|K|^{1/2}}\left(1 + 2.1 g^{4}\right)~.
\ee
if $g$ is less than $1$ then for $|K| \approx 0.01-0.1$, we
expect an enhancement factor not much larger than about $10^{3}$.
For $g$ less than $1$, most of the enhancement factor comes from
the unsuppressed tree level decays occurring at $r > r_{*}$.

%%%%%%%%%%%%%%%%%%%%%%%%%%%%%%%%%%%%%%%%%%%%%%%%%%%%%%%%

\subsection{Cosmological formation of $Q$-balls}

Cosmological formation of $Q$-balls was initially proposed by Kusenko
in~\cite{kusenko97405}, by Kusenko and Shaposhnikov in
\cite{kusenko98418}, and by Enqvist and McDonald who studied $Q$-balls
from the fragmentation of the flat direction condensates in
the gravity mediated case \cite{enqvist98,enqvist99}. Subsequently,
the problem has been attacked numerically by Kasuya and Kawasaki in
case of gravity mediation \cite{kasuya0062,kasuya01} as well as in gauge
mediation \cite{kasuya00,kasuya01}, and by Multam\"aki and Vilja in the
gravity mediated case \cite{multamaki02535}.

%%%%%%%%%%%%%%%%%%%%%%%%%%%%%%%%%%%%%%%%%%%%%%%%%%%%%%%%%
\subsubsection{In gravity mediated case}

In the gravity mediated case the size of the $Q$-ball depends
on the charge; the larger the charge, the larger is the size of a $Q$-ball,
see Fig.~(\ref{fig-1}). In a cosmological context it is natural to think
that the $Q$-balls form when the the most amplified AD condensate mode is
as large as the horizon size just after the AD field starts rotation,
such that $H^{-1}\sim\omega^{-1}\sim\varphi_0/m_{\phi}^2$.
The charge of a $Q$-ball should be given by~\cite{kasuya01}
\begin{equation}
    Q \sim H^{-D} n_{\varphi} \sim \omega^{-D} \omega\varphi_0^2
      \sim m_{\phi}^{3-D}\left(\frac{\varphi_0}{m_{\phi}}\right)^{1+D}\,,
\end{equation}
where $D=1,2,3$ stands for the number of spatial dimensions, and 
$\omega\approx m_{\phi}$.
This naive expectation has been verified numerically on lattice
by Kasuya and Kawasaki \cite{kasuya01}, although the formation
turns out to be slightly delayed \cite{kasuya0062}. The
maximum charge of a thick wall $Q$-ball can be written as
\begin{equation}
\label{Qest0}
Q_{max}= \beta_{D}\varphi_{0}^{1+D}\,,
\end{equation}
where $\beta_D$'s are some numerical factors with
$\beta_1\approx 0.1$, $\beta_2 \approx 0.02$, and
$\beta_3 \approx 6\times 10^{-4}$ \cite{kasuya01}.

Kasuya and Kawasaki~\cite{kasuya01} also noticed that the charge of a
$Q$-ball depends on the helical motion of the AD condensate, and it is
proportional to $\varepsilon$
\begin{equation}
\label{helical}
\varepsilon=\frac{n_{\varphi}(t_{osc})}{n^{max}_{Q}(t_{osc})}=
\frac{m_{3/2}\varphi^2}{\omega\varphi^2}=\frac{m_{3/2}}{\omega}
\approx\frac{m_{3/2}}{m_{\phi}}\,,
\end{equation}
where $\varepsilon=1$ corresponds to a circular motion, and
$\varepsilon =0$ for the radial motion. The numerical
calculation in~\cite{kasuya0062,kasuya01} indicates that $Q_{max}$ is
constant for small $\varepsilon$ where both positive and
negative $Q$-balls with charges of the same order of magnitude
are produced, while linearly dependent on $\varepsilon$, around
$\varepsilon \sim 1$. It was noticed that dominantly positive $Q$ balls
were formed. Numerical simulations also reveal the presence of small
negatively charged $Q$ balls~\cite{kasuya0062,kasuya01}.

%%%%%%%%%%%%%%%%%%%%%%%%%%%%%%%%%%%%%%%%%%%%%%%%%%%%%%%%%%%%%%%%%

\subsubsection{In gauge mediated case}

The potential for the condensate $\Phi$ forming a $Q$-ball is given by
Eq.~(\ref{potgmsb}). If the AD condensate couples directly to particles
in a thermal bath, the potential receives a thermal mass correction of
order $\sim T^2|\Phi|^2$ while integrating out the heavy modes yields a
contribution $\sim \pm T^4\log(|\Phi|^2/T^2)$. In the latter case the
actual sign depends upon the integrated modes. If the integrated heavy
modes mainly belong to the matter multiplet then the sign comes out to
be positive, otherwise if it were dominated by the heavy gauge degrees of
freedom then it turns out to be negative. In a particular case of
$LH_{u}$ flat direction this sign is positive as shown by Fujii,
Hamaguchi and Yanagida in \cite{fujii0163}. Combining these gives the
relevant part of the effective potential (without non-renormalizable terms)
as \cite{kasuya01}
\begin{equation}
    \label{pot-gen}
    V(\Phi) \approx m_{\phi}^4(T) \log \left( 1+\frac{|\Phi|^2}
              {m_{\phi}^2(T)}\right).
\end{equation}
Here we have assumed that the flat direction obtains a positive
contribution to the thermal potential. The effective mass can be
written as \cite{kasuya01}
\begin{equation}
    m_{\phi}(T) = \left\{
    \begin{array}{ll}
        m_{\phi} & (T<m_{\phi}) \\[2mm]
        T & (T>m_{\phi}) \\
    \end{array} \right.,
\end{equation}
Note that $T\propto t^{-1/4}$ during inflaton oscillations dominated phase.

At very large amplitudes of the AD condensate, the gravity mediation
effects for supersymmetry breaking dominates and a stable
$Q$-ball of hybrid type, as discussed in Sect.~$6.2$, can form
 \cite{kasuya00,kasuya01}. In this case the AD potential is
dominated by the terms
\begin{equation}
    \label{grav-pot}
    V(\Phi) \approx m_{3/2}^2\left[1+K\log
        \left(\frac{|\Phi|^2}{M^2}\right)\right]|\Phi|^2.
\end{equation}
Since here the curvature of the potential depends weakly on the
amplitude, the AD condensate starts rotating when
$H \simeq m_{3/2} \ll 1$~TeV. In \cite{kasuya01},
the authors have simulated the dynamics of the AD condensate on
$1,2,3$-d lattices, and verified the formation of Q balls
for $K=-0.01$.

In a cosmological context one may estimate the maximum charge of a $Q$-ball
from the fragmentation of the AD condensate. The analysis is similar to
the gravity mediated case, except for the fact that the times when the
AD condensate starts oscillating and the most amplified mode
(or instability band) enters the horizon do not coincide.
The charge of a $Q$-ball is then given by \cite{kasuya00,kasuya01}
\begin{equation}
\label{Qest1}
    Q \sim H_f^{-D} m_{\phi}\varphi_f^2
       \sim  (|K|^{1/2}m_{\phi})^{-D} m_{3/2}|K|\varphi_0^2
       \sim  |K|^{1-D/2}m_{\phi}^{3-D}
            \left(\frac{\varphi_0}{m_{\phi}}\right)^2,
\end{equation}
where the subscript `$f$' denotes the time when the $Q$-ball forms, and
we have assumed $m_{3/2}\sim m_{\phi}$ in our final expression.
Note that $Q \propto \varphi_0^{2}$, with the proportionality constant
to be determined numerically; one finds
$Q_{max}\sim \widetilde \beta_3\varphi^2$~\cite{kasuya01}, where
$\widetilde{\beta}_3 \approx 6\times10^{-3}$.

%%%%%%%%%%%%%%%%%%%%%%%%%%%%%%%%%%%%%%%%%%%%%%%%%%%%%%%%%%%%

\subsection{$Q$-ball collisions}

The dynamics of any extended object carrying charge is quite different
from the dynamics of charged point-like objects. In this respect studying
the $Q$-ball collision is important in order to understand the charge
distribution of $Q$-matter in the Universe.  $Q$-ball collisions have
been studied by  a number of authors \cite{makhanov79,kusenko98418,kasuya0062,axenides00,battye00,multamaki00482,multamaki00484}.
In \cite{kasuya01,multamaki00482}, collisions have been considered
in the context of gravity mediated supersymmetry breaking, while in
\cite{multamaki00484}, $Q$-ball collisions have been tackled within
gauge mediated supersymmetry breaking. An animation
of $Q$-ball collisions can be found at
{\it www.utu.fi/$\sim$tuomul/collision.mpg}.

The usual ansatz is to take initially two spatially well separated
$Q$-balls so that initial field configuration (in 1d) is~\cite{battye00}
\begin{equation}
\label{mqb0}
\Phi(t,x)=e^{i\omega_{1}t+i\alpha}\varphi_{\omega_{1}}(|x+a|)+e^{i\omega_{2}t}
\phi_{\omega_{2}}(|x-a|)\,.
\end{equation}
The two
$Q$-balls have the profiles $\varphi_{\omega_{1}}$ and $\varphi_{\omega_{2}}$,
correspondingly to the frequencies $\omega_{1}$ and $\omega_{2}$, and they are
separated by a distance $2a$. The total charge of the configuration
(in 1d) is given by \cite{battye00}
\begin{equation}
\label{mqb1}
Q=Q_{\omega_{1}}+Q_{\omega_{2}}+(\omega_{1}+\omega_{2})\cos(\alpha)
\int_{-\infty}^{+\infty}\varphi_{\omega_{1}}(|x+a|)\varphi_{\omega_{2}}
(|x-a|)dx\,.
\end{equation}
Note that the last term is exponentially small in the separation
parameter $a$, because the profiles of the individual $Q$-balls die
off exponentially outside $Q$-matter. Usually, for a single $Q$-ball
the phase is unimportant because of the global $U(1)$ symmetry, but
for multi-$Q$-ball case the relative phase $\alpha$ plays an important
role and affects the total charge of the configuration.

There are two extreme cases: the two charges are equal; or they have
opposite sign. The key parameters are the relative phase, the incident
velocity, and the charge \cite{axenides00,battye00}. The generic
interaction for two $Q$-balls of equal charges is attractive if the
relative phase $\alpha =0$, and repulsive when $\alpha =\pi$ . In
case of attraction the two $Q$-balls coalesce to form one larger $Q$-ball
with a resultant charge less than the sum total of the individual charges.
A loss in charge also occurs when the $Q$-ball suffers a large distortion.

If the initial phase $\alpha \neq 0,\pi$ or if the charges of the
$Q$-balls are not equal, then the dynamics of the $Q$-ball collisions
result in charge transfer. $Q$-balls tend to repel each other, which
happens even after charge transfer. If the incident velocity is extremely
high (relativistic), then $Q$-balls simply pass through each other without
losing much charge~\cite{battye00,axenides00}.

A collision of a $Q$-ball and an anti-$Q$-ball exhibits several
interesting features. A naive expectation would have them annihilating.
Instead, they bounce back or pass through each other. Charge is partially
annihilated, though. The main reason is the fact that generically
$Q$-balls  can transfer their charges only very slowly. The charge transfer
is very seldom complete~\cite{battye00}.

These conclusions hold mainly for thin-wall $Q$-balls. There is not much
difference between gauge and gravity mediated cases. Because in gravity
mediation the $Q$-ball size is smaller, in a fixed volume the AD condensate
tends to break into larger number of $Q$-balls \cite{kasuya0062}. As a
result the $Q$-balls can have larger peculiar velocities than in the gauge
mediated case. Multam\"aki and Vilja have studied the gravity mediated case
in~\cite{multamaki00482} for $2$-d lattice for a range of velocities between
$v=10^{-3}$ and $v=10^{-2}$. The authors allowed all possible values of
the relative
phase and found that the fusion cross section of two $Q$-balls appears to
be smaller than the geometric cross section, whereas the cross section for
the charge exchange is larger than the geometric cross section. The probability
for charge exchange processes increases with increasing $\omega$.
The actual velocity of the $Q$-ball is an open issue, which however,
to some extent, can be determined in a cosmological context~\cite{kasuya0062}.
In the gauge mediated case the peculiar velocity of the $Q$-balls is small
and the main interactions are either elastic scattering or partial charge
exchange \cite{multamaki00484}.

%%%%%%%%%%%%%%%%%%%%%%%%%%%%%%%%%%%%%%%%%%%%%%%%%%%%%%%%%%%%%%%
\subsection{$Q$-balls in a thermal bath}

There are several effects which one must take into account  when
$Q$-balls are immersed in a thermal bath at a temperature higher than
the $Q$-ball formation scale. The temperature dependent effective potential
might not even allow for a $Q$-ball solution, but if it does, then for a large
$Q$ thermal corrections are negligible as argued by Kusenko and
Shaposhnikov~\cite{kusenko98418}, and Laine and Shaposhnikov~\cite{laine98}.
Finite temperature effects always lead to an erosion of the condensate
and therefore to a loss of charge from the $Q$-ball into the ambient
plasma. As a result a chemical potential $\mu_{plasma}\sim\delta Q/(VT^2)$
arises in the surrounding plasma, where $V$ is  the Hubble volume
and $T$ is the ambient temperature. As pointed out in~\cite{laine98},
the process of $Q$-ball evaporation will stop when the chemical potential
$\mu_{Q} \sim m_{\phi}(T)(Q-\delta Q)^{-1/4}$ associated with the $Q$-ball
becomes equal to $\mu_{plasma}$. The conclusion is strictly valid
for the flat potential studied in \cite{laine98}, but should hold
qualitatively for all kinds of $Q$-balls.

In most cases of interest $Q$-balls are produced non-adiabatically with a
formation time scale  much shorter than the evaporation scale.
Then it is natural to ask how $Q$-balls come in thermal and chemical
equilibrium. Since the $Q$-ball does not lose its charge
significantly, chemical equilibrium between $Q$-balls and thermal
plasma may never be reached, in which case $\mu_{Q}$ always dominates over
$\mu_{plasma}$. It is possible to obtain a thermal equilibrium
at least between the soft edge of the $Q$-ball and the hot plasma.

In general there are three different thermally induced effects:
dissociation, diffusion, and evaporation of the $Q$-ball.

%%%%%%%%%%%%%%%%%%%%%%%%%%%%%%%%%%%%%%%%%%%%%%%%%%%%%%%%%%%%%%%%%%%
\subsubsection{Dissociation}

If $Q$-balls never reach thermal equilibrium with the plasma,
then $Q$-ball dissociation by the bombardment of thermal particles
as discussed in ~\cite{kusenko98418,enqvist99} is important.
In case of charge dissociation, the thermal particles in the plasma
collide with the $Q$-ball and may even penetrate inside. The penetration
width depends on the kinetic energy of the particles. For definiteness,
let us focus on the gauge mediated case with thick wall $Q$-balls.
A particle $\psi$ which interacts with $\varphi$ receives
a mass contribution $m_{\psi}\sim g\langle\varphi(x)\rangle$, while outside
$m_{\psi} \sim gT$. At an ambient temperature $T$, the particle cannot
penetrate the $Q$-ball beyond $x_{st}$, known as stopping radius and
determined by $g\varphi(x_{st})\approx 3T$. If $\psi$  imparts sufficient
energy to the $Q$-ball in order to overcome its binding energy within
the dynamical time scale then the $Q$-ball may simply break up.
On the other hand, if the energy is delivered to the $Q$-ball is below
the dissociation limit, the $Q$-ball will be able to radiate the excess
energy away adiabatically and will not dissociate.

The rate of dissociation depends on the flux of the incoming thermal
particles $f=(g_{\ast}(T)/\pi^2)4\pi x^2_{st}T^3$, and the energy per thermal
particle transferred to the $Q$-ball $\sim \gamma_{T}T$, where
$\gamma_{T}\leq 3$. Then the rate of energy imparted to the $Q$-ball
is given by \cite{enqvist99}
\begin{equation}
\label{QT0}
\frac{dE}{dt}=\frac{4g_{\ast}(T)\gamma_{T}T^4\beta^2R^2}{\pi}\,,
\end{equation}
where $x_{st} =\beta R$ is defined by the Gaussian thick wall profile
$\varphi(r)=\varphi(0)\exp(-r^2/R^2)$, and therefore
$\beta=\sqrt{\log(g\varphi(0)/3T)}$. In order to evade complete dissociation
$\Delta E \ll \Delta m_{\phi}Q$, where $\Delta m_{\phi}\approx |K|m_{\phi}$.
The dissociation will not be completed provided the temperature of
the thermal bath is given by \cite{enqvist99}
\begin{equation}
T\leq \left[\frac{\pi |K|^2}{4g_{\ast}(T)k_{r}\gamma_{T}\beta^2}\right]^{1/4}
m_{\phi}Q^{1/4}\,,
\end{equation}
where the dynamical time scale is assumed to be $\sim k_{r}/m_{\phi}$,
with $k_{r}>1$.

In a realistic case  dissociation alone cannot erode the $Q$-ball
completely. The $Q$-ball will rather come into thermal equilibrium with the
ambient plasma. In an expanding Universe the minimum energy is
configured in such a way that the $Q$-charges are always present
in the Universe along with the $Q$-balls. In this case the energy
of the $Q$-balls decreases as the temperature of the Universe decreases.

%%%%%%%%%%%%%%%%%%%%%%%%%%%%%%%%%%%%%%%%%%%%%%%%%%%%%%%%%%%%
\subsubsection{Diffusion}

Diffusion takes place only through the soft edge of the $Q$-ball at a
distance over which $\varphi$ does not change much. There are two
factors which determine the diffusion rate; firstly, how efficiently
the edge of the $Q$-ball diffuses, and secondly how fast the core
of the $Q$-ball readjusts itself in order to compensate for the loss
of charge. At large temperatures the diffusion rate is large as the
$Q$-ball tries to relax into chemical equilibrium with a
thermal plasma. The net diffusion rate is given by
\cite{banerjee00,kasuya00,kasuya01}
\begin{equation}
\label{QT1}
\Gamma_{diff}\equiv \frac{dQ}{dt}\sim -4\pi DR_{Q}\mu_{Q}T^2\sim -4\pi AT\,,
\end{equation}
where $D=A/T$ is the diffusion coefficient with $A={\cal O}(1)$, and
$\mu_{Q}\sim \omega$ is the chemical potential of a $Q$-ball. When the
temperature of a thermal bath drops due to expansion of the Universe,
surface evaporation rate takes over.

If there is a thermal bath already prior to reheating, the
instantaneous temperature of the plasma would be large:
$T \sim (M_{\rm P}^2\Gamma_IH)^{1/4} \sim (M_{\rm P} T_{rh}^2 H)^{1/4}$,
where $\Gamma_I$ is the decay rate of the inflaton field. During this
period, before reheating, $t \propto T^{-4}$, while after
reheating $t\propto T^{-2}$. The diffusion rate can be
obtained from Eq.~(\ref{QT1}) \cite{kasuya01}
\begin{equation}
    \left(\frac{dQ}{dT} \right)_{diff} \sim \left\{
      \begin{array}{ll}
          \ds{10 \frac{M_{\rm P}T_{rh}^2}{T^4}} & \ds{(T>T_{rh}),} \\[2mm]
          \ds{10 \frac{M_{\rm P}}{T^2}} & \ds{(T<T_{rh}).}
      \end{array}
      \right.
\end{equation}

%%%%%%%%%%%%%%%%%%%%%%%%%%%%%%%%%%%%%%%%%%%%%%%%%%%%%%%%%%%%%%%%%%%%%

\subsubsection{Evaporation at finite $T$}

In a thermal bath the surface evaporation rate is no longer given by
Eq.~(\ref{qevap0}), since one must take into account of  thermal
corrections. At finite $T$ the evaporation rate of a $Q$-ball has
been found to be \cite{laine98,kasuya00,kasuya01}
\begin{equation}
\label{QT2}
\Gamma_{evap}\equiv \frac{dQ}{dt}=-\zeta(\mu_{Q}-\mu_{plasma}) T^2 4\pi
R_{Q}^2 \sim e\pi \zeta\frac{T^2}{m(T)}Q^{1/4}\,,
\end{equation}
where $\mu_{plasma} \ll \mu_{Q}\sim \omega \sim m(T)Q^{1/4}$, while
$\zeta$ and $m(T)$ are given by
\begin{equation}
 m(T)=\left\{
      \begin{array}{l}
          \ds{m_{\phi}} \\[3mm]
          \ds{T}
      \end{array}
    \right.,
    \quad
    \zeta=\left\{
      \begin{array}{cl}
          \ds{\left(\frac{T}{m_{\phi}}\right)^2}
                    & \ds{(T<m_{\phi}),} \\[2mm]
          \ds{1}  & \ds{(T>m_{\phi}).}
      \end{array}
      \right.
\end{equation}
The evaporation rate can be calculated as \cite{kasuya01}
\begin{equation}
    \label{evap-rate-2}
    \Gamma_{evap} = \frac{dQ}{dt} = \left\{
      \begin{array}{ll}
          \ds{-4\pi T Q^{1/4}} & \ds{(T>m_{\phi}),} \\
          \ds{-4\pi \frac{T^4}{m_{\phi}^3}Q^{1/4}} &
                      \ds{(T<m_{\phi}).}
      \end{array}
      \right.
\end{equation}
In order to determine which rate is dominating, let us
consider the ratio $R_{diff}\equiv \Gamma_{diff}/\Gamma_{evap}$. For
$T>m_{\phi}$, the ratio is given by $R_{diff} = A Q^{-1/4}$. If
$R_{diff}<1$, the diffusion rate is the bottle-neck for the charge
transfer. This condition is met when the $Q$-ball charge is large enough:
\begin{equation}
    Q > 10^2 \left(\frac{A}{4}\right)^2.
\end{equation}
On the other hand when $T<m_{\phi}$, the condition $R_{diff}<1$ corresponds to
\begin{equation}
    \label{crit-T}
    T > T_* \equiv A^{1/3} m_{\phi} Q^{-1/12},
\end{equation}
and the transition temperature $T_*$ is lower than $m_{\phi}$ for
large enough $Q$-ball charge.

If there exists a thermal bath prior to reheating then there are additional
complications regarding the evaporation rate, which now exhibits four
different possibilities depending on how the temperature compares with
the reheating temperature and the mass of the AD particle
\cite{kasuya01,enqvist99}. The time-temperature relationship changes at
$T=T_{rh}$, while the rate $dQ/dt$ changes at $T=m_{\phi}$. Combining all
the effects, one obtains~\cite{kasuya01}
\begin{equation}
     \left(\frac{dQ}{dT} \right)_{evap} \sim \left\{
      \begin{array}{ll}
          \ds{10 \frac{M_{\rm P}T_{rh}^2}{T^4}Q^{1/4}}
                       & \ds{(T>T_{rh},m_{\phi}),} \\[2mm]
          \ds{10 \frac{M_{\rm P}T_{rh}^2}{m_{\phi}^3T}Q^{1/4}}
                       & \ds{(T_{rh}<T<m_{\phi}),} \\[2mm]
          \ds{10 \frac{M_{\rm P}}{T^2}Q^{1/4}}
                       & \ds{(m_{\phi}<T<T_{rh}),} \\[2mm]
          \ds{10 \frac{M_{\rm P}T}{m_{\phi}^3}Q^{1/4}}
                       & \ds{(T<m_{\phi},T_{rh}).}
      \end{array}
      \right.
\end{equation}
The estimate for the loss  charge turns out to be of
similar magnitude in all possible regimes with \cite{kasuya01}
\begin{equation}
    \label{evap-Q}
    \Delta Q \sim 10 \frac{M_{\rm P}}{m_{\phi}}Q^{1/12}
       \sim 2.4\times 10^{18}
              \left(\frac{m_{\phi}}{{\rm TeV}}\right)^{-1}
              \left(\frac{Q}{10^{24}}\right)^{1/12},
\end{equation}
for any case.

In the gravity mediated hybrid case the evaporation and diffusion
rates have the same forms in terms of $Q$-ball parameters
$R_Q \sim |K|^{-1/2} m_{3/2}$, and $\omega \sim m_{3/2}$.
The transition temperature at which $\Gamma_{evap}=\Gamma_{diff}$,
reads $T_*\equiv A^{1/3}|K|^{1/6}(m_{3/2}m_{\phi}^2)^{1/3}$. As in
the 'usual' type of $Q$ balls where the potential is dominated by the
logarithmic term, the charge evaporation near $T_*$ is dominant and
the total evaporated charge is found to be \cite{kasuya00,kasuya01}
\begin{equation}
\label{evap-Qg}
    \Delta Q \sim 10^{20} \left(\frac{m_{3/2}}{\rm MeV}\right)^{-1/3}
       \left(\frac{m_{\phi}}{\rm TeV}\right)^{-2/3}.
\end{equation}

%%%%%%%%%%%%%%%%%%%%%%%%%%%%%%%%%%%%%%%%%%%%%%%%%%%%%%%%%%%%%%%

\newpage

%%%%%%%%%%%%%%%%%%%%%%%%%%%%%%%%%%%%%%%%%%%%%%%%%%%%%%%%%
%%%%%%%%%%%%%%%%%%%%%%%%%%%%%%%%%%%%%%%%%%%%%%%%%%%%%%%%%
%%%%%%%%%%%%%%%%%%%%%%%%%%%%%%%%%%%%%%%%%%%%%%%%%%%%%%%%%

\section{Cosmological consequences of $Q$-balls}

In this section we discuss various issues concerning $Q$-ball
cosmology. If the MSSM flat direction carries some combination of
$B$ and/or $L$, the charge will be stored in $Q$-balls
created by the fragmentation of the initial condensate. If $Q$-balls
are stable, like in the case of gauge mediated supersymmetry breaking,
then they will be a good candidate for dark matter which can be searched
directly. If $Q$-balls eventually decay, the charge will be released
in a form of baryonic quanta, providing an interesting alternative
mechanism for baryogenesis which does not necessarily depends on sphalerons
transitions. The evaporation of $Q$-ball also gives rise to supersymmetric
dark matter. This is an added advantage of $Q$-ball cosmology: it provides
a physical mechanism for relating the dark matter and the baryon densities.

%%%%%%%%%%%%%%%%%%%%%%%%%%%%%%%%%%%%%%%%%%%%%%%%%%%%%%%%%

\subsection{$L$-ball cosmology}

$Q$-balls only carrying a leptonic charge are known as $L$-balls.
They emerge from the $LH_{u}$ flat direction which is quite different
from the rest of the MSSM flat directions. Even though $LH_{u}$ flat direction
might not fragment as already mentioned in Sect.~$5.3.1$., there
are choices for the initial conditions at $M_{GUT}$ which make
it possible to obtain a decreasing mass with decreasing $\mu$ in the
RG equations given in Eq.~(\ref{rge}). It has been noticed that
$m_{LH_{u}}^{2}$ becomes negative for scales typically smaller
than $10^{8}$ GeV or so \cite{enqvist98,enqvist99,enqvist9881}. Depending
on the choice of parameters there can be a "hill" in the plot of
$m_{LH_{u}}^{2}$ versus $|\varphi|$ (representing $LH_{u}$),
such that $m_{\phi}^{2}$ starts decreasing with increasing
$|\varphi|$ for sufficiently large values of $|\varphi|$. The
effect of negative $m_{\phi}^{2}$ at small enough $|\varphi|$ will
generate a minimum for $U(|\phi|)/|\phi|^{2}$ as required for $L$-ball
formation, typically at $|\varphi_{0}| \approx 1\TeV$.
Effectively such a potential can be given by \cite{enqvist99}
\be{l-pot}
U(\phi) \approx \frac{m_{\phi}^{2}}{2}(2 e^{-s \varphi} - 1)\varphi^{2} ~,
\ee
where $s \approx 1~{\rm TeV}^{-1}$. This gives rise to thick wall
$L$-balls with radius $ R \approx m_{\phi}^{-1}$. The charge of a
$L$-ball is bounded, i.e. $L \lae m_{\phi} s^{-2} V$, which
tends to become zero once $r \gae m_{\phi}^{-1}$. Since
$L \lae (sm_{\phi})^{-2}$, the $L$-balls will have a maximum charge,
which  for typical values of $s$ and $m_{\phi}$ cannot be larger
than $10^{3}$ (with an essentially fixed radius). This stands in
contrast to other thin wall $Q$-balls for which the charge is
proportional to the volume.

Inside $L$-balls the field strength is of order $\TeV$, which
is much smaller than the initial amplitude for $d=4$ AD condensate at
$H \approx m_{\phi}$. This suggests that $L$-balls cannot  form
by the collapse of an unstable condensate at $H \approx m_{\phi}$.
Therefore the AD baryogenesis along the $LH_{u}$ direction will be
essentially unaltered from the conventional scenario. Even if there were
a primordial formation of $L$-balls, these objects would decay at
$T_{d} \approx 10^{7}\GeV$. It is therefore unlikely that such $L$-balls
could have any cosmological consequences. Though it is possible that
$L$-balls, which have a field strength of order $1$~TeV or less,
could play a role in the physics of the electroweak phase transition
(we will discuss about phase transition aided by solitons in Sect.~$8.3.1$).

%%%%%%%%%%%%%%%%%%%%%%%%%%%%%%%%%%%%%%%%%%%%%%%%%%%%%%%%%
\subsection{$B$-ball cosmology}

Apart from $LH_{u}$ flat direction, there are purely
baryonic directions such as $\bar u\bar d\bar d$. $Q$-ball
forming along this direction carries only baryonic charge and
is dubbed as $B$-ball. There are also the $\bar dQL$ and
$\bar eLL$ directions. One may expect these directions
to be phenomenologically similar to $\bar u\bar d\bar d$ direction.
$R$-parity conservation allows $d=4$ non-renormalizable superpotential
term $(H_{u}L)^{2}$ and $d=6$ term $(\bar u\bar d\bar d)^{2}$.
In addition, there is also the $d=4$,
$B-L$ conserving $\bar u\bar u\bar d\bar e$
direction (and phenomenologically similar $QQQL$ direction).
Although this will not produce a $B$ asymmetry via a direct decay
of the AD condensate in case there is subsequent anomalous $B+L$
violation (i.e. sphalerons, see Sect.~$2.3.4$.), but it can generate a baryon
asymmetry via $Q$-ball decays occurring after the electroweak phase
transition \cite{enqvist98,enqvist99,enqvist9881}.

For large enough $B$, a $B$-ball cannot decay into the lightest
$B$-carrying fermions (the nucleons), and so it is completely stable.
Stable $B$-balls could have a wide ranging astrophysical
\cite{kusenko98418,kusenko98423} and experimental implications
\cite{dvali98}, as will be discussed in Sects.~$7.4$ and $7.6$.

In the gravity mediated case the $B$-balls may decay at temperatures below
$T_{ew}$, whence the observed baryon number will be a combination of
baryon number originating from the decay of the $B$-balls
and baryon number from free squarks left over after the
break-up of the squark condensate. Since the $B$-balls are composed
of squarks, when they decay they will naturally produce a $number$
density of neutralinos which is of the same order of magnitude as the
number density of baryons, as was first pointed out in \cite{enqvist98}.
If the $B$-balls decay sufficiently below the freeze-out
temperature of LSPs, and if the number density of thermal relic
neutralinos is less than that from the $B$-ball decay, then the dark
matter density and baryon number densities in the Universe
will be naturally related.

The actual ratio of baryons to dark matter will essentially be
determined by two variables: (i) the mass of the neutralino LSP,
and (ii) the proportion of baryon number trapped in B-balls as compared to
baryon number in free squarks;
this is usually referred to as the efficiency of $B$-ball formation.

When a $B$-ball decays, for each unit of $B$ produced,
corresponding to the decay of $3$ squarks to quarks, there will be at least
three units of $R$-parity produced, corresponding to at least $3$ neutralino
LSPs (depending on the nature of the cascade produced by the squark decay
and the LSP mass, more LSP pairs could be produced). Let $N_{\chi} \gae 3$
be the number of LSPs produced per baryon number and $f_{B}$ be the fraction
of the total $B$ asymmetry contained in  $B$-balls. Then the
baryon to dark matter ratio is given by~\cite{enqvist98,enqvist99},
see also \cite{fujii02aj},
\be{i1}
r_{B} = \frac{\rho_{B}}{\rho_{DM}} = \frac{m_{n}}{N_{\chi} f_{B} m_{\chi}}~,
\ee
where $m_{n}$ is the nucleon mass and $m_{\chi}$ is the neutralino
LSP mass. It is rather natural to have $r_{B} < 1$. The present LEP
lower bound on the neutralino mass in the MSSM (assuming no constraints
on the scalar masses) is $17$~GeV \cite{ellis97}. If we were to assume
radiative electroweak symmetry breaking and universal masses for the
squarks and Higgs scalars at the unification scale, then the lower
bound would become $m_{\chi}\gae 40\GeV$ for $\tan\beta \lae 3$
\cite{ellis97}. For $N_{\chi}\geq 3$, and with $m_{\chi} \gae 17 \:(40)\GeV$,
we find that $r_{B} < 1$ occurs for $f_{B} \gae 0.02 \;(0.008)$.
As long as more than $2\%$ of the baryon asymmetry is trapped in $B$-balls,
the observed dominance of dark matter in the Universe can be naturally
explained.

Primordial nucleosynthesis \cite{olive00333} bounds the density of
baryons in the Universe to satisfy $ 0.0048 \lae \Omega_{B}h^{2} \lae 0.013$,
where $0.4 < h < 1$ (we adopt the bound based on "reasonable" limits on
primordial element abundances \cite{olive00333,olive02}). The observed
baryon to dark matter ratio, $r_{B} \approx \Omega_{B}/(1-\Omega_{B})$
(assuming a flat Universe),  satisfies $0.005 \lae r_{B} \lae 0.09$.
This can be accounted from $B$-ball baryogenesis, provided
\cite{enqvist98,enqvist99}
\be{i2}
3.7\GeV \lae \left(\frac{N_{\chi}}{3}\right) f_{B}m_{\chi} \lae 67\GeV  ~.
\ee
For example, if the LSP mass satisfies
$17 \; (40)\GeV \lae m_{\chi} \lae 500\GeV$, then the
observed baryon to dark matter ratio can be achieved by a wide range of
$f_{B}$, i.e. $0.007 \lae f_{B} (N_{\chi}/3)$ $ \lae 3.9 \; (1.7)$.
Note that if $m_{\chi} \gae 67\GeV$, then we must have $f_{B} < 1$,
implying that the observed baryon asymmetry must come from a $mixture$
of decaying $B$-balls and free baryons.

This all assumes that the asymmetry not trapped in the $B$-balls
can survive down to temperatures below $T_{ew}$. If we
were to consider a $B-L$ conserving condensate or additional $L$
violating interactions in thermal equilibrium above $T_{ew}$, then
the only $B$ asymmetry which could survive anomalous $B+L$ violation
is the one associated with the $B$-balls. In this case $f_{B}$ would be
effectively equal to $1$ (we refer to this case as "pure" $B$-ball
baryogenesis (BBB)), so $m_{\chi}$ would have to be less than $67$~GeV.

A crucial assumption in all these is that there is effectively no
subsequent annihilation of LSPs coming from $B$-ball decays.

%%%%%%%%%%%%%%%%%%%%%%%%%%%%%%%%%%%%%%%%%%%%%%%%%%%%

\subsection{$B$-balls in gravity mediated supersymmetry breaking}

The attractive force due to logarithmic radiative correction term in
the condensate scalar potential is given by Eq.~(\ref{pot}). Of particular
interest is the $d=6$ $\bar u\bar d\bar d$ squark direction with a
non-renormalizable superpotential term of the form
$(\bar u\bar d\bar d)^{2}$ and the $d=4$ $\bar u\bar u\bar d\bar e$ direction,
which conserves $B-L$. The magnitude of $K$ is important for numerical
estimates. From the $1$-loop effective potential \cite{nilles84}, for the
$\bar u\bar d\bar d$ direction, the correction due to gauginos
with supersymmetry breaking masses $M_{susy}$ is given by (see earlier
discussion in Sect.~$5.3.1$.) \cite{enqvist98,enqvist99}
\be{e1a}
K \approx -\frac{1}{3} \sum_{\alpha, \; gauginos}
\frac{\alpha_{g_{\alpha}}}{8 \pi} \frac{M_{susy}^{2}}{m_{\phi}^{2}}~,
\ee
where the sum is over those gauginos which gain a mass from the
condensate scalar $\varphi$. The main contribution will come from the
three gluinos which gain masses from the squark expectation values.
For $\alpha_{g_{3}} \approx 0.1$ we obtain $|K| \approx 0.004 (M_{3}/m)^{2}$.
Depending on the ratio of the supersymmetry breaking gluino mass to the
squark mass, we expect $|K|$ to be typically in the range $0.01$ to $0.1$
(see discussion in Sect.~$5.3.1$.).

%%%%%%%%%%%%%%%%%%%%%%%%%%%%%%%%%%%%%%%%%%%%%%%%%%%%%%%%%%%%%

\subsubsection{$B$-ball Baryogenesis}

Recall that the perturbations of the AD condensate at some scale $\lambda$
go non-linear once $t \gae m_{s}^{-1}\sim m_{\phi}^{-1}$ (see Sect.~$5.10.1$),
causing the AD condensate to collapse into fragments of size
$\lambda$ and trapping inside a baryon  density. The fragments then relax into
the state of lowest energy $B$-balls with a charge of order $B$.
The charge of $B$-ball prevents the soliton from further collapsing
and hence the perturbations do not grow on length scales smaller than
the $B$-ball radius. Once the length scale going non-linear is larger
than the final $B$-ball radius, one expects $B$-balls to form quite
efficiently.

The time at which a perturbation of scale $\lambda$ goes non-linear is
then given by~\cite{enqvist99}
\be{tnl}
t \approx \frac{\alpha_{k}}{2 \pi} \left(\frac{2}{|K|}\right)^{1/2} \lambda~,
\ee
where
\be{alpha}
\alpha_{k} = \log\left(\frac{\phi_{i}}{\delta \phi_{i\;k}}\right)~.
\ee
One finds that $\alpha_{k} \approx 34 \; (44)$, for $d=4 (d=6)$
directions. In practice $B$-balls will typically turn out to have
thick walls with radius $R \approx (|K|^{1/2} m_{\phi})^{-1}$.
Perturbations on this scale which have the largest possible growth
in time $H^{-1}$ will go non-linear at
\be{tq}
t \approx \frac{10}{|K| m_{\phi}}   ~,
\ee
corresponding to $H \approx 0.1 |K| m_{\phi}$. Assuming
that the charge asymmetry corresponds to the presently observed
baryon asymmetry; $\eta_{B} \approx 10^{-10}$ prior to reheating,
then the baryon density is given by
\be{qasy}
n_{B} \approx \left( \frac{\eta_{B}}{2 \pi}\right)  \frac{H^{2}
 M_{\rm P}^{2}}{T_{R}} ~.
\ee
The charge contained inside a region within a radius
$R \approx (|K|^{1/2}m_{\phi})^{-1}$ is given by
\cite{enqvist98,enqvist9881,enqvist99}
\be{chasy}
B \approx 10^{15} |K|^{1/2}
\left(\frac{\eta_{B}}{10^{-10}}\right)
\left(\frac{10^{9}\GeV}{T_{R}}\right)
\left(\frac{100\GeV}{m_{\phi}}\right)~.
\ee
Hence with $|K| \gae 0.01$, we see that $B$-balls of charge larger
than $10^{14}$ are likely to form.

It is quite natural for the MSSM $B$-balls to decay after the electroweak
phase transition, see Eq.~(\ref{ttw}). It is therefore possible to
generate the observed $B$ asymmetry from the decay of $B$-balls occurring
at relatively low temperatures. This is true even if there are rapid
$L$ violating interactions or $B-L$ conservation. In a thermal bath $B$-balls
could lose their charges by the processes discussed in Sect.~$6.5$.
Therefore we naturally have some kind of constraint on the ambient thermal
temperature. Combining Eqs.~(\ref{chasy},\ref{evap-Qg}), one obtains
an upper bound on the final reheat temperature of the Universe
\cite{enqvist99}, which is given by
\begin{equation}
T_{\rm rh} \leq 10^{5} |K|^{1/2}\left(\frac{\eta_{B}}{10^{-10}}\right)
\left(\frac{1~{\rm TeV}}{m_{\phi}}\right)^{1/3}~{\rm GeV}\,.
\end{equation}
The reheating temperature therefore cannot be larger than $10^{5}$~GeV,
in order that the $B$-balls have enough charge left over for the purposes
of baryogenesis.

%%%%%%%%%%%%%%%%%%%%%%%%%%%%%%%%%%%%%%%%%%%%%%%%%%%%%%%%%

\subsubsection{LSP dark matter from $B$-ball decay}

$B$-ball formation from even dimensional $d=4,6$ operators is a good
candidate for generating dark matter in the Universe through $B$-ball
decay into LSPs \cite{enqvist98,enqvist99,fujii02,fujii02b,mcdonald01}.
When the $B$-balls decay there will be $N_\chi\gae 3$ LSPs produced
per baryon number or equivalently
\begin{equation}
\label{lsp1}
\Omega_{\chi}\approx 3f_{B}\left(\frac{m_{\chi}}{m_{n}}\right)
\Omega_{B}\,,
\end{equation}
where $f_{B}$ is the efficiency parameter which denotes the fraction of
the total charge stored in the $B$-ball, which ought to be less than one
in a realistic situation, and $m_{n}$ is the nucleon mass. Considering
the bounds on baryon number density from nucleosynthesis
$0.004 \leq \Omega_{B}h^2 \leq 0.013$~\cite{olive00333}, and the conservative
bound from CMB $0.004 \leq \Omega_{B}h^2\leq 0.023$,
one obtains the upper limits on the LSP mass~\cite{enqvist99}
\begin{equation}
\label{nmb}
m_{\chi}\leq (17.6-20.8) f_{B}^{-1}\left(\frac{\Omega_{\chi}}{0.4}\right)
\left(\frac{h}{0.8}\right)^2~{\rm GeV}\,.
\end{equation}
The direct experimental lower limit on the neutralino as
a LSP comes from ALEPH $m_{\chi}> 32.2$~GeV \cite{hutchcroft00}, which requires
$f_{B} \leq 0.64$. LEP and Tevatron constraints for the universal $A$-term
and gaugino masses lead to $m_{\chi}>46$~GeV, and for the universal scalar
masses $m_{\chi}>51$~GeV, which implies that $f_{B}<0.45$ and $f_{B}<0.41$,
respectively. This discourages the maximally charged condensate
hypothesis for a $B$-ball formation.

If one adopts Eq.~(\ref{lsp1}) as it stands, without taking into account of
possible LSP annihilations, then the amount of dark matter density would
be given by~\cite{enqvist99}
\begin{eqnarray}
 \left.
 \Omega_{\chi} \right|_{\rm no\,\,ann}
  &=&
  3\left(\frac{N_{\chi}}{3}\right)f_{B}\left(\frac{m_{\chi}}{m_n}\right)
    \Omega_B \nonumber\\
 &\gsim& 2.6\,f_{B}\times\left(\frac{N_{\chi}}{3}\right)
   \left(\frac{m_{\chi}}{100\GeV}\right)\left(\frac{0.7}{h}\right)^2
     \,,
\end{eqnarray}
where $m_n\simeq 1\GeV$ is the nucleon mass, and we have used the bound
$\Omega_B h^2\gsim 0.004$. In case of a bino-like LSP, the
$B$-ball formation could be a serious obstacle for the AD baryogenesis.

In this regard a $Q$-ball with a smaller charge would be beneficial.
One  option was considered in \cite{fujii0164}, where the
authors proposed a gauged $B-L$ symmetry to make $Q$-ball small enough
(for a discussion on $Q$-balls with a local gauge symmetry, see 6.2.6).
The gauged $U(1)_{B-L}$ was assumed to be broken at a scale
$\sim 10^{14}$~GeV. It was argued that in the gravity-mediated models
the D-term from $U(1)_{B-L}$ helps forming smaller $B$-balls from the
oscillations of the flat directions at weak scale. As noticed
by the authors~\cite{fujii0164}, their mechanism fails to ameliorate
the problem for the gauge-mediated models. Another solution has been
invoked in~\cite{allahverdi0265b}, where it was argued that if
one takes into account of the Hubble induced radiative
corrections to the flat direction then for a range of gaugino masses
$3H \lsim m_{1/2}\lsim 5H$, the amplitude of the AD condensate oscillations
is redshifted and leads to a formation of considerably smaller $Q$-balls
at low scales. The advantage is that the mechanism works for both
gravity and gauge mediated type AD potentials.

Otherwise, because of a large annihilation cross section, Wino and Higgsino
like LSP can be rather more promising candidates for CDM if ultimately
originating from an AD condensate, as pointed out by Fujii and Hamaguchi in
\cite{fujii02,fujii02b}. If the LSP is not very heavy, say
$m_{\chi} \lae 200\GeV$, so that their freeze-out temperature is
about $m_{\chi}/20\simeq 10$ GeV, then $B$-balls should decay at temperatures
below $10$~GeV for all the produced LSPs to survive. It is quite possible
that $B$-balls are not the only source of LSPs and that some of
the LSPs produced in $B$-ball decay will be annihilated with the LSPs in
the background. Using \eq{chasy}, one finds that if the $B$-ball decay
temperature $T_{d}\leq 10$~GeV, there is an upper limit on the
reheating temperature~\cite{enqvist99}
\be{d4}
T_{rh} \lae 5 \times 10^{8} \;
\frac{|K|^{3/2}}{f_{s}}
\left( \frac{100\GeV}{m_{\phi}} \right)^{2}
\left( \frac{40}{\alpha_{k}} \right)^{2}
\left( \frac{T_{d}}{10\GeV}
\right)^{2}\left(\frac{\eta_B}{10^{-10}}\right)
 \GeV
~.\ee
With $|K| \approx (0.01-0.1)$, and $\alpha_{k} \sim 40$,
$T_{rh} \lae (5 \times 10^{5}-2 \times 10^{7})f_{s}^{-1}\; \GeV$.

%%%%%%%%%%%%%%%%%%%%%%%%%%%%%%%%%%%%%%%%%%%%%%%%%%%%%%%%%%%%%%

\subsubsection{The LSP abundance}

The LSPs produced in $B$-ball decays will collide with themselves
and with other weakly interacting particles in the background and
settle locally into a kinetic equilibrium. Thermal contact can be
maintained until $T_f\sim m_{\chi}/20$~\cite{ellis84,jungman96}, and
a rough freeze-out condition for LSPs (if they were initially in thermal
equilibrium) will be given by \cite{jungman96}
\be{lspfreeze}
n_{\rm LSP}\langle \sigma_{\rm ann}v\rangle \approx H_f {m_\chi\over T_f}~,
\ee
where $\sigma_{\rm ann}$ is the LSP annihilation cross-section and
the subscript $f$ refers to the freeze-out values. The thermally
averaged cross section can be written as
$\langle\sigma_{\rm ann}v\rangle =a+bT/m_\chi$, where $a$ and $b$
depend on the couplings and the masses of the light fermions \cite{jungman96}.
For a light neutralino with $m_{\chi} < m_{W}$, neglecting the final
state fermion masses and assuming an efficient LSP production,
so that $f_{B} =1$, one finds for the LSP density from \eq{lspfreeze},
that $b \approx H m_{\chi}^{2} T_{f}^{-2} n_{f}^{-1}$, and
\be{eqdist}
n_f=\frac{1}{(2\pi)^{3/2}}(m_\chi T_f)^{3/2}e^{-m_\chi/T_f}\approx
1.46\times 10^{-12}m_\chi^3~.
\ee

The LSPs produced in $B$-ball decays will spread out by a random
walk with a rate $\nu$ determined by the collision frequency
divided by a thermal velocity $v_{th} \approx \sqrt{T/m_\chi}$. Since
the decay is spherically symmetric, it is very likely that the LSPs have
a Gaussian distribution around the central region of radius
as pointed out in~\cite{enqvist98}
\be{rcent}
\overline{r} \approx \left( \frac{\nu x}{\Gamma_{B}}\right)^{1/2}~
\ee
Within this central region annihilation is significant if
$\bar n_{\rm LSP}\langle \sigma_{\rm ann}v_{\rm rel}\rangle \gae H$.
In~\cite{enqvist98}, it was shown that the annihilation of LSPs is
insignificant provided the $Q$-ball decay temperature is given by
\be{noann}
T_d\ll 21 \left({m_\chi \over 100 \GeV}\right)^{3/16}\left(
{10^{20}\over N_{\rm LSP}^{tot}}\right)^{1/8}
\left(\frac{100}{g(T)}\right)^{3/16} ~\GeV~.
\ee
The main conclusion is that typically most of the LSPs will
survive if the $B$-ball decay temperature is less than a few GeVs.

The final abundance of the LSPs can be approximately expressed by a simple
analytical form \cite{fujii02,fujii02b}. Solving the Boltzmann equation
analytically for the LSP, one finds \cite{fujii02b}
\begin{eqnarray}
 Y_{\chi}(T)
  \simeq
  \left[
   \frac{1}{Y_{\chi}(T_d)}
   +
  \sqrt{
   \frac{8\pi^2 g_*(T_d)}{45}
   }
   \langle{\sigma v}\rangle
    M_{pl}
    (T_d - T)
   \right]^{-1}
   \,,
   \label{EQ-analytic-YT}
\end{eqnarray}
where $Y_{\chi}=n_\chi/s$.
In Fig.~(\ref{FIG-Boltzmann}), we present the numerical solution  of
Fujii and Hamaguchi for two cases: with and without large entropy
production from the $B$-balls. In the latter case $n_{Q}$ need not be
directly related to the present baryon asymmetry. This case can be
simulated by choosing a very small $\varepsilon$ and including the
radiation energy density generated by the $B$-ball decay in the
Boltzmann equation \cite{fujii02b}.

%%%%%%%%%%%%%%%%%%%%%%%%%%%%%%%%%%%%%%%%%%%%%%%%%%%%%%
\begin{figure}[h!]
\centering
\hspace*{-7mm}
\leavevmode\epsfysize=6cm \epsfbox{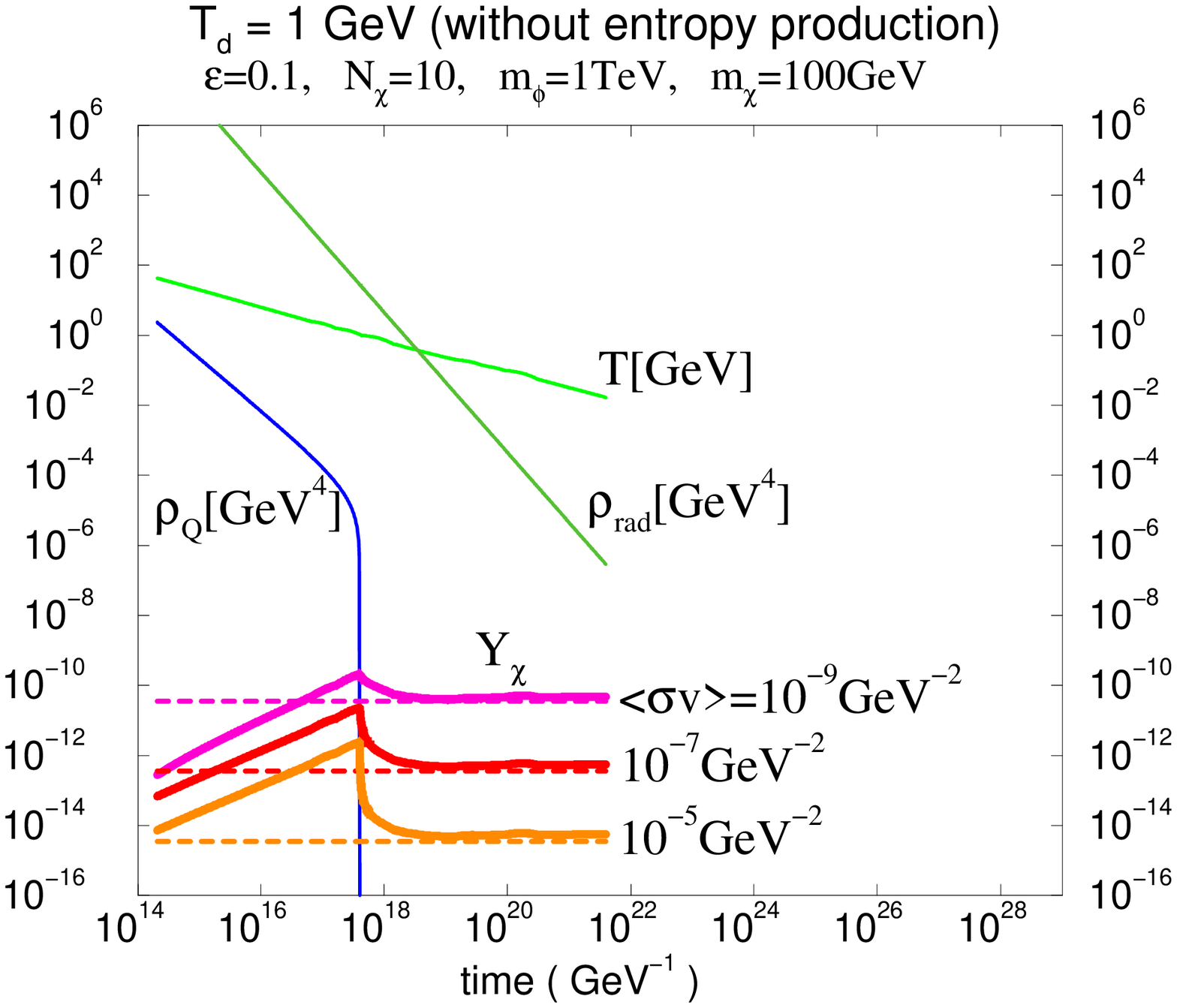}
\leavevmode\epsfysize=6cm \epsfbox{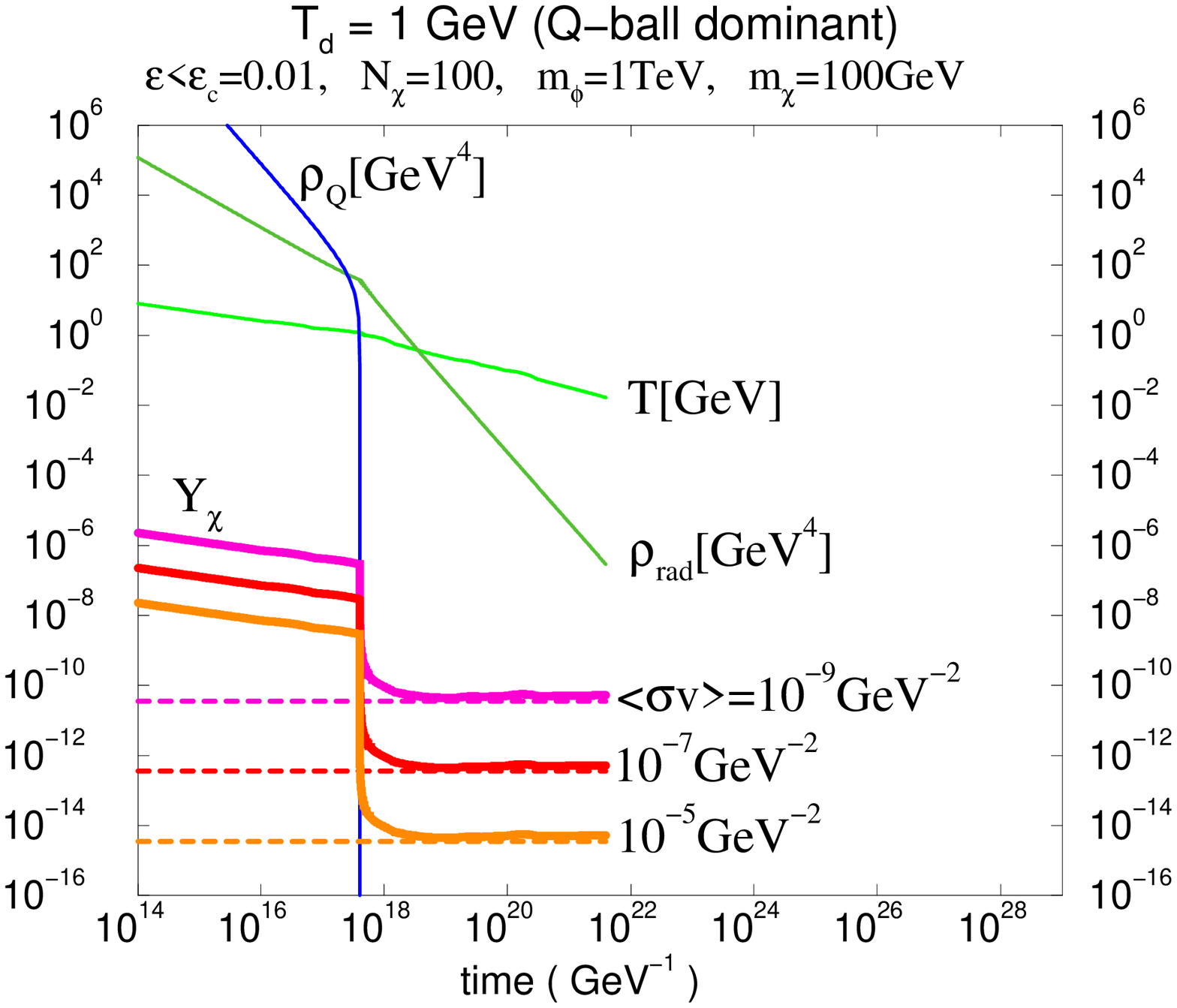}\\[2mm]
\caption{\label{FIG-Boltzmann}
\small The evolution of the neutralino dark matter abundance
generated from the $Q$-ball decay for $T_d = 1$~GeV and
$\langle{\sigma v}\rangle =10^{-9}$, $10^{-7}$, and $10^{-5}{\rm GeV}^{-2}$,
represented by thick solid lines. The abundances estimated from
the analytic formula in Eq.~(\ref{EQ-Ychi-analytic}) are shown in
dashed lines. In the first figure the energy density of the $Q$-balls
has been assumed to be small enough with respect
to radiation; the parameter values are $m_\varphi = 1$~TeV,
$m_\chi = 100$~GeV, $\epsilon = 0.1$ and $N_\chi=10$. In the
second plot $Q$-balls have been assumed to dominate the
energy density before their decay; here the parameters are
$m_\varphi = 1$~TeV, $m_\chi = 100$~GeV, $\varepsilon <0.01$
and $N_\chi= 100$, from \cite{fujii02b}.}
\end{figure}

%%%%%%%%%%%%%%%%%%%%%%%%%%%%%%%%%%%%%%%%%%%%%%%

If initial abundance $Y_{\chi}(T_d)$ is large enough, the final
abundance $Y_{\chi 0}$ for $T \ll T_d$ is given by \cite{fujii02,fujii02b}
\begin{eqnarray}
 Y_{\chi 0}
  \simeq Y_{\chi}^{\rm approx}\equiv
  \left[
   \sqrt{
   \frac{8\pi^2 g_*(T_d)}{45}
   }
   \langle{\sigma v}\rangle
   M_{pl}
   T_d
   \right]^{-1}
   \,.
   \label{EQ-Ychi-analytic}
\end{eqnarray}
In this case the final abundance $Y_{\chi 0}$ is determined only by the
$Q$-ball decay temperature $T_d$ and the annihilation cross section of
the LSP $\langle{\sigma v}\rangle$ (independently of the initial value
$Y_{\chi}(T_d)$ as long as $Y_{\chi}(T_d)\gg Y_{\chi}^{\rm approx}$).
In terms of the density parameter $\Omega_\chi$, the neutralino abundance
can be rewritten as \cite{fujii02b}
\begin{eqnarray}
 \Omega_{\chi}
 &\simeq&
  0.5
  \left(\frac{0.7}{h}\right)^2
  \times
  \left(
   \frac{m_{\chi}}{100 {\rm GeV}}
   \right)
   \left(
    \frac{10^{-7}{\rm GeV}^2}{\langle{\sigma v}\rangle}
    \right)
    \times
    \left(
     \frac{100 {\rm MeV}}{T_d}
     \right)
     \left(
      \frac{10}{g_*(T_d)}
      \right)^{1/2}
      \,.
      \label{Omega-ana}
\end{eqnarray}

In case of $Y_{\chi}(T_{d})<Y_{\chi}^{\rm approx}$, the final
abundance is given by
\begin{equation}
Y_{\chi 0}
\simeq Y_{\chi}(T_{d})\gsim \varepsilon^{-1}\left(\frac{n_{B}}{s}
\right)_{0}\,,
\end{equation}
This is the case where the LSP annihilation cross section is small
enough, which holds  for a bino-like LSP, where the relic
abundance of the LSPs is directly related to the observed baryon
asymmetry~\cite{enqvist98}.
Unfortunately, a bino-like neutralino will overclose
the Universe unless we assume an extremely light bino (which is
experimentally excluded).

%%%%%%%%%%%%%%%%%%%%%%%%%%%%%%%%%%%%%%%%%%%%%%%%%%%%%%%%%%%%%%

\subsubsection{Which direction?}

Consider the $\bar u\bar d\bar d$ and
$\bar u\bar u\bar d\bar d\bar e$ directions, which are both
lifted by $d=4$ and $d=6$ non-renormalizable operators.
For $d=6$ case the initial value of the field when the condensate oscillations
begin is given by
$\varphi_{0}=5.8\times 10^{14}\lambda^{-1/4}\left(m_{\phi}/100~{\rm GeV}\right)^{1/4}$~GeV,
and for $d=4$ case
$\varphi_{0}=3.2\times 10^{10}\lambda^{-1/2}\left(m_{\phi}/100~{\rm GeV}\right)^{1/2}$~GeV. Let us compare the lower bound on the reheating
temperature for the two cases. For $d=6$, it has been estimated
\cite{enqvist99} to be
$T_{rh}\gae 0.23 \; \lambda^{1/2} \left(m_{\phi}/100~{\rm GeV}\right)^{1/2}$
GeV (for $d=6$, $\lambda \sim 0.003$ if the strength of the
non-renormalizable interactions is set by $M_{\rm P}$). Such bound
on $T_{rh}$ fulfills all the requirements for the survival of
$B$-balls from thermal dissociation, diffusion, etc. Repeating the same
analysis for the $d=4$ case leads to an upper bound
$T_{rh} \gae 8 \times 10^{7} \lambda $~GeV (with $\lambda \sim 0.1$).
This is rather hard to satisfy. The robust conclusion appears
to be that for an efficient $B$-ball production which also gives rise to
dark matter, should involve the $d=6$ $\bar u\bar d\bar d$
direction. This particular direction is also favored in the sense
that $f_{s} = 1$ can be possible. As long as the reheating temperature
does not exceed $10^{3-5}\GeV$, the baryon to dark matter ratio can be
naturally accounted by $B$-ball decays in case of the MSSM flat
directions which are lifted by $d=6$ operators~\cite{enqvist98}.

In the case of the $d=4$ directions $B$-balls do not form
efficiently~\cite{enqvist99}. For $d=4$ and along the
$\bar u\bar u\bar d\bar e$ direction $B-L$ is actually conserved.
In this case only the baryon number trapped in the $B$-balls will
survive. Even though it might appear that the observed $B$ asymmetry
could be obtained if the initial condensate had a large $B$, this would
be a rather ambitious program because $\bar u\bar u\bar d\bar e$-balls
disappear rather quickly in a thermal bath as argued in~\cite{enqvist98}.

%%%%%%%%%%%%%%%%%%%%%%%%%%%%%%%%%%%%%%%%%%%%%%%%%%%%%%%%%%%

\subsubsection{Direct LSP searches and $B$-balls}

If dark matter is a neutralino, then it should be possible to confirm
its existence by direct and indirect searches. The direct detection
involves the interaction of neutralino with matter, which is
usually dominated by scalar couplings of relatively heavy nuclei
$A\gsim 20$~\cite{bednyakov94,jungman96}. The counting rate of the
elastic neutralino-nucleon scattering is given by \cite{drees93}
\begin{equation}
R=\left(\frac{\sigma \xi}{m_{\chi}m_{N}}\right)\left(\frac{
1.8\times 10^{11}{\rm GeV}^4}{{\rm Kg}\cdot {\rm days}}\right)
\left(\frac{\rho_{\chi}}{0.3 {\rm GeV/cm^3}}\right)\left(
\frac{v_{\chi}}{320~{\rm Km/s}}\right) ~{\rm events}\,,
\end{equation}
where $\rho_{\chi}$ is the density, $v_{\chi}$ the average
velocity of the neutralino, $m_{n}$ the mass of the target
nucleon, and $\xi$ a nuclear form factor. The cross section
$\sigma$ is the neutralino-nucleon cross-section at zero momentum
transfer. These interactions are mediated by heavy Higgs exchanges,
or by a sfermion exchange. The forthcoming experiments, such as
CDMS~\cite{CDMS}, CRESST~\cite{CRESST}, EDELWEISS II~\cite{EDELWEISS},
GENIUS~\cite{GENIUS} and ZEPLIN~\cite{ZEPLIN}, have high hopes
in reaching cross-sections up to $10^{-10}$~pb.

Indirect detection relies on astrophysics. The annihilation of
neutralino LSPs in any astrophysical sources can give rise to
fluxes of anti-protons and positrons which are usually not seen in the
cosmic rays. Other possible ways are neutralino annihilation into
$2\gamma$ final states or $Z\gamma$ final states \cite{bergstrom97}.
The monoenergetic gamma rays with energy $\sim m_{\chi}$ might not have
any competitive background from other astrophysical sources, and has a
possibility of being detected in the next generation air Cherenkov
telescopes observing the galactic center, such as VERITAS \cite{veritas},
HESS \cite{hess} and MAGIC \cite{magic}.

Fujii and Hamaguchi \cite{fujii02b}, have studied in detail
the parameter space where neutralino production from the decays of
$B$-balls gives the correct dark matter abundance. The allowed region
is found where the dominant contribution to the LSP is provided by
$\widetilde H$. A large $\widetilde H$ content of the LSP enhances the
neutralino annihilation cross section into $W$ bosons via chargino
exchange. There is a small difference in thermally produced $\widetilde H$
and non-thermally produced ones. In the latter case the mass of the
neutralino is much smaller than thermally produced ones. The annihilation
rate of neutralinos into $2\gamma$ final states is enhanced for
$\widetilde{H}$-like neutralino (this was previously shown in the context
of thermally distributed neutralinos in \cite{bergstrom97,bergstrom98}),
and seems to hold true even when the neutralinos are created
non-thermally \cite{fujii02,fujii02b}. A typical $B$-ball decay temperature
which leads to the desired CDM density
is $100~{\rm MeV} \lsim T_{d}\lsim ({\mbox{a few}})~{\rm GeV}$.
In the anomaly mediation and in the no-scale supersymmetry breaking
models, it was found that $\widetilde W$ is the most promising candidate for
the LSP in a wide region of the parameter space \cite{fujii02b}.
The conclusion is that if AD baryogenesis is successful within minimal
SUGRA, then Higgsino and Wino like neutralinos are perhaps the likely
dark matter candidates~\cite{fujii02b}.

%%%%%%%%%%%%%%%%%%%%%%%%%%%%%%%%%%%%%%%%%%%%%%%%%%%%%%%%%%%

\subsection{$Q$-balls and gauge mediated supersymmetry breaking }

As pointed out by Kusenko and Shaposhnikov, in gauge mediated
supersymmetry breaking scenarios the salient feature is that the
$Q$-balls are stable against decaying into nucleons~\cite{kusenko98418},
because the $Q$-ball energy per unit charge is given by
$E_{Q}/Q\simeq m_{\phi}Q^{-1/4}<1$ GeV for $m_{\phi}\sim 1$~TeV.
For sufficiently large $Q$, the $Q$-ball itself could then be a candidate
for CDM \cite{dvali98,kusenko98418,demir99,kasuya01,hisano01}.

%%%%%%%%%%%%%%%%%%%%%%%%%%%%%%%%%%%%%%%%%%%%%%%%%%%%%%

\subsubsection{Baryogenesis and gauge mediation}

A sufficiently large $Q$-ball in the gauge mediated supersymmetry
breaking case will be absolutely stable against decaying into nucleons.
Although baryogenesis by evaporation of a $Q$-ball is not
effective, $Q$-balls nevertheless loose some charge to the ambient
plasma by thermal dissociation and diffusion, which might create asymmetry
in the nucleons. An ambitious possibility, but one that would be hard
to realize, is to work out whether $Q$-ball evaporation could also lead
to dark matter as in the case of gravity mediated supersymmetry breaking.

In the gauge mediated case the baryon asymmetry can be related to the dark
matter density via
\begin{equation}
\label{bdm0}
\eta_{B}=\frac{n_{B}}{n_{\gamma}}\simeq \frac{\varepsilon n_{Q} \Delta Q}
{n_{\gamma}}\simeq \frac{\varepsilon \rho_{Q}\Delta Q}{n_{\gamma}M_{Q}}
\simeq \frac{\varepsilon\rho_{c,0} \Omega_Q \Delta Q}
                   {n_{\gamma,0}M_Q}\,,
\end{equation}
where $\varepsilon\ll 1$ displays the departure from a circular orbit,
$\Delta Q$ is the evaporated charge,
$\Omega_{Q}$ is the density parameter for the $Q$-balls,
$\rho_{c,0} \sim 8h_0^2\times 10^{-47} {\rm GeV}^4$ is the present critical
density, and $n_{\gamma,0} \sim 3.3 \times 10^{-39} {\rm GeV}^3$ is the
present photon number density, with $h_0\sim 0.7$. The evaporated charge
should yield the baryons while the remaining charge is in $Q$-ball
dark matter. The baryon-to-dark matter ratio is given by~\cite{kasuya01}
\begin{equation}
    \label{dQ-Q-ratio}
    r_B \equiv \frac{\Delta Q}{Q}
      \sim \eta_B \frac{m_{\phi}n_{\gamma,0}}
           {\varepsilon\rho_{c,0}\Omega_Q} Q^{-1/4}
      \sim 10^{11} \varepsilon^{-1} \eta_B \Omega_Q^{-1}
            \left(\frac{m_{\phi}}{\rm TeV}\right) Q^{-1/4}.
\end{equation}
The total evaporated charge from $Q$-balls
in a thermal bath is given by Eq.~(\ref{evap-Q}). Requiring
that $Q>\Delta Q$, the charge of the $Q$-ball should be sufficiently
large~\cite{kasuya01}:
\begin{equation}
\label{Qest2}
Q\geq 1.2 \times 10^{8}\eta_{B}^{-3/2}\varepsilon^{3/2}\Omega_Q^{3/2}
\left(\frac{m_{\phi}}{\rm TeV}\right)^{-3}\,.
\end{equation}
The value of $\varepsilon$ is not an independent quantity and can
be related to the charge of a $Q$-ball. At the beginning of the condensate
rotation $H\sim m_{\phi}^2(T)/\varphi_0 \sim T^2/\varphi_0$, where
$T\sim (M T_{rh}^2 H)^{1/4}$ is the instantaneous temperature prior
to reheating. Combining these two pieces of information, one obtains
$T\sim T_{rh}\sqrt{M/\varphi_{0}}$. The net charge of
the $Q$-ball is then related to the baryon number density through
$n_{B}\sim r_B\varepsilon n_{\varphi}\sim r_{B}\varepsilon\omega\varphi_0^2$,
where $\omega\sim m^2_{\phi}(T)/\varphi_{0}\sim T^2/\varphi_{0}$ is
the rotation frequency of the condensate. With the help of
Eq.~(\ref{dQ-Q-ratio}), one finds~\cite{kasuya01}
\begin{equation}
    \label{n-b}
    n_{B} \sim  10^{11} \eta_B \Omega_Q^{-1}
      \left(\frac{m_{\phi}}{\rm TeV}\right) T^2 \varphi_0 Q^{-1/4},
\end{equation}
Kasuya and Kawasaki also estimated the net charge of the $Q$-ball, given by
the initial amplitude of the condensate and the reheating temperature, as
\cite{kasuya01}
\begin{equation}
    \label{Q-phi-Trh}
    Q \sim 10^{44}
      \left(\frac{m_{\phi}}{\rm TeV}\right)^4
      \frac{\varphi_0^{16}}{T_{rh}^4M^{12}} \Omega_Q^{-4}.
\end{equation}
From Eq.~(\ref{Qest0}) one obtains
$Q \sim \beta \left(\varphi_0/T\right)^4$
for $\varepsilon \sim 1$ (for $\varepsilon \ll 1$, one replaces
$\beta$ by $\beta'=\gamma\beta$ with $\gamma\sim 0.1$).

From all these considerations one can estimate the amplitude of the
condensate and the charge of the $Q$-ball. The amplitude is given by
\cite{kasuya01}
\begin{eqnarray}
    \label{init-amp}
    \phi_0 \sim  4.6\times 10^{13}
    \varepsilon^{1/10} \Omega_Q^{2/5}\times
    \left(\frac{\beta}{6\times 10^{-4}}\right)^{1/10}
    \left(\frac{m_{\phi}}{\rm TeV}\right)^{-2/5} {\rm GeV}.
\end{eqnarray}
Inserting this in the expression $T\sim T_{rh}\sqrt{M/\varphi_{0}}$,
one obtains~\cite{kasuya01}
\begin{eqnarray}
    T \sim 2.3\times 10^7 \varepsilon^{-1/20} \Omega_Q^{-1/5}
    \left(\frac{T_{rh}}{10^5 {\rm GeV}}\right) \times
    \left(\frac{\beta}{6\times 10^{-4}}\right)^{-1/20}
    \left(\frac{m_{\phi}}{\rm TeV}\right)^{1/5} {\rm GeV}\,,
\end{eqnarray}
so that  the charge of the $Q$-ball reads~\cite{kasuya01}
\begin{eqnarray}
    \label{Q-limit}
    Q \sim 9.3 \times 10^{21} \varepsilon^{8/5} \Omega_Q^{12/5}
    \left(\frac{T_{rh}}{10^5 {\rm GeV}}\right)^{-4}\times
    \left(\frac{\beta}{6\times 10^{-4}}\right)^{8/5}
    \left(\frac{m_{\phi}}{\rm TeV}\right)^{-12/5}\,.
\end{eqnarray}
Now, with the help of Eqs.~(\ref{Qest2}) and (\ref{Q-limit}), one
obtains~\cite{kasuya01}
\begin{equation}
\label{sum-eps}
\varepsilon \sim 1.5 \times 10^{-2} \Omega_Q^{3/5}
    \left(\frac{\eta_B}{10^{-10}}\right)
    \left(\frac{T_{rh}}{10^5 {\rm GeV}}\right)^{-8/3}\times
    \left(\frac{\beta}{6\times 10^{-5}}\right)^{16/15}
    \left(\frac{m_{\phi}}{\rm TeV}\right)^{2/5}.
\end{equation}
The above equations determine the parameter space for $Q$-ball baryogenesis
to coexist with $Q$-ball dark matter in gauge mediated models.

%%%%%%%%%%%%%%%%%%%%%%%%%%%%%%%%%%%%%%%%%%%%%%%%%%%%%%%

\subsubsection{Generic gauge mediated models}

Now the task is to consider the parameter space for a realistic MSSM
flat direction.  A detailed analysis can be found in
\cite{kasuya01}, where it was pointed out that the $d=5,6$ directions are
favorable. For $d=4$, the charge does not accumulate enough to survive as
a dark matter relic while $d=7$ requires unnaturally small
values of $\varepsilon$. It is worth pointing out that as long as
the ambient temperature of the plasma is sufficiently high with
$T\geq (m_{3/2}\varphi_{0})^{1/2}\left[\log(\varphi_{0}^2/T^2)\right]^{-1/4}$,
the logarithmic term dominates over the gravity-mediation term.
For $d=5,6$, the requirement on temperature is thus $T\geq 10^{5-6}$~GeV.

On the other hand, if the $Q$-ball forming scalar field has a large vev,
or if the reheating temperature is extremely low, then gravity mediation
effects should also be taken into account. However,
as kasuya and Kawasaki pointed out in \cite{kasuya01},  in order
to produce enough baryon asymmetry and simultaneously produce surviving
$Q$-balls for dark matter, one requires a considerably large gravitino mass
$m_{3/2} \geq 10^{3}$~GeV, which is an
unacceptable value for the gravitino mass within the gauge mediated
supersymmetry breaking~\cite{giudice98}.

One could also consider a generic model for gauge mediation where the scale
of the logarithmic potential is larger than $m_{\phi}$, such that
\cite{kasuya00,kasuya01}
\begin{equation}
    \label{pot-app}
    V \sim \left\{
      \begin{array}{ll}
          \ds{M_F^4 \log \left(\frac{\varphi^2}{M_S^2}\right)}
          & (\varphi \gg M_S), \\
          m_{\phi}^2 \varphi^2 & (\varphi \ll M_S), \\
      \end{array}
      \right.
\end{equation}
where $M_S$ is the messenger mass scale. In this particular case the
condensate will start oscillating at large field amplitudes. The
$Q$-ball will form at a large vev. The mass and
the size of the $Q$-ball are now given by \cite{kasuya00,kasuya01}
\begin{equation}
    M_Q \sim M_F Q^{3/4}, \quad
    R \sim M_F^{-1}Q^{1/4}, \quad
    \omega \sim \frac{M_F^2}{\varphi},.
\end{equation}
If $M_F \gsim T$, then the $Q$-ball could be stable against decaying
into the nucleons, provided the $Q$-ball mass per unit charge is smaller
than $1$~GeV. This condition holds for
\begin{equation}
    \label{stable-app}
    Q \gsim 10^{24} \left(\frac{M_F}{10^6 {\rm GeV}}\right)^4.
\end{equation}
The ambient temperature is smaller than $M_{F}$, but could be larger
than $m_{\phi}$. In this case the $Q$-ball would rather
evaporate. The total evaporated charge can be estimated to be \cite{kasuya01}
\begin{equation}
\label{bdm1}
\Delta Q\sim 10^{15}\left(\frac{m_{\phi}}{10^{2}~{\rm GeV}}\right)^{-2/3}
\left(\frac{M_{F}}{10^{6}~{\rm GeV}}\right)^{-1/3}\times Q^{1/12}\,.
\end{equation}
The survival condition: $Q\geq \Delta Q$ then imposes a bound on an
initially accumulated charge to be
\begin{equation}
Q\geq 10^{17}\left(\frac{m_{\phi}}{10^{2}~{\rm GeV}}\right)^{-8/11}
\left(\frac{M_{F}}{10^{6}~{\rm GeV}}\right)^{-4/11}\,.
\end{equation}
While combining with Eq.~(\ref{bdm0}), one obtains a relationship between the
baryon number and the amount of dark matter as \cite{kasuya01}
\begin{equation}
\label{bdm2}
Q\sim 10^{17}\Omega_{Q}^{3/2}\varepsilon^{3/2}\left(\frac{\eta_{B}}{10^{-10}}
\right)^{-3/2}\left(\frac{m_{\phi}}{10^{2}~{\rm GeV}}\right)^{-1}
\left(\frac{M_{F}}{10^{6}~{\rm GeV}}\right)^{-2}\,.
\end{equation}
The consistency condition requires the initial amplitude of the AD
condensate to be $\varphi_{0}\sim 10^{11}$~GeV. Note that when
all the above conditions are taken into account, then there is
hardly any region in the parameter space which allows for the required
baryon asymmetry from evaporation of $Q$-balls together with enough
surviving $Q$-balls to provide the dark matter, as was concluded by
Kasuya and Kawasaki in~\cite{kasuya01}.

%%%%%%%%%%%%%%%%%%%%%%%%%%%%%%%%%%%%%%%%%%%%%%%%%%%%%%%%%%%%%%%

\subsubsection{Late formation of gauged $Q$-balls}

In~\cite{kasuya01}, the authors have also considered the late
formation of $Q$-ball at a scale when gravity mediation and thermal
logarithmic correction terms dominate the potential at large and small
scales, respectively. The $Q$-ball forms when the instability band
enters the horizon with a wavelength $k_{eq}^{-1}$ and the angular velocity
$\omega_{eq}\sim (T^2_{eq}/\varphi_{eq})^{-1}\sim m_{3/2}^{-1}$.
The number density of the $Q$-balls
is found to be $n_{eq}\sim  T^4_{eq}/m_{3/2}$, while the charges are
$Q\sim (\varphi/T_{eq})^4$~\cite{kasuya01}. The exact relationship
depends on the helicity of the condensate. In this late formation scenario
the only realistic flat directions turn out to be $d=6$ with
$Q\sim 10^{24}-10^{21}$, $T_{rh}\sim 1.0\times 10^{7}-30$~GeV,
and $M_{F}\sim 10^{2}-10^{4}$~GeV; and $d=7$ with $Q\sim 10^{25}-10^{22}$,
$T_{rh}\sim 60~{\rm GeV}-10~{\rm MeV}$, and $M_{F}\sim 10^{2}-10^{3}$~GeV
\cite{kasuya01}.

Kasuya and Kawasaki also repeated their analysis when temperature effects
are negligible. In this case the $Q$-ball forms when
$V\sim M_F^4\log\left(\varphi_{eq}^2/M_s^2\right)\sim m_{3/2}^2\varphi_{eq}^2$,
where $\varphi_{eq} \sim M_{F}^2/m_{3/2}$. The AD condensate fragment just
after its amplitude becomes smaller than $\varphi_{eq}$. Consistent
scenarios arise only in $d=6,7,8$ cases. For the allowed parameter
space, one should require $m_{3/2} \sim 0.1$~GeV, $M_{F}\sim 10^{4}$~GeV,
$T_{rh}\sim 5$~GeV, and $Q\sim 10^{20}$.

These many  proposals indicate very clearly that in order to find
a consistent cosmological scenario where $Q$-ball evaporation leads to
baryon asymmetry while the survived $Q$-balls act as a dark matter
candidate requires a stringent condition on $M_{F}\le 10^{6}$~GeV.
Note that $M_{F}\sim 10^{6}$~GeV is required to provide the
right spectrum for sparticle masses, see~\cite{dine96a,giudice98}.

%%%%%%%%%%%%%%%%%%%%%%%%%%%%%%%%%%%%%%%%%%%%%%%%%%%%%%%%%%%%%%%%%%%

\subsection{$Q$-balls as self-interacting dark matter}

Recently, $Q$-ball has been proposed as a candidate for
self-interacting dark matter by Kusenko and Steinhardt \cite{kusenko01}.
Such a consideration is motivated by the fact that collisionless CDM
appears to have certain discrepancies between numerical simulations
and observations. The halo density profiles, and the number density
of satellite galaxies, do not match well with the observations
\cite{moore94,spergel00}. A possible remedy is dark matter that has
fairly strong self-interactions \cite{spergel00}, a situation which
however is not easily achieved in the standard particle physics models.

The self-interaction cross section and the mass of the dark matter
particles undergoing elastic scattering should satisfy the relation
$s=\sigma_{DD}/m_{DM} \sim 2\times 10^{3}-3\times 10^{4}~{\rm GeV}^{-3}$.
This might change if one considers other processes such as
dark matter annihilation.

$Q$-balls, being extended objects, can certainly have a large
geometric cross-section, but it was found that in order to match
the required cross-section to mass ratio $s$, the AD particles
should have a very low mass scale $\sim {\cal O}(1)$~MeV. This
requires that the charge of the $Q$-ball should also be very small,
i.e. $Q\leq 10^{5}$. Such $Q$-balls would not originate within
MSSM but could be possible in some extended theories. There are
however a number of issues concerning production of such a
small charged $Q$-balls, their thermal distribution, their evaporation,
annihilation and scattering should be taken into account consistently.
It has been argued that thick wall $Q$-balls do not
seem to have much admissible parameter space~\cite{enqvist02},
while thin wall $Q$-balls have a slightly better chance to succeed with a
vev $\varphi_{0}\geq {\cal O}({\rm MeV})$. In any case, such a
small AD condensate amplitude seems hard to obtain, although not
completely impossible, in order to reconcile with the existing
particle physics models.

Recently, it has been pointed out~\cite{gelmini02} that future
experiments should be able to discern the spatial extent of the
dark-matter particle. It was noticed that the extended
objects such as $Q$-balls leave its distinct imprint on the spectrum
which falls off very fast with increasing energy. The signal is
primarily dominated by the low-energy events near the threshold.

%%%%%%%%%%%%%%%%%%%%%%%%%%%%%%%%%%%%%%%%%%%%%%%%%%%%%%%%

\subsection{Direct searches for gauge mediated $Q$-balls}

If the initial charge of a $Q$-ball is larger than the
evaporated charge, the $Q$-ball survives and contributes to
the energy density of the Universe. Let us take as
an example $B$-balls, such as $\bar u_{2}\bar d_{1}\bar d_{2}$, which
in gauge mediated case are unstable because of the
presence of baryon number violating operators (required in the
first place to charge up the condensate) with dimensions
larger than $5$ \cite{hisano01}. Nevertheless
for $Q\geq 10^{20}$, the lifetime of the $Q$-ball is in fact greater than
the age of the Universe ($t_{0}\sim 10^{10}$~years)~\cite{hisano01}.
Such $Q$-balls are potential candidates for CDM.

From Eq.(\ref{evap-Q}), we can read the limit on the initial charge of
a $Q$-ball
\begin{equation}
   \label{survive}
   Q_{{\rm init}} \geq 7.4\times 10^{17}
          \left( \frac{m_{\phi}}{{\rm TeV}} \right)^{-12/11}\,.
\end{equation}
On the other hand, in order for the $Q$-ball to be stable against
decaying into nucleons, i.e. $E_Q/Q\leq 1$~GeV, one finds~\cite{kasuya01}
\begin{equation}
   \label{stable}
   Q \geq 10^{12} \left( \frac{m_{\phi}}{{\rm TeV}} \right)^4 \,.
\end{equation}
The condition relating baryon number and dark matter is given by
Eq.(\ref{Qest2}), and can be written as
\begin{equation}
    \label{dm}
     Q \leq 10^{23} \varepsilon^{3/2}
              \left( \frac{m_{\phi}}{{\rm TeV}} \right)^{-3}\,,
\end{equation}
where we have taken $\eta_B \sim 10^{-10}$ and $\Omega_Q \leq 1$.

If we do not impose the condition that the evaporated charge
accounts for the baryons in the Universe, the only constraint is
that the energy density of $Q$-balls must not exceed the critical
density. As mentioned in the previous Section, this condition is
given by Eq.(\ref{Q-limit}).

Let us now assume that the gauge mediated $Q$-balls indeed make
up the dark matter in the galactic halo. Then the corresponding number
density is given by~\cite{kusenko9880}
\begin{equation}
n_{Q} \sim \frac{\rho_{dm}}{M_{Q}}\sim 5\times 10^{-5}Q_{B}^{-3/4}
\left(\frac{1~{\rm TeV}}{m_{\phi}}\right)~{\rm cm}^{-3}\,,
\end{equation}
and the $Q$-ball flux will be determined by
${\cal F} \sim (1/4\pi)n_{Q}v\sim 10^{2}Q_{B}^{-3/4}(1~{\rm TeV}/m_{\phi})$
~${\rm cm}^{-2}~{\rm s}^{-1}~{\rm sr}^{-1}$, where we have taken
$v\sim 10^{-3}$c. The number of events will obviously depend on
the surface area of the detector. In case of Super-Kamiokande
the number of events could be estimated to be
$N \sim (10^{24}/Q_{B})^{3/4} (1~{\rm TeV}/m_{\phi})~{\rm yr}^{-1}$
\cite{kusenko9880}, where the surface area of the water tank is
$\sim 7.5 \times 10^{7}~{\rm cm}^2$~\cite{kamiokanda}.

The $Q$-balls can be detected via
Kusenko-Kuzmin-Shaposhnikov-Tinyakov (KKST) process \cite{kusenko9880}.
Note that in the absence of any fundamental singlet a $B$-ball would
necessarily break $SU(3)$ while the electroweak symmetry could be restored
inside, provided the Higgses do not obtain vevs or if the flat direction
is already a $SU(2)$ singlet such as $\bar u\bar d\bar d$.
The instanton mediated baryon number violation will be much weaker in
these cases because of the limitation on the size of the instanton,
which cannot be larger than the size of the $Q$-ball. When nucleons
collide with a $Q$-ball, they enter the surface layer of the $Q$ ball,
and dissociate into quarks, which are converted into squarks via gluino
exchange. In this process, $Q$-balls release $\sim 1$~GeV energy per
collision by emitting soft pions.

For an electrically neutral $Q$-ball the absorption cross-section is
quite large and  is determined by the soliton size to be~\cite{kusenko9880}
$\sigma_{B}\sim 10^{-33}Q_{B}^{1/2}(1{\rm TeV}/m_{\phi})^2~{\rm cm}^2$.
The estimated mean free path in matter is given by
$\lambda\sim 10^{-3}A(10^{24}/Q_B)^{1/2}(m_\phi/1~{\rm TeV})(1~{g/cm^3}/\rho)$
cm \cite{kusenko9880,arafune00},
where $A$ is the weight of the atomic nucleus with a density distribution
$\rho$. Obviously absorption of quarks takes place at a higher rate than the
collisions of $Q$-balls with nuclei. The BAIKAL experiment~\cite{baikal}
sets a limit on the monopole flux which also gives a lower bound on the charge
of a $Q$-ball as $Q_{B}\geq 10^{22}$ for $m_{\phi}\sim 1$~TeV.

If the $Q$-balls are electrically charged, for instance when the
selectrons obtain a large vev along the $QQQLLL\bar e$ flat direction,
the detection prospects worsen, because of two reasons. First, there
will be a Coulomb barrier which will prevent absorption of the incoming
nuclei. Secondly, the absorption cross section will be determined by the Bohr
radius $\sigma \sim \pi r_{B}^2\sim 10^{-16}~{\rm cm}^{-2}$, and therefore
the corresponding mean free path length will decrease to
$\lambda \sim 10^{-8}A(1~{\rm g}/{\rm cm^3}/\rho)$~cm \cite{kusenko9880}.
The present limit on electrically charged $Q$-ball comes from the
MACRO search \cite{ahlen92} with a flux
${\cal F}\leq 1.1\times 10^{-14}~{\rm cm}^{-2}~{\rm s}^{-1}~{\rm sr}^{-1}$,
which constraints the charge of a $B$-ball to $Q_{B}\geq 10^{21}$.

A wide range of charges are already ruled out, see Fig.~(\ref{z0com})
from~\cite{arafune00}. The only allowed range of charges is
$Q\sim 10^{22}-10^{26}$ for $m_{\phi}\sim 1~{\rm TeV}-100~{\rm GeV}$.
The range however depends on $\varepsilon$, which one tacitly assumes to
be of order one in Eq.~(\ref{dm}). For smaller values of $\varepsilon$
the range  of allowed charges will further reduce. In a generic logarithmic
potential a charge $Q\sim 10^{24}$ and $M_{F}\sim 10^{2}$~GeV is still allowed
\cite{kasuya01}. In many cases such as in delayed $Q$-ball formation
\cite{kasuya01}, the charge of a $Q$-ball is large $\sim 10^{26}$. Such
$Q$-balls could be detectable in the Telescope Array Project or
the OWL-AIRWATCH detector.

%%%%%%%%%%%%%%%%%%%%%%%%%%%%%%%%%%%%%%%%%%%%%%%%%%%%%%%%%%
\begin{figure}
\setlength{\epsfxsize}{4in}
\epsfclipon
\centerline{\epsfbox{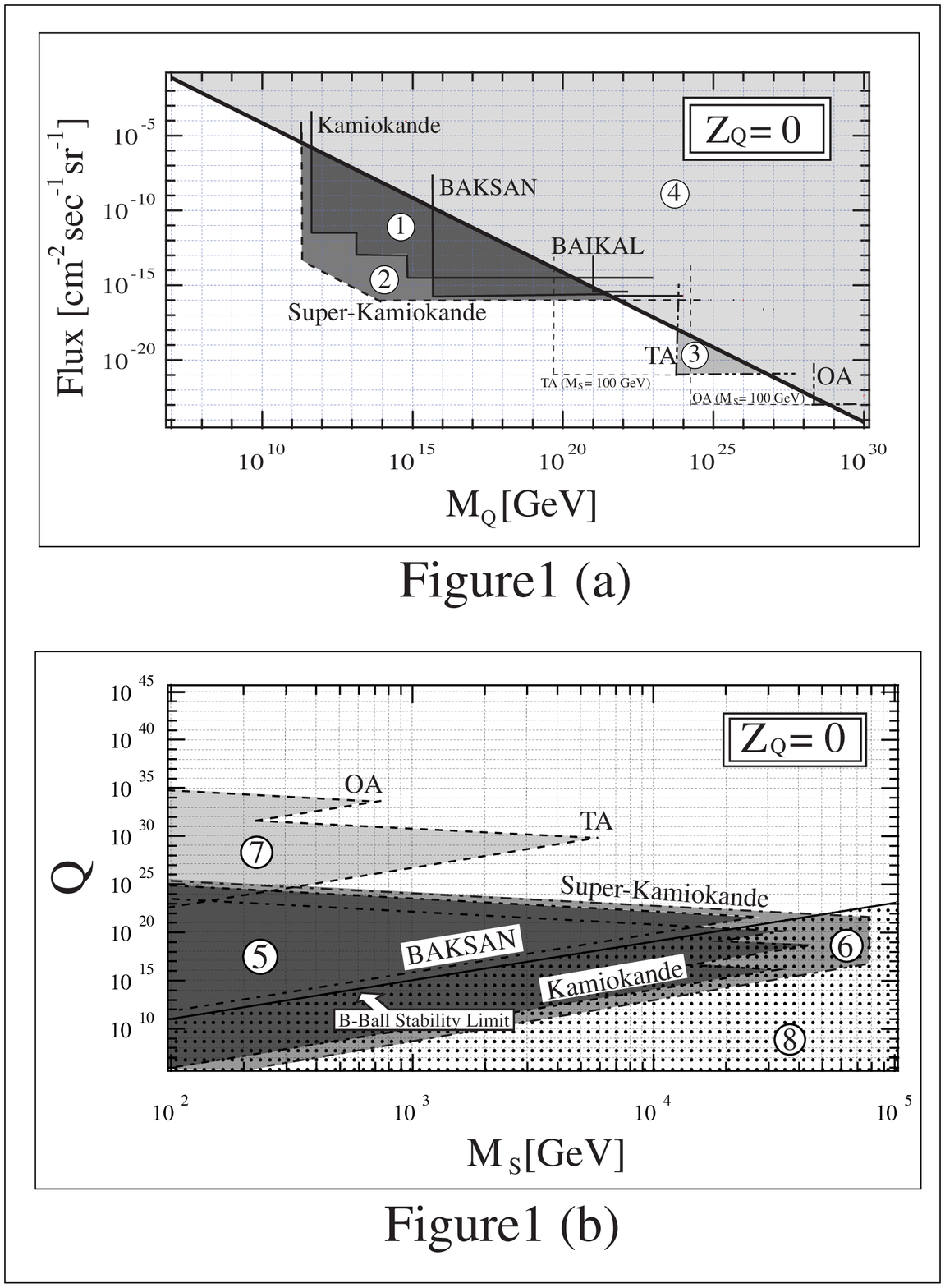}}
\epsfclipoff
\caption{\label{z0com}
\small The bounds on the charge of an electrically neutral $Q$-ball
with respect to the supersymmetry breaking scale. Various shaded
regions of the parameter space are already ruled out. The line
"B-Ball Stability Limit" marks the region below which
$B$-balls are not stable. The allowed region
$Q > 10^{22}$ can be found only in the upper part of the figure, from
\cite{arafune00}.}
\end{figure}

%%%%%%%%%%%%%%%%%%%%%%%%%%%%%%%%%%%%%%%%%%%%%%%%%%%%%%%%%

There are also astrophysical ways for detecting stable $Q$-balls. Charged
$Q$-balls can dissipate a large amount of energy quite efficiently before
they can be detected. The $Q$-balls might leave a track behind whose
stopping range is  roughly $1000$~m for a charge
$Q_{B} \leq 10^{13}(m_{\phi}/1~{\rm TeV})^{-4/3}$ \cite{kusenko9880}.
Electrically neutral  $Q$-balls would hardly have any impact when they pass
through a planet like ours. The loss in $Q$-ball kinetic energy would
be almost undetectable; the velocity would decrease by
$\delta v/v \sim 10^{-2}Q_{B}^{-1/4}(1~{\rm TeV}/m_{\phi})^3$.

$Q$-balls could also be captured in sufficiently dense stars such
as inside the core of neutron stars. This will result in an
increase in the temperature of the neutron star studied in
~\cite{kusenko9880}, but an insignificant enhancement
$\sim 0.01 (Q_{B}/10^{24})^{-1/16}$ may not result in
any observable consequences. There is still a possibility that the
gradual accumulation of $Q$-balls may seal the fate of a neutron star
by decreasing its mass and thereby reaching the critical condition which
may lead to a supernovae explosion~\cite{kusenko9880}.

%%%%%%%%%%%%%%%%%%%%%%%%%%%%%%%%%%%%%%%%%%%%%%%%%%%%%%%%%%%%
\newpage

%%%%%%%%%%%%%%%%%%%%%%%%%%%%%%%%%%%%%%%%%%%%%%%%%%%%%%%%%%%%

\section{Flat directions other than MSSM}

Flat directions, fragmentation of scalar condensates and $Q$-balls
are generic features that could be encountered in many cosmological
models with scalar fields. Examples include such cornerstones of modern
cosmology as inflation, as well as the currently popular particle
physics models with extra dimensions. It has also been suggested
that the motion of a complex flat direction condensate field,
spinning in a $U(1)$-symmetric potential,
could give a dynamical explanation for the
dark energy \cite{bckamionkowski}, although the situation could be
complicated by copious $Q$-ball formation \cite{shinta01}.

%%%%%%%%%%%%%%%%%%%%%%%%%%%%%%%%%%%%%%%%%%%%%%%%%%%%%%%%%%%%%

\subsection{Fragmentation of the inflaton condensate}

By definition, the inflaton is a homogeneous scalar condensate with
a small quantum induced spatial fluctuations. Reheating of the Universe,
which is a consequence of the inflaton condensate break-up and decay,
may take place via ordinary smooth perturbative decay of the condensate
\cite{albrecht82,kolbturner90} or via a non-perturbative
process dubbed as preheating~\cite{traschen90,kofman94}, which
typically involves an amplification in some of the fluctuation modes
as well as the fragmentation of the inflaton condensate. Reheating dynamics
depends very much on the form assumed for the inflaton potential,
and for some choices, the inflaton condensate may also form $Q$-balls,
see \cite{enqvist02a,enqvist02b}.

%%%%%%%%%%%%%%%%%%%%%%%%%%%%%%%%%%%%%%%%%%%%%%%%%%%%%%%

\subsubsection{Reheating as a surface effect}

Usually the process of reheating is taken to be entirely a volume
effect. This can be problematic, however, especially if the scale
of inflation is high, i.e. $H_{inf}\sim 10^{15}-10^{16}$~GeV, as it
is sometimes assumed in order to provide the right magnitude for the
density perturbations, and also for reasons that have to do with
non-thermal heavy dark matter production or exciting right-handed
Majorana neutrinos for leptogenesis, etc. \cite{giudice99}.
As it is well known, the entropy thus dumped into the Universe
may pose a problem for big bang nucleosynthesis by overproducing
gravitinos from a thermal bath; an often quoted bound on the reheat
temperature is $T_{\rm rh} \leq 10^9~{\rm GeV}$~\cite{ellis85b}.
Obtaining such a low reheat temperature is a challenge for high scale
inflation models. (One way to solve the gravitino problem is
to dilute them via a brief period of late thermal inflation
\cite{lythstewart}).

A novel way to avoid the gravitino and other moduli problems is reheating
via the surface evaporation of an inflatonic soliton. Compared with the
volume driven inflaton decay, the surface evaporation naturally
suppresses the decay rate by a factor
\begin{equation}
\label{sur}
\frac{{\it area}}{{\it volume}} \propto L^{-1}\,,
\end{equation}
where $L$ is the effective size of an object whose surface is evaporating.
The larger the size, the smaller is the evaporation rate, and therefore
the smaller is the reheat temperature.

Reheating as a surface phenomenon has been considered
\cite{enqvist02a,enqvist02b} in a class of chaotic inflation models
where the inflaton field is not real but complex. As the inflaton
should have coupling to other fields, the inflaton mass should in
general receive radiative corrections \cite{liddle-lyth00}, resulting
in a running inflaton mass and in the simplest case in the inflaton
potential that can be written as
\begin{equation}
    \label{qpotr}
    V = m^2 |\Phi|^2
    \left[ 1 + K\log\left(\frac{|\Phi|^2}{M^2}\right)\right ]\,,
\end{equation}
where the coefficient $K$ could be negative or positive, and $m$ is
the bare mass of the inflaton. The logarithmic correction to
the mass of the inflaton is something one would expect
because of the possible Yukawa and/or gauge couplings to other fields.
Though it is not pertinent, we note that the potential Eq.~(\ref{qpotr})
can be generated in a supersymmetric theory if the inflaton has a
gauge coupling \cite{enqvist99,kasuya0061,enqvist0163} where
$K\sim -(\alpha/8\pi)(m_{1/2}^2/m_{\widetilde{\ell}}^2)$,
where $m_{1/2}$ is the gaugino mass and $m_{\widetilde{\ell}}$ denotes the
slepton mass and $\alpha$ is a gauge coupling constant. It is also
possible to obtain the potential
Eq.~(\ref{qpotr}) in a non-supersymmetric (or in a broken supersymmetry)
theory, provided the fermions live in a larger representation than the
bosons. In this latter situation the value of $K$ is determined by the
Yukawa coupling $h$ with $ K =-C({h^2}/{16\pi^2})$, where $C$ is
some number.

As long as $|K| \ll 1$, during inflation the dominant contribution to the
potential comes from $m^2|\Phi|^2$ term, and inflationary slow roll
conditions are satisfied as in the case of the standard chaotic model.
COBE normalization then implies $m\sim 10^{13}$~GeV. If $K < 0$,
the inflaton condensate feels a negative pressure (see, Sect.~$5.9.1$, for
the discussion on negative pressure of the AD condensate) and it is
bound to fragment into lumps of inflatonic matter. Moreover, since the
inflation potential Eq.~(\ref{qpotr}) respects a global $U(1)$ symmetry
and since for a negative $K$ it is shallower than $m^2|\Phi|^2$, it
admits a $Q$-ball solution (see Sect.~$5.9.3$).
Comparing with $Q$-balls along the MSSM flat directions, here
the major difference is that the inflatonic condensate
has no classical motion along the imaginary direction as usually required
for a $Q$-ball solution.

%%%%%%%%%%%%%%%%%%%%%%%%%%%%%%%%%%%%%%%%%%%%%%%%%%%%%%%%
\subsubsection{$Q$-balls from the inflaton condensate}

As pointed out in~\cite{enqvist02a}, there are quantum fluctuations
along both the real and imaginary directions which may act as the initial
seed that triggers on the condensate motion in a whole complex plane.
The fluctuations in the real direction grow and drag the imaginary direction
along via mode-mode interactions, as illustrated by $2$ dimensional
lattice simulation in \cite{enqvist02b}, see Figs.~(\ref{mode-osc}).
The first plot shows the linear fluctuations without rescattering effects;
scattering effects are accounted for in the second plot. The late time
formation of inflatonic solitons is shown in Fig.~(\ref{3Dinf}). $Q$-balls
were observed to form with both positive and negative charges, as can be
seen in the first plot of Fig.~(\ref{3Dinf}), while keeping the net
global charge conserved. Inflatonic $Q$-balls are of same size because
the running mass potential resembles the MSSM flat direction potential
in the gravity mediated case, where the $Q$-ball radius is independent
of the charge.

%%%%%%%%%%%%%%%%%%%%%%%%%%%%%%%%%%%%%%%%%%%%%%%%%%%%%%%%%
\begin{figure}[t!]
\centering
\hspace*{-7mm}
\leavevmode\epsfysize=5cm \epsfbox{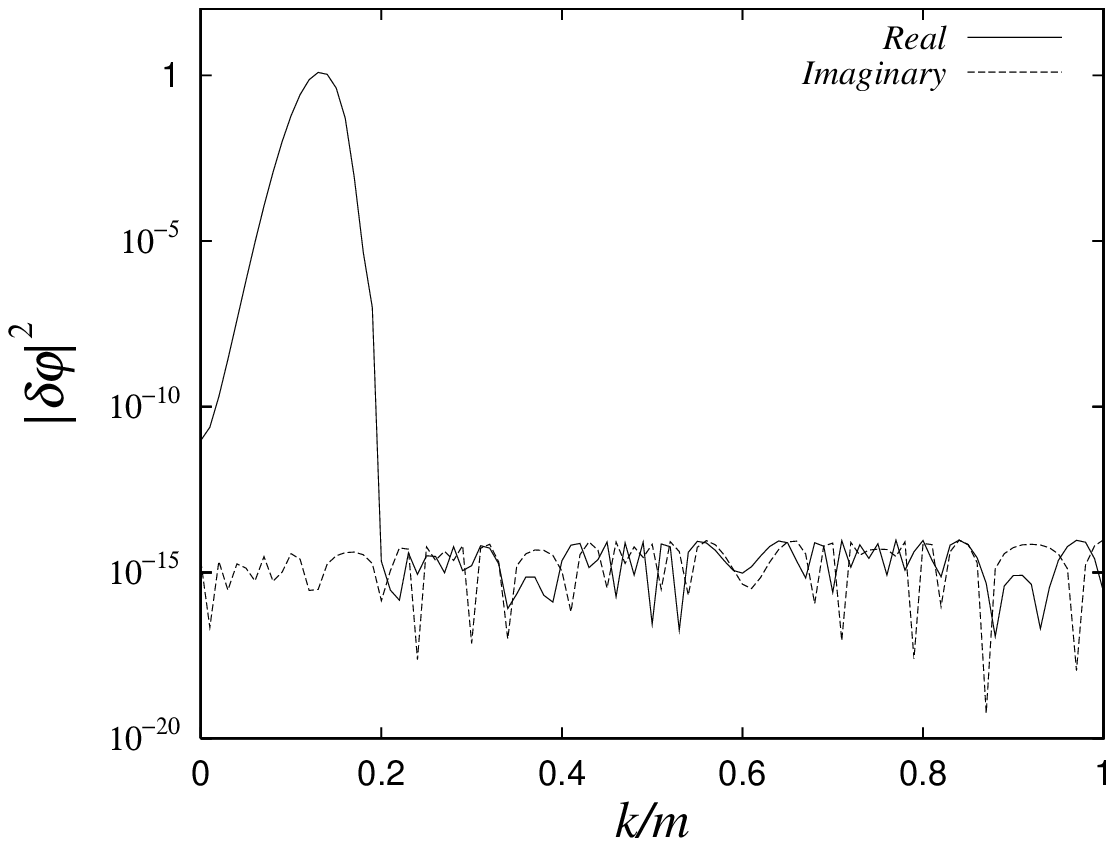}
\leavevmode\epsfysize=5cm \epsfbox{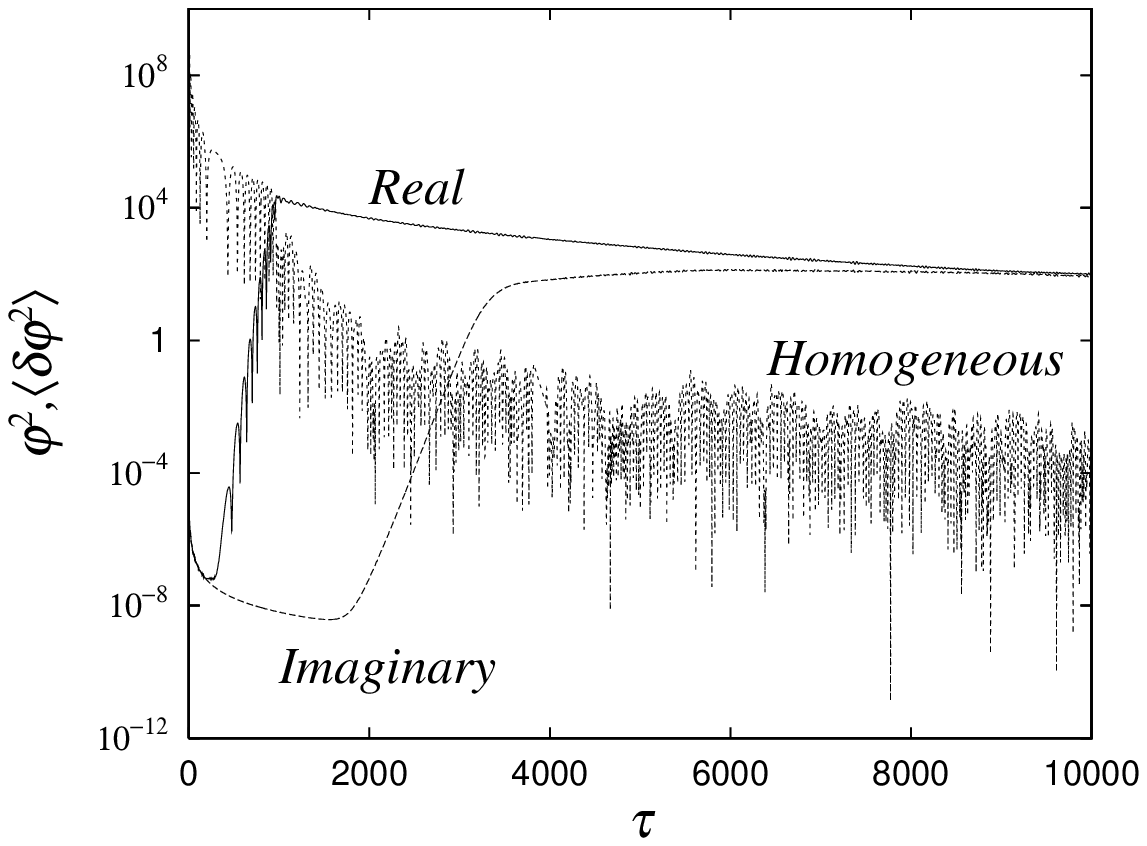}\,\\
\leavevmode\epsfysize=5cm \epsfbox{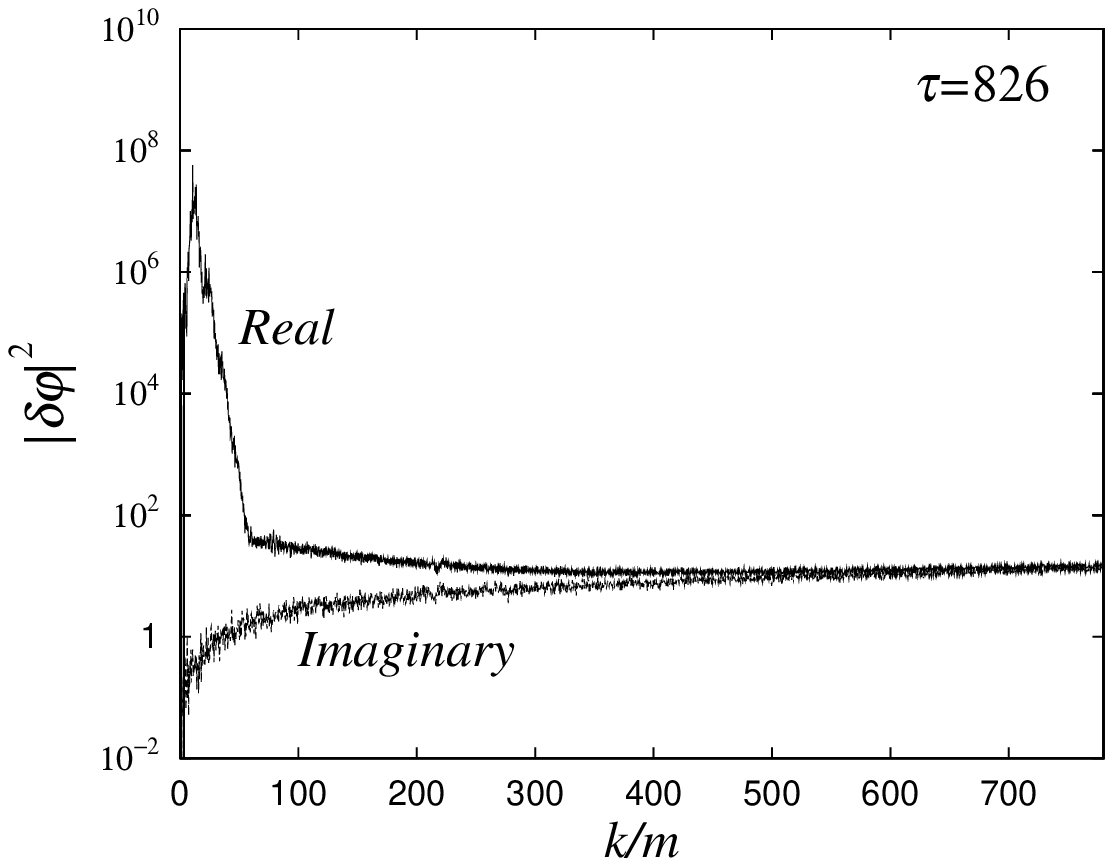}\\[2mm]
\caption{\label{mode-osc}
\small The first plot from left shows the instability bands
of the homogeneous mode of the inflaton along the real (solid) and imaginary
(dotted) directions. The second plot shows the result of
lattice simulation in the real and imaginary
directions, together with the evolution of the homogeneous  mode.
The third plot shows the power spectra of fluctuations at late times.
All plots assume $K=-0.02$, from \cite{enqvist02b}.}
\end{figure}

%%%%%%%%%%%%%%%%%%%%%%%%%%%%%%%%%%%%%%%%%%%%%%%%%%%%%%%%%%%%%

\begin{figure}[t!]
\centering
\hspace*{-7mm}
\leavevmode\epsfysize=6cm \epsfbox{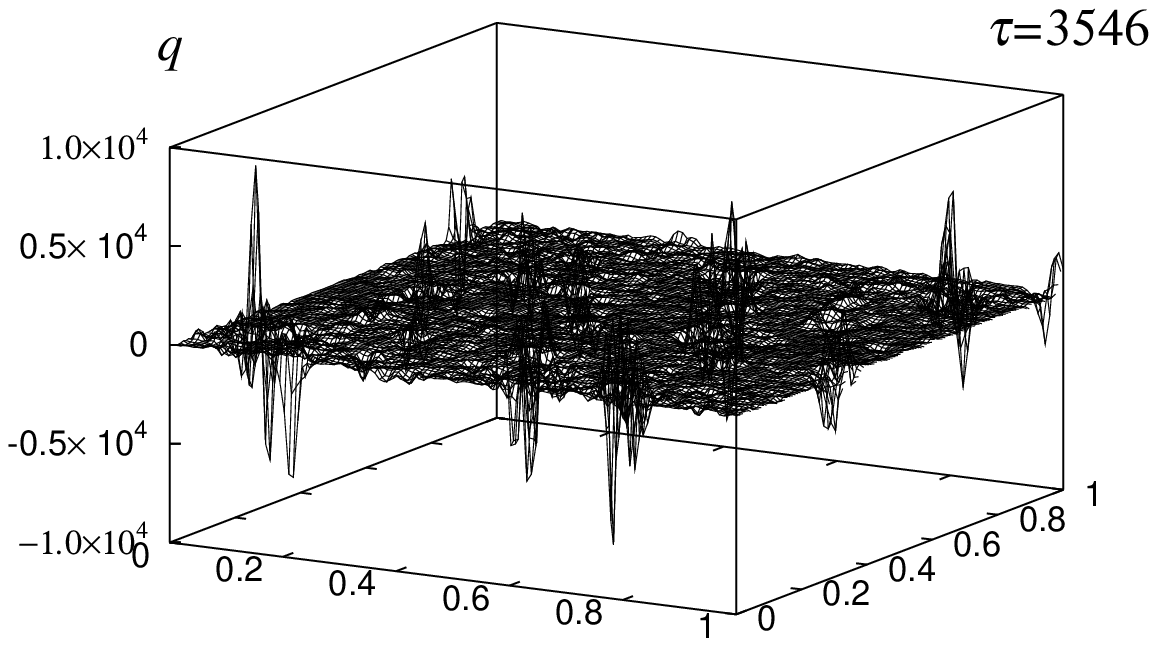}
\leavevmode\epsfysize=6cm \epsfbox{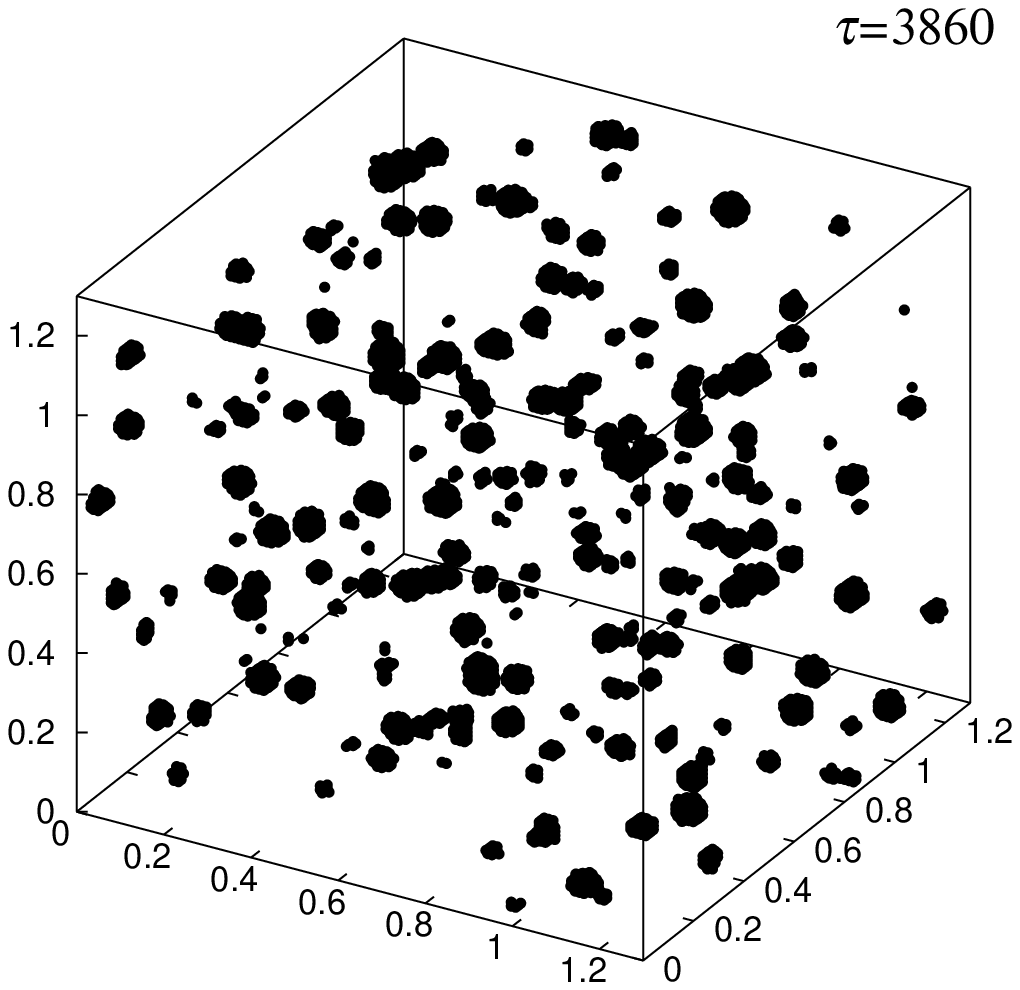}\\[2mm]
\caption{\label{3Dinf}
\small The first plot shows the charge density distribution in a small
sub-lattice at late times. The second plot shows inflatonic solitons
forming in $3D$ lattice.  Here $K=-0.02$, from
\cite{enqvist02b}.}
\end{figure}

%%%%%%%%%%%%%%%%%%%%%%%%%%%%%%%%%%%%%%%%%%%%%%%%%%%%%%%%%%%%%

$Q$-balls of size $R \sim |K|^{-1/2}m^{-1}$ form when the
fluctuations grow nonlinear (see Eq.~(\ref{a5}) \cite{enqvist02a,enqvist02b}).
Since the growth rate of fluctuation is $\sim |K|m$, the Hubble parameter at
 the formation time can be estimated as $H_{f} \sim \gamma |K|m$,
where $\gamma$ is a numerical coefficient less than one. For $|K|\ll 1$, we can
approximate the decrease in the amplitude of the oscillations by
$\phi_f \sim \phi_i(H_f/H_i)$ as in the matter dominated era, where
$\phi_i\simeq M_{\rm P}$ denotes the amplitude at the
end of inflation in the chaotic model, and $H_i\sim m$ when the
oscillations begin. The total charge of a $Q$-ball is given
by $Q\sim(4\pi/3)R^3n_q\sim(1/9)\beta\zeta^2\gamma^2|K|^2R^3mM_{\rm P}^2$,
where $n_q = \beta \omega\phi_0^2$, $\phi_0 \simeq \zeta \phi_f$, and
$\beta \ll 1$ and $\zeta \geq 1$ are numerical factors.

Given an inflaton coupling to fermions of the type
$ h\phi\bar\psi\psi$ it has been shown that \cite{enqvist02a,enqvist02b}
reheating is driven by surface evaporation of inflatonic $Q$-balls for
relatively large Yukawa couplings $h\le 1$. In general $K$ and $h$ are
not independent quantities but are related to each other by
$|K| \sim C(h^2/16\pi^2)$. If the inflaton sector does not
belong to the hidden sector, it is very natural that the inflaton
coupling to other matter fields is relatively large, i.e.
$h\geq (m/M_{\rm P})$. In this regime the evaporation rate is
saturated by Eq.~(\ref{qevap1}) and~\cite{enqvist02b}
\begin{equation}
    \label{evap2}
    \Gamma_Q = \frac{1}{Q}\frac{dQ}{dt}
        \simeq \frac{3}{16\pi \beta\zeta^2\gamma^2|K|^{3/2}}
                \left(\frac{m}{M_{\rm P}}\right)^2 m\,.
\end{equation}
Note that the decay rate is determined by the ratio
$m/M_{\rm P} \simeq 10^{-6}$, which is fixed by the anisotropies
seen in the cosmic microwave background radiation. Even
though we are in a relatively large coupling limit, the decay rate
mimics that of a Planck suppressed interaction of the inflatonic
$Q$-ball with matter fields. Fermionic preheating~\cite{heitmann99}
is not a problem in this case because the whole inflaton energy is
not transferred in this process and the energy density stored in
the fermions remains small compared to the inflaton energy density,
as argued in ~\cite{enqvist02b}. Fermions cannot scatter inflaton
quanta off the condensate~\cite{heitmann99}, unlike in the case of
bosonic preheating~\cite{kofman94}. Inflatonic $Q$-ball formation 
is reminiscent of bosonic preheating~\cite{copeland0265} due to the 
presence of attractive self coupling of the inflaton, which stems 
from the logarithmic term in the potential.

%%%%%%%%%%%%%%%%%%%%%%%%%%%%%%%%%%%%%%%%%%%%%%%%%%%%%

\subsection{Affleck-Dine baryogenesis without MSSM flat directions}

%%%%%%%%%%%%%%%%%%%%%%%%%%%%%%%%%%%%%%%%%%%%%%%%%%%%

\subsubsection{Leptogenesis with sneutrino}

It is interesting to note that it is possible to mimic the AD
baryogenesis even without  MSSM flat directions. This can happen
in F-term hybrid inflation with a superpotential~\cite{zurab01}
\begin{equation}
\label{newsuper}
W=-\Lambda^2 S+\lambda S \Phi^2+\kappa \Phi\Psi^2\,,
\end{equation}
where $\Lambda\approx 6.5\cdot 10^{16}\epsilon^{1/4}$~GeV,
or equivalently $\Lambda\approx 1.3\cdot 10^{15} |\eta|\lambda^{-1/2}$~GeV,
where $\epsilon,\eta \ll 1$ are the slow roll parameters.
The superpotential \eq{newsuper} extends the usual hybrid case
Eq.~(\ref{suppot1}) in order to provide a natural initial condition
for $\Phi$, which should be zero within the accuracy $10^{-5}$ \cite{zurab98},
while the inflaton should have an initial value close to the Planck scale.
$\Psi$ is identified with the $U(1)_{B-L}$ carrying right-handed neutrino
which alleviates the problem of initial conditions through the dynamics of
\eq{newsuper}~\cite{zurab98}. During inflation the system is trapped
in a false vacuum with $\phi=0$. Once $\phi$
decays, sneutrino $\widetilde\Psi$ obtains a vev of order
of the Hubble parameter at the end of inflation. It then starts
oscillating around the origin and due to dynamical breaking of
$U(1)_{B-L}$, a net $B-L$ asymmetry is generated~\cite{zurab01}.
Right after inflation the real and imaginary component of $\Psi$,
$\psi_{1}$ and $\psi_{2}$ have a relative
sign difference which induces a helical motion on
$\Psi$ similar to the MSSM flat direction condensate. In this model
the neutrino mass is generated by the ordinary see-saw
mechanism~\cite{see-saw}
\begin{equation}
\label{numass}
m_{\nu} = \frac{g^2 \langle\varphi_2\rangle^2}{ M_{\Psi}}
\simeq \frac{g^2\lambda}{\kappa \eta} \times
1.2 \cdot 10^{-2} ~{\rm eV}\,,
\end{equation}
(where we have taken $\langle\varphi_2\rangle \simeq 170$ GeV).
The Hubble parameter during inflation is
\begin{equation}
H_{inf} \simeq \frac{\Lambda^2}{\sqrt{3}M_{\rm P}} \simeq
\frac{\eta^2}{\lambda}\times 4 \cdot 10^{11} ~{\rm GeV}.
\end{equation}
Consideration of the detailed dynamics and decay of the
condensate leads to an estimation of the lepton asymmetry,
which is given by~\cite{zurab01}
\begin{equation} \label{final}
B-L \sim X \frac{T_{\rm r}}{M_{\rm P}} \simeq X
\left(\frac{T_{\rm r}} {10^{9} ~{\rm GeV}}\right) \times 10^{-10} \,,
\end{equation}
where the numerical factor is $X = \min\{1, x\}C\kappa^{-1}$, with $C\sim 1$
coefficient if the field $\widetilde\Psi$ gets a supergravity
induced mass term during inflation, otherwise $C = 3/N_{\rm e}$.
The reheat temperature appears to respect the
gravitino bound.

%%%%%%%%%%%%%%%%%%%%%%%%%%%%%%%%%%%%%%%%%%%%%%%%%%%%%%%%%%%%%%%%%%

\subsubsection{AD baryogenesis in theories with low scale quantum gravity}

In low scale gravity models the weakness of gravity
arises because of extra spatial dimensions, the scale of which could
be as large as $mm$,~\cite{nima0}. In the simplest models, the
SM fields live on a three dimensional brane while gravity can also
permeate the bulk~\cite{nima0,early}. The fundamental scale is
$M_{*}\sim {\cal O}({\rm TeV})$ in $4+d$ dimensions, which is
related to the the volume suppression $V_{d}$ and to the effective
four dimensional Planck mass $M_{\rm p}$ by a simple relationship~\cite{nima0}
\begin{equation}
M_{\rm p}^2 =M_{*}^{2+d}V_{d}\,.
\end{equation}
This automatically sets a {\it common size} for all the extra
dimensions at $b_0$. For two extra dimensions with $M_{*}=1$ TeV, the
size is $0.2$ mm, just below the current experimental limits from the
search for deviations in Newtonian gravity~\cite{exp2}.
Recent astrophysical bounds based on neutron stars suggests
$M_\ast\geq 500$~TeV  for two extra dimensions~\cite{hannestad02}.
Naturally, low scale gravity models have important implications for
collider experiments~\cite{exp1} and for cosmology \cite{davidson99,many}
(for a review on large extra dimensions, see \cite{lorenzana00}).

Although quite attractive from a particle physics point of view,
large extra dimensions bring along a host of cosmological problems.
There are dynamical questions regarding the  stabilization of the
size of the extra dimension(s), or equivalently the vev of the
radion field, whose mass can be as small as ${\cal O}(\rm eV)$
(for two large extra dimensions)~\cite{nima0}.
Cosmologically stabilization should take place very early by
some trapping mechanism as discussed in~\cite{abdel0}.
Another challenge is how to realize inflation in these models.
There have been many proposals \cite{many}, such as invoking a
SM singlet scalar living in the bulk \cite{abdel1}. There is
the problem of the Kaluza Klein (KK) states of the graviton and
any other fields residing in the bulk, which above a certain
temperature known as the {\em normalcy temperature}, should
fill the Universe. The normalcy temperature is constrained by
cosmological  considerations to lie in the range $1$ MeV to $100$ MeV
\cite{nima0,davidson99,abdel2,abdel3,green02}, and the reheat temperature
should be lower than the normalcy temperature.

In addition, one must not only forbid dangerous higher order operators which
can mediate  proton decay, but also ensure that such operators are not
being reintroduced by whatever mechanism is responsible for baryogenesis.
This tends to make baryogenesis in low scale gravity models quite difficult.
Moreover, it has been argued \cite{abdel2,abdel3} that leptogenesis
is not a viable option; obviously at very low reheat temperatures required
by the normalcy temperature, the sphalerons cannot be activated
(a possible way out of this problem could be to increase the reheat 
temperature by increasing the number of large extra dimensions to six 
\cite{pilaftsis00b}).

One solution to this predicament was provided in \cite{abdel3}
(see also \cite{maz01}). There $U(1)_{\chi}$ carried by a gauge
singlet $\chi$ was broken dynamically in order to provide a small
asymmetry in the current density. Baryon asymmetry is produced by
the decays of $\chi$ and $\bar \chi$ into SM quarks and leptons,
analogously to the old AD baryogenesis. The decay channels are
constrained because quarks and leptons must carry a non zero global
$\chi$ charge. This prevents $\chi$-$\bar\chi$ asymmetry to be
transferred into non-baryon-number violating interactions such as
interactions involving the Higgses.

A consistent model for baryogenesis can be constructed \cite{abdel3}
along the lines of sneutrino leptogenesis discussed in Sect.~$8.2.1$.
It requires a hybrid inflaton sector, described by the fields $\phi$ and
$N$, and the flat direction field $\chi$, which are promoted to the bulk.
This ensures the right amplitude for the density perturbations and provides
enough baryon asymmetry towards the end of reheating~\cite{abdel2,abdel3}.
The potential for the zero modes in $4$ dimensions can be written
as~\cite{abdel3}
\begin{eqnarray}
\label{adchi}
V_{\rm AD}(\phi,N,\chi_1,\chi_2) =
\kappa_1^2 \left(\frac{M_{\ast}}{M_{\rm p}}\right)^2 N^2(\chi_1^2+\chi_2^2)
 +\frac{\kappa_2^2}{4}\left(\frac{M_{\ast}}{M_{\rm p}}\right)^2
 (\chi_1^2+\chi_2^2)^2\, \nonumber \\
+\kappa_3 ^2\left(\frac{M_{\ast}}{M_{\rm p}}\right)^2\phi N
(\chi_1^2-\chi_2^2)\,,
\end{eqnarray}
where $\kappa_1,\kappa_2,\kappa_3$ are order one constants, and $\chi_1$
and $\chi_2$ are the real and imaginary components of the complex
field $\chi$. Note that all the terms are Planck mass  suppressed
because \eq{adchi} is an effective 4d potential derived from a higher
dimensional Lagrangian by integrating out the extra spatial dimensions.
Since during inflation the auxiliary field  $N=0$, the flat direction
condensates  $\chi_1$ and $\chi_2$ are massless and the potential
is almost flat.

The final $\chi$ asymmetry was found to be given by \cite{abdel3}
\begin{eqnarray}
\label{final11}
\frac{n_{\chi}}{s} \approx
{2\kappa_3^2\over 27\lambda^2} \left({|\chi(0)|\over M_{\rm p}}\right)^2
 \left({T_{\rm r}\over H_0}\right) \leq
\frac{2\sqrt{6\pi}}{27}\left(\frac{\kappa_3}{\kappa_2}
\right)^2\left(\frac{1}{\lambda N_{\rm e}}\right)
\left(\frac{T_{\rm r}}{M_{\ast}}\right)\, ,
\end{eqnarray}
which at the same time provides an upper bound on the baryon to
entropy ratio. As an example, taking $T_{\rm r} \sim 100$~MeV,
$M_{\ast}\sim 100$~TeV, the number of e-foldings $N_{\rm e}\sim 100$,
and couplings of order one, yields an asymmetry of order $\sim 10^{-10}$.

%%%%%%%%%%%%%%%%%%%%%%%%%%%%%%%%%%%%%%%%%%%%%%%%%%%%%%%%%%%%

\subsection{Solitosynthesis}

Solitosynthesis is a mechanism of charge accretion with the help of
pre-existing small $Q$-balls in a charge asymmetric background
\cite{frieman88,frieman89,griest89,kusenko97406,postma02}. The
accumulation of charge and forming a large charged $Q$-ball has
been shown to be quite efficient especially in a finite temperature
thermal bath.

%%%%%%%%%%%%%%%%%%%%%%%%%%%%%%%%%%%%%%%%%%%%%%%%%%%%%%%%%%%%%
\subsubsection{Accretion of charge by $Q$-balls}

Consider thermodynamics of $Q$-balls surrounded by massive
non-relativistic fermions $\psi$. The number densities of
$Q$-balls and $\psi$ particles are governed by the Boltzmann
distributions $n_{Q}(T)=g_{Q}(M_{Q}T/2\pi)^{3/2}$ $\exp[(\mu_{Q}-M_{Q})/T]$,
and $n_{\psi}=g_{\psi}(m_{\psi}T/2\pi)^{3/2}\exp[(\mu_{Q}-m_{\psi})/T]$,
where $g_{Q}$ is the partition function for the $Q$-ball and $g_{\psi}=2$.
The respective chemical potentials are denoted by $\mu$. In
chemical equilibrium the absorption and evaporation of charge is
equally possible, i.e. $(Q)+\psi\leftrightarrow (Q+1)$, which relates
the two chemical potentials through $\mu_{Q}=Q\mu_{\psi}$. The number
density of $Q$-balls can be expressed in terms of the number density of
the fermions \cite{frieman89,postma02}
\begin{equation}
\label{soli0}
n_{Q}=\frac{g_{Q}}{g^{Q}_{\psi}}n^{Q}_{\psi}\left(\frac{M_{Q}}{m_{\psi}}
\right)^{3/2}\left(\frac{2\pi}{m_{\psi}T}\right)^{3(Q-1)/2}e^{B_{Q}/T}\,,
\end{equation}
where $B_{Q}\equiv Qm_{\psi}-M_{Q}$ is the binding energy per charge of
a $Q$-ball.
When $B_{Q}$ grows with $Q$, the formation of large charged
$Q$-ball is likely. The interactions between $Q$-ball and $\psi$ quanta
leads to a chemical equilibrium when \cite{frieman89,postma02}
\begin{eqnarray}
n_{\psi}v_{\psi}\sigma_{abs}(Q)=n_{Q+1}r_{evap}(Q+1)\,.
\end{eqnarray}
The accretion and evaporation rate from a charge $Q$-ball is
given by Saha equation:
\begin{eqnarray}
\label{saha}
\frac{dQ}{dt}&=&r_{abs}(Q)-r_{evap}(Q) \,, \nonumber \\
&=&n_{\psi}v_{\psi}\left[\sigma_{abs}(Q)-\frac{n_{Q}-1}{n_{Q}}\sigma_{abs}
(Q-1)\right]\,,
\end{eqnarray}
where $v_{\psi}=(T/2\pi m_{\psi})^{1/2}$ is the mean velocity of
$\psi$ particles.
The charge of a $Q$-ball grows when $r_{abs}(Q)>r_{evap}(Q)$.
The necessary charge asymmetry in $\psi$ quanta affects
the $Q$-ball abundance. Charge conservation requires
\begin{equation}
N=n_{\psi}-n_{\psi^{\ast}}+\sum Qn_{Q}+\sum Q^{\ast}n_{Q^{\ast}}=\eta_{\psi}
n_{\gamma}\,,
\end{equation}
where $\eta_{\psi}=n_{\psi}-n_{\psi^{\ast}}/n_{\gamma}$ is the charge
asymmetry. If $\eta$ is close to zero,  annihilations of $\psi$ quanta will
be effective and there will be virtually no solitosynthesis. The
number density of the stable $\psi$ quanta is $n_{\psi}=\eta_{\psi}n_{\gamma}$,
and the charge asymmetry is given by \cite{frieman89}
\begin{equation}
\eta_{\psi}=2.5\times 10^{-8}\Omega_{\psi}h^2\frac{{\rm GeV}}{m_{\psi}}\,.
\end{equation}
Obviously the over-closure limit $\Omega_{\psi}h^2 \leq 1$ yields an upper
bound on $\eta_{\psi}$ in terms of $m_{\psi}$. At high temperatures
large $Q$-balls are suppressed by a small asymmetry factor
$\sim \eta_{\psi}^{Q-1}$. At
lower temperatures the abundance is dominated by $Q^{5/2}$, where
$M_{Q} \propto m_{\psi}Q$ \cite{frieman89}.

The $Q$-ball starts growing at a temperature $T_{s}$ when
$r_{abs}(Q)>r_{evap}(Q)$. This can be estimated from the Saha equation
\cite{kusenko97406,postma02}
\begin{equation}
T_{s}=\frac{m_{\psi}+M_{Q-1}-M_{Q}}{-\frac{3}{2}\ln\left(\frac{T_{s}}{m_{\psi}}
\right)-\ln(c\xi^3\eta_{\psi})}\,,
\end{equation}
where $c\sim {\cal O}(1)$ number for large enough $Q$.

When the absorption of  $Q$-charge from the surroundings freezes
out, solitosynthesis stops. There are two distinct era which one may
consider; solitosynthesis freezes out during the radiation dominated epoch,
or during the matter dominated epoch. The demarcation temperature  is
$T_{eq}\approx 5.5(\Omega_{0}h^2)^{-1}\xi^{-1}$~eV. The freeze-out temperature
is defined by $\Gamma[(Q)+\psi\rightarrow (Q+1)]\leq H(T)$. For a
geometric cross section $\sigma_{abs} \sim \pi R_{Q}^2$, the freeze-out
temperature is given by \cite{postma02}
\begin{equation}
\frac{T_{F}}{m_{\psi}} \leq \left\{
      \begin{array}{ll}
          \ds{\left[\frac{10^{-9}}{\xi\beta_{Q}Q^{2/3}}\left(\frac{m_{\psi}}
{{\rm GeV}}\right)^2\left(\frac{0.3}{\Omega_{\psi}h^2}\right)\left(
\frac{g_{\ast}^{1/2}}{10}\right)\right]^{2/3}}  & \ds{(T> T_{eq})} \\
          \ds{\left[\frac{10^{-13}}{\xi\beta_{Q}Q^{2/3}}\left(\frac{m_{\psi}}
{{\rm GeV}}\right)^{3/2}\left(\frac{0.3}{\Omega_{\psi}h^2}\right)^{1/2}\left(
\frac{g_{\ast}^{1/2}}{10}\right)\right]^{1/2}} &
                      \ds{(T<T_{eq}).}
      \end{array}\right.
\end{equation}
where $\xi \equiv (g_{\ast}(T_{D})-g_{\ast}(T))^{1/3}$, the temperature $T_{D}$
signifies the decoupling temperature of a $Q$-ball with a rest of the
plasma, and $\beta_{Q}=(3m_{\phi}^3/4\pi \omega\varphi_{0}^2)^{1/3}$.
One should require that $T_{F} < T_{s}$, which is
satisfied for a very low $\psi$-mass. In order to have some feeling for
the numbers involved, we note that for the smallest charged $Q$-ball
with $Q=2$, the freeze-out temperature is smaller than $T_s$ only when
$m_{\psi}\leq{\cal O}(1)$~GeV. During the matter dominated epoch, due to
the gravitational clustering, $Q$-ball synthesis is favorable but for
ambient temperature $T_{\gamma} \leq {\cal O}(1)$~eV,
$m_{\psi}\leq {\cal O}(1)$~MeV.

%%%%%%%%%%%%%%%%%%%%%%%%%%%%%%%%%%%%%%%%%%%%%%%%%%%

\subsubsection{Phase transition via solitogenesis}

An interesting observation has been made in \cite{ellis89225,kusenko97406},
where it was pointed out that solitosynthesis may also lead to a first
order phase transition. The gradual absorption of charge may lead to a
critical charge $Q=Q_{c}$ for which the false vacuum inside the $Q$-ball
becomes unstable and expands as the
the whole space is filled with the true vacuum. The value of the critical
charge is determined by minimizing the energy of the $Q$-ball with respect
to the radius when $dE/dR=0$ and $d^2E/dR^2=0$ are satisfied simultaneously.
In a thin wall limit when $Q_{c}\gg 1$, the critical charge is given by
\cite{kusenko97406}
\begin{equation}
Q_{c}=\frac{100\pi\sqrt{10}}{81}\frac{\varphi_{0}S_{1}^3}{U(\varphi_0)^{5/2}}
\,,
\end{equation}
where
$S_1=\int_{0}^{\varphi_{0}}d\varphi\sqrt{2(U(\varphi)-(\omega^2/2)\varphi^2)}$.

One can also enquire how large the critical charge should be in order
to facilitate the destabilization of the otherwise cosmologically
stable false vacuum. It has been found that the decay of a metastable
false vacuum at zero temperature requires a small charge
$Q_{C} \sim 28(\varphi_{0}/U(\varphi_0)^{1/4})$ for $A\sim (100~{\rm GeV})^4$
\cite{kusenko97406}. Similar considerations at finite temperature result
in a larger critical charge
$Q>Q_{c}\sim 146\varphi_{0}T/U(\varphi_{0},T)^{1/2}$ for the
phase transition to proceed.

Solitosynthesis-catalyzed phase transitions in supersymmetric models
would require a local violation of lepton or baryon number.
As an example (see~\cite{kusenko97406}), one could consider a
lepton number violating local minimum along
$H_{d}\widetilde L_{L}\widetilde L_{R} \neq 0$. This false
vacuum can decay into the standard true vacuum but the sphaleron induced
transition rate is  almost negligible compared to the cosmological time
scale because $SU(2)\times U(1)$ is broken. If the
typical  mass scale of squarks and sleptons is considerably heavier
than $1$~TeV, $L$-balls will accumulate  when
$T_{s}>T_{F}\approx m_{\phi}/40$, and can catalyze a phase transition within
one Hubble time at temperatures $T_{s}$ \cite{ellis89225,kusenko97406}.

\newpage
%%%%%%%%%%%%%%%%%%%%%%%%%%%%%%%%%%%%%%%%%%%%%%%%%%%%%

%%%%%%%%%%%%%%%%%%%%%%%%%%%%%%%%%%%%%%%%%%%%%%%%%%%%%%%%%%
%%%%%%%%%%%%%%%%%%%%%%%%%%%%%%%%%%%%%%%%%%%%%%%%%%%%%%%%%%
\section*{Acknowledgements}

The authors are thankful to Rouzbeh Allahverdi, Mar Bastero-Gil,
Zurab Berezhiani, Ed Copeland, Masaaki Fujii, Katrin Heitmann, Asko Jokinen,
Shinta Kasuya, Alex Kusenko, Mikko Laine, Andrew Liddle, John McDonald,
Tuomas Multam\"aki, Altug \"Ozpineci, Abdel P\'erez-Lorenzana and Iiro Vilja
for useful discussions.

A.M. was partially supported by {\it The Early Universe Network}:
HPRN-CT-2000-00152. A.M. is also thankful to the Helsinki Institute of
Physics where part of the project has been carried out. K.E.
acknowledges the Academy of Finland grant no. 51433.

%%%%%%%%%%%%%%%%%%%%%%%%%%%%%%%%%%%%%%%%%%%%%%%%%%%%%%%%%%%%
\pagebreak
%%%%%%%%%%%%%%%%%%%%%%%%%%%%%%%%%%%%%%%%%%%%%%%%%%%%%%%%%%%%

%%%%%%%%%%%%%%%%%%%%%%%%%%%%%%%%%%%%%%%%%%%%%%%%%%%%%%%%%%%%%%%%%
%%%%%%%%%%%%%%%%%%%%%%%%%%%%%%%%%%%%%%%%%%%%%%%%%%%%%%%%%%%%%%%%%
%%%%%%%%%%%%%%%%%%%%%%%%%%%%%%%%%%%%%%%%%%%%%%%%%%%%%%%%%%%%%%%%%
%%%%%%%%%%%%%%%%%%%%%%%%%%%%%%%%%%%%%%%%%%%%%%%%%%%%%%%%%%%%%%%%%

\end{document}